%% file: main.tex
\documentclass[10pt]{book}

\usepackage[a4paper,margin=1.5in,heightrounded]{geometry}

\usepackage{mathrsfs}
\usepackage{amssymb}

\usepackage{multirow}
\usepackage{latexsym}

\usepackage[T1]{fontenc}

\usepackage[font=footnotesize,labelfont=bf]{caption}

\usepackage{graphicx}
\usepackage{enumitem}
\usepackage{amsmath}
\usepackage[dvipsnames]{xcolor}
\definecolor{color1}{RGB}{100,200,255}
\usepackage[linktocpage]{hyperref}
\hypersetup{
    colorlinks=true,
    linkcolor=blue,
    filecolor=magenta,      
    urlcolor=color1,
    citecolor=color1,
}
\hypersetup{pdfstartview=}
\usepackage{tikz}
\usetikzlibrary{mindmap}
\newcommand{\rh}{r_h}
\newcommand{\gb}{\mathcal{G}}

\newcommand{\meff}{m_{\text{eff}}}
\usepackage[%
    style=phys,%
    articletitle=false,biblabel=brackets,%
    chaptertitle=false,pageranges=false,%
    eprint=true
]{biblatex}
\newcommand\tablex{2.5mm}
\newcommand\tabley{1mm}

\newcommand{\GB}{\mathscr{G}}

\definecolor{color2}{rgb}{1,0.3,0.5}
\definecolor{color1}{rgb}{0.4,0.25,1}

\usepackage{titlesec}
\titleformat{\chapter}[display]
  {\bfseries\Large}
  {\filright\MakeUppercase{\chaptertitlename} \Huge\thechapter}
  {1ex}
  {\titlerule\vspace{2mm}}
  [\vspace{1ex}\titlerule]

\usepackage{fancyhdr}
\usepackage{lipsum}

\allowdisplaybreaks

\begin{document}

\pagenumbering{gobble}

\begin{titlepage}

\newgeometry{left=1in,right=1in,top=1in,bottom=1in}

\begin{center}
    \vspace{3cm}
    {\huge\textbf{New perspectives on scalar fields\\[5mm]
    in strong gravity}}\\
    \vspace{2cm}
    {\Large\textbf{Georgios Antoniou}}\\[5mm]
    \vfill
    {\large Thesis submitted to the University of Nottingham\\[2mm]
    for the degree of \\[10mm]
    \Large\textsc{Doctor of Philosophy}}\\[3cm]
    \includegraphics[width=0.3\textwidth]{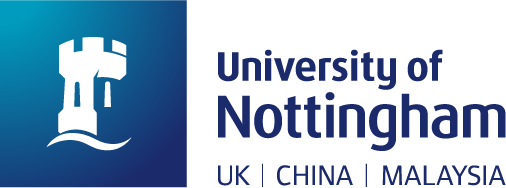}\\[2cm]
    {\large University of Nottingham\\[2mm]
            School of Mathematical Sciences}\\[2cm]
    {\Large April 2023}
\end{center}
    
\end{titlepage}
\newgeometry{left=1.5in,right=1.5in,top=1.5in,bottom=1.5in}

\null
\thispagestyle{empty}
\newpage

\begin{flushright}
    \vspace*{\stretch{1}}
    \textit{To my brother Niko}
\end{flushright}
\vspace{\stretch{3}}
\thispagestyle{empty}
\newpage

\null
\thispagestyle{empty}
\newpage

\frontmatter

\chapter{Acknowledgements}
\input{Chapters/acknowledgements}

\chapter{Abstract}
\input{Chapters/abstract}

\chapter{Publication List}
\input{Chapters/publications}

\chapter{Notation}
\input{Chapters/Notation}


\chapter{Preface}
\input{Chapters/preface}

\tableofcontents

\listoffigures

\listoftables

\mainmatter

\chapter{Introduction}
\label{ch:Introduction}
\input{Chapters/introduction}


\chapter{Evasion of no-hair theorems in scalar-tensor gravity}
\label{ch:Evasions}
\input{Chapters/evasion_of_no_hair_theorems}

\chapter[Minimum size and scalar charge in shift symmetry]{Minimum black hole size and scalar charge in shift symmetry}
\label{ch:Mass-Charge}
\input{Chapters/minimum_bh_mass}

\chapter{Cosmological attractors and scalarization}
\label{ch:Cosmology}
\input{Chapters/cosmological_attractors_and_scalarization}

\chapter[Scalarized black holes in EsRGB]{Scalarized black holes in EsRGB}
\label{ch:Black holes}
\input{Chapters/sRGB_bh}

\chapter[Scalarized neutron stars in EsRGB]{Scalarized neutron stars with Ricci and Gauss-Bonnet couplings}
\label{ch:Neutron Stars}
\input{Chapters/sRGB_ns}

\chapter{Stability and Quasinormal modes}
\label{ch:Stability}
\input{Chapters/stability_qnms}

\chapter{Shadows of compact objects with scalar fields}
\label{ch:Shadows}
\input{Chapters/shadows}

\chapter{Conclusions}
\label{ch:conclusions}
\input{Chapters/conclusions}

\appendix

\chapter{Black hole thermodynamics}
\label{ch:Appendix-Thermodynamics}
\input{Chapters/appendix_thermodynamics}

\chapter{Other ``hairy'' solutions}
\label{ch:Appendix_othersol}
\input{Chapters/appendix_other_solutions}

\chapter{Equations of motion and other lengthy expressions}
\label{ch:Appendix_eom}
\input{Chapters/appendix_eom}

\chapter{Quasinormal mode techniques}
\label{ch:Appendix_qnms}
\input{Chapters/appendix_qnms}


\printbibliography

\end{document}

%% file: Chapters/acknowledgements.tex
First and foremost, I would like to thank my supervisor Thomas Sotiriou for his guidance and support throughout my Ph.D. Thoma your scientific insight had a massive impact on my development as a researcher. I will never forget that and I hope that our collaboration will continue to flourish throughout spacetime.
I also want to thank Panagiota Kanti, who has been there since the very beginning of my physics endeavours, supporting me and helping me as if I were her own student.
In alphabetical order, I want to thank all my collaborators: Athanasios Bakopoulos, Lorenzo Bordin, Panagiota Kanti, Burkhard Kleihaus, Jutta Kunz, Antoine Leh\'ebel, Caio Macedo, Ryan McManus, Alexandros Papageorgiou, Thomas Sotiriou, Farid Thaalba, and Giulia Ventagli.
Many thanks to my viva examiners Christos Charmousis and Tony Padilla for their insightful comments and suggestions.

I am very grateful to the School of Mathematical Sciences at the University of Nottingham for the many opportunities it gave me to travel abroad, share my research with the world, and develop as a researcher overall.
Moreover, from the start of my graduate studies I have been a fellow of the Onassis Foundation, for which I am deeply grateful.

On a personal level, I would like to thank my parents for their immense love and support. I will forever be grateful to have you as parents.
Finally, thank you Niko and Eleni. Niko, apart from my brother you are the best friend anyone could ask for.
Eleni you are you.

%% file: Chapters/abstract.tex
Recent developments in the field of gravitational physics, including the emergence of gravitational wave astronomy, black hole images, and more accurate telescopes, have allowed us to probe the strong-field character of gravity in a novel and revolutionary manner. This accessibility related to strong gravity brings into the foreground discussions about potential modifications to General Relativity (GR) that are particularly relevant in high curvature regimes. The most straightforward way to generalise GR is to consider an additional degree of freedom, in the form of a scalar field. In this thesis, we study generalised scalar tensor theories that predict interesting strong-gravity phenomenology. First, we review scalar no-hair theorems and the conditions under which they can be evaded. Next, we study solutions of black holes with scalar hair and the way in which higher derivative terms alter their properties. We then move our discussion to the spontaneously scalarized solutions, which only deviate from GR in the strong-field regime. We propose a model consistent with compact object scalarization, that allows for a GR attractor at late times, without fine-tuning (EsRGB model). Then, we proceed to study properties of black holes and neutron stars in this theory, revealing the interesting phenomenology of the solutions. We also study the radial stability of black holes in EsRGB and perform a preliminary analysis of the hyperbolicity of the problem. Finally, we take a look at the shadows of black holes and wormholes in theories with scalar fields, in light of recent observations of black hole shadows.

%% file: Chapters/publications.tex
The work presented in this thesis is based on the following publications made during my PhD \cite{Antoniou:2020nax, Antoniou:2021zoy, Ventagli:2021ubn, Antoniou:2022agj, Antoniou:2022dre, Thaalba:2022bnt}:

\begin{enumerate}[label={[\arabic*]}]

    \item \textbf{``Black hole minimum size and scalar charge in
    shift-symmetric theories''}\\
    F.~Thaalba, \textbf{G.~Antoniou}, T.~P.~Sotiriou\\
    \href{https://arxiv.org/abs/2211.05099}{arXiv:2211.05099 [gr-qc]}.

    \item \textbf{``Constraining modified gravity theories with scalar fields using black-hole images''}\\
    \textbf{G.~Antoniou}, A.~Papageorgiou, P.~Kanti\\
    \href{https://arxiv.org/abs/2210.17533}{arXiv:2210.17533 [gr-qc].}

    \item \textbf{``Stable spontaneously-scalarized black holes in generalized scalar-tensor theories''}\\
    \textbf{G.~Antoniou}, C.~F.~B.~Macedo, R.~McManus, T.~P.~Sotiriou\\
    \href{https://journals.aps.org/prd/abstract/10.1103/PhysRevD.106.024029}{Phys. Rev. D \textbf{106} (2022) 2, 024029},
    \href{https://arxiv.org/abs/2204.01684}{arXiv:2204.01684 [gr-qc]}.

    \item \textbf{``Neutron star scalarization with Gauss-Bonnet and Ricci scalar couplings''}\\
    G.~Ventagli, \textbf{G.~Antoniou}, A.~Leh\'ebel, T.~P.~Sotiriou\\
    \href{https://journals.aps.org/prd/abstract/10.1103/PhysRevD.104.124078}{Phys. Rev. D \textbf{104} (2021) 12, 124078},
    \href{https://arxiv.org/abs/2111.03644}{arXiv:2111.03644 [gr-qc]}.

    \item \textbf{``Black hole scalarization with Gauss-Bonnet and Ricci scalar couplings''}\\
    \textbf{G.~Antoniou}, A.~Leh\'ebel, G.~Ventagli, T.~P.~Sotiriou\\
    \href{https://journals.aps.org/prd/abstract/10.1103/PhysRevD.104.044002}{Phys. Rev. D \textbf{104} (2021) 4, 044002}, 
    \href{https://arxiv.org/abs/2105.04479}{arXiv:2105.04479 [gr-qc]}.

    \item \textbf{``Compact object scalarization with general relativity as a cosmic attractor''}\\
    \textbf{G.~Antoniou}, L.~Bordin, T.~P.~Sotiriou\\
    \href{https://journals.aps.org/prd/abstract/10.1103/PhysRevD.103.024012}{Phys. Rev. D \textbf{103} (2021), 024012}, \href{https://arxiv.org/abs/2004.14985}{arXiv:2004.14985 [gr-qc]}.

\vspace{5mm} 

\end{enumerate}

%% file: Chapters/Notation.tex
For this thesis, the metric signature is chosen to be $(-,+,+,+)$. Greek letters take values $0,1,2,3$ denoting the indices of the spacetime coordinates. We will be working in the natural system of units, where we set $c=\hbar\equiv 1$. Let us also note that unless explicitly said otherwise, we will be using the terms ``spontaneous scalarization'' and ``scalarization'' interchangeably. Following, is a list of symbols, tensors, and conventions we employ:\\[5mm]

\begin{description}[leftmargin=!,labelwidth=3cm]

\item[$c$] speed of light, we set $c\equiv 1$

\item[$\hbar$] reduced Planck's constant, we set $\hbar\equiv 1$

\item[$G$] Newton's gravitational constant
 
\item[$M_\odot$] solar mass
 
\item[$\kappa$] $8\pi G/c^4$

\item[$g_{\mu\nu}$] spacetime metric

\item[$g$] determinant of the metric

\item[$\partial_\mu$] partial derivative

\item[$\Gamma^\rho_{\mu\nu}$] Christoffel symbol

\item[$\nabla_\mu$] covariant derivative

\item[$R^\rho_{\mu\sigma\nu}$] Riemann tensor

\item[$R_{\mu\nu}$] Ricci tensor

\item[$R$] Ricci scalar

\item[$G_{\mu\nu}$] Einstein tensor

\item[$\mathscr{G}$] $R^2-4\,R_{\mu\nu}R^{\mu\nu}+R_{\mu\nu\rho\sigma}R^{\mu\nu\rho\sigma}$, Gauss-Bonnet invariant

\item[$\phi$] scalar field
 
\item[$X$] $-\partial_\mu\phi\,\partial^\mu\phi/2$, scalar kinetic term

\item[$T^{(\phi)}_{\mu\nu}$] scalar stress-energy tensor

\item[$S_M$] matter action
 
\item[$\psi_M$] matter fields
 
\item[$T^{(m)}_{\mu\nu}$] matter stress-energy tensor
 
\item[$\epsilon$] energy density of a perfect fluid
 
\item[$p$] pressure of a perfect fluid
 
\item[$u_\mu$] 4-velocity of a perfect fluid
 
\item[$T^{PF}_{\mu\nu}$] stress-energy tensor of a perfect fluid
 
\item[$M$] ADM mass of the compact object
 
\item[$Q$] scalar charge of the compact object

\item[$\hat{M},\hat{Q}$] normalized scalar charge and mass

\item[$(\mu\nu)$] symmetrization with respect to $\mu\nu$
  
\end{description}
\vspace{10mm}
For the Fourier transformations we will employ throughout this thesis we will assume the notation

\begin{align*}
    f(x)=&\;\frac{1}{\sqrt{2\pi}}\int F(k)\,e^{-i k x}\, dk\\
    F(k)=&\;\frac{1}{\sqrt{2\pi}}\int f(x)\,e^{+i k x}\, dx
\end{align*}

%% file: Chapters/preface.tex
Through the centuries, the field of gravity has provided the ground for momentous debates, groundbreaking research, and epochal discoveries, intimately associated with our understanding of the cosmos and the human perspective.
From Aristotle's ideas on gravity's fundamental connection with earth, to Kepler and Galileo, and from Newton's law of universal gravitation to Einstein's mathematical revolutionization of the field, gravity has been an area of ongoing development.
The theory of General Relativity (GR) has been one of the cornerstones of modern physics, having made accurate predictions and been repeatedly verified observationally. GR is considered the conventional theory describing the gravitational interaction and has been extensively studied over the last century.
In its framework, spacetime is described by the metric tensor, which encodes its casual structure and constitutes the main object of study.

Physicists, however, have been looking into ways of generalising GR even from its early stages, driven either by pure scientific curiosity, or (more recently) by concrete observational data, including observations of dark matter/energy, LSS etc.
The easiest way to generalise GR is by adding a degree of freedom in the form of a scalar field, resulting in the so-called scalar-tensor theories of gravity.
In this thesis we will study such theories, mainly motivated by our current unprecedented access to the strong field regime of gravity.
Over the last years, we have been experiencing the beginning of a new era in astrophysics, in the face of gravitational wave astronomy, that allows us to probe never-before accessed regions of spacetime, in the very neighbourhood of black holes and neutron stars. From a theoretical standpoint, these regions delve into the high-energy regime of gravity, where new physics might lurk, being associated with the quantum nature of the interaction.
Even more recently, we have been granted ``visual'' access to the vicinity of black holes, by having the images of the supermassive black holes M87$^*$ and Sagittarius A$^*$ taken by the Event Horizon collaboration.
Such observations open up a direct visual path onto potential deviations from the predictions of GR in the high-curvature regime.

As mentioned earlier, black holes and very compact neutron stars are ideal candidates for carrying beyond-GR characteristics. However, in the case of black holes, there have been a number of no-hair theorems preventing solutions carrying (scalar) hair. As is usually the case, from their conception, the various no-hair theorems have been challenged, and eventually solutions with nontrivial hair were found in several theories.
Recently a class of theories belonging in the Horndeski framework, predicting black holes evading the existing no-hair theorems, has attracted a lot of attention.
The type of nonminimal coupling between the scalar field and the metric tensor that acts as the source for the scalar hairs in this model, is the scalar-Gauss-Bonnet (sGB) term.
Hairy black holes, neutron stars, and wormholes with interesting phenomenology, all appear within the sGB framework.

Of particular interest is a subclass of the sGB scalar-tensor models that allow for what we call \textit{spontaneous scalarization} of compact objects.
These scalarized solutions are spontaneously endowed by hair in a process resembling a phase transition, that is triggered at the linear level by a tachyonic instability.
Spontaneously scalarized objects only present deviations from their GR counterparts in the strong-field region of spacetime, \textit{i.e.} near the black hole horizon.
Far away from the compact object, on the other hand, spacetime approaches the flat one just as the GR solution.

In this work we study compact objects carrying scalar hair in an overall attempt to discover a theory, consistent with observations and predicting deviations from GR at those regions where the latter is still unconstrained.
The outline of the thesis is the following: in Chapter~\ref{ch:Introduction} we introduce briefly the concepts of GR, and summarise some of its main successes and shortcomings. We discuss potential modifications of GR, focusing particularly on the Horndeski theory, and explain the existence of no-hair theorems concerning scalar fields.
In Chapter~\ref{ch:Evasions} we examine how the no-hair theorems presented in the introduction may be evaded and once again focus on the evasion of the no-hair theorem regarding Horndeski gravity. In the last section of this chapter we introduce the concept of spontaneous scalarization of compact objects, a theme that will be on the foreground for the largest part of this thesis.
In Chapter~\ref{ch:Mass-Charge} we focus on hairy black holes in shift-symmetric theories, in an attempt to verify how the properties of the solutions depend on the Lagrangian terms that are not directly related with the sourcing of the scalar hair. We derive the existence conditions for scalarized black holes to exist and determine the dependence of the black hole mass and scalar charge on the terms appearing in the Lagrangian.
In Chapter~\ref{ch:Cosmology} we present a model within the spontaneous scalarization framework, that allows for a late-time cosmological attractor, a characteristic that is crucial from an observational standpoint. In this model, which we call EsRGB model and exists in the more general Horndeski theory, a scalar field couples nonminimally with gravity through the Ricci and Gauss-Bonnet geometrical invariants.
Next, in Chapter~\ref{ch:Black holes} we discuss the properties of the black holes arising in the EsRGB model demonstrating an interesting dependence of them on the particular synergy between the two couplings.
In Chapter~\ref{ch:Neutron Stars} we do a similar analysis but for scalarized neutron stars emerging in EsRGB. We examine a few different neutron star configurations corresponding to different choices for the energy density. We also discuss our results in light of the very strong binary pulsar constraints that have thus far significantly restricted models predicting scalarization.
In Chapter~\ref{ch:Stability} we turn our attention once again to the spontaneously scalarized black holes found in Chapter~\ref{ch:Black holes}. We now perform a stability analysis for the solutions, perform a quasinormal mode analysis to demonstrate the transition from GR to scalarized solutions, and finally investigate the effects the sRGB synergy has on the hyperbolicity formulation of the problem.
Finally, in Chapter~\ref{ch:Shadows} we look into the impact that potential modifications to GR with scalar fields may have on the shadows of the hairy solutions. We do that especially, considering the recent images of black holes captured by the EHT. We analyse shadows of black holes and wormholes in Gauss-Bonnet gravity, curvature-induced spontaneous scalarization models, and the Einstein-Maxwell-scalar theory.
We present our final conclusions in Chapter~\ref{ch:conclusions}.\\[10mm]
\textbf{Nottingham, 14 April 2023 \hfill Georgios Antoniou}

%% file: Chapters/introduction.tex
\section{General Relativity: should we go beyond?}
In the beginning of the previous century Albert Einstein revolutionized the way the gravitational interaction is perceived. According to the General theory of Relativity (GR), gravity is inherentlty connected to the geometry of space and time, in an equally impressive and nontrivial way.

The Einstein-Hilbert action
\begin{equation}
    S_{\text{EH}}= \frac{1}{2\kappa}\int{d^4x\,\sqrt{-g}\,R}\,,
    \label{eq:EH_action}
\end{equation}
is the action that after variation with respect to the metric tensor yields the Einstein field equations, which elegantly describe the correspondence between the spacetime and the energy and momentum contained in it.
The constant appearing in the Einstein-Hilbert action is given by $\kappa=8\pi G c^{-4}$, where $G$ is Newton's gravitational constant and $c$ is the speed of light; $g\equiv\det(g_{\mu\nu})$ is the determinant of the metric tensor, and $R$ is the Ricci curvature invariant. In the context of this thesis, as is also usually assumed for GR, the connection and the metric are related through

\begin{equation}
    \Gamma_{\mu\nu\sigma}=\frac{1}{2} (g_{{\mu \sigma},\nu}+g_{{\nu \sigma},\mu}-g_{{\mu \nu},\sigma})\,,
\end{equation}
while the Riemann tensor, expressing the spacetime curvature, is defined as

\begin{equation} \label{1.6}
    R^{\rho}_{\mu \sigma \nu}=\Gamma ^{\rho}_{\mu \nu, \sigma}-\Gamma ^{\rho}_{\mu \sigma, \nu}+\Gamma^{\rho}_{\lambda \sigma}\Gamma^{\lambda}_{\mu \nu}-\Gamma^{\rho}_{\lambda \nu}\Gamma^{\lambda}_{\mu \sigma}\,.
\end{equation}
After variation of the Einstein-Hilbert action with respect to the metric tensor, we find that the vacuum Einstein field equations are given by:

\begin{equation}
    \frac{\delta S_{EH}}{\delta g^{\mu\nu}}=0 \;\Rightarrow \; G_{\mu\nu}\equiv R_{\mu\nu}-\frac{1}{2}R g_{\mu\nu}=0\,,
\end{equation}
where $G_{\mu\nu}$ is the Einstein tensor. If a matter distribution is assumed to be present, and the presence of a cosmological constant is accounted for, the action for the theory reads
\begin{equation}
    S=\frac{1}{2\kappa}\int{d^4x\,\sqrt{-g}\,(R-2\Lambda)+\mathcal{L}_{\text{M}}}\,,
\end{equation}
where $\mathcal{L}_{\text{M}}$ is the matter Lagrangian and $\Lambda$ is the cosmological constant.
Then, the Einstein equations are modified by the addition of the $\Lambda$ contribution and the energy-momentum tensor, the latter of which appears in the right-hand side of the equation below
\begin{equation}
    G_{\mu\nu}+\Lambda g_{\mu\nu}=\kappa T_{\mu\nu}\,\quad\text{where} \quad T_{\mu\nu}=-\frac{2}{\sqrt{-g}}\frac{\delta (\sqrt{-g} \mathcal{L}_M)}{\delta g^{\mu\nu}}\,.
\end{equation}
As of now, GR has managed to pass a variety of experimental tests and has proven to be one of the most successful physical theories. Gravitational lensing provides one of the most direct observationally confrontable predictions of GR. During the solar eclipse of May 1919, Dyson, Eddington, and collaborators made the first experimental observation of light deflection as predicted by GR, by noticing the change of the position of stars around the sun in the celestial sphere \cite{Dyson:1920cwa}. Another important prediction of GR concerned the gravitational redshift of light. In 1954 Popper performed the first accurate measurement of the redshift for the light of a white dwarf \cite{1954ApJ...120..316P}. Moreover, the equivalence principle, is considered to be one of the basic pillars of GR, essentially stating the equivalence between the inertial and gravitational mass. Since the 1960s a number of tests have managed to verify the equivalence principle to very high accuracy. Recently, a lot of tests have managed to confront GR in the strong field regime, with the field of gravitational-wave astronomy evolving rapidly. In 1993, Hulse and Taylor won the Nobel Prize in Physics for their discover of a binary pulsar and the measurement of the effects of the gravitational wave emission on its orbit \cite{Hulse:1974eb}. More recently, in 2016 the first direct measurement of gravitational waves was announced \cite{LIGOScientific:2016aoc}, officially marking a new era in the field of gravitational waves and opening a never before accessible window onto the strong gravitational regime.
These waves were predicted by Einstein \cite{Einstein:1918btx} and their direct detection marks perhaps the most important experimental verification of GR. A few years later, in 2018, the first ever images of a black hole shadow to be taken were published \cite{EventHorizonTelescope:2019dse,EventHorizonTelescope:2019uob,EventHorizonTelescope:2019jan,EventHorizonTelescope:2019ths,EventHorizonTelescope:2019pgp,EventHorizonTelescope:2019ggy,EventHorizonTelescope:2021bee,EventHorizonTelescope:2021srq}, in agreement (for the most part!) with black hole shadows predicted by models using the principles of GR. We will return to the two last points in the next section.

Despite its numerous successes, however, GR fails to address a number of issues related to gravity as a mathematically-consistent physical theory, as well as within the context of cosmology.
To begin with, GR cannot be reconciled with quantum physics. In many scenarios we can safely rely on classical GR to describe strong gravitational fields at large scales, or QFT to explain small-scale physical processes on effectively flat backgrounds. If one, however wants to investigate strong gravitational interactions at small scales, the importance of a quantum description of gravity becomes obvious.

In QFT, a perturbative quantization process, inevitably leads to ultraviolet divergences and nonsensical results for the sought-after quantities. These divergences, however, can be cleverly removed by the use of renormalization techniques, which effectively allow us to absorb the seemingly infinite self-interactions into re-definitions. A similar approach in GR seems rather bleak as the dimensions of the gravitational coupling would lead to an infinite number of terms, with increasingly high curvature, rendering the theory non-renormalizable \cite{tHooft:1974toh, Goroff:1985th}.

A quantum gravitational theory would allow us to predict the behaviour of gravity at all scales, but as of now the achievement of that goal has not been fully realized.
GR has been extensively tested and verified in the weak-field limit as already mentioned, but in the strong-gravity regime it remains relatively unconstrained \cite{Berti:2015itd}.
In these strong-field regions new physics may emerge while evading our attention due to the limitations introduced by the use of GR.

Perhaps most famously, GR fails to incorporate the physics of dark matter and dark energy into its scheme, and also to address the cosmological constant problem.
By now it is well accepted among the scientific community that regular matter constitutes no more than a small part of the universe around us. Discussions about unseen matter affecting in some way astrophysical observables occurred at many different points of the 20th century \cite{1922ApJ....55..302K, 1937ApJ....86..217Z, 1940ApJ....91..273O, 1970ApJ...159..379R, 1970ApJ...160..811F}. The latest Planck data \cite{Planck:2015fie, Planck:2018vyg} suggests that around 31\% of the matter-energy distribution in the universe is in the form of matter, with ordinary matter contributing around 5\% and dark matter contributing 26\% of that. Through the years there have been several candidates for dark matter ranging from  neutrinos to axions, supersymmetric particles, and primordial black holes \cite{Preskill:1982cy, Ellis:1983ew, Feng:2010gw, Marsh:2015xka, Carr:2016drx, Adams:2022pbo}. Despite the plethora of potential explanations, however, dark matter has not yet been identified.

The rest, which happens to be the largest contribution to the total energy of the universe is in the form of dark energy, comprising an astonishing 69\%.
In the  $\Lambda$-CDM (Lambda cold dark matter) model which is primarily used to describe the cosmological history of our universes, all three of the above contributions are considered, with the dark energy expressed through the cosmological constant $\Lambda$-term.
The cosmological constant was first introduced by A. Einstein in 1917 in an attempt to retrieve a model predicting a static universe \cite{1917SPAW.......142E}, and can be perceived as the energy density of the vacuum itself. From a QFT point of view, this energy density emanates from zero-point fluctuations of the quantum fields spread across the quantum vacuum. A reasonable assumption for the cut off threshold for the vacuum energy is the Planck scale. Then we can estimate for the vacuum energy density the following
\begin{equation}
    \langle\rho\rangle=\int_0^{M_{Pl}}\frac{d^3p}{(2\pi)^3}\frac{1}{2}\sqrt{p^2+m^2}\sim 10^{72}\,\text{GeV}^4\, .
\end{equation}

On the other hand the numerical value of the cosmological constant as it is estimated from observations and cosmological models, is found to be 120 orders of magnitude smaller. Because of Lorentz invariance for the vacuum we have that
\begin{equation}
    \langle T_{\mu\nu} \rangle = - \langle \rho \rangle g_{\mu\nu}\,,
\end{equation}
which in turn contributes to the Einstein equations in the form of a cosmological constant
\begin{equation}
    \Lambda= 8\pi G \langle\rho\rangle\, .
\end{equation}
From cosmological observations we know that the expectation value for the vacuum energy density is
\begin{equation}
    \langle \rho \rangle \sim 10^{-47}\,\text{GeV}^4\,.  
\end{equation}
By requiring Lorentz invariance of the vacuum state, this vast difference in orders of magnitude can be brought down to some 50 orders of magnitude \cite{Koksma:2011cq, Martin:2012bt}, but remains mysteriously large either way. We are confronted therefore with a huge discrepancy that would require extremely precise fine-tuning.

Various solutions have been suggested to the cosmological constant problem.
For instance, applying one interpretation of the \textit{Anthropic Principle}, which was first proposed by Dicke \cite{Dicke:1957zz, 1961Natur.192..440D}, we may claim that we just happen to live in a universe characterized by the observed value of the cosmological constant. Our universe is only one of many regions (universes) with different vacuum energies. In string theory, there is supposedly a humongous landscape ($\gtrsim 10^{500}$) of solutions \cite{Bousso:2000xa, Douglas:2003um, Douglas:2004kp} allowing for many different cosmological constants.
The cosmological problem has also been considered alongside supersymmetry.
When supersymmetry holds, the fermionic and bosonic contributions cancel out and the vacuum energy is exactly zero, while breaking supersymmetry induces a nonzero vacuum energy \cite{Weinberg:1988cp, Weinberg:2000yb}. Considering the supersymmetric scale to be $\mathcal{M}_{\text{SUSY}}\sim 10^3\,\text{GeV}$, would lead to a discrepancy $\mathcal{M}_{\text{SUSY}}/\mathcal{M}_{\text{vacuum}}\sim 10^{15}$ which is significantly smaller than $\mathcal{M}_{\text{Planck}}/\mathcal{M}_{\text{vacuum}}\sim 10^{30}$, but still not small enough for the problem to be resolved.

In the process of tackling the variety of different issues emerging in the context of GR, we might have to resort to the basics of the theory itself \cite{Deser:1969wk}.
There are three main (non-mutually exclusive) ways to define alternatives theories of gravity \cite{Ezquiaga:2018btd}: (i) breaking one of the assumptions related to the fact that GR is defined on a 4-dimensional pseudo-Riemannian manifold, where local Lorentz invariance holds, (ii) adding extra fields (interacting with the metric tensor), and (iii) considering the graviton to be massive.
Following either one of these three paths is equivalent to modifying gravity, which could presumably resolve one or more of the aforementioned problems associated with GR. In the following section we discuss various ways in which one can reasonably modify GR.

Before we close this section let us explain the unit system we will be using throughout this thesis. Unless stated otherwise we will be using geometrized units where all units are expressed in terms of powers of length or mass. We also set $c\equiv\hbar \equiv G = 1$. In these units the metric tensor is dimensionless, the Ricci invariant $R$ had units $L^{-2}$, and unless stated otherwise we will define the scalar fields appearing in our theories so that they are is dimensionless.

\section{Modifying GR}

Going from our current understanding of gravity, which is mostly limited to low energies, to the one that probes extremely high energy regimes, does not have to be direct. In other words modifying gravity as we currently know it, may be enough to capture physics inaccessible by GR, potentially realized at long distances, while still remaining within an effective theory framework that only breaks down at some high energy limit.
In technical terms, these modifications that one can introduce to GR would appear as additional terms in the Einstein-Hilbert action \eqref{eq:EH_action}. So the question that follows naturally is, how can one reasonably modify GR?

First of all, as pointed out in the previous section, GR is properly fixed on a 4-dimensional pseudo-Riemannian manifold where locality and Lorentz invariance are respected. A possible avenue that allows for GR modifications is the introduction of extra spatial dimensions permitting the construction of new operators using the metric tensor. Kaluza-Klein theory, Lovelock gravity, brane-world models and string theory \cite{1926ZPhy...37..895K, Lovelock:1971yv, Lovelock:1972vz, Scherk:1974jj, Siegel:1988yz, Overduin:1997sri, Arkani-Hamed:1998jmv, Antoniadis:1998ig, Randall:1999ee, Randall:1999vf} are examples of theories considering more than four spacetime dimensions in an attempt to tackle issues not addressed by standard GR. Moreover, theories violating Lorentz invariance can be considered. Arguing in favor of a preferred foliation of spacetime can result in that, as in the case of Ho\u{r}ava gravity \cite{Horava:2009uw, Horava:2009if, Sotiriou:2010wn, Blas:2014aca}. Another famous Lorent-violating example is that of Einstein-aether gravity \cite{Gasperini:1987nq, Jacobson:2000xp, Arkani-Hamed:2002bjr}, where an additional to the metric, unit vector field is assumed, and the existence of a preferred reference frame breaks Lorentz invariance.

\begin{figure}[t]
\centering
\begin{tikzpicture}
  \path[mindmap,concept color=black!50!red!80,text=black,
    level 1/.append style={level distance=5cm,sibling angle=35},
    level 2/.append style={level distance=3cm,sibling angle=50},
    level 3/.append style={level distance=2cm,sibling angle=45},
    level 4/.append style={level distance=2cm,sibling angle=35}]
    node[concept] {General\\Relativity}
    [clockwise from=10]
    child[concept color=black!50!red!50] {
    node[concept] {Massive\\Gravity}
      [clockwise from=5]
      child[concept color=black!25!red!25] { node[concept] {dRGT} }
}
    child[concept color=teal!70] {
      node[concept] {Extra\\fields}
      [clockwise from=20]
      child[concept color=teal!50] { node[concept] {Tensors}
                    [clockwise from=20]
                    child[concept color=blue!50] { node[concept] {Bigravity}
                    }
                    child[concept color=blue!50] { node[concept] {Multi-\\Gravity}
                    }
      }
      child[concept color=teal!50] { node[concept] {Vectors}
                    [clockwise from=0]
                    child[concept color=blue!50] { node[concept] {TeVeS}
                    }
                    child[concept color=blue!50] { node[concept] {Proca}
                    }
      }
      child[concept color=teal!50] { node[concept] {Scalars}
              [clockwise from=-40]
              child[concept color=teal!50!blue!50] { node[concept] {Beyond\\Horn-\\deski}
              [clockwise from=0]
                    child[concept color=blue!70] { node[concept] {$C(X)$}
                    }
                    child[concept color=blue!70] { node[concept] {$D(X)$}
                    }
              }
              child[concept color=teal!50!blue!50] { node[concept] {Horn-\\deski}
              [clockwise from=-14]
                    child[concept color=blue!70] { node[concept] {Brans-Dicke}
                    }
                    child[concept color=blue!70] { node[concept] {Quint-essence}
                    }
                    child[concept color=blue!70] { node[concept] {Gali-leon}
                    }
                    child[concept color=blue!70] { node[concept] {GB}
                    }
                    child[concept color=blue!70] { node[concept] {$f(R)$}
                    }
              }
              child[concept color=blue!50] { node[concept] {Parity\\Viola-\\tion}
              }
    }
}
    child[concept color=black!50!green!70] {
    node[concept] {Broken\\Assump-\\tions}
      [clockwise from=-100]
      child[concept color=black!50!green!50] { node[concept] {Extra\\Dimen-\\sions}
      }
      child[concept color=black!50!green!50] { node[concept] {Lorentz\\
      Violation}
            child[concept color=black!50!green!25] { node[concept] {Horava}
            }
            child[concept color=black!50!green!25] { node[concept] {Einstein\\Aether}
            }
      }
      child[concept color=black!50!green!50] { node[concept] {Non-\\Locality}
      }
}
;
\end{tikzpicture}
\caption[Roadmap of modified gravitational theories]{Roadmap of modified gravitational theories. This figure is based on a similar one presented in \cite{Ezquiaga:2018btd}, where a tour through the avenues of modified gravity and the arising observational constraints can be found.}
\label{fig:roadmap}
\end{figure}
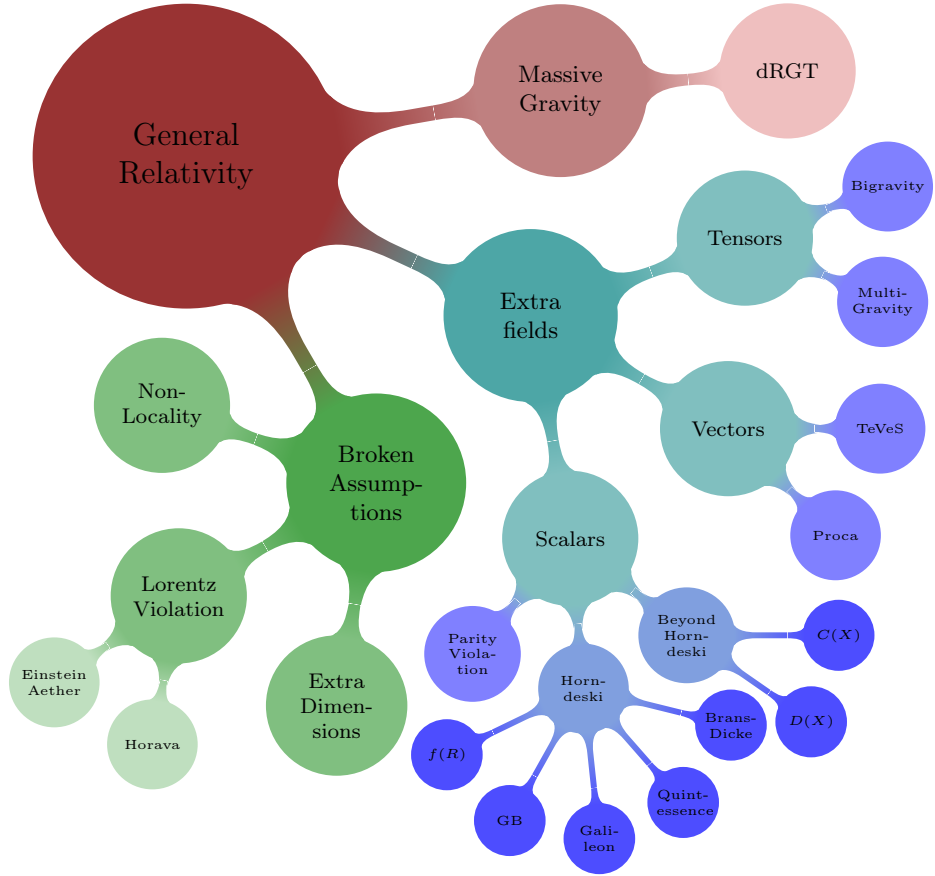

Another relatively obvious way to modify GR is to consider the graviton to be massive. Despite the simplicity of the theoretical consideration, making the graviton massive has proven to be a task followed by a number of issues that are not trivial to address. They pertained to inconsistencies between the zero-mass limit of the massive theory with GR, and more importantly to the appearance of ghosts \cite{Fierz:1939ix, vanDam:1970vg, Zakharov:1970cc, Vainshtein:1972sx}. A decade ago the dRGT model was proposed, managing to present a massive gravity model free of ghosts \cite{deRham:2010kj, deRham:2014zqa}.

Lorentz invariance and locality are properties that we would like to preserve. Additionally, from our previous discussion, especially on the issue pertaining to the cosmological constant, the need to explore GR in the infrared is important.
Consequently, it is reasonable to investigate whether changing the gravitational degrees of freedom could potentially be helpful. In that respect, we can distinguish these additional degrees of freedom to tensorial, vectorial, and scalar.
Characteristic examples of modified gravitational theories with additional tensor fields are those of \textit{Bigravity} \cite{Hassan:2011zd}, where two spin-2 particles exist, and \textit{Multigravity} \cite{Hinterbichler:2012cn} where more than two interacting metrics are considered.
Theories introducing vector fields have been considered in the context of dark energy. Perhaps the most well-known theory in the framework of vector-tensor theories is the Proca theory including a kinetic and a mass term for the vector field \cite{Proca:1936fbw}. Generalizations of the Proca theory \cite{Heisenberg:2014rta, Allys:2015sht, BeltranJimenez:2016rff} have also been made resembling the Horndeski theory which we will analyze in detail in the following pages. On this matter, maybe the most straightforward generalization of GR is the one considering one or multiple additional scalar fields.
The most general theory considering a metric tensor and a scalar field that leads to second order equations of motion is the Horndeski theory \cite{Horndeski:1974wa}. Scalar fields are studied both in the context of strong gravitational effects and of cosmology. Especially in the latter case, an important advantage scalars offer in contrast to the vectors scenario, is that they do not affect the isotropy of the universe if they are only time-dependent, as they do not have a preferred direction. Furthermore, one can introduce extra high-order terms in the Lagrangian formulation of GR, in such a way that the resulting equations do not yield unphysical results despite being of higher order. These extended theories are usually referred to as DHOST (Degenerate-Higher-Order-Scalar-Tensor) theories \cite{Langlois:2015cwa, BenAchour:2016cay, BenAchour:2016fzp, Motohashi:2018pxg}. There also exist models including both additional vector and scalar degrees of freedom \cite{Bruneton:2007si}. A road map of the possible avenues that one can follow in order to modify gravity can be seen in Fig.~\ref{fig:roadmap}.

\subsection{Early scalar-tensor theories}

The origins of scalar-tensor theories trace back to Jordan's work in 1955, who introduced the concept of a scalar field, in a attempt to embed a curved 4-dimensional manifold in a flat 5-dimensional spacetime. The Jordan action is given by
\begin{equation}
    S_{\text{J}}= \int{ d^4x\,\sqrt{-g}\;\phi^\gamma \bigg[R -\frac{\omega}{\phi^2}(\partial\phi)^2
    +L_{\text{M}}
    (g_{\mu\nu},\phi,\psi_M)   \bigg]}  \,.
    \label{eq:J_action}
\end{equation}
where $\psi_{\text{M}}$ correspond to the matter distribution, which is coupled with the metric tensor and the scalar field.
When the matter Lagrangian does not depend on $\phi$ there exists an invariance with respect to the constant $\gamma$. This invariance was taken into account by Robert H. Dicke and Carl H. Brans in 1961, when they redifined the scalar field, in a way that absorbs $\gamma$ into the definition of the field and developed the \textit{first} scalar-tensor theory as a competitor to Einstein's GR \cite{Brans:1961sx}. In addition to the metric tensor the gravitational interaction is mediated by a scalar field, nonminimally coupled with curvature. The action for the Brans-Dicke model reads:
\begin{equation}
    S_{\text{BD}}= \int{d^4x\,\sqrt{-g}\,\bigg[\phi\,R -\frac{\omega}{\phi}(\partial\phi)^2+L_M
    (g_{\mu\nu},\psi_M)\bigg]}\,,
    \label{eq:BD_action}
\end{equation}
where $\omega$ is a parameter of the theory that can be chosen in order to satisfy observational constraints, and is called the Brans-Dicke coupling constant.
Now the matter is not coupled with the scalar field.
This is important when considering the \textit{weak equivalence principle} (WEP) which has been so far verified to a high level of precision \cite{Baessler:1999iv, Blaser:2001}. The absence of the scalar-matter coupling at the level of the Lagrangian, implies that the inertial mass appearing in the the action for the geodesics of a point particle, is independent of the scalar field and can therefore be factored out, \textit{i.e.}
\begin{equation}
    S_M=-m\int d\tau\, ,
\end{equation}
where $\tau$ is the proper time. Therefore, the trajectories are not affected by it, and the WEP holds.
It is worth pointing out that the kinetic term appears to be singular due to the presence of the $\phi^{-1}$ factor in front of it. It is however easy to employ the following redefinition
 \begin{equation}
 \phi\to \frac{1}{2} \xi \tilde{\phi}^2\quad , \quad \xi^{-1}\equiv 4\omega\, ,
 \end{equation}
that allows us to bring this term into the standard canonical form, so that the action can then be written as
\begin{equation}
    S_{\text{BD}}= \int{d^4x\,\sqrt{-g}\,\bigg[\frac{\xi}{2}\tilde{\phi}^2 R -\frac{1}{2}(\partial\tilde{\phi})^2+L_{\text{M}}
    (g_{\mu\nu},\psi_{\text{M}})\bigg]}\,,
\end{equation}
where now the coupling between the scalar and the Ricci invariant is quadratic. We will be calling couplings of this form \textit{nonminimal} couplings. In particular, if a term of the action cannot be retrieved from the flat limit by applying the \textit{comma-to-semicolon rule}, which substitutes $\eta_{\mu\nu}$ with $g_{\mu\nu}$ and promotes partial derivatives to covariant ones, then we will be calling it nonminimally coupled.

In the Brans-Dicke theory the Einstein equations are modified by the presence of an effective energy-momentum tensor for the scalar, and a scalar coefficient in front of the Einstein tensor, eventually leading to
\begin{equation}
\begin{split}
    G_{\mu\nu}=&\;\frac{\kappa\pi}{\phi} T^M_{\mu\nu} + \frac{\omega}{\phi^2}\left[\partial_\mu\phi \partial_\nu\phi - \frac{1}{2} g_{\mu\nu} (\partial \phi)^2\right]\\
    &+ \frac{1}{\phi}(\nabla_\mu\nabla_\nu\phi - g_{\mu\nu} \Box\phi)\,,
\end{split}
\end{equation}
while the scalar equation of motion reads:
\begin{equation}
    \Box\phi = \frac{\kappa}{3+2\omega}T^{M}\,.
\end{equation}
It is noteworthy that despite the fact that the matter Lagrangian does not contain couplings with the scalar field, the latter appears to be sourced from the energy momentum tensor as a result of the nonminimal coupling, and it strengthens the detectability prospects for the scalar.
It is straightforward to deduce that GR is retrieved in the limit $\phi\rightarrow constant$. One of the main points of the Brans-Dicke model concerns the appearance of an effective gravitational constant proportional to the inverse of the scalar field, \textit{i.e.} $G_{\text{eff}}\sim 1/\phi$, provided that the time-variation of the scalar is slow. Like GR, the Brans-Dicke model predicts gravitational lensing and precessions in a way that is dependent on the exact value of the Brans-Dicke parameter. Therefore, observations can be used to place constraints on this parameter. The most recent and larger lower bound on the Brans-Dicke parameter has been set by the  Cassini–Huygens experiment, and is of the order of $\omega>40,000$ \cite{Alsing:2011er}.

\subsection{Horndeski gravity}\label{subsec:Horndeski}
As we mentioned already, Horndeski's theory is the most general four-dimensional diffeomorphism invariant theory involving a metric tensor and a scalar field that leads to second-order field equations upon variation \cite{Horndeski:1974wa}. It is important for a theory to yield second-order equations of motion in order to avoid a type of instability know as Ostrogradsky' instability \cite{Woodard:2015zca}. Whenever the Lagrangian satisfies a condition known as \textit{nondegeneracy}, theories yielding equations of motion of order higher than two, are usually associated with linear contributions of canonical momenta in the Hamiltonian formulation of the theory. This leads to unphysical modes that are unbounded from below, otherwise known as ghosts. The action of the Horndeski theory is given by
\begin{equation}
    S_{\text{H}}=\frac{1}{2\kappa}\sum_{i=2}^{5}\int \mathrm{d}^4x\,\sqrt{-g}\mathcal{L}_i,
\label{eq:Horndeski}
\end{equation}
with each sub-Lagrangian $\mathcal{L}_i$ given by
\begin{align}
    \label{L2}
    \mathcal{L}_2 = & \, G_2(\phi,X),\\
    \label{L3}
    \mathcal{L}_3 = & \, -G_3(\phi,X)\Box\phi,\\
    \label{L4}
    \mathcal{L}_4 = & \, G_4(\phi,X){R}+G_{4X}[(\Box\phi)^2-(\nabla_\mu\nabla_\nu\phi)^2],\\
    \begin{split}
        \mathcal{L}_5 = & \, G_5(\phi,X)G_{\mu\nu}\nabla^\mu\nabla^\nu\phi\\
        &- \frac{G_{5X}}{6}\left[\left(\Box\phi\right)^3 -3\Box\phi(\nabla_\mu\nabla_\nu\phi)^2+2(\nabla_\mu\nabla_\nu\phi)^3\right],
    \end{split}
\end{align}
where the canonical kinetic term is defined as $X=-(\partial \phi)^2/2$ and $G_{iX}\equiv \partial_X G_i$. The action of the full theory is completed by the matter contributions, namely $S_{\text{M}}[g_{\mu\nu},\psi_{\text{M}}]$. In order to derive the equations of motion a variation of the action with respect to the metric and the scalar must be taken \cite{Kobayashi:2011nu}:
\begin{equation}
\begin{split}
    \delta\left(\sqrt{-g}\sum_{i=2}^5\mathcal{L}_i\right)
    =&\, \sqrt{-g}\left[\sum_{i=2}^5{\cal G}^i_{\mu\nu}\delta g^{\mu\nu}
    +\sum_{i=2}^5\left(P_\phi^i-\nabla^\mu J_\mu^i\right)\delta\phi
    \right]\\
    &+\, \text{total derivative}\, .
\end{split}
\end{equation}
Minimizing the action then yields the metric and scalar field equations

\begin{equation}
    \sum_{i=2}^5{\cal G}_{\mu\nu}^i=0,\quad \nabla^\mu\left(\sum_{i=2}^5 J_\mu^i \right)= \sum_{i=2}^5 P^i_\phi.
\label{eq:Horndeski-eom}
\end{equation}
with some of the specific expressions for the quantities defined above given in appendix~\ref{ch:Appendix_eom}. In this format the scalar equation has been put in a useful way, as the quantities $P^i_{\phi}$ are absent when shift symmetry is considered.

\section{No-hair theorem}
Despite the mystery associated with black holes, they are in general rather simple physical objects characterized by a small number of independent parameters. Let us consider a stationary spacetime, i.e. one admitting an asymptotically timelike Killing vector field. According to the \textit{no-hair theorem} stationary black holes derived as solutions of the Einstein-Maxwell Lagrangian of general relativity, are characterized by three observables, namely their mass, electric charge, and angular momentum.

Since black holes can in principle be characterized by non-trivial electromagnetic fields, it is reasonable to question whether more general black hole classes exist, that are characterized by more degrees of freedom. The simplest extension to the three "charges" mentioned earlier would be an additional scalar degree of freedom associated with a scalar field. In the following chapters we will be referring to the property quantifying the scalar hair of a black hole solution as the \textit{scalar charge} of the black hole.

\subsection{No scalar hair theorems}

In the Brans-Dicke theory, and in the Jordan frame (the physical one where matter is coupled minimally to gravity and therefore particles have constant mass), the action for the theory reads
\begin{equation}
    S=\frac{1}{2\kappa}\int d^4x\left\{\sqrt{-\hat{g}} \left(\varphi \hat{R} - \frac{\omega}{\varphi}\hat{\nabla}_\mu\varphi\hat{\nabla}^\mu\varphi\right) + \mathcal{L}_m(\hat{g}_{\mu\nu},\psi)\right\}\,,
\end{equation}
where the second contribution corresponds to the matter distribution. Stationarity implies that the solution admits an \textit{asymptotically timelike killing vector} $\xi^\mu$. We will be using the symbol $T_{(\phi)}$ for the contribution to the energy-momentum tensor coming from the scalar field. Then, in the Einstein frame, $T_{(\phi)}$ does not contain second derivatives for the scalar and satisfies the \textit{weak energy condition}. The weak energy condition essentially states that the energy density for any field as seen by any observer should not be negative, namely:
\begin{equation}
    T^{(\phi)}_{\mu\nu}\xi^\mu\xi^\nu \ge 0 \, ,
\end{equation}
where the vector $\xi^\mu$ is timelike (e.g. 4-velocity).
In \cite{Hawking:1971vc} it was shown that stationary and asymptotically flat spacetimes that satisfy the WEC are axisymmetric or static (globally timelike killing vector), and therefore admit an \textit{asymptotically spacelike killing vector} $\zeta^\mu$ (associated with rotational symmetry). The event horizon occurs wherever the time translational killing vector becomes null.
By defining the bivector $\xi^{[\mu}\zeta^{\nu]}$, it can be claimed that following to Carter's analysis \cite{Carter:1969zz}, the horizon is located at $h\equiv\xi^{[\mu}\zeta^{\nu]}\xi_{[\mu}\zeta_{\nu]}=0$.
Then, following Hawking's analysis \cite{Hawking:1972qk}, outside of the horizon
$$h=\xi^{[\mu}\zeta^{\nu]}\xi_{[\mu}\zeta_{\nu]}<0\, .$$
In the Einstein frame, and when no matter distribution is present, the scalar equation simply reads
\begin{equation}
    g^{\mu\nu}\nabla_\mu\nabla_\nu \phi=0\, .
\end{equation}
This can be straightforwardly integrated to yield
$$\int_\mathcal{V} \phi \;g^{\mu\nu}\nabla_\mu\nabla_\nu \phi=0\Rightarrow \int_{\mathcal{V}}g^{\mu\nu}\,\nabla_\mu \phi \,\nabla_\nu \phi=\int_{\partial \mathcal{V}}\phi\, n^\mu \nabla_\nu \phi\,,$$
where $\partial\mathcal{V}$ bounds $\mathcal{V}$ and is made from the horizon hypersurface, the spacelike infinity, the future and the past timelike infinities.
Bekenstein's ``old'' no-hair theorem \cite{Bekenstein:1971hc} yields the same result following a slightly different analysis. There, it was shown that for an arbitrary number of minimally coupled scalar fields ${\phi_k},\,k=1,2...$, the scalar equation of motion yields
\begin{equation}
    \sum_k\int_\mathcal{V} d^4x\,\sqrt{-g}\left[\partial_\mu\phi_k\,\frac{\partial\mathcal{L}}{\partial\phi_\mu}+\phi_k\,\frac{\partial\mathcal{L}}{\partial\phi_k} \right]-\int_{\partial\mathcal{V}} b^\mu\,dS_\mu=0\, ,
\label{eq:many_scalars}
\end{equation}
where $dS_\mu$ is the hypersurface element bounding the domain $\mathcal{V}$. The following quantity was also defined
\begin{equation}
    b^\mu=\sum_k \phi_k\,\frac{\partial \mathcal{L}}{\partial(\partial_\mu\phi_k)}.
\end{equation}
The reader is referred to \cite{Bekenstein:1971hc} for the details where it is shown that for static and asymptotically flat spacetimes the surface integral appearing in \eqref{eq:many_scalars} should vanish (at the horizon the resulting Schwarz inequality bounds the quantity $b^\mu dS_\mu=(g_{ij}dS^i b^j)^2$ to be less than or equal to zero, while the contribution at infinity vanishes due to the fall-off of physically relevant scalar fields). Then it is shown that the only allowed solution for the scalar fields is the trivial one.

The theorem was generalized to scalar-tensor theories where $\omega\rightarrow \omega(\phi)$ and a scalar potential may be included \cite{Sotiriou:2011dz}
\begin{equation}
    S=\frac{1}{2\kappa}\int \hspace{-1mm} d^4x\sqrt{-\hat{g}} \left\{\left[\varphi \hat{R} - \frac{\omega(\varphi)}{\varphi}(\hat{\nabla}\varphi)^2-\hspace{-1mm}V(\varphi)\right] \hspace{-1mm}+ \hspace{-1mm}\mathcal{L}_m(\hat{g}_{\mu\nu},\psi)\right\}\, .
\end{equation}\\
By employing the transformation
\begin{equation}
    g_{\mu\nu}=\varphi\,\hat{g}_{\mu\nu}\quad \text{and}\quad d\phi = \frac{d\varphi}{\varphi}\sqrt{\frac{2\omega(\varphi)+3}{2\kappa}}\, ,
\end{equation}
we can go to the Einstein frame, and like before we can show with a slight modification that
\begin{equation}
\begin{split}
    g^{\mu\nu}\nabla_\mu\nabla_\nu \phi=U'&\Rightarrow \int_\mathcal{V} U' \;g^{\mu\nu}\nabla_\mu\nabla_\nu \phi-U'^2=0\\
    &\Rightarrow \int_{\mathcal{V}}U''\,g^{\mu\nu}\,\nabla_\mu \phi \,\nabla_\nu \phi+U'^2=\int_{\partial \mathcal{V}}U'\, n^\mu \nabla_\nu \phi\, ,
\end{split}
\end{equation}\\
where $U(\phi)=V(\varphi)/\varphi^2$. As long as $U''>0$, the same reasoning as before leads to $\phi$ being constant. At this point let us stress that we can identify the quantity $U''$ with the effective mass of the scalar field. As long as the effective mass is not tachyonic the no-hair theorem holds and the only solution is the constant scalar.

\subsection{``Novel'' no-hair theorem}
A different approach to formulating a no scalar hair theorem for black holes was followed in \cite{Bekenstein:1995un}, with Bekenstein's ``novel'' no-hair theorem. It concerns static and asymptotically flat solutions, and theories where the scalar field(s) are only minimally coupled to gravity. It is proved by showing that for hairy black holes, the behaviour of the $T_{rr}$ component of the energy-momentum tensor near the horizon and at asymptotic infinity does not allow for a smooth matching.

The action considered by Bekenstein was the following 
\begin{equation}
S_{\phi,\chi,\ldots}=\int d^4x\sqrt{-g}\bigg[R+\mathcal{E}(\mathcal{J},\mathcal{F},\mathcal{K},\ldots,\phi,\chi,\ldots)\bigg]\, ,
\end{equation}
where $\phi,\,\chi,\ldots$ are scalar fields, and the quantities $\mathcal{J},\, \mathcal{F},\,\mathcal{K},\ldots$ are invariant terms made of first derivatives of the scalar fields, \textit{i.e.} for the $\phi$ and $\chi$ scalars we have
\begin{equation}
\mathcal{J}=\partial_\mu\phi\,\partial^\mu\phi\,,\quad
\mathcal{F}=\partial_\mu\chi\,\partial^\mu\chi\,,\quad\text{and}\quad
\mathcal{K}=\partial_\mu\phi\,\partial^\mu\chi\, . 
\end{equation}
The energy momentum tensor for the case of two scalar fields is given by
\begin{equation}
\begin{split}
    T_{\mu\nu}=
    & -\mathcal{E}\,g^{\mu\nu}
    +2\frac{\partial\mathcal{E}}{\partial\mathcal{J}}\partial_\mu\phi\partial_\nu\phi
    +2\frac{\partial\mathcal{E}}{\partial\mathcal{F}}\partial_\mu\chi\partial_\nu\chi +\frac{\partial\mathcal{E}}{\partial\mathcal{K}}\big( \partial_\mu\phi\partial_\nu\chi\\
    & +\partial_\mu\phi\partial_\nu\chi\big).
\end{split}
\end{equation}
By employing the energy momentum conservation for a static and spherically symmetric spacetime, one can derive the components of the energy momentum tensor near the horizon and at asymptotic infinity, where the spacetime becomes flat. The following hold:
\begin{itemize}
    \item The spacetime is static, spherically symmetric, and asymptotically flat. The scalar fields respect the symmetries, \textit{i.e.} they have a timelike Killing vector.

    \item For an observer with 4-velocity $u^\mu$ ($u^2=-1$) moving along the timelike Killing vector $\partial_\mu\phi\, u^\mu=0,\ldots$, so then from the weak energy condition
    $$\mathcal{E}\ge 0\, .$$
    
    \item Because the spacetime is static it is straightforward to see that $T^\theta_{\;\theta}=T^\varphi_{\;\varphi}=-\mathcal{E}$.

    \item Near the black hole horizon it is found that $T^r_{\;r}<0$ and $(T^r_{\;r})'<0$.

    \item At asymptotic infinity it is found that $T^r_{\;r}>0$ and $(T^r_{\;r})'<0$.
\end{itemize}
It is obvious then, that there has to be some intermediate point where the monotonicity of the spatial derivative of the $rr$ component of the energy momentum tensor must change. By using the Einstein equations however, it becomes clear that $(T^r_{\;r})'$ retains its monotonicity throughout the black hole exterior, and therefore the only reasonable conclusion is that the scalar fields should be trivial.

\subsection{No-hair theorem in Horndeski}
In \cite{Hui:2012qt}, a no-hair theorem was proposed for generalized Galileons. In \cite{Kobayashi:2011nu} the equivalence between generalized Galileons and Horndeski gravity was demosntrated. This new no-hair theorem for Galileons is formulated under a number of assumptions that we will analyze below, but it allows for arbitrarily defined functions $G_2,\,G_3,\,G_4$ and $G_5$ provided they do not depend explicitly on $\phi$, i.e. $\partial_\phi G_i=0$. The main assumptions of the theorem are that  (i) the spacetime is static and spherically symmetric, (ii) the scalar field respects the same symmetries, (iii) the Lagrangian includes a canonical kinetic term for the scalar, (iv) $G_{iX}\sim X^n$ with $n\ge 0$ when $X\to 0$, and (v) the current does not diverge. We then present the steps followed to establish the theorem:

\begin{itemize}
    \item For the generalized Galileon, in the covariant formulation, a Noether current can be derived originating from the shift symmetry $\phi\rightarrow \phi +constant$.
        \begin{equation*}
        \label{Jcons}
        \text{For} \quad \nabla_\mu J^\mu = 0, \;\;\text{where}\;\; J^\mu=\frac{1}{\sqrt{-g}}\frac{\delta S[\phi]}{\delta (\partial_\mu \phi)}\quad
        \text{we \; find:}
        \hspace{5cm}
        \end{equation*}
        \vspace{-1mm}
        \begin{equation}
            \hspace{0mm}\begin{split}
                &J^\mu =\, -\partial^\mu\phi \bigg\{ G_{2X} - G_{3X} \Box \phi +G_{4X} {R}
                + G_{4XX} \big[ (\Box \phi)^2 -\\
                &(\nabla_\rho\nabla_\sigma\phi)^2 \big] +G_{5X}G^{\rho\sigma}\nabla_{\rho}\nabla_{\sigma}\phi-\frac{G_{5XX}}{6} \big[ (\Box \phi)^3\\
                &-3\Box \phi(\nabla_\rho\nabla_\sigma\phi)^2 + 2(\nabla_\rho\nabla_\sigma\phi)^3 \big] \bigg\}
                -\partial^\nu X \big( - \delta^\mu_\nu G_{3X}\\
                &+2 G_{4XX} (\Box \phi \delta^\mu_\nu-\nabla^\mu\nabla_\nu \phi)+G_{5X} G^\mu{}_\nu
                -\frac12 G_{5XX} \big[ \delta^{\mu}_{\nu}(\Box\phi)^2\\
                &-\delta^{\mu}_{\nu}(\nabla_\rho\nabla_\sigma\phi)^2 -2\Box\phi \nabla^\mu\nabla_\nu\phi
                +2\nabla^\mu \nabla_\rho \phi \nabla^\rho \nabla_\nu \phi \big]   \big)\\
                &+2G_{4X} {R}^{\mu}{}_{\rho} \nabla^\rho \phi + G_{5X} \big( -\Box \phi {R}^\mu{}_\rho \nabla^\rho\phi
                +{R}_{\rho\nu}{}^{\sigma\mu} \nabla^\rho\nabla_\sigma\phi \nabla^\nu\phi
                \\
                &+{R}_\rho{}^\sigma \nabla^\rho\phi  \nabla^\mu\nabla_\sigma\phi \big).
            \end{split}
        \end{equation}

    \item In static spherical coordinates it turns out that the only nonzero component of the current is the $J^r$, because the current has to respect the symmetries of the configuration, i.e. $J=(0,J^r,0,0)$. It is then easy to calculate the norm of the current, $J^2=g_{rr}(J^r)^2$. The norm should remain regular everywhere, and therefore, since at the horizon $g_{rr}\rightarrow\infty$, we must have $J^r\big|_h=0$.

    \item From the scalar equation of motion it then follows that $J^r$ has to vanish everywhere. In order to see that let us introduce the static and spherically symmetric ansatz
    \begin{equation}
    \label{eq:metric}
    ds^2=-A(r)dt^2+\frac{1}{B(r)}dr^2+r^2 d\Omega^2\,,
    \end{equation}
    so then the scalar equation yields
    \begin{equation}
        J^r=\frac{\tilde{c}}{r^2}\sqrt{\frac{B}{A}}\Rightarrow J^2=\frac{\tilde{c}^2}{r^4 A}\,.
    \end{equation}
    where $\tilde{c}$ is a constant. Evaluating the expression above at the horizon, while demanding that the current remains finite, leads to $\tilde{c}=0$, so that $J^r=0$ everywhere.

    \item 
    For the final part of the proof of the theorem it is useful to express the current in the specific metric components. In this case the current can be written in the following form:
    \begin{align}
    \begin{split}
        &J^r= -B\,G_{2X}\phi' + \frac{r A' + 4 A}{2rA}B^2(\phi')^2 G_{3X}\\
        &+\frac{2A -2A B + 2r A'}{r^2A}B^2\phi'G_{4X}- \frac{2A +2rA' }{r^2 A}B^3(\phi')^3 G_{4XX}\\
        &+\frac{ B A' - 3(\phi')^2 A'}{2r^2A}B^3(\phi')^2 G_{5X}+\frac{ A' }{2r^2A}B^4(\phi')^4 G_{5XX}\\[3mm]
        & \hspace{3mm}\equiv \;B\,\phi'\,\mathcal{P}(r;\phi',A'/A,B)
        \,.
    \end{split}
    \label{eq:Jr_spherical}
    \end{align}
    where the prime denotes differentiation with respect to the radial coordinate. The spacetime is assumed to be asymptotically flat which translates to the asymptotic conditions
    $$(A,B,A',B',\phi')\rightarrow (1,1,0,0,0)\,.$$
    Then, it follows that asymptotically $\mathcal{P}\rightarrow -1$, provided that $G_2 \subseteq X$.
    Let us assume that the scalar field has a non-trivial profile that asymptotically approaches a constant, since $\phi'_\infty = 0$ (the value of the constant is irrelevant since shift-symmetry is assumed anyway). Then as we start moving to smaller radii we expect that $\phi'$ will start acquiring nonzero values. The only way for $J^r$ to yield zero then, is if $\mathcal{P}=0$ whenever $\phi'\ne 0$. However, for every nondegenerate Galileon theory featuring a kinetic term, $\mathcal{P}$ should approach a constant asymptotically. By continuity as we start moving from infinity inwards $\mathcal{P}$ will still be nonzero when $\phi'$ picks up a nonzero value, which means that at some point $J^r\ne 0$ violating one of our earlier conclusions. Therefore, $\phi'=0$ should be the only viable option, which means that GR is retrieved.

\end{itemize}

%% file: Chapters/evasion_of_no_hair_theorems.tex
In the previous chapter we discussed some well-known no-hair theorems governing black holes with scalar fields and the conditions under which they apply. As it has turned out however, it has been possible to evade them, either by breaking one of the assumptions concerning them, or simply by gaining a deeper understanding of the underlying theories. The old no-hair theorems for instance were outdated by the discovery of black holes with Yang-Mills \cite{Volkov:1989fi, Bizon:1990sr, Greene:1992fw}, Skyrme fields \cite{Luckock:1986tr, Droz:1991cx, Heusler:1991xx}, or a conformal coupling to gravity \cite{Bekenstein:1974sf}. The novel formulation of the no-hair theorem proposed in 1995 by Bekenstein was evaded within a year with the discovery of the dilatonic black holes found in the context of the Einstein-dilaton-Gauss-Bonnet theory (EdGB) \cite{Kanti:1995vq}  (for some earlier studies that paved the way, see \cite{Gibbons:1987ps, Callan:1988hs, Campbell:1990ai, Duncan:1992vz, Mignemi:1992nt, Kanti:1995cp}). In the EdGB theory the motivation emanates from the lower energy limit of superstring theory \cite{Metsaev:1987zx}.
The colored black holes were found next in the context of the same theory completed by the presence of a Yang-Mills field \cite{Torii:1996yi, Kanti:1996gs, Kanti:1999sp}, and higher-dimensional models \cite{Bamba:2007ef, Maeda:2009uy, Ohta:2010ae} or rotating versions \cite{Kleihaus:2011tg, Kleihaus:2015aje, Pani:2011gy, Pani:2011xm, Herdeiro:2014goa, Ayzenberg:2014aka} were also constructed.

Over the last years, the construction of generalized gravitational theories was significantly enlarged via the revival of the Horndeski theory and led to new formulations of the no-hair theorems, with the most famous one presented in the last section of the previous chapter. However, these recent forms were also evaded \cite{Sotiriou:2013qea} and concrete black-hole solutions were constructed \cite{Sotiriou:2014pfa, Babichev:2013cya, Benkel:2016rlz, Benkel:2016kcq}.

In the first section of this chapter, we show how the presence of the scalar-GB coupling in the action, which is equivalent to $G_5\sim \ln|X|$, can lead to the evasion of the no-hair theorems formulated for the shift-symmetric Horndeski theory. We then generalize the analysis by abandoning the shift-symmetry requirement and discuss how both the old \cite{Bekenstein:1971hc} and the ``novel'' no-hair theorem \cite{Bekenstein:1995un,Hui:2012qt} by Bekenstein are evaded, allowing for a large class of hairy black hole solutions with interesting phenomenology. Finally, we discuss a particular class of hairy black hole solutions that evade the no-hair theorems, which belong to the framework of \textit{spontaneous scalarization}, which is a process that spontaneously endows black holes and compact neutron stars with scalar hair.

\section{Evasion of the no-hair theorem for the generalized Galileon}

We will begin our discussion on evasions of the black-hole no-hair theorems by considering the case of shift-symmetric Horndeski gravity. Let us briefly discuss why this case is a justifiably good starting point for the analysis. Considering the length scale of very compact objects, strong gravity observations probe lengths of the order of kilometres. If one assumes that the no-hair theorems for scalar fields can somehow be evaded, the experiments probing strong-field gravity are expected to be relevant for massless or ultiralight scalar fields. That is because, massive scalars in this case, produce nontrivial profiles around compact objects that are expected to decay exponentially at a timescale set by the inverse of the mass. Shift symmetry, \textit{i.e.} invariance under $\phi\to \phi+constant$, protects scalars from acquiring a mass. Consequently, strong gravity observations effectively probe scalars that exhibit either this symmetry or very small violations of it.

Let us begin by examining the no-hair theorem postulation for the shift-symmetric generalized Galileon (ssgG) theory presented in the introduction.
If we examine closely the expression for the current we see that there is one general case that violates one of the main assumptions of the ssgG no-hair theorem, which states that $\mathcal{P}$ approaches a finite value asymptotically. A way to achieve that is if $J^r$ is of order $n\ge 0$ with respect to $\phi'$, with at least one term being of zeroth order \cite{Sotiriou:2014pfa}. In other words, if at least one of the terms appearing in \eqref{eq:Jr_spherical} is independent of $\phi'$, the condition $J^r=0$ forces $\phi'$ to acquire a nonzero value. This can be achieved by specific choices for the functions $G_i$ appearing in the Lagrangian, \textit{e.g.} $G_2,G_4\sim \sqrt{-X}$ and $G_3,G_5\sim \ln|X|$  \cite{Babichev:2017guv}.

Taking a step back and reviewing the problem at the Lagrangian level, we can see that the only way through which $\phi$-dependent terms do not appear in the equation of motion for the scalar, and the gravitational equations of motion remain second-order and divergence-free, is if we add a term of the form $\phi\,\mathcal{A}[g]$ in the action, where from the Lovelock theorem the scalar quantity $\mathcal{A}[g]
$ is the Gauss-Bonnet invariant $\GB$, defined as
\begin{equation}
    \GB={R}^{\mu \nu \rho \sigma}{R}_{\mu \nu \rho \sigma}-4 {R}^{\mu \nu}{R}_{\mu \nu}+{R}^{2}\,.
\end{equation}
At first glance, the GB invariant seems to be absent from the Horndeski Lagrangian. However, it was shown \cite{Minamitsuji:2018xde} that it simply corresponds to the choice $G_2=G_3=G_4=0$ and $G_5=-4\ln|X|$. Let us make this discussion a bit more tangible by considering as an example the following action
\begin{equation}
\label{eq:lgbaction}
     S= \frac{1}{2 k}\int \mathrm{d}^4 x\,\sqrt{-g}\bigg[R+X+\alpha\,\phi\,\GB\bigg]\,.
\end{equation}
By varying the above action with respect to the scalar we get the corresponding scalar equation
\begin{equation}
\label{eq:lgbeq}
    \Box\phi =\alpha \GB.
\end{equation}
As we explained earlier the shift-symmetric property of the theory allows one to write the scalar equation as the conservation of a current, i.e. $\nabla_\mu (\nabla^\mu\phi-\alpha {\GB}^\mu)=0$, where $\GB=\nabla_\mu\GB^\mu$, and $\GB^\mu$ considers the contributions to the current by the GB term. It is apparent that in this scenario the theory does not admit any constant $\phi$ solutions unless $\GB=0$. This however is not the case for black holes and hence they have to be endowed with scalar hair. As explained earlier in a generic context, the contradiction with the no-hair theorem of~\cite{Hui:2012qt} relates to the assumption that every single term in the current depends on the gradient of $\phi$. The contribution of the linear coupling to $\GB$ clearly violates this assumption and is indeed unique in this respect \cite{Sotiriou:2013qea}. Interestingly, hairy black hole solutions in this theory violate another assumption of the theory of the Galileon no-hair theorem \cite{Hui:2012qt}:  $\left(J^{r}\right)^2/g^{rr}$ diverges on the horizon \cite{Babichev:2017guv}. It was subsequently shown in \cite{Creminelli:2020lxn} that this quantity is not an invariant when it receives a contribution from a linear coupling with $\GB$ and hence there is no reason to impose that it is finite in this case. It is worth noting that the linear coupling term $\phi\, \GB$ would arise in a small coupling or small $\phi$ expansion of the exponential coupling $e^\phi \GB$ that had already been known to introduce black hole hair \cite{Mignemi:1992pm,Kanti:1995vq,Yunes:2011we}.

For black hole solutions with a non-divergent scalar field to emerge in this theory, it was shown that the derivative of the scalar field at the horizon should satisfy a particular condition which dictates the existence of a lower mass limit for the solutions, i.e. $r_h>r_{h(\text{min})}$. Hairy black hole solutions in the $\phi\, {\GB}$ but also in the $e^\phi \GB$ cases share this property, with the minimum mass being in principle different for the different coupling functions. Moreover, in both cases, the scalar charge is not an independent parameter, but it is instead fixed with respect to the black hole mass and spin by a regularity condition on the horizon. In the next chapter we will focus on the linear coupling case and determine how these two properties are affected by the presence of additional shift-symmetric (derivative) interactions in the action.

In~\cite{Saravani:2019xwx}, theories described by the shift-symmetric Horndeski action \eqref{eq:Horndeski} were classified as follows: 
\begin{align}
    \textbf{Class-1: } &\mathcal{E}_\phi[\phi=0,\,g]=0,\qquad \forall g,\\[2mm]
    \textbf{Class-2: } &\lim_{g\rightarrow \eta}\mathcal{E}_\phi [\phi=0,\, g]=0.\\[0mm]
    \textbf{Class-3: } &\text{All the rest}.
\end{align}
Class-1 theories are defined by their property to accept $\phi=0$ as a solution for any general background; hence, they admit all possible GR solutions. Class-2 theories allow for $\phi=0$ to be realized only for flat spacetimes.  The third class is defined as the complement of the other two. Therefore, class-3 theories admit a non-trivial scalar configuration in flat spacetime as a solution, or flat spacetime is not a solution, and hence they violate Local Lorentz symmetry.  

 At first sight, it appears that classes 1 and 2 are unrelated. On the contrary, it was shown in~\cite{Saravani:2019xwx} that a class-2 Lagrangian can always be expressed as a class-1 Lagrangian plus a contribution from the Gauss-Bonnet invariant, namely:
\begin{equation}
    \mathcal{L}_{(2)}=\mathcal{L}_{(1)}+\alpha\phi\,\GB.
\end{equation}
Consequently, all shift-symmetric non-Lorentz violating Horndeski theories admit all GR solutions, provided that a linear coupling between the scalar and the Gauss-Bonnet invariant is not present. Using this result, it was then shown that the scalar charge $Q$ of a stationary black hole in any theory in classes 1 and 2 is given by
\begin{equation}
\label{eq:charge}
4\pi Q= \alpha \int_\mathcal{H} n_a \GB^a,
\end{equation}
where $\GB^a$ is the Gauss-Bonnet contribution to the shift-symmetric current, $\mathcal{H}$ denotes the Killing horizon and $n^a$ its normal. We will discuss the scalar charge property of black holes later on.


\section{General couplings}
\label{sec:no-hair_theorems}

The no-hair theorem for the generalized Galileon \cite{Hui:2012qt} pertains to shift-symmetric Horndeski theories. Now we will show that hairy black holes can be found in the Horndeski framework even when one abandons shift symmetry, leading to the evasion of the old and new no-hair theorems proposed by Bekenstein. Consider the following action
\begin{equation}
\label{eq:action_GB}
S=\frac{1}{16\pi}\int{d^4x \sqrt{-g}\bigg\{ R+X+f(\phi)\,\GB\bigg\} }\,,
\end{equation}
so then the dilatonic and shift-symmetric models are found for $f(\phi)=\alpha e^{\gamma\phi}$ and $f(\phi)=\alpha\phi$ respectively. In order to retrieve \eqref{eq:action_GB} from the Horndeski Lagrangian \eqref{eq:Horndeski} the following choices have to be made
\begin{align}
\begin{split}
    G_2=&\, X+8f^{(4)}X^2(3-\ln|X|)\,,\\
    G_3=&\, 4f^{(3)}X(7-3\ln|X|)\,,\\
    G_4=&\, 1+4f^{(2)}X(2-\ln|X|)\,,\\
    G_5=&\,= -4f^{(1)}\ln|X|\,,
\end{split}
\end{align}
where $f^{(n)}\equiv \partial_{\phi^n} \equiv \partial^n f/\partial \phi^n$. The scalar equation of motion and the Einstein field equations in this theory are given by
\begin{align}
    & \Box\phi+(\partial_\phi f)\,\GB=0\,,
    \label{eq:scalar_GB}\\
    \begin{split}
        & G_{\mu\nu}=
        T^\phi_{\mu\nu}\equiv
        -\frac{1}{4}g_{\mu\nu}(\nabla\phi)^2+\frac{1}{2}\nabla_\mu\phi\nabla_\nu\phi\\
        &\hspace{2.5cm}-\frac{1}{g}g_{\mu(\rho}g_{\sigma)\nu}\epsilon^{\kappa\rho\alpha\beta}\epsilon^{\sigma\gamma\lambda\tau}R_{\lambda\tau\alpha\beta}\nabla_{\gamma}\partial_{\kappa}f(\phi)\, ,
    \end{split}
\end{align}
where $\epsilon^{\mu\nu\rho\sigma}$ is the 4-dimensional antisymmetric symbol.

In order to demonstrate the evasion of the no-hair theorems, and especially Bekenstein's novel one, we will need the near-horizon and asymptotic expansions for the functions appearing in our analysis. Casting the equations of motion in the specific metric ansatz \eqref{eq:metric} we end up with three independent equations, that after careful manipulation allow for $B(r)$ to be eliminated and reduce to the following two ordinary differential equations of second order for the functions $A$ and $\phi$
\begin{equation}
    A''=\frac{P\big(r, \phi', A', \partial_\phi f, \partial_{\phi\phi} f\big)}{S\big(r, \phi', A', \partial_\phi f, \partial_{\phi\phi} f\big)}\quad\,,\quad
    \phi''=\frac{Q\big(r, \phi', A', \partial_\phi f, \partial_{\phi\phi} f\big)}{S\big(r, \phi', A', \partial_\phi f, \partial_{\phi\phi} f}\big)\,, \label{eq:Aphi-system}
\end{equation}
where $P,\,Q,\,S$ are functions of the quantities in the parentheses.
We will now demonstrate that our set of equations, with a general coupling function $f(\phi)$, allows for the construction of a black-hole solution with a regular horizon  provided that $f$ satisfies certain constraints. For a spherically-symmetric spacetime, the presence of a horizon is realised for $A \rightarrow 0$, as $r \rightarrow r_h$, or equivalently for $A'/A \rightarrow \infty$ -- the latter will be used in our analysis as an assumption but it can be shown to follow from the former. On the other hand, the regularity of the horizon amounts to demanding that $\phi$, $\phi'$ and $\phi''$ remain finite in the limit $r \rightarrow r_h$. Then, the assumption that $A' \rightarrow \infty$ while $\phi'$ remains finite, results in \eqref{eq:Aphi-system} taking the following form to leading order
\begin{align}
    \begin{split}
        &\left(\frac{A'}{A}\right)'\approx-\frac{r^4+4r^3\phi'(\partial_\phi f)+4r^2\phi'^2(\partial_\phi f)^2-24(\partial_\phi f)^2}{r^4+2r^3\phi'(\partial_\phi f)-48(\partial_\phi f)^2}\left(\frac{A'}{A}\right)^2\, ,
    \end{split}
    \label{eq:A-approx-h}\\
    \begin{split}
        &\quad\phi''\approx-\frac{\big[2\phi'(\partial_\phi f)+r\big]\big[r^3\phi'+12(\partial_\phi f)+2r^2\phi'^2(\partial_\phi f)\big]}{r^4+2r^3\phi'(\partial_\phi f)-48(\partial_\phi f)^2}\left(\frac{A'}{A}\right)\, .
    \end{split}
    \label{eq:phi-approx-h}
\end{align}
Focusing on the second of the above equations, we observe that $\phi''$ diverges at the horizon if $f(\phi)$ is either zero or left unconstrained. However, $\phi''$ may be rendered finite if either one of the two expressions in the numerator of  Eq. \eqref{eq:phi-approx-h} is zero close to the horizon. 

If we assume that $(2\phi'\partial_\phi f+r)=0$ close to the horizon, then a careful inspection of our equations reveals that, in that case, $\phi'' \simeq \sqrt{A'/A}/(\partial_\phi f)$.
Thus, for $\phi''$ to remain finite, we must demand that $\partial_\phi f \rightarrow \pm\infty$, near the horizon. This may be easily shown to lead to either a divergent or a trivial scalar field near the horizon, for every elementary form of $f(\phi)$ we have tried. 
Therefore, for the construction of a regular horizon in the presence of a non-trivial scalar field, we are led to consider the second choice:
$r^3\phi'+12(\partial_\phi f)+2r^2{\phi'}^2(\partial_\phi f)=0$ when $r\rightarrow r_h$. This may be easily solved to yield
\begin{equation}
\label{eq:con-phi'}
\phi'_h=\frac{r_h}{4(\partial_\phi f)_h}\left(-1\pm\sqrt{1-\frac{96(\partial_\phi f)_h^2}{r_h^4}}\right),
\end{equation}
where all quantities have been evaluated at $r_h$.
To ensure that $\phi'_h$ is real, we must impose a constraint on the coupling function, which translates to a constraint regarding the black hole size
\begin{equation}
r_h^4>96\big( \partial_\phi f \big)_h^2\,.
\label{eq:rh_min}
\end{equation}
For a given coupling function $f(\phi)$, and fixed ($\alpha$, $\phi_h$) - where $\alpha$ is the coupling constant introduced in the function $f(\phi)$, the constraint \eqref{eq:rh_min} imposes that there is a lower bound for the horizon radius of the derived black hole solutions given by $r_h^2 > 4\sqrt{6}\,|( \partial_\phi f )|$, and thus a lower-bound on their mass, in agreement with previous works, including the dilatonic and the shift-symmetric cases \cite{Kanti:1995vq, Kanti:1997br, Sotiriou:2013qea, Sotiriou:2014pfa}.

Turning now to Eq. \eqref{eq:A-approx-h}, and using the constraint \eqref{eq:con-phi'}, we find that the coefficient of $A'^2$ simplifies to $-A^2$. Then, upon integration with respect to $r$, we obtain $A'/A=(r-r_h)^{-1}+\mathcal{O}(1)$ which, in accordance to our initial assumption, diverges close to the horizon. Integrating once more and putting everything together, we may write the near-horizon solution as
\begin{align}
    &A=a_1 (r-r_h) + ... \,, \label{eq:A-rh}\\
    &B=b_1 (r-r_h) + ... \,, \label{eq:B-rh}\\
    &\phi =\phi_h + \phi_h'(r-r_h)+ \phi_h'' (r-r_h)^2+ ... \,. \label{eq:phi-rh}
\end{align}
The above describes a regular black-hole horizon in the presence of a scalar field provided that $\phi'$ and the coupling function $f$ satisfy the constraints \eqref{eq:con-phi'}-\eqref{eq:rh_min}.

A general coupling function $f(\phi)$ for the scalar field does not interfere with the requirement for the existence of an asymptotically flat limit for the spacetime (\ref{eq:metric}). We assume the following power-law expressions for the metric functions and scalar field, in the limit $r \rightarrow \infty$
\begin{equation}
    (A,B)=1+\sum_{n=1}^\infty{\frac{(p_n,q_n)}{r}}\,,\quad 
    \phi =\phi_{\infty}+\sum_{n=1}^{\infty}{\frac{d_n}{r}}\,.
    \label{eq:far}
\end{equation}

\subsection{Evading the old no-hair theorem}
Let us now turn our attention to the no-hair theorems. The older version of the no-hair theorem for scalar fields \cite{Bekenstein:1971hc}, that employs the scalar equation of motion, fails to exclude the existence of
black-hole solutions in our theory (\ref{eq:action_GB}): multiplying the scalar equation (\ref{eq:scalar_GB}) by $f(\phi)$ and integrating over the black-hole exterior region, we obtain the integral constraint
\begin{equation}
    \int_{\mathcal{V}} d^4x \sqrt{-g} \,f(\phi) \left[\nabla^2 \phi + (\partial_\phi f) \GB \right] =0\,.
\end{equation}
Integrating by parts the first term, the above becomes
$$
\int_{\mathcal{V}}  d^4x \sqrt{-g}\,(\partial_\phi f) \left[\,2X  + 
f(\phi)  \GB \, \right] =-\int_{\partial\mathcal{V}} d^3x\sqrt{-h}f(\phi)n^\mu\partial_\mu \phi\,,
\label{eq:old-con}
$$
where $h$ is the induced metric on the hypersurface $\partial\mathcal{V}$ serving as the boundary of $\mathcal{V}$. The boundary term $[\sqrt{-h} f(\phi) \partial_\mu \phi]_{r_h}^\infty$ vanishes both at the horizon (due to the $\sqrt{A\,B}$ factor) and at infinity, for an asymptotically vanishing scalar field.
Since $\phi=\phi(r)$, the first term in Eq. (\ref{eq:old-con}) gives $2X = -g^{rr}(\partial_r \phi)^2<0$ 
throughout the exterior region.  Also, for the metric (\ref{eq:metric}), the GB term has the explicit form
\begin{equation}
    \GB =\frac{2 B}{r^2}\bigg[(3B-1)\,\frac{A' B'}{A B} -(1-B)\,\frac{2A''A-A'^2}{A^2}\bigg]z, .
    \label{eq:GB}
\end{equation}
Employing the asymptotic solutions near the horizon \eqref{eq:A-rh}-\eqref{eq:B-rh} and at infinity \eqref{eq:far}, we may easily see that the GB term takes on a positive value at both regimes. Therefore, in the simplest possible case where both $f(\phi)$ and $\GB$ are sign-definite, Eq. \eqref{eq:old-con} allows for black-hole solutions with scalar hair for every choice of the coupling function that merely satisfies $f(\phi)>0$.

The boundary term does not necessarily vanish in the more generic scenario where $f(\phi_\infty)\ne 0$. Therefore, the condition $f(\phi)>0$ is not absolutely derived, and solutions with $f(\phi)<0$ can also be recovered provided that the field does not vanish asymptotically.
It is important to stress out that the constraint $f(\phi)>0$ is not unique and is based on the steps we followed in the analysis above, and particularly on the fact that we chose to multiply the scalar equation with $f(\phi)$ before integrating by parts. For example, if we had instead multiplied with $\phi$, then we would have found:
$$\int_{\mathcal{V}} d^4x \sqrt{-g}\,\left[\,2X + \phi(\partial_\phi f)\GB\, \right] = - \int_{\partial\mathcal{V}} d^3x\sqrt{-h}\, \phi \, n^\mu\partial_\mu \phi\,\,,$$
and so when the field vanishes asymptotically, the constrain now becomes $\phi(\partial_\phi f)>0$.

\subsection{Evading the ``novel'' no-hair theorem}

We will now continue with the ``novel'' no-hair theorem developed by Bekenstein \cite{Bekenstein:1995un}. As we saw in the introduction, assuming positivity and conservation of energy, it was demonstrated that the asymptotic forms of the $T^r_{\;\,r}$ component of the energy-momentum tensor near the horizon and at infinity could never be smoothly matched. That argument proved beyond doubt that there were no black-hole solutions in the context of a large class of minimally-coupled-to-gravity scalar field theories.  Here, we will show that the coupling of the scalar field to the quadratic GB term causes the complete evasion of Bekenstein's theorem. This may be once again be realised for a large class of scalar field theories, with the previously-studied exponential \cite{Kanti:1995vq,Kanti:1997br} and linear \cite{Sotiriou:2013qea,Sotiriou:2014pfa} GB couplings comprising special cases of our argument.

The energy-momentum tensor $T_{\mu\nu}$ satisfies the equation of
conservation $\nabla_\mu T^\mu_{\;\,\nu}=0$ due to the invariance of the action
(\ref{eq:action_GB}) under coordinate transformations. Its $r$-component may take the explicit form
\begin{equation}
    (T^r_{\;\,r})'=\frac{A'}{2A}\,(T^t_{\;\,t} -T^r_{\;\,r}) +
    \frac{2}{r}\,(T^\theta_{\;\,\theta}-T^r_{\;\,r})\,,
\label{eq:Trr-deriv}
\end{equation}
where the relation $T^\theta_\theta=T^\varphi_\varphi$ has been used due
to the spherical symmetry. The non-trivial components of the energy momentum
tensor $T_{\mu\nu}$ for our theory (\ref{eq:action_GB}) with a generic coupling function
$f$ are:
\begin{align}
    \begin{split}
        T^t_{\;\,t}=
        &-\frac{B^2}{4r^2}\bigg\{\frac{\phi'^2}{B}\big[r^2+16f^{(2)}(1-B)\big]\\
        &+8f^{(1)}\bigg[\frac{B'}{B^2}\phi'(1-3B)
        +\frac{2\phi''}{B}(1-B)\bigg]\bigg\},\label{eq:Ttt}
    \end{split}
    \\[2mm]
        T^r_{\;\,r}=\;
        &\frac{B\phi'}{4}\Bigl[\phi'-
        \frac{8\left(1-3B\right)f^{(1)}A'}{r^2 A}\Bigr],\label{eq:Trr}\\[2mm]
    \begin{split}
        T^{\theta}_{\;\,\theta}=
        &\frac{B^2}{4 r}\bigg\{4f^{(1)}\bigg[\frac{\phi'(2A''A-A'^2)}{A^2} +
        \frac{A'}{A}\bigg(2\phi''+\frac{3B'\phi'}{B}\bigg)\bigg]\\
        &-\phi'^2\bigg(\frac{r}{B}-8f^{(2)}\frac{A'}{A}\bigg)
        \bigg\}. \label{eq:Tthth}
    \end{split}
\end{align}
We will first investigate the profile of $T^r_{\;\,r}$ at infinity. We find that
$-T^t_{\;\,t} \simeq-T^\theta_{\;\,\theta} \simeq T^r_{\;\,r} \simeq \phi'^2/4
+ {\mathcal O}\left(1/r^6\right)$.
Since the metric function $A$ there adopts a constant value ($A \rightarrow 1$), the dominant contribution to the right-hand-side of Eq. (\ref{eq:Trr-deriv}) is
\begin{equation}
    (T^r_{\;\,r})' \simeq \frac{2}{r}\,(T^\theta_\theta-T^r_r) \simeq -\frac{1}{r}\,\phi'^2 +....\,.
\end{equation}
Therefore, at asymptotic infinity, the $T^r_{\;\,r}$ component is positive and decreasing, in agreement with \cite{Bekenstein:1995un} since the GB term is insignificant in this regime. 

In the near-horizon regime, $r \rightarrow r_h$, the $T^r_{\;\,r}$ component (\ref{eq:Trr})
takes the approximate form:
\begin{equation}
    T^r_{\;\,r}=-\frac{2A'B}{r^2 A}\phi'f^{(1)}+\mathcal{O}(r-r_h)\,.\label{eq:Trr-rh}
\end{equation}
The above combination is finite but not negative-definite as in \cite{Bekenstein:1995un}. In fact, the $T^r_{\;\,r}$ component is positive-definite since, close to the black-hole horizon, $A'>0$, and $\dot f\,\phi'<0$. Therefore, the existence of a regular black-hole horizon in the context of the
theory (\ref{eq:action_GB}) automatically evades one of the two requirements of the
novel no-hair theorem.

Turning finally to the expression for $(T^r_{\;\,r})'$ and employing the energy-momentum components (\ref{eq:Ttt})-(\ref{eq:Tthth}) in Eq. (\ref{eq:Trr-deriv}), we obtain the following expression, in the limit $r \rightarrow r_h$
\begin{equation}
\begin{split}
    (T^r_r)' = \;&\frac{A' B}{A}\Bigl[-\frac{r \phi'^2}{4Z} -\frac{2 (f^{(2)} \phi'^2
    +f^{(1)} \phi'')}{r Z}+\frac{4 f^{(1)} \phi'}{r^2}\,\bigg(\frac{1}{r}+ B' \bigg)
    \Bigr]\\
    &+ {\mathcal O}(r-r_h)\,,     
\end{split}
\label{eq:Trr'-rh}
\end{equation}
where we have defined $Z \equiv r+2 (\partial_\phi f) \phi'$. Close to the black-hole
horizon, Eq. (\ref{eq:con-phi'}) guarantees that $(\partial_\phi f) \phi'<0$ while $Z>0$.
Employing also the metric functions behaviour $A'>0$ and $B'>0$, we conclude
that $(T^r_{\;\,r})'$ is negative in the near-horizon regime if a sole additional
constraint, namely $(\partial_{\phi\phi} f) \phi'^2+\dot f \phi''>0$, is satisfied. This may
be alternatively written as $\partial_r[(\partial_\phi f) \phi']|_{r_h}>0$, and merely
demands that the negative value of the quantity $(\dot f \phi')|_{r_h}$,
necessary for the existence of a regular black-hole solution, should be 
constrained away from the horizon. This is in fact the only way for the
matching of the two asymptotic solutions (\ref{eq:A-rh})-(\ref{eq:phi-rh}) and
(\ref{eq:far}) to be realised. Therefore, for $\partial_r (\dot f \phi')|_{r_h}>0$,
the $T^r_{\;\,r}$ component is positive and decreasing also near the horizon
regime. As a result, both requirements of the ``novel'' no-hair theorem 
\cite{Bekenstein:1995un} do not apply in this theory, and thus it can be evaded.

\subsection{Numerical demonstration}

In this subsection we briefly discuss a few representative numerical results as a means to graphically confirm that the behavior explained approximately in the last subsection, is independent of the particular coupling function. We also want to demonstrate the complete profile of the $rr$ component of the energy momentum tensor.

The derivation of exact solutions, valid over the entire radial domain, demands the numerical integration of the equations of motion. Our integration starts at a distance very close to the horizon of the black hole, \textit{i.e}. at $r\approx r_h+\mathcal{O}(10^{-5})$, and for simplicity, we set $r_h=1$. We use as boundary conditions the near-horizon solution \eqref{eq:A-rh}-\eqref{eq:phi-rh} together with Eq.~\eqref{eq:con-phi'} for $\phi'_h$ upon choosing a particular coupling function $f(\phi)$. The integration proceeds towards large values of the radial coordinate until the form of the derived solution matches the asymptotic solution.

\begin{figure}[ht]
    \centering
    \includegraphics[width=0.47\textwidth]{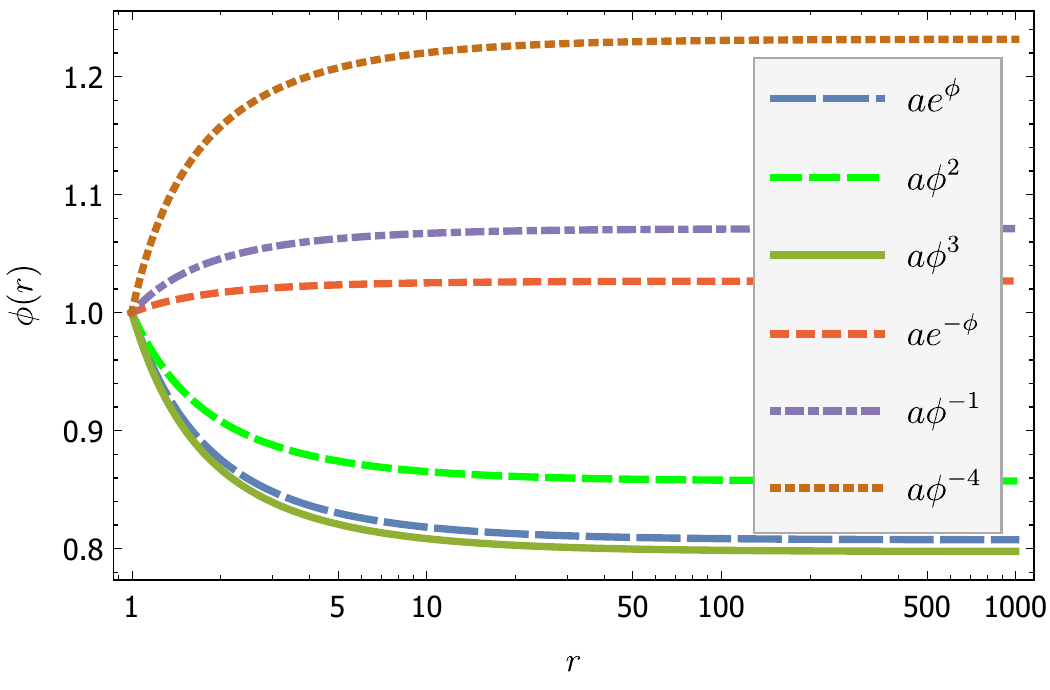}\hspace{5mm}
    \includegraphics[width=0.47\textwidth]{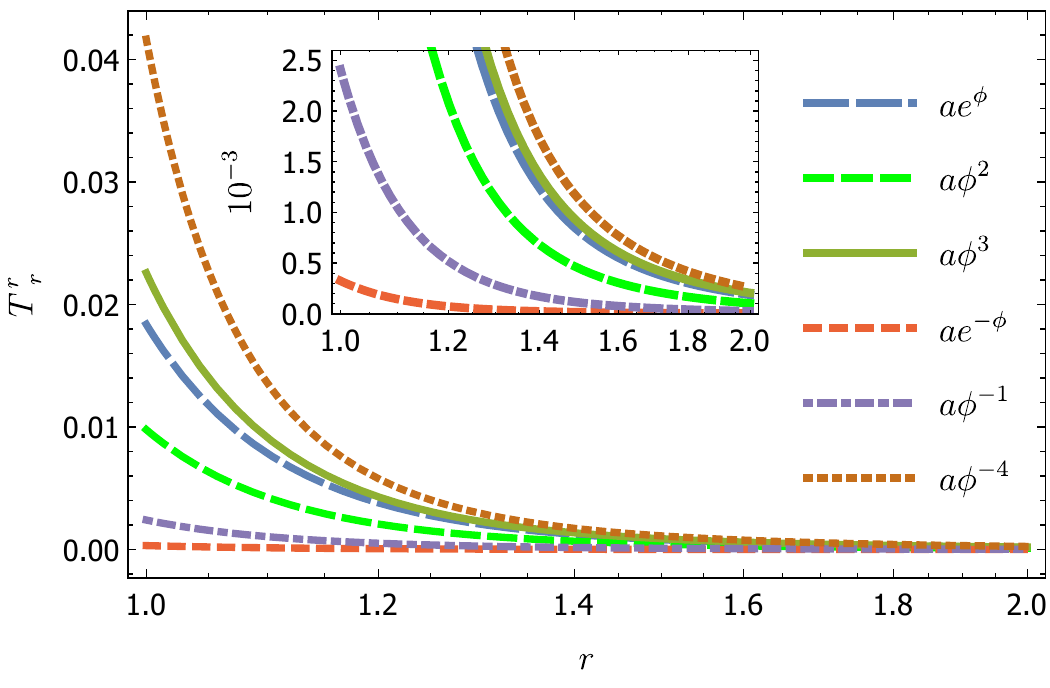}\hspace{2mm
    }
    \caption[Scalar profile and energy-momentum tensor in EsGB]{The scalar field $\phi$ and the $T^r_{\;\,r}$ component for different coupling functions $f(\phi)$, for $a=0.01$ and $\phi_h=1$. The plots are taken from \cite{Antoniou:2017acq}.}
    \label{fig:field_emt}
\end{figure}

The solutions for the scalar field are depicted in Fig. \ref{fig:field_emt}, for a variety of forms of the coupling function $f(\phi)$: exponential, odd and even power-law, odd and even inverse-power-law. For easy comparison, the coupling constant in all cases has been set to $\alpha =0.01$ and the near-horizon value of the field to $\phi_h=1$. For $f(\phi)=(\alpha e^\phi, \alpha \phi^2, 
\alpha \phi^3)$, which all have $f'_h>0$, our constraint \eqref{eq:con-phi'} leads to a negative $\phi'_h$; for
$f(\phi)=(\alpha e^{-\phi}, \alpha \phi^{-1}, \alpha \phi^{-4})$, which all have $f'_h<0$, Eq. \eqref{eq:con-phi'} demands a positive $\phi'_h$ -- the decreasing and increasing, respectively, profiles are clearly depicted in Fig. \ref{fig:field_emt}. In all cases, for a given value of $\phi_h$, Eq. \eqref{eq:con-phi'} uniquely determines the quantity $\phi'_h$. The integration of the system \eqref{eq:Aphi-system} with initial conditions $(\phi_h, \phi'_h)$ then leads to the presented solutions. The positivity and decreasing profile of the $T^r_{\,\,r}$ component,  necessary features for the evasion of the novel no-hair theorem, are clearly seen in Fig. \ref{fig:field_emt}.
It is worth mentioning that the second constraint, $\partial_r [(\partial_\phi f)] \phi')|_{r_h}>0$, we derived in the discussion about the evasion of Bekenstein's ``novel'' theorem, is automatically satisfied for every solution found and does not need any further action or fine-tuning of the free parameters.

\section{Spontaneous scalarization}
In the previous sections we discussed how the evasion of the scalar no-hair theorems occuring for theories belonging in the Horndeski framework can lead to solutions described by non-trivial scalar configurations. Here we discuss a subclass of these theories that allow for a very interesting phenomenology, predicting severe deviations from GR in the strong-field regime while retrieving GR in the asymptotic weak-field one. The general idea can be expressed in the form of the following question: is it possible that GR only becomes unstable beyond some threshold related to the strength of the gravitational field, while remaining perfectly stable elsewhere?

\emph{Spontaneous  scalarization} is perhaps the most direct manifestation of new physics that stays dormant in the weak field regime and yet leads to large deviations from GR in the strong field regime. The first model that exhibits the spontaneous scalarization phenomenon was proposed by Damour and Esposito-Far\`ese (DEF) in \cite{Damour:1993hw}. 
Here, a direct coupling between a scalar field $\phi$ and the Ricci scalar, $R$ (or equivalently the matter in a different conformal frame), generates at linear level an effective mass for $\phi$. As the compactness of an object increases, this effective mass can become negative and trigger a tachyonic instability. The scalar field then grows until nonlinear effects kick in and quench the instability, thereby leading to a ``scalarized'' object: a neutron star that is dressed with a scalar configuration and, hence, has different structure than its GR counterpart. In the original work the DEF action expressed in the Einstein frame was given by
\begin{equation}
    S= \frac{1}{2\kappa}\int d^4x \sqrt{-g}\bigg[R-2(\partial\phi)^2\bigg]+S_\text{M}[\psi_m,A^2(\phi)g_{\mu\nu}]\, ,
\label{eq:DEF_action}
\end{equation}
It was shown that the nonminimal coupling between the scalar and matter was responsible for these nonperturbative strong field effects in neuron stars. In order to demonstrate this, we need to consider the scalar field equation of motion
\begin{equation}
    \Box \phi=-\frac{\kappa}{2}\,\alpha(\phi)\, T\;, \quad \alpha(\phi)\equiv\frac{\partial \ln A(\phi)}{\partial \phi}\, .
\end{equation}
One of the key points of spontaneous scalarization is the fact that GR can be retrieved in the weak field limit. That is ensured by demanding that for some $\phi_0$, the condition $\alpha(\phi_0)=0$ is satisfied. A simple choice for the coupling function allowing that, is the one made by the authors in their original work, namely
\begin{equation}
    A(\phi)=\exp\bigg(\frac{\beta \phi^2}{2}\bigg)\,.
\end{equation}
As mentioned earlier the onset of the scalarization can be captured at the linear level as the manifestation of a tachyonic instability. To see that consider
\begin{equation}
    \phi=\phi_0+\delta\phi\,,
\end{equation}
so then the scalar equation for the particular choice of the coupling function, yields
\begin{equation}
    \Box \delta\phi+\frac{1}{2}\,\kappa\,\beta T\,\delta\phi=0\, ,
\end{equation}
where we identify the effective mass as
\begin{equation}
    m_{\text{eff}}^2=-\frac{1}{2}\,\kappa\,\beta T\, .
\end{equation}
For very compact neutron stars, it turns out that (considering the energy momentum tensor of a perfect fluid) the trace of the energy momentum tensor is negative. Therefore a tachyonic instability, corresponding to negative values of the effective scalar mass, develops for negative values of the coupling parameter $\beta$. It was found that scalarized neutron star configurations emerge for $\beta<-4$ \footnote{In contrast with flat spacetime, a negative effective mass in curved spacetime does not guarantee the development of an instability. The nontrivial way in which curvature changes may actually stabilize the configuration. In curved spacetimes, in principle, some threshold needs to be exceeded.}.

GR solutions of black holes in this theory do not exhibit instabilities. In order to see that clearly, it is useful to move to the Jordan frame \cite{Damour:1996ke}, so then action \eqref{eq:DEF_action} transforms to one containing the following nonminimal coupling
\begin{equation}
    \sqrt{-g}\,\zeta\,\varphi^2 R\, .
\end{equation}
where $\zeta$ is a coupling constant and $\varphi$ is the scalar field expressed in the Jordan frame.
The effective mass then receives the following contribution $m_{\text{eff}}^2 \subset \zeta\, R$, which cannot attribute to a tachyonic instability in the Ricci flat geometries of black holes in GR.

\begin{figure}[ht]
    \centering
    \includegraphics[width=0.9\textwidth]{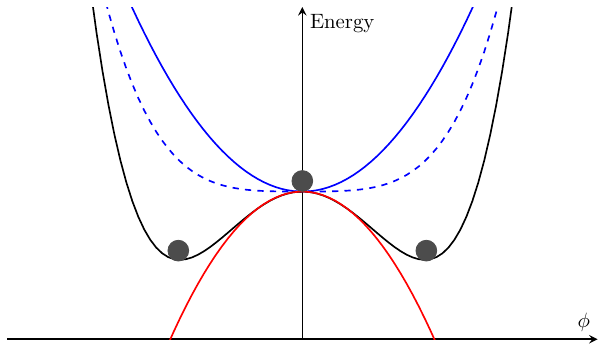}
    \caption[Qualitative description of spontaneous scalarization]{Qualitative description of spontaneous scalarization. The blue line corresponds to stable GR solution where a positive effective mass keeps the scalar settled in its GR configuration. The red line depicts the tachyonic instability driving the scalar away from GR, and the grey line corresponds to the situation after non-linearities have managed to stabilize the configuration.}
    \label{fig:instability}
\end{figure}

In the previous chapter we discussed how the no-hair theorem can be evaded within the context of the Horndeski theory when a coupling between the Gauss-Bonnet invariant and the scalar field is assumed. We will now show how this general framework allows for the spontaneous scalarization of both black holes and neutron stars.
Let us consider the model presented in \eqref{eq:action_GB}, which as explained leads to hairy solutions for an arbitrary coupling function $f(\phi)$. If we turn our attention to the scalar equation \eqref{eq:scalar_GB} we can see that the only way for the theory to admit $\phi=\phi_0$ constant solutions is if $f^{(1)}(\phi_0)=0$ for some $\phi_0$. It is instantly clear that this can never be the case for the shift-symmetric case $f(\phi)\sim\phi$ \cite{Sotiriou:2013qea,Sotiriou:2014pfa} or the dilatonic black holes with $f(\phi)\sim e^{\alpha\phi}$ \cite{Kanti:1995vq,Kanti:1995cp}. If we accept that in order for GR to be a solution of our theory, the coupling function must satisfy $f^{(1)}(\phi_0)=0$, then the next step is to examine under what conditions GR becomes unstable and when the system transitions to the scalarized version.
Let us rewrite the scalar equation of motion \eqref{eq:scal_eq} after taking linear perturbations around the GR solution
\begin{equation}
    \left[\Box+f^{(2)}(\phi_0)\GB\right]\delta\phi=0\,.
\end{equation}
We can now identify the term involving the Gauss-Bonnet invariant as the negative of the effective mass for the scalar field, \textit{i.e.}
\begin{equation}
    m_{\text{eff}}^{2}\equiv -f^{(2)}(\phi_0)\,\GB\, .
\end{equation}
Once the effective mass of the scalar becomes suffieciently negative a tachyonic instability develops driving the scalar to an exponential growth. That growth is quenched by non-linearities which are however irrelevant when one studies the onset of the scalarization. The phenomenology of the full non-decoupled physical system was first studied in \cite{Silva:2017uqg,Doneva:2017bvd} for the quadratic coupling function\footnote{In \cite{Doneva:2017bvd} the quadratic coupling function was given in an exponential format but the two functions are equivalent in the small $\phi$ limit. Therefore, the onset of the instability is effectively the same since it is derived in the linear limit.}, for which the special value for the scalar that allows for spontaneous scalarization is $\phi_0=0$.
In \cite{Silva:2017uqg} it was shown that black holes but also neutron stars can undergo spontaneous scalarization in this model.
For static and spherical black hole solutions, as we saw in the previous section, the Gauss-Bonnet invariant is positive definite. Therefore, in the quadratic model, spontaneous scalarization is only allowed when the coupling constant is positive.

For rotating black hole spacetimes, on the other hand, the parameter space of solutions widens and a new class of scalarized black holes emerges, that of \textit{spin-induced} spontaneously scalarized solutions \cite{Dima:2020yac, Herdeiro:2020wei}. More specifically, the Gauss-Bonnet invariant takes the following form for the Kerr spacetime
\begin{equation}
    \GB_{\text{Kerr}}=\frac{48M^2}{(r^2+\chi^2)^6}\big(r^6-15r^4\chi^2+15r^2\chi^4-\chi^6\big)\, ,
\end{equation}
where $\chi\equiv \alpha\cos\theta$, with $\alpha=J/M$ being the spin parameter of the black hole and $J$ its angular momentum. For rapidly rotating black holes it is now possible for the quantity in the parenthesis to become negative rendering the Gauss-Bonnet invariant negative too. Therefore, in this scenario spontaneous scalarization is possible even for negative values of the coupling constant in the quadratic coupling case, provided that the black hole is spinning fast enough. The study of spin induced scalarization has revealed interesting phenomenology but will not concern us it the context of this thesis, since we will focus on static spacetimes.

%% file: Chapters/minimum_bh_mass.tex
It the first part of Chapter~\ref{ch:Evasions} we discussed the evasion of the no-hair theorems governing the shift-symmetric Horndeski Lagrangian. As we explained there, a theory that respects shift symmetry does not allow the scalar to acquire a mass which would lead to an exponential decay. Therefore, it is of particular interest considering observations probing the strong-field gravitational regime.
We saw in the last chapter that black holes carrying scalar hair, that emerge in Horndeski through a nonminimal coupling of the scalar with the Gauss-Bonnet invariant share two key properties:
\begin{itemize}
    \item For the general coupling between the scalar and the Gauss-Bonnet invariant (which includes the linear one) a condition for the minimum black hole size emerges \eqref{eq:rh_min}. In other words, for any fixed value of the coupling constant that controls these nonminimal terms ({\textit i.e.}~any given theory within the class), black holes have a minimum mass, controlled by the value of that coupling constant(s).

    \item The scalar charge of the hairy solutions is not an independent quantity, but is instead fixed with respect to the black hole mass and spin by a regularity condition on the horizon.
    This was first shown in \cite{Saravani:2019xwx}, were the authors proved that for the most general shift symmetric action that leads to second-order equations upon variation (shift-symmetric Horndeski theory) and respects local Lorentz symmetry, the scalar charge $Q$ is given by $4\pi Q= \alpha \int_{\mathcal H} n_a {\GB}^a$, as discussed in the previous chapter. 
\end{itemize}
Indeed, hairy black hole solutions in the  $\phi\, {\GB}$ and in the $e^\phi {\GB}$ scenarios which are subcases of the generic $f(\phi)$ case, share these two properties \cite{Sotiriou:2014pfa}.
The scope of this chapter is to focus on the linear coupling case and determine how these two properties are affected by the presence of additional shift-symmetric (derivative) interactions in the action \cite{Thaalba:2022bnt}.

It is worth emphasising that having a minimum mass for black holes in the $\phi\, {\GB}$ model leads to a strong constraint on the coupling constant of this term, coming from the lightest black hole observed \cite{Fernandes:2022kvg, Charmousis:2021npl}. 
Most other observations are sensitive to the scalar charge but this gets converted to a constraint on the same coupling constant using the relation that fixes the scalar charge in terms of the black hole mass (and spin) {\textit e.g.~}\cite{Lyu:2022gdr,Perkins:2021mhb,Maselli:2020zgv,Maselli:2021men}. In the case of extreme mass ratio inspirals (EMRIs), a scaling of the scalar charge with respect to the black hole mass inspired by the $\phi\, {\GB}$ theory proved to be a crucial ingredient to drastically simplify the modelling for a vastly broader class of nonminimally coupled scalars \cite{Maselli:2020zgv}. Note that, although there are good reasons to believe that such terms can remain subdominant when modelling binary dynamics and gravitational wave radiation \cite{Witek:2018dmd,Maselli:2020zgv}, they could still have a crucial effect on the properties of the sources, including their quasi-normal ringing \cite{Hui:2021cpm}, and their dependence on the coupling constants of the theory \cite{Saravani:2019xwx}. This can  affect how observations get translated to bounds for these coupling constants. 
Considering also that Effective field theory (EFT) strongly suggests that additional terms should be present, it is rather pertinent to understand their contributions to relations between mass and charge and to check whether they affect the minimum size of black holes.
 
As expected from the discussion above, the charge vanishes if the $\phi {\GB}$ term is absent. The value of ${\GB}^a$ does however depend on any additional couplings, where corrections with respect to GR are suppressed by the mass scales that correspond to these couplings. If one assumes continuity as these couplings are driven to zero and that their characteristic energy scales are similar to that of the Gauss-Bonnet one which sources the hair ({\textit i.e.}~no hierarchy of scales), then one expects corrections to the charge and its scaling with the mass of the black hole to be subdominant. How much so is a matter of further exploration. Moreover, the aforementioned expression for the charge does not give any information on the minimum size of black holes.

\section{A theory with higher order operators}
As we explained before, our broader goal in this chapter is to understand how adding additional shift symmetric terms to action \eqref{eq:lgbaction} would affect the two key properties of the black hole solutions, namely the minimum size and the scalar charge. In this section we will avoid using numerical methods, and we will constrain our analysis in analytical or semi-analytical methods to track and study the aforementioned properties.

\subsection{The high-order action}
To make the calculations more tractable, we will not consider all of the terms appearing in the Horndeski Lagrangian \eqref{eq:Horndeski}, but we will instead restrict ourselves to the following theory
\begin{equation}
\begin{split}
    \mathcal{L}\sim\bigg[
    R+X\big(1+\sigma\Box\phi+\kappa X\big)
    +\alpha \phi\GB+\gamma \,G_{\mu\nu}\nabla^\mu\phi\nabla^\nu\phi\bigg]\, .
\end{split}
\label{eq:action}
\end{equation}
This action can be obtained from action \eqref{eq:Horndeski} by selecting  \footnote{Up to a total derivative the term $X R+(\Box\phi)^2-(\nabla_\mu\nabla_\nu\phi)^2$ is equivalent to $\,G_{\mu\nu}\nabla^\mu\phi\nabla^\nu\phi$. Since, $\int d^4x \sqrt{-g}\,(\Box\phi)^2=\int d^4x \sqrt{-g}\,\big[(\nabla_\mu\nabla_\nu\phi)^2+R_{\mu\nu}\nabla^\mu\phi\nabla^\nu\phi\big]+\int \text{total derivative}\;.$} 
\begin{equation}
    \begin{split}
    G_2(X)= & X+\kappa X^2, \\
    G_3(X)= & -\sigma X, \\
    G_4(X)= & 1/2 + \gamma X, \\ 
    G_5(X)= & -4\alpha\ln|X|.
    \end{split}
\end{equation}
In units where $G=c=1$ the scalar field is  dimensionless while $\alpha, \gamma, \sigma, \kappa$ have dimensions of a length squared. We will work on the spherical symmetric and static background we introduced in \eqref{eq:metric} when we were discussing the no-hair theorem for the Galileon. There, we also introduced the conserved current \eqref{eq:Jr_spherical} for the shift-symmetric Horndeski. Once again, the only non-vanishing component of the current $J^\mu$ is $J^r$, which for the theory \eqref{eq:action} yields
\begin{align}
\label{eq:spherical_current}
\begin{split}
    J^r=& 
    -B\phi'(1-\kappa B\phi^{\prime 2}) -\sigma \phi^{\prime 2} \frac{r A' + 4 A}{2rA}B^2\\[2mm]
    &
    +\gamma \phi' \frac{2A B -2A + 2rA'B}{r^2A}B+ \alpha \frac{4(1-B) B A'}{r^2 A}\,,
\end{split}
\end{align}\\
where a prime denotes a derivative w.r.t. the radial coordinate $r$. As discussed earlier, and according to the classification of \cite{Saravani:2019xwx}, the current can be separated into a part in which every term contains $\phi'$ and a contribution by the coupling of the scalar field with ${\GB}$, as follows
\begin{equation}
    J^r=\tilde{J}^r-\alpha\GB^r,\qquad \GB^r=\frac{4(B-1) B A'}{r^2 A},
    \label{eq:current}
\end{equation}
The conservation of the current can be straightforwardly integrated:
\begin{equation}
    \nabla_\mu J^\mu=0\Rightarrow J^r=\frac{\tilde{c}}{r^2}\sqrt{\frac{B}{A}}. 
\end{equation}
Using \eqref{eq:spherical_current}, one can then determine $\phi'$ and then integrate once more to obtain $\phi(r)$. Due to shift symmetry, $c$ in \eqref{eq:current} is the only meaningful integration constant. Hence, considering also the mass parameter of the black hole, one would have a two-parameter family of solutions. The scalar charge would then be independent.

However, $\phi'$ evaluated on the horizon of a black hole, $r=r_h$, denoted as $\phi'_h$, generically diverges. 
If we impose that $\phi'_h$ is finite, which is required for regularity, and we take into account that at $r=r_h$ we have $A(r_h)=B(r_h)\rightarrow0$, then  \eqref{eq:spherical_current} implies that $\tilde{J}^r(r_h)=0$. Evaluating \eqref{eq:current} on the horizon then fixes the value of $c$. As a consequence, the scalar charge ceases to be an independent parameter. After substituting $c$ back in \eqref{eq:current}, and solving with respect to $\tilde{J}^r$ we find
\begin{equation}
    \tilde{J}^r=\frac{4 \alpha}{r^2}\sqrt{\frac{B}{A}}\bigg[\lim_{r\rightarrow r_h}\bigg(\sqrt{A'B'}\text{ sgn}(B')\bigg)+\left(B-1\right)\sqrt{\frac{B}{A}}A'\bigg]. 
\end{equation}
This is the equation we will be using in the following subsection.

\subsection{Decoupling limit}
As a warm-up,  we consider the scalar in a fixed Schwarzschild background. We look for  solutions that are regular at the horizon and approach a constant value asymptotically. For simplicity, we fix that value to zero, as the value of the constant is irrelevant due to shift symmetry. The $\gamma$ term is not expected to contribute anything to the decoupled equations since it multiplies the Einstein tensor which vanishes at decoupling. The scalar equation on a GR-Schwarzschild background is:
\begin{equation}
\begin{split}
      m r^3 \kappa (2 m-r) \phi'^3
      &+m r^2 \sigma  (2 r-3 m) \phi'^2\\
      &+m r^4 \phi'+2 \alpha  \left(4 m^2+2 m r+r^2\right)=0 \, ,
\end{split}
\label{eq:sc_GR}
\end{equation}
where $m$ is the black hole mass.
First, let us note that if only the $\alpha$-term is present, one can find an analytic solution for the scalar field \cite{Sotiriou:2013qea}:
\begin{equation}
    \phi_\alpha=\frac{2 \alpha  \left(4 m^2+3 m r+3 r^2\right)}{3 m r^3}\, .
\end{equation}
We note that, in this case, no specific restriction on the choices of $\alpha$ is suggested. We then consider the $\sigma$-term in addition to the $\alpha$ one, and we solve for the derivative of the scalar field:
\begin{equation}
    \phi'_{\alpha\sigma}=\frac{r^3-\sqrt{r^6 + 8\alpha\sigma \left(12m^2-2m r - r^2 -\frac{2r^3}{m}\right)}}{6 m r \sigma -4 r^2 \sigma }\; ,\quad \alpha\sigma<\frac{2m^4}{3}\, .
\label{eq:sc_sigma_GR}
\end{equation}
The quantity under the square root needs to be non-zero and positive, therefore, an existence condition emerges. It is straightforward to see that
\begin{equation}
    \lim_{\sigma\rightarrow 0}\phi'_{\alpha\sigma}=\,\phi'_\alpha\quad , \quad\phi'_{\alpha\sigma}(r\gg r_h)=\phi'_{\alpha}(r\gg r_h)\approx-\frac{2\alpha}{m r^2}\,.
\end{equation}
The inequality condition appearing in \eqref{eq:sc_sigma_GR} imposes an upper bound on the product $\alpha\sigma$. This, in turn, yields an upper bound on $\alpha$ when $\sigma>0$, and a lower one when $\sigma<0$. It is also possible to employ a near-horizon expansion, i.e. $r=r+\epsilon$, for the two cases discussed above. This yields 
\begin{equation}
    \phi'_{\alpha} = -\frac{3\alpha}{2 m^3} + \ldots \quad , \quad
    \phi'_{\alpha\sigma} = -\frac{2 m^2 - \sqrt{4 m^4-6 \alpha  \sigma }}{m \sigma } + \ldots\, .
\end{equation}
When $\alpha$ is positive, from the near-horizon expressions we can deduce that for $\sigma>0$ ($\sigma<0$) the scalar field fall-off is larger (smaller) than in the $\sigma=0,\,\alpha\ne 0$ case, while in the limit $\sigma\rightarrow -\infty$ we retrieve the trivial solution $\phi_{\alpha\sigma}=0$ for the near-horizon expansion. When $\alpha$ is negative the aforementioned properties are reversed.

The case where $\kappa\ne 0$ is more subtle. By examining the $\phi'^3$ coefficient in \eqref{eq:sc_GR}, we see that when the $\kappa$-term is present, the derivative of the scalar at the horizon does not depend on $\kappa$, therefore, we deduce that $\kappa$ does not enter an existence condition analogous to \eqref{eq:sc_sigma_GR}. If one attempts to solve the equation in the region $[r_h+\epsilon,\infty)$, for $\epsilon \ll 1$, it turns out that a regular solution can be found $\forall\,\kappa\in R$. However, not all of those solutions have the desired asymptotic behaviour, and for large positive values of $\kappa$,  $\phi'_{\alpha\kappa}(\infty)\ne 0$.

\subsection{Existence conditions}
In the previous subsection, we saw how the existence conditions for the scalar equation are affected by the extra terms in the decoupling limit. Here, we derive the existence conditions for black holes beyond decoupling, for the full system of equations. To do so, we assume the existence of a horizon located at $r=r_h$, so that $A_{h^\pm}\rightarrow 0^\pm$, where the $+$ sign corresponds to approaching the horizon from the outside, while the $-$ to approaching it from the inside. By employing the near-horizon expansions, we find the following expression for the second derivative of the scalar at the horizon
\begin{equation}\label{eq:phidd}
    \phi''\approx\frac{\left(4 \alpha  \phi'+r\right) \left\{24 \alpha +\phi'^2 \left[24 \alpha  \gamma+r^2 (8 \alpha +\sigma )\right]+2 r^3 \phi'\right\}}{-2 \left\{r^4-96 \alpha ^2+\phi' \left[r^3 (4 \alpha +\sigma )+24 \alpha \gamma r \right]\right\}}\left(\frac{A'}{A}\right)
\end{equation}
We note that in order to get a black hole solution $\left(4 \alpha  \phi'+r\right)\ne 0$. Since $(A'/A)_h=\infty$, in order for $\phi$ to be regular at the horizon, it is required that 
$$[24 \alpha +\phi'^2 \left[24 \alpha  \gamma +r^2 (8 \alpha +\sigma)\right]+2 r^3 \phi']_{r_h}=0,$$ therefore, 
\begin{equation}\label{eq:phidh}
    \phi_h'=\frac{\sqrt{r_h^6-576 \alpha ^2 \gamma-24 \alpha  r_h^2 (8 \alpha +\sigma )}-r_h^3}{24 \alpha  \gamma +r_h^2 (8 \alpha +\sigma )}.
\end{equation}
It is possible to derive from equations \eqref{eq:phidd}, and \eqref{eq:phidh} two conditions for black hole solutions to exist. The first is for the quantity under the square root of equation \eqref{eq:phidh} to be positive; the other condition imposes a nonzero denominator in both equations. 
\begin{align}
    \label{eq:condition_1}
    \textbf{I:}\quad & r_h^6-576 \alpha ^2 \gamma -24 \alpha  r_h^2 (8 \alpha +\sigma ) \ge 0\,,\\
    \label{eq:condition_2}
    \begin{split}
        \textbf{II:}\quad & \left[24 \alpha  \gamma  r_h+(4 \alpha +\sigma ) r_h^3\right] \sqrt{r_h^6-576 \alpha ^2 \gamma -24 \alpha  (8 \alpha +\sigma ) r_h^2}\\
        &+4\alpha\left[r_h^6-576 \alpha ^2 \gamma-24 \alpha  (8 \alpha +\sigma ) r_h^2 \right]\ne 0\,.
    \end{split}
\end{align}
It is worth pointing out that when we consider only the GB term, conditions I and II reduce to the existence condition appearing in \cite{Sotiriou:2014pfa}. Thus, we see that in the non-perturbative approach $\sigma$ but also $\gamma$ enter the existence conditions. In particular, in the case $\gamma=0$, the parameter $\alpha$ has both an upper and a lower bound for either sign of $\sigma$.

\subsection{Perturbative treatment}
We will now attempt to gain insight into the dependece of the scalar charge of the solutions on these higher order terms by employing a perturbative approach with respect to the coupling constant  $\alpha$, where the latter is associated with the term that sources the hair. To do that we define the dimensionless parameter $\tilde{\alpha}\equiv \alpha/r_h^2\ll 1$, where the horizon radius $r_h$ is the length scale we associate with our solution. In a similar manner we can define $\tilde{\gamma}=\gamma/r_h^2,\, \tilde{\sigma}=\sigma/r_h^2$ and $\tilde{\kappa}=\kappa/r_h^2$.  For nonzero, small values of $\tilde{\alpha}$ we expect to acquire perturbative deformations to the Schwarzschild solution. Those are expressed through the following expansions for the metric elements
\begin{align}
    A(r)=&\left(1-\frac{2M}{r}\right)\left[1+\sum_{n=1}^{\infty}A_n(r)\tilde{\alpha}^n\right]^2,\label{eq:pertA}\\
    B(r)=&\left(1-\frac{2M}{r}\right)\left[1+\sum_{n=1}^{\infty}B_n(r)\tilde{\alpha}^n\right]^{-2}\label{eq:pertB}\, ,\\
    \phi(r)=&\;\phi_0+\sum_{n=1}^\infty \phi_n \tilde{\alpha}^n \label{eq:pertphi} \,.
\end{align}
For $\tilde{\alpha}=0$, we retrieve GR minimally coupled to a scalar field, which for the spherically symmetric static configurations yields the Schwarzschild solution. These expansions are substituted in the equations of motion, which are then solved order by order for the unknown coefficients $\{A_n$, $\,B_n$, $\phi_n\}$. We work out the calculations up to the fifth order in the perturbative parameter $\mathcal{O}(\tilde{a}^5)$. The solutions become very lengthy beyond $2^{\text{nd}}$-order and, therefore, are omitted. However, the expressions for the scalar charge and the mass of the black hole can be written in the following compact form
\begin{align}
    Q=&\,Q_1\,\tilde{\alpha}+Q_3\,\tilde{\alpha}^3+Q_4(\sigma)\,\tilde{\alpha}^4+Q_5(\gamma,\sigma,\kappa)\,\tilde{\alpha}^5,\label{eq:Qa}\\[4mm]
    M=&\,m+M_2\,\tilde{\alpha}^2+M_3(\sigma)\tilde{\alpha}^3+M_4(\gamma,\sigma,\kappa)\,\tilde{\alpha}^4
    +M_5(\gamma,\sigma)\,\tilde{\alpha}^5\label{eq:Ma},
\end{align}
where the coefficients $Q_n$ and $M_n$ can be found in Appendix~\ref{ch:Appendix_eom}. Notice that we have to extend to the $5^{\text{th}}$ order in $\tilde{\alpha}$ before we see $\kappa$ contributing to the scalar charge.

Perturbative treatment will break down at some radius. To trace when that happens we simultaneously scan the following expressions
\begin{align}
    A(r)=\,& A_0(r)+A_2(r)\,{\tilde{\alpha}}^2+A_3(r,\sigma)\,{\tilde{\alpha}}^3\, ,\label{eq:A_pert}\\
    B(r)=\,& B_0(r)+B_2(r,\gamma)\,{\tilde{\alpha}}^2+B_3(r,\gamma,\sigma)\,{\tilde{\alpha}}^3,\, ,\label{eq:B_pert}\\
    \phi(r)=\,& \phi_0+\phi_1(r)\,{\tilde{\alpha}}+\phi_2(r, \sigma)\,{\tilde{\alpha}}^2+\phi_3(r,\gamma,\sigma,\kappa)\,{\tilde{\alpha}}^3\, ,\label{eq:phi_pert}\\
    \GB(r)=\,& \GB_0(r)+\GB_2(r,\gamma)\,{\tilde{\alpha}}^2+\GB_3(r,\gamma,\sigma)\,{\tilde{\alpha}}^3\, ,\label{eq:GB_pert}
\end{align}
for perturbative inconsistencies.
If at some radius $r_{\text{np}}$ terms of different orders of $\tilde{\alpha}$ become comparable in size, the perturbative treatment can no longer be trusted. The coefficients $A_n,\,\GB_n$ are also given in Appendix~\ref{ch:Appendix_eom}. From \eqref{eq:A_pert}-\eqref{eq:GB_pert} we see that even at second-order in $\tilde{\alpha}$, terms involving $\gamma$ appear. We note that in the case where  $\tilde{\gamma},\,\tilde{\sigma},\,\tilde{\kappa}=0$ it was shown in \cite{Sotiriou:2014pfa} that loss of perturbativity occurred at roughly the same radius at which the non-perturbative solutions exhibited a finite area singularity. We will return to this issue in the next section.

\begin{figure}[t]
    \centering
    \includegraphics[width=.49\linewidth]{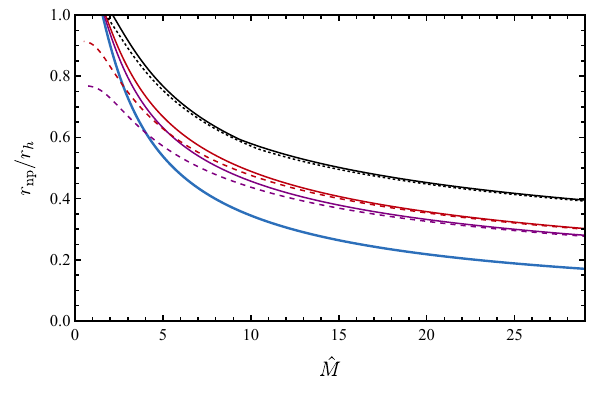}
    \hspace{0mm}
    \includegraphics[width=.49\linewidth]{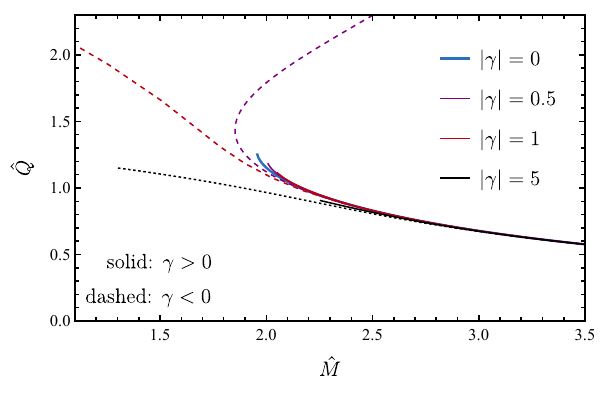}\\
    
    \includegraphics[width=.49\linewidth]{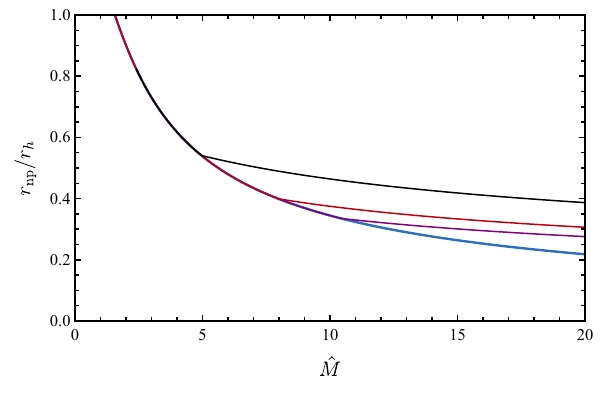}
    \hspace{0mm}
    \includegraphics[width=.49\linewidth]{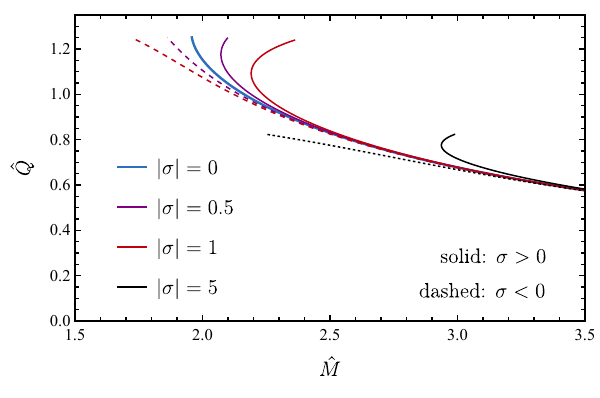}
    
    \caption[Perturbative approach in the high-order shift-symmetric model]{\textit{Top left}: Singular radius derived from the perturbative analysis, for $\tilde{\sigma}=0$ and $\tilde{\gamma}=0,\,\pm 0.5,\,\pm 1,\, \pm 5$. \textit{Top right}: Normalized scalar charge and mass derived from the perturbative analysis. \textit{Bottom left}: Singular radius derived from the perturbative analysis, for $\tilde{\gamma}=0$ and $\tilde{\sigma}=0,\,\pm 0.5,\,\pm 1,\, \pm 5$.\textit{Bottom right}: Normalized scalar charge and mass derived from the perturbative analysis.}
    \label{fig:perturbative_treatment}
    
\end{figure}
In the top-left panel of Fig.~\ref{fig:perturbative_treatment} we present the radius $r_{\text{np}}$ denoting the point where the perturbative analysis breaks down, for $\tilde{\gamma}=0$, $\pm 0.5$, $\pm 1$, $\pm 5$ and $\tilde{\sigma}=0$. In the top-right panel of Fig.~\ref{fig:perturbative_treatment} we present the $\hat{Q}$-$\hat{M}$ solution-existence curves for the same choices of $\tilde{\gamma}$. The quantities $\hat{Q}$ and $\hat{M}$ are defined as
\begin{equation}
    \hat{M}=\,M/\alpha^{1/2}\,,\;\; \hat{Q}=Q/\alpha^{1/2}\,,
    \label{eq:scaled_quantities}
\end{equation}
In the bottom panels, we present the analogous results for the case $\tilde{\sigma}=0,\,\pm 0.5,\,\pm 1,\, \pm 5$ and $\tilde{\gamma}=0$. The horizontal axis in these plots corresponds to the normalized mass with respect to the GB coupling, $\hat{M}=m/\alpha^{1/2}$, with $m=0.5$.

When $r_{\text{np}}$ exceeds $r_h$ part of the exterior cannot be described by the perturbative solution.  One can read of the corresponding mass from Fig.~\ref{fig:perturbative_treatment}. From both top panels, we deduce that this mass increases (decreases) for positive (negative) values of $\tilde{\gamma}$. We also see that for $\hat{M}\gtrapprox 2.5 $ all curves start merging, as the $\gamma$-term becomes significantly subdominant with respect to the GB one. One other interesting property we notice occurs for the larger value of $\tilde{\gamma}=5$ and it pertains to more than one solutions existing for the same mass. These solutions are different from one another however as they describe black holes with different scalar charges.

From the bottom panels, we notice a similar trend regarding the effects of $\tilde{\sigma}$ and the mass for which perturbativity is lost already at the exterior. For $\hat{M}\gtrapprox 3.5$ all solutions in the $\hat{M}$-$\hat{Q}$ plots begin to merge as the $\sigma$-term becomes subdominant. One of the main differences with respect to the top-panel plots, however, has to do with the maximum scalar charge. In the $\tilde{\sigma}=0,\tilde{\gamma}\ne 0$ scenario we manage to get solutions with substantially larger scalar charges in comparison to the $\tilde{\sigma}=\tilde{\gamma}=0$ (blue line). In the $\tilde{\gamma}=0,\tilde{\sigma}\ne 0$ case this does not happen. Let us also point out that the values for $r_{\text{np}}$ depicted in the bottom left panel, are the same for positive and negative values of $\sigma$. To understand why this occurs it is helpful to discuss how $\sigma$ enters the perturbative expansions. Specifically, $\sigma$ appears at second order in the expansion of the scalar field as a multiplicative constant, which explains why changing its sign yields the same solution for $r_{\text{np}}$.

It should be noted, however, that these conclusions have to be drawn with care, as they correspond to a region of the parameter space where $\hat{M}<5$, or $\tilde{\alpha}>0.01$. This is a region that can in principle render the perturbative approach problematic in general and only a proper numerical analysis can either confirm or disprove the aforementioned effects.

\section{Numerical results}
\label{sec:numerical}
We now move to solving the full system of equations numerically. This is a system of ODEs  of the form $\{\phi'',A',B'\}=f(r,\phi',\phi,A,B)$. We separate the analysis into two regions: the black-hole exterior and the black-hole interior. In both cases, the  integration starts at the horizon. The theoretical parameter space is made by $(\gamma,\,\sigma,\,\kappa,\,r_h)$, where $r_h$ is the black-hole horizon radius. Since $r_h$ appears in the existence condition \eqref{eq:phidh}, the allowed values for the coupling parameters are expected to be affected if we variate $r_h$. We can straightforwardly reduce the dimension of the parameter space by one if we normalize the coupling parameters with the horizon radius as we did in the previous section.

For a given theory defined by $(\tilde{\gamma},\,\tilde{\sigma},\,\tilde{\kappa})$ we allow the values of $\tilde{\alpha}$ to scan the parameter space starting from small $\tilde{\alpha}$ and gradually increasing until the existence conditions are saturated, We, therefore, need the set of values $\{\phi',\phi,A,B\}_{r_h}$. Despite appearing to constitute ``initial data'', this set of values is not entirely free to choose. In practice, in order to apply the existence conditions \eqref{eq:condition_1}-\eqref{eq:condition_2} with reasonable numerical accuracy, 
we use a perturbative expansion near the horizon and we numerically solve the system of algebraic equations for the first coefficients appearing in the expansions, up to order $\mathcal{O}(r-r_h)^2$.  This process reduces the number of the free initial conditions to two, namely the value of the scalar field at the horizon and that of the first-order coefficient of $A$. The latter one, however, is fixed by asymptotic flatness, leaving $\phi_h$ as the only free-to-chose initial condition. The asymptotic value of the scalar field should be constant but otherwise unconstrained since our model is shift-symmetric. For simplicity, we choose $\phi_h$ so that $\phi_\infty=0$. To achieve that we employ a shooting method while integrating outwards, demanding $\phi$ vanishing to a part in $10^4$. The remaining free parameter is the horizon radius $r_h$.
We then start the numerical integration outwards (inwards) from $r=r_h\pm\mathcal{O}(10^{-5})$. In the exterior, we typically integrate up to $r/r_h\approx 10^{5}$.

In the following subsections, we present plots corresponding to different cases of couplings. In each case we numerically calculate the scalar charge and ADM mass of the black hole using the following expressions:
\begin{equation}
    Q=-\lim_{r\rightarrow\infty}\left(r^2\phi'\right)\,,\quad M=\lim_{r\rightarrow \infty}\left[\frac{r \left(2-2 B+r^2 \phi'^2\right)}{r^2 \phi'^2-4}\right].
\end{equation}
We were also able to verify the emergence of a finite-radius singularity, consistent with the existence conditions. While integrating from the horizon and inwards we noticed the following general trend: Starting from GR ($\alpha\rightarrow 0$) and gradually increasing the couplings, the geometric invariants diverge and the solutions become singular at some radius $r_s$. The larger GR deviations become the more $r_s$ approaches $r_h$. When one of the existence conditions is saturated the singularity radius approaches the horizon radius, \textit{i.e.} $r_s\rightarrow r_h$.

\subsubsection{Charge, mass and scalar profile}
\begin{figure}[ht]
    \centering
    \includegraphics[width=.47\linewidth]{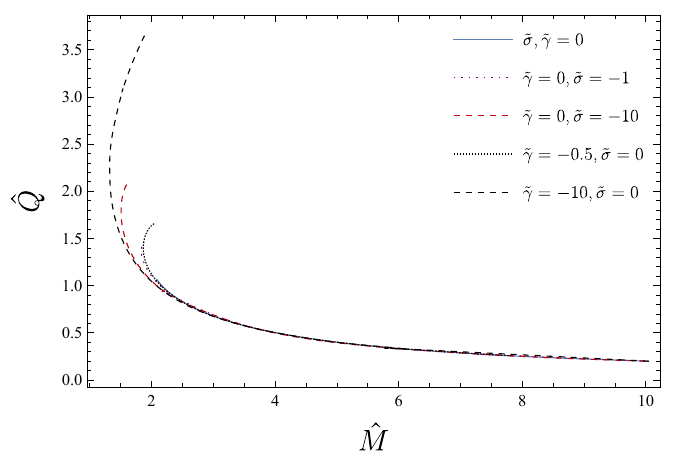}
    \hspace{5mm}
    \includegraphics[width=.47\linewidth]{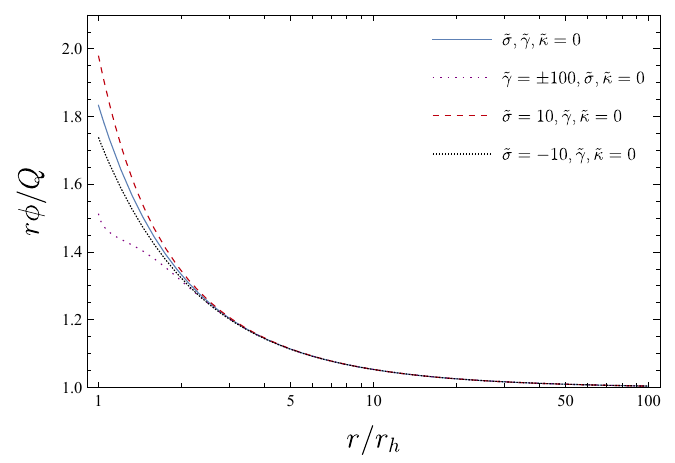}
    \caption[Scalar charge and scalar profiles for the high-order shift-symmetric model]{\textit{Left:} The relation between the normalized mass and charge. \textit{Right:} The scalar field profile for black hole solutions with $\hat{M} \sim 10$.}
    \label{fig:MQphi}
\end{figure}
In the following paragraphs we attempt to present the overall generic trend that the black-hole properties follow, if one considers action \eqref{eq:action}. We discuss the charge, mass and scalar profile for a few examples corresponding to different scenarios, which motivate the more thorough analysis that follows in the next subsections.

On the left panel of Fig.~\ref{fig:MQphi}, we show the $\hat{M}$-$\hat{Q}$ plot for different negative values of the coupling constants $\gamma$ and $\sigma$. The corresponding plots for positive couplings are not presented here, since -at these scales- they are overshadowed by the $\tilde{\gamma} = \tilde{\sigma} = 0$ curve, as will be explained later in more detail. In all positive-coupling cases, the minimum black-hole mass is larger than the one corresponding to $\tilde{\gamma} = \tilde{\sigma} = 0$. Furthermore, non-zero $\tilde{\kappa}$ curves are almost indistinguishable from the $\tilde{\gamma} = \tilde{\sigma} = 0$ one, since as we saw $\kappa$ does not enter the existence conditions. Consequently, the corresponding $\tilde{\kappa}$-plots are not presented here.
We see that for large $\hat{M}$ which corresponds to small GB couplings, the charge in all cases drops off to zero and GR is retrieved. This is of course associated with the fact that the GB term is the one sourcing the hair. In the small $\hat{M}$ regime significant deviations are observed, which are explained in the following coupling-specific subsections. On the right panel of Fig.~\ref{fig:MQphi}, we show the profile of the scalar field, properly normalized with the distance and the scalar charge. All curves exhibit a $1/r$ fall-off and asymptotically approach $1$. For large radii, the scalar field profiles are indiscernible for different couplings. In the near horizon regime, however, there are apparent deviations in accordance with the non-trivial deviations shown in the left panel.

To make things easier for the reader, in what follows, we consider the GB coupling $\tilde{\alpha}$ in combination with $\tilde{\gamma},\,\tilde{\sigma}$ and $\tilde{\kappa}$ separately. In this work, we consider $\alpha>0$ as this is consistent with most of the bibliography. However, it is worth pointing out, that action \eqref{eq:action} is invariant under the simultaneous transformation $\alpha\rightarrow-\alpha$, $\phi\rightarrow-\phi$ and $\sigma\rightarrow-\sigma$ and that in the case of $\sigma=\gamma=\kappa=0$, the sign of $\phi$ is determined by the sign of $\alpha$ for solutions that are continuously connected to Schwarzschild as $\alpha \to 0$. In what follows we consider both positive and negative values for $\sigma$. $\gamma$, and $\kappa$, and hence our analysis should effectively cover the $\alpha<0$ case as well, at least for configurations that are continuously connected to  Schwarzschild.

\subsubsection{The \texorpdfstring{$\tilde{\gamma}$}{TEXT} and \texorpdfstring{$\tilde{\sigma}$}{TEXT} terms}

\begin{figure}[t]
    \centering
    \includegraphics[width=.48\linewidth]{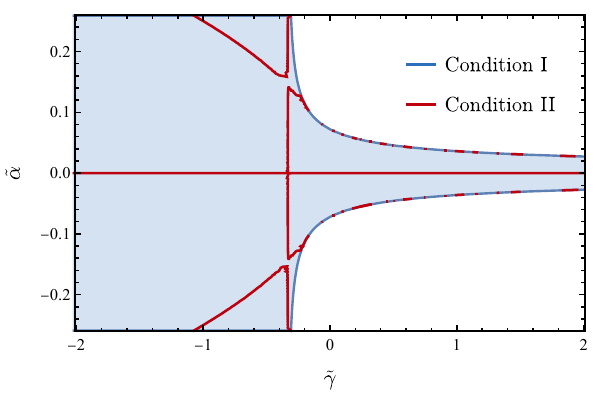}\hspace{5mm}
    \includegraphics[width=.47\linewidth]{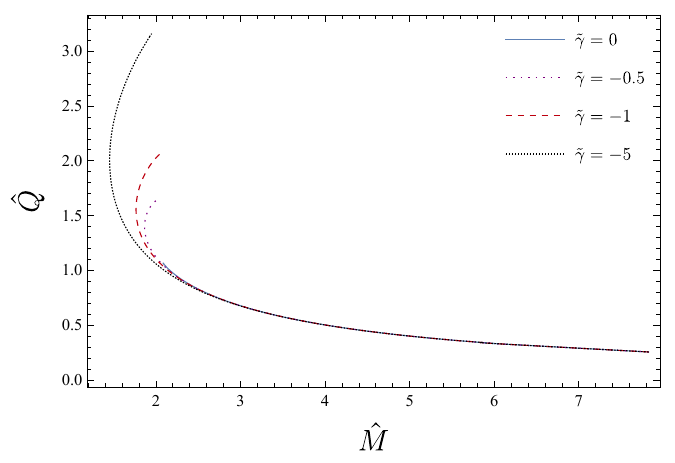}\\

    \includegraphics[width=.47\linewidth]{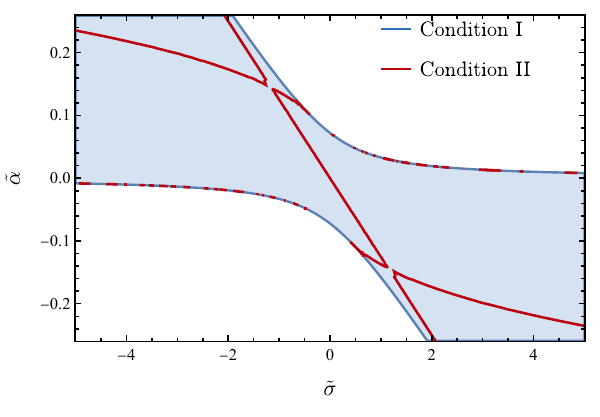}\hspace{5mm}
    \includegraphics[width=.47\linewidth]{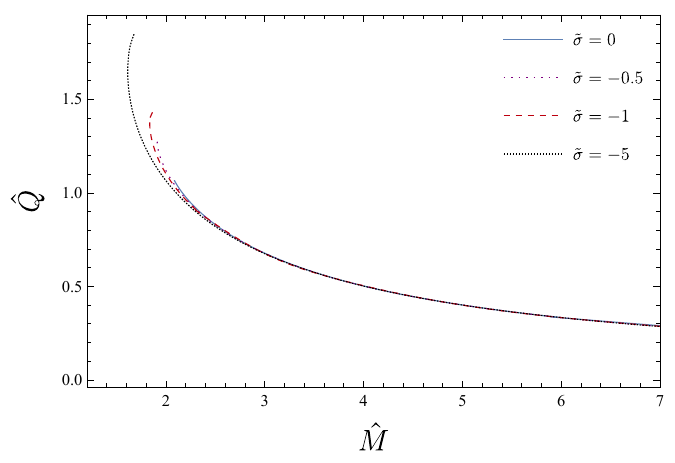}
    
    \caption[Existence conditions and scalar charge in the high-order shift-symmetric model]{\textit{Top Left:} Existence conditions in the case of $\tilde{\sigma}=\tilde{\kappa}=0$. The blue shaded region corresponds to the area of the parameter space allowed by condition I \eqref{eq:condition_1}. The red line corresponds to the values within the allowed blue region that are excluded by condition II \eqref{eq:condition_2}. \textit{Top Right:} Mass-Charge plots for $\tilde{\sigma} = \tilde{\kappa} = 0$ and $\tilde{\gamma}=\{-5,-1,-0.5,0\}$. \textit{Bottom Left:} Existence conditions in the case of $\tilde{\gamma}=\tilde{\kappa}=0$. The blue shaded region corresponds to the inequality of condition I \eqref{eq:condition_1}, while the red lined corresponds to the inequality of condition II \eqref{eq:condition_2}. \textit{Bottom Right:} Mass-Charge plots for $\tilde{\gamma} = \tilde{\kappa} = 0$ and $\tilde{\sigma}=\{-5,-1,-0.5,0\}$.}
    \label{fig:existence_QM}
\end{figure}

First we consider the case $\tilde{\sigma}=\tilde{\kappa}=0$. From \eqref{eq:condition_1} and \eqref{eq:condition_2} we find the conditions on $\tilde{\gamma}$ necessary for regularity at the horizon. The existence conditions I-II are in general non-trivial and the easiest way to track them is to examine the corresponding region plot. In the left panel of Fig.~\ref{fig:existence_QM} we see the aforementioned plot with $\tilde{\gamma}$ being on the horizontal axis and $\tilde{\alpha}$ occupying the vertical one. The first obvious observation relates to the apparent asymmetry about the vertical axis. Therefore, we expect the sign of $\tilde{\gamma}$ to influence the black-hole solutions and properties. In particular, for negative values of $\tilde{\gamma}$ the parameter space of allowed values for $\tilde{\alpha}$ increases, and so we expect negative values of $\tilde{\gamma}$ to allow for hairy solutions with smaller masses. On the other hand when $\tilde{\gamma}>0$ the parameter space of $\tilde{\alpha}$ shrinks and we expect the black-hole-mass range to also decrease. Regarding the GB-coupling $\tilde{\alpha}$, extending the plot to negative values of $\tilde{\alpha}$ is trivial as, for $\tilde{\sigma}=0$, the action \eqref{eq:action} is invariant under the simultaneous transformation $\tilde{\alpha}\rightarrow-\tilde{\alpha}$ and $\phi\rightarrow-\phi$.

\begin{figure}[t]
    \centering
    \includegraphics[width=.47\linewidth]{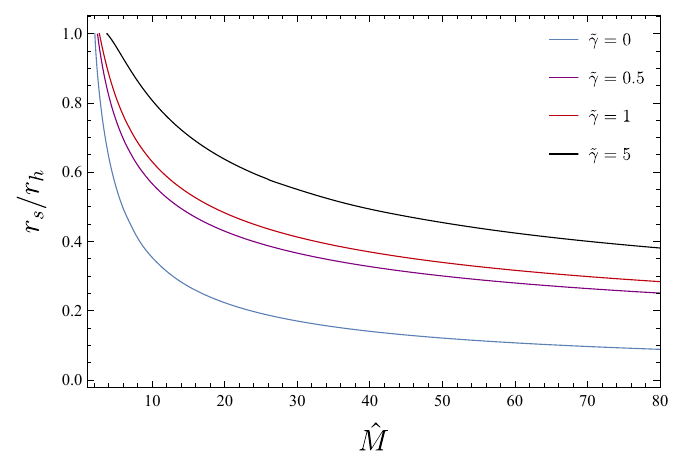}\hspace{5mm}
    \includegraphics[width=.47\linewidth]{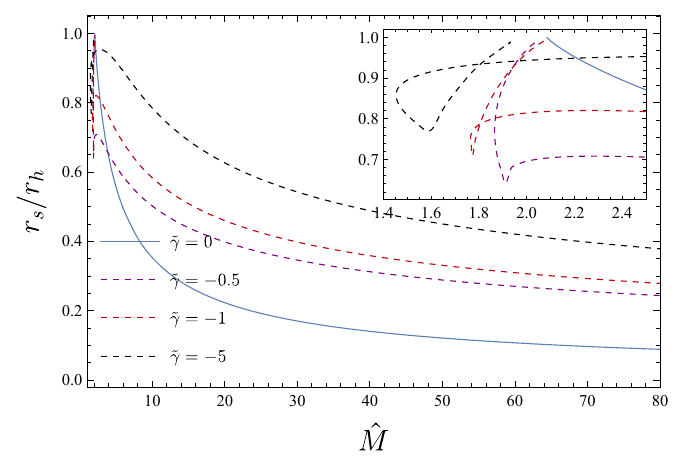}\\

    \includegraphics[width=.47\linewidth]{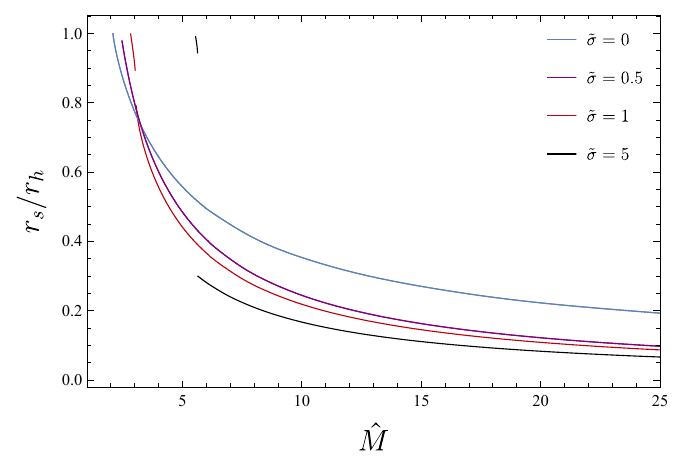}\hspace{5mm}
    \includegraphics[width=.47\linewidth]{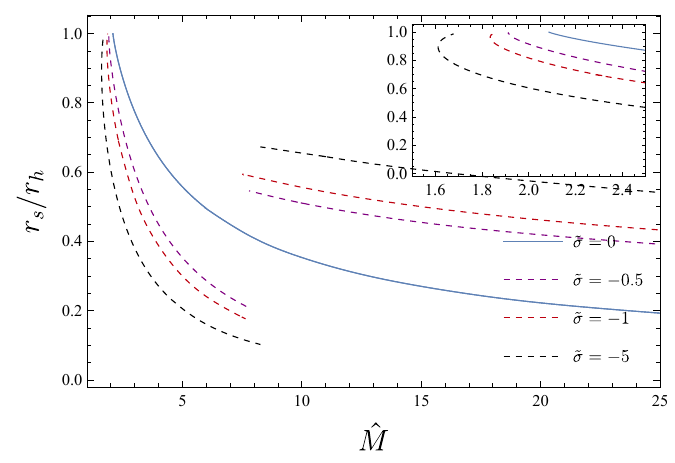}
    
    \caption[Finite area singularity for the $\gamma$ and $\sigma$ cases]{\textit{Top:} The finite singularity radius $r_{s}$ as a function of the normalized mass $\hat{M}$ for $\tilde{\sigma} = \tilde{\kappa} = 0$ and $\tilde{\gamma}=\{-5,-1,-0.5,0,0.5,1,5\}$. \textit{Bottom:} The finite singularity radius $r_{s}$ as a function of the normalized mass $\hat{M}$, for $\tilde{\gamma} = \tilde{\kappa} = 0$ and $\tilde{\sigma}=\{-5,-1,-0.5,0,0.5,1,5\}$. The horizon radius is $r_h = 1$. }
    \label{fig:singularity}
\end{figure}

These are indeed verified in Fig.~\ref{fig:singularity}, where the emergence of a finite-radius singularity is demonstrated in the interior of the black hole. The left panel shows the singularity radius of the black-hole mass for $\tilde{\gamma}=\{0,0.5,1,5\}$ while the right panel shows the corresponding results for $\tilde{\gamma}=\{0,-0.5,-1,5\}$. The values are chosen to be of order $\sim 1-10$ with respect to $\tilde{\alpha}_{\text{max}}$, where $\tilde{\alpha}_{\text{max}}$ corresponds to the largest allowed value for $\alpha$ satisfying the existence conditions.  For the choices of $\tilde{\gamma}$ made, we present the results for the minimum hairy black hole mass in the following table:

\begin{table}[b]
    \centering
    \begin{tabular}{c||c|c|c|c|c|c|c}
    \multicolumn{8}{c}{Minimum mass for $\tilde{\sigma}=\tilde{\kappa}=0,\; \tilde{\gamma} \ne 0,\,\tilde{\alpha}>0$}\\ \hline
    \raisebox{-1mm}{$\tilde{\gamma}$} & \raisebox{-1mm}{$-5.0$} & \raisebox{-1mm}{$-1.0$} & \raisebox{-1mm}{$-0.5$} & \raisebox{-1mm}{\;\,$0.0$\;\,} & \raisebox{-1mm}{$+0.5$} & \raisebox{-1mm}{$+1.0$} & \raisebox{-1mm}{$+5.0$} \\[2mm] \hline
    \raisebox{-1mm}{$\hat{M}$} & 1.45 & 1.77 & 1.87 & 2.08 & 2.45 & 2.71 & 3.75 \\[2mm] \hline
    \multicolumn{8}{c}{Minimum mass for $\tilde{\gamma}=\tilde{\kappa}=0,\; \tilde{\sigma} \ne 0\,,\tilde{\alpha}>0$}\\ \hline
    \raisebox{-1mm}{$\tilde{\sigma}$} & \raisebox{-1mm}{$-5.0$} & \raisebox{-1mm}{$-1.0$} & \raisebox{-1mm}{$-0.5$} & \raisebox{-1mm}{\;\,$0.0$\;\,} & \raisebox{-1mm}{$+0.5$} & \raisebox{-1mm}{$+1.0$} & \raisebox{-1mm}{$+5.0$} \\[2mm] \hline
    \raisebox{-1mm}{$\hat{M}$} & 1.61 & 1.84 & 1.91 & 2.08 & 2.47 & 2.83 & 5.55  \\[2mm] \hline
    \end{tabular}
    \caption[Minimum mass in the high-order shift-symmetric model]{Minimum mass in the high-order shift-symmetric model.}
\end{table}

For a negative $\tilde{\gamma}$, we notice another interesting property of the solutions: at small masses, the apparent change in monotonicity in the $\hat{M}$-$\hat{Q}$ and $\hat{M}$-$r_s$ (see the inset) plots indicates that black holes with the same mass can correspond to different scalar charges and singularity radii. Therefore, one would expect that the black hole with the larger scalar charge, would shed some of it to reach a more favourable scalar configuration with a smaller charge. Finally, from Fig.~\ref{fig:singularity} it is pointed out that in the larger mass regime, the sign of $\tilde{\gamma}$ becomes unimportant and the cases with opposite signs merge.

In the left panel of Fig. \ref{fig:existence_QM} we present the allowed and excluded regions of the parameter space according to the existence conditions in the case of $\tilde{\gamma} = 0$, with $\tilde{\sigma}$ being on the horizontal and $\tilde{\alpha}$ on the vertical axis. For negative values of $\tilde{\alpha}$ the region plot we retrieve demonstrates an origin symmetry which was anticipated since the action \eqref{eq:action} is invariant under the simultaneous transformation $\tilde{\alpha}\rightarrow-\tilde{\alpha}$, $\tilde{\sigma}\rightarrow-\tilde{\sigma}$, and $\phi\rightarrow-\phi$.
For $\tilde{\alpha}>0,\,\tilde{\sigma}<0$ the allowed values for $\tilde{\alpha}$ increase and therefore the mass range also increases, and hairy black holes with smaller masses are found. At the same time for $\tilde{\alpha}>0,\,\tilde{\sigma}>0$, the parameter space of $\tilde{\alpha}$ shrinks and the black-hole mass range should also decrease. If we considered $\tilde{\alpha}<0$ the above conclusions would be reversed.

In Fig. \ref{fig:singularity} we display the singularity radius in this scenario and its dependence on the value of $\tilde{\sigma}$. Verifying the above, positive(negative) $\tilde{\sigma}$ leads to a larger(smaller) minimum black-hole mass.

In the $\tilde{\sigma}\ne 0$ scenario, the relation between the finite singularity radius and the normalized mass exhibits discontinuous behaviour, which is evident from the vertical jumps shown in Fig. \ref{fig:singularity}. As already explained, we identify the singularity radius as the one for which a geometric invariant (\textit{e.g.}, the Gauss-Bonnet or equivalently the Kretschmann invariant) diverges. To explain the discontinuity let us imagine that we start from some large $\hat{M}$ moving inwards towards smaller masses. At $r=r_s$ the GB invariant diverges and we identify $r_s$ as the singularity radius. There exists a second special point at $r=r_s'>r_s$ where the metric functions and the scalar field appear to lose differentiability.
The differential equations however can still be integrated for $r_s'>r>r_s$. If we plotted $r_s'$ instead of $r_s$, then the vertical jump would no longer be present and the lines would be continuous. In all cases, however, we chose to plot the singularity corresponding to the divergence of the geometric invariants. On the other hand, for positive $\tilde{\sigma}$, we do not encounter any other ``singularities'' than the ones we plot, which correspond once again to the geometric invariants diverging.

Similar discontinuities have also been encountered in Einstein-scalar-Gauss-Bonnet gravity with a quadratic exponential coupling \cite{Fernandes:2022kvg}. In the zoomed-in part of the right panel of Fig. \ref{fig:singularity}, similar behaviour to the negative $\tilde{\gamma}$ case is exhibited, where same-mass black holes have different singularity radii. This can also be understood from the $\hat{M}$-$\hat{Q}$ plot, in the right panel of Fig.~\ref{fig:existence_QM}, where a turning point appears at small masses.

\subsubsection{The \texorpdfstring{$\tilde{\kappa}$}{TEXT} term}
\begin{figure}[t]
    \centering
    \includegraphics[width=.47\linewidth]{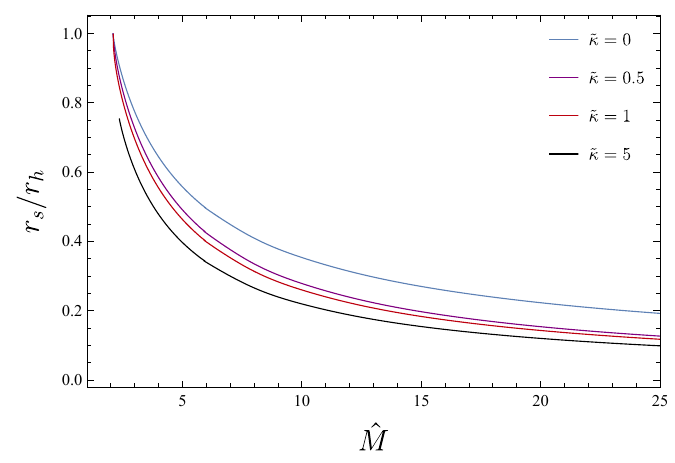}
    \hspace{5mm}
    \includegraphics[width=.47\linewidth]{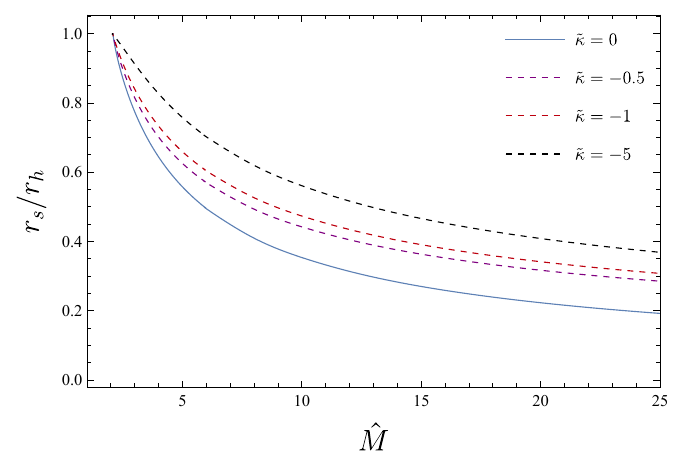}\\
    
    \includegraphics[width=.47\linewidth]{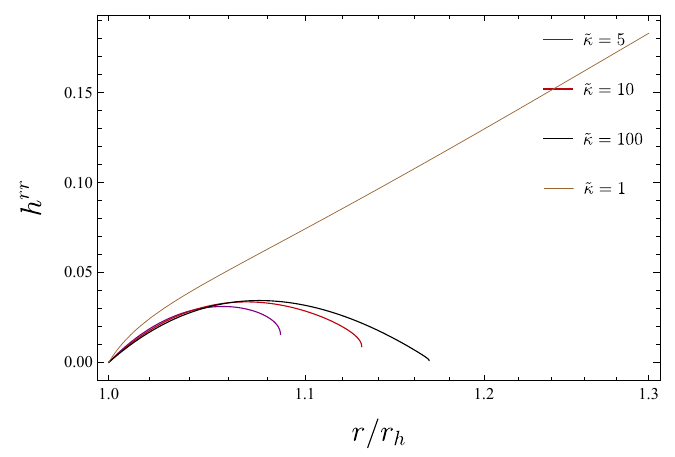}
    \hspace{5mm}
    \includegraphics[width=.47\linewidth]{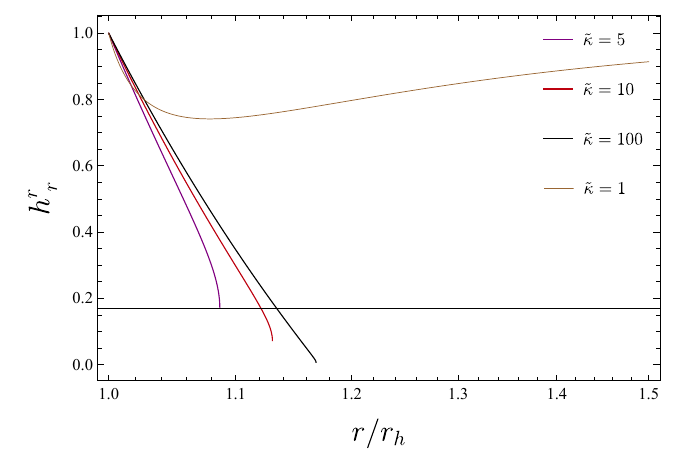}
    \caption[Plots for the $\kappa$ case]{\textit{Top:} The finite singularity radius $r_{s}$ as a function of the normalized mass $\hat{M}$, with $\tilde{\sigma} = \tilde{\gamma} = 0$. The horizon radius $r_h = 1$. \textit{Bottom Left:} $h^{rr} = B(1-3\kappa B \phi'^{2})$. \textit{Bottom Right:} $h^{r}_{\,\,r} = 1-3\kappa B \phi'^{2}$.}
    \label{fig:kappa}
\end{figure}

It is evident from equations \eqref{eq:condition_1} and \eqref{eq:condition_2} that  $\kappa$ does not enter the existence conditions. As a result, one might naively conclude that black hole solutions exist irrespective of the value that $\kappa$ takes, given that the remaining parameters satisfy the existence conditions. Contrarily, that is not the observed behaviour. If $\kappa$ is taken to be positive, then we cannot  find solutions for all values of $\alpha$ that are allowed by the existence conditions; however, if $\kappa$ is negative, then solutions could be found for all values of $\alpha$ allowed by the conditions. This behaviour is illustrated in Fig. \ref{fig:kappa}, where for negative values of $\kappa$ it is possible to saturate the existence condition and have solutions with a naked singularity, but for $\kappa > 0$, in general, that cannot be achieved.  
To better understand this trend, it is useful to rewrite the scalar equation for $\gamma = \sigma = 0$ as
\begin{align}
h^{\mu \nu} \nabla_{\mu}\nabla_{\nu}\phi \equiv \left[g^{\mu \nu}\left(1-\kappa(\nabla\phi)^{2}\right) -2\kappa \nabla^{\mu}\phi \nabla^{\nu}\phi\right] \nabla_{\mu}\nabla_{\nu}\phi
 = \alpha \GB.
\label{eq:scalar_kappa}
\end{align}

In practice, we see that when $\tilde{\kappa}>1$ not all values of $\tilde{\alpha}$, allowed by the existence conditions, yield black-hole solutions. In order to give an explanation to this issue, we numerically examine the value of the quantity inside the square brackets in eq.~\eqref{eq:scalar_kappa}, namely $h^{\mu\nu}$. Due to the symmetry of our problem only $h^{rr}$ will be nonzero.
In the bottom panels of Fig. \ref{fig:kappa}, we plot $h^{rr}$ (and $h^r_{\; r}$) for values of $\tilde{\kappa}$ spanning a few orders of magnitude, i.e. $\mathcal{O}(1)-\mathcal{O}(10^2)$. We see that for $\mathcal{O}(\tilde{\kappa})>\mathcal{O}(1)$, the quantity $h^{rr}$ approaches zero at some intermediate radius, which seems to increase as we increase the value of $\kappa$. Beyond that point, the ODE system can no longer be integrated.
This bares similarities with the behaviour of $\phi''$ at the horizon, the regulation of which yielded the existence conditions. Thus, it appears that imposing regularity for the scalar field at the horizon results in divergences appearing elsewhere for large positive values of $\tilde{\kappa}$.

\subsection{Numerical solutions vs perturbative solutions}

\begin{figure}[t]
    \centering
    \includegraphics[width=.47\linewidth]{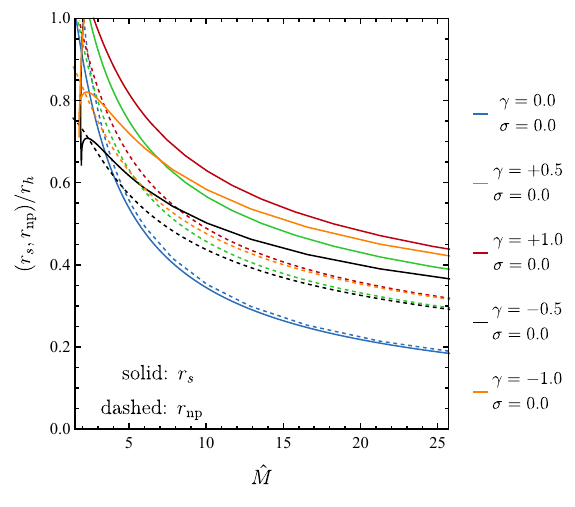}\hspace{2mm}
    \includegraphics[width=.47\linewidth]{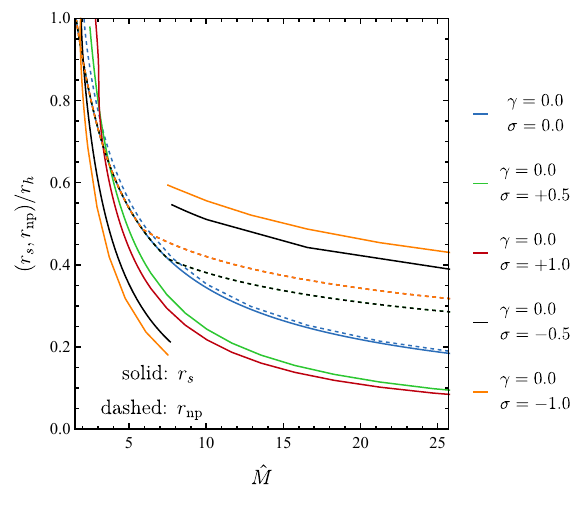}
    \label{fig:singularity_comparisons}
    \caption[Comparing the perturbative and the numerical approach]{\textit{Left}: Plot for the radii $r_{\text{np}}$ and $r_s$ for different values of $\gamma$ with $\sigma=0$. \textit{Right}: Same as left but for different values of $\sigma$ with $\gamma=0$.}
    \label{fig:comparisons}
\end{figure}

As mentioned earlier, it has already been demonstrated that in the case $\gamma=\sigma=0$ loss of perturbativity is associated with the appearance of a finite-radius singularity in the black-hole interior. 
Here we discuss the relation between the perturbative treatment breakdown radius $r_{\text{np}}$ and the finite-radius singularity $r_s$ in the general case where $\tilde{\gamma},\,\tilde{\sigma}$ are nonzero.
We present the comparative plots in Fig.~\ref{fig:comparisons}.
Verifying the results of \cite{Sotiriou:2014pfa}, we see that the radius of the singularity in the black-hole interior in the case $\tilde{\gamma}=\tilde{\sigma}=0$, is traced almost perfectly by the perturbative analysis. However, this is not the case, at least to the same level of success, when one considers the $\gamma$ and $\sigma$ contributions. From the left panel of Fig.~\ref{fig:comparisons}, we see that when $\tilde{\gamma}\ne 0$ the $r_{\text{np}}$ curve sits below the singularity radius $r_s$. From the right panel, we notice that in the $\tilde{\sigma}\ne 0$ case on the other hand the $r_{\text{np}}$ curve sits between the disconnected branches of the numerical solutions.

We can also compare the results regarding the scalar charge and mass of the black holes, by inspecting Figs.~\ref{fig:perturbative_treatment} and \ref{fig:existence_QM}, \ref{fig:existence_QM}. Specifically, for the $\tilde{\gamma}\ne 0$ case, by comparing Figs.~\ref{fig:perturbative_treatment} and \ref{fig:existence_QM} we deduce that the perturbative approach captures at least quantitatively the two main conclusions drawn by the numerical analysis: (i) For $\tilde{\gamma}>0$ the minimum black-hole mass for each $\tilde{\gamma}$ increases and the mass-parameter space of solutions shrinks. (ii) For $\tilde{\gamma}>0$ the minimum black-hole mass for each $\tilde{\gamma}$ decreases, and for small masses it is possible to retrieve black holes of the same mass with different scalar charges. Similar conclusions were also drawn in the numerical analysis $\tilde{\sigma}\ne 0$ scenario (with the addition of the discontinuities), where the qualitative trends were also captured by the perturbative analysis.

\section{Discussion}

In this chapter we studied hairy black holes in generalized scalar-tensor theories that exhibit a range of shift-symmetric derivative interactions, in addition to  the linear coupling to the Gauss-Bonnet invariant that is known to introduce black hole hair. We found that, although these additional interactions cannot introduce hair themselves, they can significantly influence the behaviour of the scalar fields near the horizon of the black hole and hence affect the configuration in general, including the value of the scalar charge for a given mass.

Interestingly $G_{\mu\nu}\nabla^\mu\phi\nabla^\nu\phi$ and
 $X\Box\phi$ modify the regularity condition on the horizon that determines the scalar charge of the black hole with respect to its mass and affects the regularity condition that determines the minimum black hole mass, whereas  $X^2$ leaves both conditions unaffected. All terms affect the scalar configuration however and a large positive coupling for $X^2$ can compromise the existence of black holes altogether. 
 
 Our two key findings are the following: (i) additional shift-symmetric interactions affect the minimum size of hair black holes, and hence the constraints one can derive from that ({\em e.g.}~\cite{Fernandes:2022kvg, Charmousis:2021npl}), but their effect is rather moderate for dimensionless couplings (with respect to the scale of the black hole in geometric units) of order 1 or less. (ii) additional shift-symmetric interactions have an effect on the scaling of the charge per unit mass versus the mass of the black hole only for masses that are fairly close to the minimum mass. Hence, sufficiently large black holes in shift-symmetric theories will not carry a significant charge per unit mass, irrespective of the presence of additional shift-symmetric interactions [c.f. \cite{Saravani:2019xwx,Maselli:2020zgv,Maselli:2021men}].

%% file: Chapters/cosmological_attractors_and_scalarization.tex
As explained in the Chapter~\ref{ch:Evasions}, spontaneous scalarization models rely on the fact that the effective mass for the scalar field $\meff^2$ depends on curvature. This allows for objects characterized by high curvature to scalarize, while objects charactirized by low curvature will be described by GR solutions with $\phi=\phi^{(0)}$. There is a subtlety though: if one treats these objects as isolated and hence asymptotically flat, as usual, then one can always assume that  $\phi=\phi^{(0)}$ asymptotically. However, in a more realistic setup the value of $\phi$ far away from the object is actually determined by cosmological considerations. As it turns out, when the coupling constant of the DEF model is such that scalarization can occur for neutron stars, GR solutions with $\phi=\phi^{(0)}$ are not attractors in late time cosmology \cite{Damour:1992kf}, see also \cite{Anderson:2016aoi} for a more recent detailed analysis. Similarly, models that exhibit black hole scalarization due to a coupling between the scalar and the Gauss-Bonnet invariant also exhibit exponential growth of the scalar during cosmological constant domination \cite{Franchini:2019npi}. Hence, without severely fine-tuning initial conditions in cosmology, localized matter configurations in the late universe could not be described by GR with $\phi=\phi^{(0)}$ and scalarization models would be effectively ruled out. 

In this chapter, we point out that (generalized) scalar-tensor theories that have GR as a cosmological attractor and still exhibit scalarization at large curvatures actually exist. We first demonstrate this by means of a simple (perhaps the simplest) example and argue intuitively about why this is expected. We then proceed to discuss the cosmology of such models a bit more thoroughly, examine how generic our results are, and explain how they would change in more general classes of scalarization models. 

\section{Framework of the model}

Let us consider the following action,
\begin{align}
\label{eq:theory}
S  =  & \frac{1}{2\kappa}  \int   d^4x \sqrt{-g}\;\bigg\{
R + X - \bigg(\frac{\beta}{2} R -\lambda L^2 \GB\bigg)\phi^2 \bigg\}
\end{align}
 where $X=-(\partial\phi)^2/2$ is once again the kinetic term of the scalar field, $\beta$, $\lambda$ are coupling constants and $L$ is an additional lengthscale that one needs to choose. We assume that the metric is minimally coupled to matter.
 The corresponding scalar equation of motion is 
\begin{equation}\label{sc}
\Box\phi + \left( \lambda L^2 \GB-\frac{\beta}{2} R\right)\phi =0.
\end{equation}
The couplings with $R$ and $\GB$ generate an effective mass for the scalar field, 
\begin{equation}\label{eq:eff_mass}
\meff^2 = \frac{\beta}{2} R-\lambda L^2  \GB\,.
\end{equation}
We are interested in models that exhibit spontaneous scalarization around compact objects so we need to demand that $\meff^2$ becomes negative at high curvature in order to trigger a tachyonic instability. For the time being our goal is to just demonstrate that this simple model can exhibit spontaneous scalarization for some type of compact objects and still have GR as a cosmological attractor. So, we restrict our attention to spherical black holes. Our GR solution will then be the Schwarzschild solution, for which we have $R=0$ and $\GB = 12 r_S^2/r^6$.

As mentioned earlier, $\GB$ is then sign-definite and the condition for having a negative $\meff^2$ becomes $\lambda > 0$ \cite{Silva:2017uqg}. For scalarization to be relevant to astrophysical black holes we need to choose $L$ to be of the order of the characteristic lengthscale of the compact object, so we choose $L\sim 10 \, {\rm km}$. Finally, we stress that for GR solutions to be admissible in the model under consideration, one should have $\phi=\phi^{(0)}=0$. Hence, this is the asymptotic value that $\phi$ would need to take for unscalarized configurations.

\section{Cosmological scales}
Next, we turn our attention to studying this theory on cosmological scales. Assuming a flat Friedman-Lema\^itre-Robertson-Walker metric:
\begin{equation}
    ds^2 = -dt^2 + \alpha(t) \left[dr^2 +r^2 \left(d\theta^2 +\sin^2\theta \,d\varphi^2 \right) \right] \,,
\end{equation}
the equation of motion for $\phi$ can be expressed as,
\begin{equation}
\label{eq:cosmo_scalar_eom}
\ddot \phi + 3H\dot \phi + \meff^2(t)\phi=0\,,
\end{equation}
where $\meff^2(t)$ is given by eq.~\eqref{eq:eff_mass} and depends on the cosmological background.
To get the evolution of the scale factor $a(t)$ we study the $tt$ component of the modified Einstein e quations
\begin{equation} \label{eq:tt_ee}
G_{tt} = \, \kappa \, \left(\rho_\phi + \rho_a \right)\,,
\end{equation}
where  $\rho_a$ denotes the energy densities of the various conventional components of the cosmic fluid and $\rho_\phi$ is an {\em effective}  energy density associated with the scalar field, given by
\begin{equation}\label{eq:sc_density}
\rho_\phi = \kappa^{-1} \left[ \dot\phi^2 +6\beta H^2 \phi^2 +12 H \phi\, \dot\phi \left( \beta - 8 \lambda L^2 H^2 \right) \right]\,.
\end{equation} 
The cosmic fluid is well approximated by a barotropic fluid whose pressure is given by $p_a = w_a \rho_a$, with the index $a=r,m, de$ and $w_a=1/3,\,-1,\,0$ for radiation domination (RD), matter domination (MD) and dark energy domination (DED) respectively.

We do not require $\phi$ to play any role in late universe cosmology, so we will assume that it is subdominant with respect to $\rho_a$.
This assumption helps avoid the gravitational wave constraints on Dark Energy theories (see \cite{Creminelli:2017sry,Ezquiaga:2017ekz,Sakstein:2017xjx,Baker:2017hug,Creminelli:2018xsv}), as discussed in detail in \cite{Franchini:2019npi}. Under the condition $\rho_\phi \ll \rho_a$
Eq.~\eqref{eq:tt_ee} simplifies to the usual Friedmann equation, $H^2\approx \kappa \rho_a/3$.
This, together with the continuity equation, $\dot{\rho}_a + 3H\rho_a(1+w_a) = 0$ allows us to simplify the expressions for the curvature terms
\begin{align}
    R = & \ 6(2H^2+\dot H) = \kappa \, \rho_a\,(1-3{w_a}), \\
    \GB = & \ 24H^2 (H^2+\dot H) = -\frac{4}{3} (\kappa \, \rho_a)^2(1+3w_a),
\end{align}
and hence the expression for the effective mass.

Let us now return to  \eqref{eq:cosmo_scalar_eom} and consider the behaviour of the scalar in different cosmological eras. Table \ref{tab:signs} summarizes the signs of the Ricci scalar, $R$, and the Gauss-Bonnet invariant, $\GB$, during each era. Note that these, together with the signs of the coupling constants $\beta$ and $\lambda$, control the sign of the effective mass. It is also worth emphasising that $R$ and $\GB$ have different dimensions and hence different scaling with time, with $\GB$ being clearly dominant at earlier times. 

\begin{table}[htb]
\centering
\begin{tabular}{c c c c}
& \hspace{3mm}Radiation\hspace{3mm} & \hspace{3mm}Matter\hspace{3mm} & \hspace{3mm}Dark Energy\hspace{3mm} \\
\hline\hline
$\GB$ & $<0$ & $<0$ & $>0$ \\
$R$ & $0$ & $>0$ & $>0$ \\
\hline
\end{tabular}
\caption{Signs of the Ricci scalar and the Gauss-Bonnet invariant during  different cosmological eras. \label{tab:signs}}
\end{table}

During RD, $R$ effectively vanishes and, hence, the mass of the scalar field is entirely controlled by the ${\GB}$ term, with $\meff^2 \simeq - \lambda L^2 \GB \approx 24 \lambda H^4 L^2$ or $\meff^2 \propto 1/t^4$, since $H\propto 1/t$. 
At very early times $\meff^2$ will dominate over the friction term in Eq.~\eqref{eq:cosmo_scalar_eom}. However, $\meff^2$ decays much faster than the Hubble friction and the latter will rapidly take over and drive $\phi$ to a constant.
The time that $\phi$ takes to freeze is approximatively given by the time at which the potential is comparable with the Hubble friction. 
After this point, it only takes a few Hubble times for $\dot\phi$ to effectively vanish. More concretely, $\meff \lesssim H \Rightarrow H(z) \times L \lesssim 1$, which happens very early, around the redshift $z\approx10^{11}$ for our choice of $L$. As a result, the scalar field is already frozen to a constant solution well before MD.

At the onset of MD, $\phi$ starts evolving again. 
This is because $R$ no longer vanishes on cosmological scales and thus it provides a non negligible contribution to $\meff^2$. 
The contribution of the ${\GB}$ term in $\meff^2$ has actually become largely subdominant to that of the $R$ term because of their different scaling.
During MD, $H\ll L^{-1}$.

As has been pointed out in \cite{Andreou:2019ikc}, action \eqref{eq:theory} with $\lambda=0$ is related by a simple field redefinition to a linearized version of the DEF model. In fact, we have defined $\beta$ such that $8\beta=\beta_{\rm DEF}$ in the appropriate limit. Nonlinearities are not important in our regime. As a result, one expects that once the ${\GB}$ term in our theory has become negligible, cosmological evolution will match that of the DEF model. Interestingly, the latter actually exhibits our desired cosmological behaviour for $\beta>0$ \cite{Damour:1992kf}: GR is a cosmological attractor! Hence the scalar field will naturally be driven to $\phi=0$. 
The transition to DED does not chance the dynamics of the scalar qualitatively and GR with $\phi=0$ continues to be the attractor.

\section{Dynamical evolution}
All of the above can be verified by studying the scalar dynamics quantitatively.
We can chose to either express everything with respect to the time parameter $\tau=M \,t$, where $M$ is the mass scale of the problem, or the redshift $1+z=1/a$. In the former case it takes the following form
\begin{equation}
\label{cosmo_scalar_eom2}
\frac{d^2\phi_{(a)}}{d\tau^2}+f_a^{(\tau)}\frac{d\phi_{(a)}}{d\tau}+q_a^{(\tau)}\phi_a=0,
\end{equation}
where the coefficients in the different epochs are given by
\begin{equation}
f_r^{(\tau)}=\frac{3}{2\tau},\quad q_r^{(\tau)}=\frac{3\lambda}{2\tau^4}, \quad
f_m^{(\tau)}=\frac{2}{\tau},\quad q_m^{(\tau)}=\left(\frac{128\lambda}{27\tau^4}-\frac{4\beta}{3\tau^2} \right).
\end{equation}
In this work, however, we chose to work with respect to the redshift. Then, Eq.~\eqref{eq:cosmo_scalar_eom} in terms of the redshift takes the following form:
\begin{equation}
    \phi_{(a)}''+f_a\phi_{(a)}'+q_a\phi_{(a)}=0,
    \label{eq:cosmo_scalar_eom2}
\end{equation}
where prime denotes differentiation with respect to $z$, with
\begin{equation}
f_a(z)=\frac{H'(z)}{H(z)}-\frac{2}{z+1}\,,\quad
q_a(z)=\frac{24 L^2 H(z)^2 (\lambda +3 \lambda  w_a)+3 \beta (3 w_a-1)}{(z+1)^2}\, .
\end{equation}
We begin our numerical analysis at $z_{i} = 10^{10}$,  just before Big Bang Nucleosynthesis (BBN). 
To set the initial conditions for the scalar field and its derivative, we assume that $\phi$ is just coupled with the thermal bath.
Therefore a natural initial value is $\phi_{i} \simeq H(z_{i}) / \kappa \ll 1$.  
The initial value $\phi'_{i}$ can be, instead, derived from $\dot \phi_{i}$: we expect $\dot \phi_{i} \simeq H(z_{i}) \phi_{i} \Rightarrow \phi'_{i} \simeq \phi_i / z_{i}$, which is, again, much smaller than unity.
These two conditions ensure that $\rho_{\phi}(z_{i}) \ll \rho_r(z_{i})$ and are hence consistent with the assumption that $\phi$ is cosmologically subdominant. 
We stress that $\phi \sim 1$ would imply Planckian energy scales in our units and hence initial conditions with $\phi_{i} \ll 1$ do not constitute fine tuning.

\begin{figure*}[t]
\begin{tabular}{cc}
\includegraphics[width=0.9\textwidth]{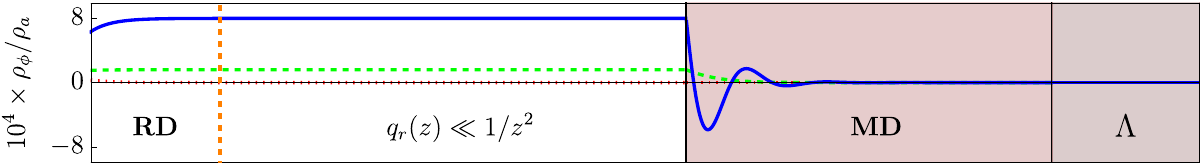} & \\[-2mm]
\includegraphics[width=0.9\textwidth]{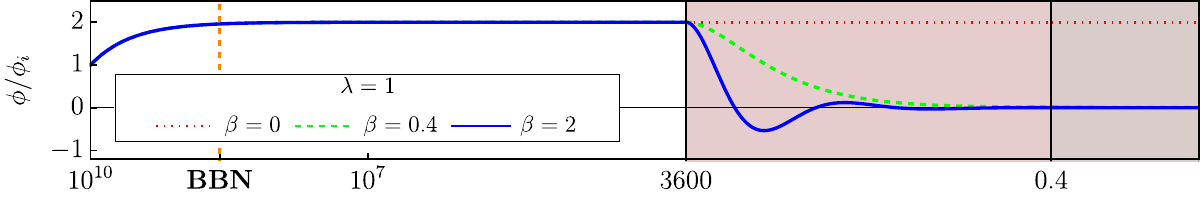} & \\
\end{tabular}
\caption[Cosmological evolution of the scalar field and the scalar energy density]{\label{fig:fig1} {\em Top panel:} Effective energy density of the scalar $\rho_\phi$ over  the energy density of the cosmic fluid $\rho_a$ as a function of redshift. {\em Bottom panel:} Evolution of the scalar field $\phi$ in units of its a reference value $\phi_i$, fixed at $z=10^{10}$.}
\end{figure*}

Fig.~\ref{fig:fig1} shows the evolution of the scalar and of the ratio $\rho^\phi/\rho_a$
for $z<z_i$. $\rho_\phi$ remains subdominant  as expected and the plots confirm the qualitative behaviour described previously. In particular, $\phi$ remains constant throughout, with the exception of transitions between cosmological eras.  The ratio of the energy density with respect to the redshift, is given by the following expression
\begin{equation}\label{eq:density_ratio}
\begin{split}
    \rho^\phi_a/\rho_a=\frac{1}{3}\bigg[12 (z+1) \phi_{(a)}  \phi_{(a)}' &\left(\beta -8 \lambda L^2 H^2  \right)\\
    &+(z+1)^2 \phi_{(a)}'^2+6 \beta  \phi_{(a)} ^2\bigg]\, .
\end{split}
\end{equation}

The value that $\phi$ takes at late times does depend crucially on $\beta$. For $\beta=0$, $\phi$ effectively remains frozen to the value it has in the early RD era. Unless this value is set to be extremely close to zero by fine tuning initial data, any local configuration in the late universe will have to be scalarized because cosmological asymptotics will be incompatible with having unscalarized configurations. As discussed in the begining of this chapter, this would clash with weak field constraints. For $\beta>0$ instead, $\phi\to 0$ during MD and GR with $\phi=0$ becomes a cosmological attractor. To approach this attractor fast enough, $\beta$ should be of order unity so that the oscillations seen in Fig.~\ref{fig:fig1}
at the onset of MD are nearly \emph{critically damped}. 
These oscillations correspond to changes on the effective Newton's constant  that will, in principle, affect the formation of Large Scale Structures. 
However, the time scale of the oscillations is very large, of order of the Hubble rate. Moreover, the corrections to Newton's constant would be $\propto |\beta| \, \Delta \phi^2$, and hence negligible.
In summary, cosmic evolution is expected to be almost identical to GR for late times.

\begin{figure}[t]
\begin{tabular}{c}
\includegraphics[width=0.47\textwidth]{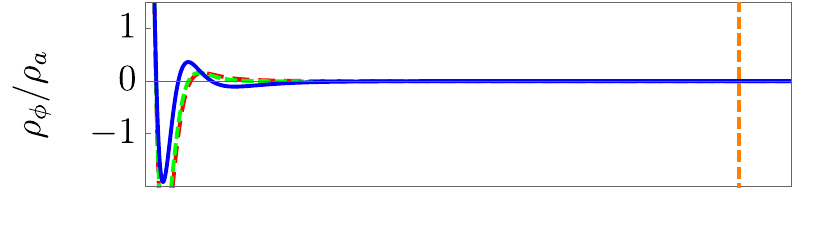} \\[-7.3mm]
\includegraphics[width=0.47\textwidth]{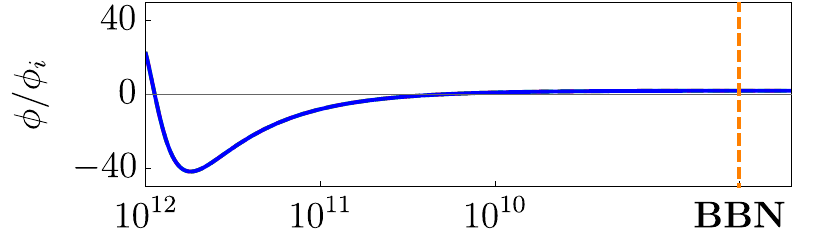}
\end{tabular}\hspace{-5mm}
\begin{tabular}{cc}
\includegraphics[width=0.47\textwidth]{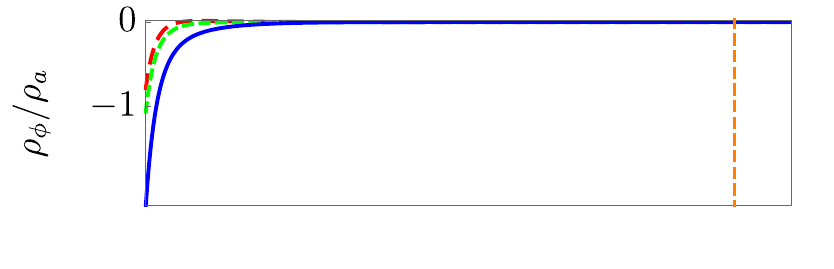} & \\[-10mm]
\includegraphics[width=0.47\textwidth]{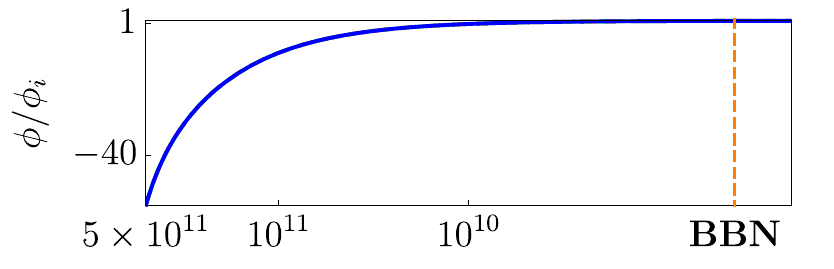} & \\
\end{tabular}
\caption[Early-time cosmic evolution for a positive coupling]{\textit{Left:} Same as Fig.~\ref{fig:fig1} but for high redshifts. \textit{Right:} Same as Fig.~\ref{fig:fig1} but for high redhsifts and $\lambda=-1$.}
\label{fig:fig2}
\end{figure}

Fig.~\ref{fig:fig2} shows the evolution of $\rho_\phi$ and $\phi$ for $z>z_i$ and the very early epochs before recombination. As anticipated in our qualitative analysis, for $z \gg 10^{11}$, the significant contribution of the ${\GB}$ term to the effective mass results in a sinusoidal behaviour. The oscillation is damped by Hubble friction when we move forward in time. $\rho_\phi$ also shows oscillatory behaviour and, moving to higher redshift, the oscillations are amplified. Eventually, our approximation that $\phi$ is subdominant ceases to be valid. It is worth emphasising that $\rho_\phi$ does not need to remain positive, as it is just an effective energy density.

All of the above referred to $\lambda, \beta>0$. Next we discuss the case $\beta>0$, $\lambda<0$. 
On astrophysical scales, $\lambda < 0$ leads to spontaneous scalarization triggered by a tachyonic instability in the interior of neutron stars \cite{Silva:2017uqg}. 
As the previous analysis has already shown, on cosmological scales the $\lambda$ term has an impact only at very early times, before $z\simeq10^{11}$. Indeed, numerical analysis confirms that flipping the sign of $\lambda$ makes no difference during BBN and at later times. However, as seen from Table \ref{tab:signs}, for $\lambda<0$, the $\lambda {\GB}$ contribution to $\meff^2$ will be negative and will lead to exponential growth of $\phi$ once one reaches sufficiently $z$ for the mass contribution to dominate over Hubble friction. 
As shown in Fig.~\ref{fig:fig2}, $\rho_\phi/\rho_a$ grows exponentially fast and reaches $1$ a lot earlier than in the $\lambda>0$ scenario. 

\begin{figure}[t]
\begin{tabular}{c}
\includegraphics[width=0.47\textwidth]{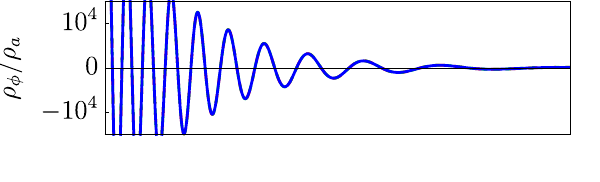} \\[-7.6mm]
\includegraphics[width=0.47\textwidth]{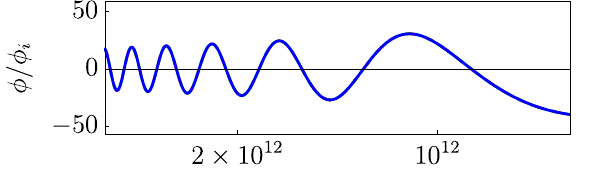}
\end{tabular}\hspace{-5mm}
\begin{tabular}{cc}
\includegraphics[width=0.47\textwidth]{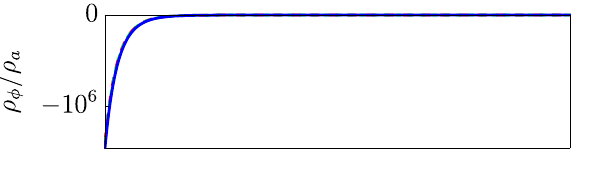} & \\[-9.2mm]
\includegraphics[width=0.47\textwidth]{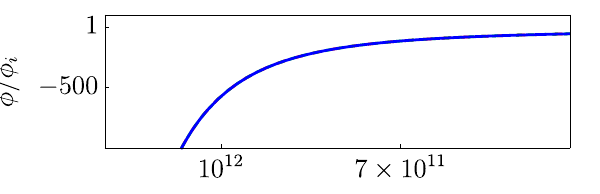} & \\
\end{tabular}
\caption[Early-time cosmic evolution for a negative coupling]{\textit{Left:} Same as Fig.~\ref{fig:fig1} but for very high redshifts. \textit{Right:} Same as Fig.~\ref{fig:fig1} but for very high redshifts and $\lambda=-1$.}
\label{fig:very_early}
\end{figure}

Note that, since scalarization relies on curvature couplings, it is rather intuitive that the terms that trigger it will become relevant in the very early universe. The coupling with the Gauss-Bonnet invariant is the dominant one at large curvatures and its coupling constant is dimensionful. As such, it controls the curvature scale at which departure from standard cosmology would appear. This would happen when the universe is of the size of a few kilometres, well before BBN, for values of the coupling that are compatible with compact object scalarization. At earlier times, departures from standard cosmology would be significant, as our results show, and as has been pointed out in the literature \cite{Anson:2019uto}. However, it is quite a stretch to consider these models as good effective field theories, and hence take their predictions seriously, all the way to energy scales where the universe is the size of kilometres. Instead, it seems sensible to try to embed them in a theoretical framework valid at those scales, and eventually in a suitable UV completion with the appropriate inflationary cosmology. For completeness in Fig.~\ref{fig:very_early} we depict the behaviour of our solutions at very early times. For positive values of $\lambda$ we notice that the energy density blows up in a sinusoidal fashion. For negative $\lambda$ this happens monotically.

Finally, we consider $\beta<0$. For $\lambda=0$, one expects to recover the results of Refs.~\cite{Anderson:2016aoi}. In fact, for any value of $\lambda$ one will have a tachyonic instability  on cosmological scales at late times. This instability will be very slow, so it is not particularly threatening in its own right. However, without an attractor mechanism at late times, severe tuning of initial conditions would be needed to have GR configurations locally (see $\beta=0$ case) and the instability would only make things worse.

 \section{Discussion}

 To conclude this chapter, we have demonstrated by using a specific model as an example, that the phenomenon of spontaneous scalarization around compact objects is compatible with having 
 an attractor mechanism to GR on cosmological scales. In fact, our results show that fairly simple scalarization models can track GR cosmology over a vast range of redshift and all the way back to BBN. The key feature that leads to the desired behaviour is that the scalar can couple in two different ways to curvature --- through the Gauss-Bonnet invariant and through the Ricci scalar --- with one coupling triggering scalarization locally and the other providing a late time attractor cosmologically.
 
The action we have considered is rather minimal, as it only includes terms that contribute to linearized perturbations around GR solutions with constant scalar. It is sufficient for discussions about the onset of scalarization and whether GR is a cosmological attractor. However, the properties of scalarized solutions will be controlled by the nonlinear (self)interactions of the scalar that one can add to our action \cite{Macedo:2019sem,Andreou:2019ikc}. Hence, there is actually a wide variety of scalarization models with the desired cosmological behaviour at late time and different properties for compact objects. In the next chapters we will look into the properties of the scalarized compact objects arising in this theory.

%% file: Chapters/sRGB_bh.tex
In Chapter~\ref{ch:Cosmology} we examined a simple scalar-tensor model that is motivated from compact object scalarization and is consistent with the appearance of a late time cosmological attractor. This model is slightly more complicated than the ones proposed initially in \cite{Silva:2017uqg, Doneva:2017bvd} as a coupling of the scalar with the Ricci scalar is also assumed in addition to the one with the GB invariant. The compact objects in this model are therefore expected to present differences in comparison with the EsGB model solutions.

As has been stressed in Ref.~\cite{Silva:2018qhn}, although the onset of scalarization is determined by terms that are linear (in the equations) in the scalar, the properties of the scalarized object depend crucially on nonlinear interactions, as these are the ones that quench the linear instability and determine its endpoint. Non-linearities can originate from scalar self-interactions \cite{Macedo:2019sem}, from the coupling function to $\GB$ \cite{Doneva:2017bvd}, and from the backreaction of the scalar onto the metric.
The potential coupling between the scalar field and the Ricci scalar, $R$, has mostly been disregarded in the case of black holes.

This is entirely justified when studying the onset of scalarization, as GR black holes have a vanishing $R$. However, it is bound to have an effect on the properties of scalarized objects, as it will contribute to the nonlinear quenching of the tachyonic instability that leads to scalarization. Indeed, as soon as the scalar becomes nontrivial, $R$ will cease to be zero and it will contribute directly to the effective mass of the scalar. From an effective field theory (EFT) perspective there seems to be no justification to exclude such a coupling. As we saw in the previous chapter this coupling makes GR a cosmological attractor and hence reconciles Gauss-Bonnet scalarization  with late-time cosmological observations. It has also been pointed out in Ref.~\cite{Ventagli:2020rnx} that this coupling can help suppress scalarization of neutron stars and hence evade the relevant constaints. This will be further explored in the next chapter.

In this section we examine the role a coupling with the Ricci scalar can have on scalarized black holes and their properties. We consider the simplest model that contains a coupling with both the Ricci scalar and Gauss-Bonnet invariant, and we  study static, spherically symmetric black holes. We explore the region of existence of scalarized solutions when varying both couplings and the black hole mass. We examine the influence of the Ricci coupling  on the scalar charge of the black holes, which is the quantity that controls the deviations from GR in the observation of binaries. We also discuss the role this coupling can play in stability considerations even though a more detailed analysis will follow in Chapter~\ref{ch:Stability}. 

\section{Setup}
We will consider the same action as in the previous chapter but now we absorb the length scale into the definition of the $\GB$ coupling parameter, \textit{i.e.}
\begin{equation}
\label{eq:ActionGeneric}
    S=\frac{1}{2\kappa}\int d^4x\sqrt{-g\,}\bigg\{R-\dfrac12(\partial\phi)^2
    -\left(\frac{\beta}{2}R-\alpha\GB\right)\frac{\phi^2}{2} \bigg\}\,,
\end{equation}
where $\kappa = 8\pi G/c^4$, $\beta$ is a dimensionless parameter, while $\alpha$ has dimensions of length squared.  The normalization of $\beta$ is chosen to match the standard DEF literature. From now on we will be referring to this model as the Einstein-scalar-Ricci-Gauss-Bonnet model (EsRGB).

One can consider the above action as part of an EFT in which the scalar enjoys $\phi \to -\phi$ symmetry while shift symmetry is broken only by the coupling to gravity. For linear perturbations around solutions that solve the vacuum Einstein equations, the $\phi^2 \GB$ coupling will be the leading correction to GR \cite{Andreou:2019ikc}. However, more generally, the $\phi^2 R$ term comes with a lower mass dimension and provides a direct contribution to the effective mass. Hence it is expected to play a crucial role in the non-linear quenching of the tachyonic instability that one associates with scalarization, and in determining the properties of scalarized black holes. The complete EFT would include more terms that can contribute to the effective mass (nonlinearly), such as the operators $R\phi^4$ and $G^{\mu\nu}\partial_\mu\phi\partial_\nu\phi$. These operators would enlarge the parameter space, while they are characterized by a higher mass dimension than $\phi^2 R$. We will neglect them in our analysis and we do not expect them to change significantly the final results qualitatively. The modified Einstein equation is
\begin{equation}
\label{eq:grav-BH}
\begin{split}
    G_{\mu\nu}=T^{(\phi)}_{\mu\nu}\equiv &-\frac{1}{4}g_{\mu\nu}(\nabla\phi)^2+\frac{1}{2}\nabla_\mu\phi\nabla_\nu\phi+\frac{\beta\phi^2}{4}G_{\mu\nu}+\frac{\beta}{4}\big(g_{\mu\nu} \nabla^2\\
    &-\nabla_\mu\nabla_\nu \big)\phi^2
    -\frac{\alpha}{2g}g_{\mu(\rho}g_{\sigma)\nu}\epsilon^{\kappa\rho\alpha\beta}\epsilon^{\sigma\gamma\lambda\tau}R_{\lambda\tau\alpha\beta}\nabla_{\gamma}\nabla_{\kappa}\phi^2\, ,
\end{split}
\end{equation}
where $\epsilon^{\mu\nu\rho\sigma}$ is the ant-symmetric symbol in four dimensions, and $T^{(\phi)}_{\mu\nu}$ is the energy momentum tensor contribution that comes from the variation of the $\phi$-dependent part of the action with respect to the metric. The scalar field equation reads
\begin{equation}\label{eq:scal_eq}
    \Box \phi =m_\text{eff}^2\,\phi\,,\quad \text{where} \quad \meff^2=\frac{\beta}{2}R-\alpha \gb.
\end{equation}
The effective mass clearly receives two contributions, one from the Ricci and one from the Gauss-Bonnet coupling as already explained in the previous chapter.

\section{Scalarization threshold}\label{subsec:threshold}

\begin{figure}[ht]
    \centering
    \includegraphics[width=\textwidth]{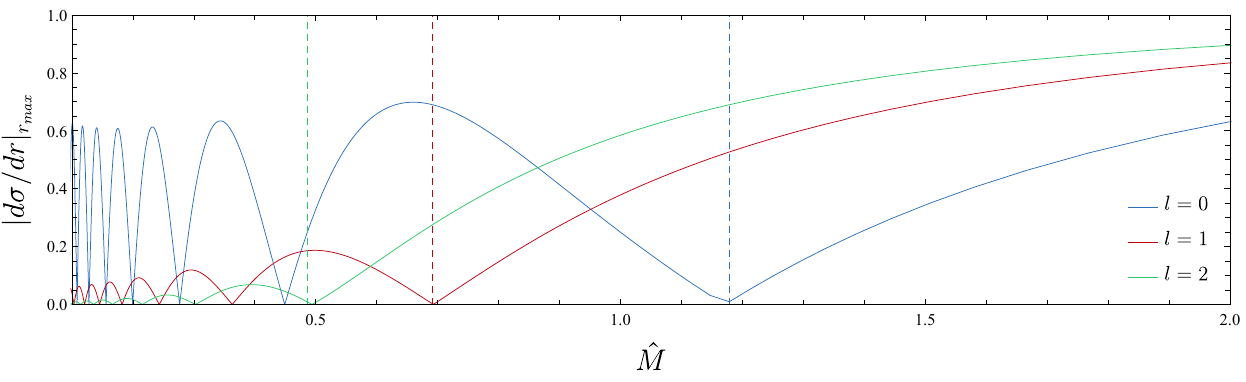}
    \caption[Scalarization thresholds]{Numerical solution of the decoupled scalar equation on a GR background. The points where the line touches the horizontal axis correspond to the scalarization thresholds $\hat{M}^{(n,l)}_{\text{th}}$.}
    \label{fig:fig_Mth}
\end{figure}

Let us begin the analysis by deriving the scalarization onset for black holes in this model. In order to do that, we look at the linearized equation for scalar perturbations in the scalar equation of motion. Assuming spherical symmetry we are allowed to decompose the scalar in the following way
\begin{equation}
    \phi(t,r,\theta,\varphi)=\int_{-\infty}^{+\infty} e^{-i\omega t}\left[\;\sum_{m=-l}^{+l}\;\sum_{l=0}^{\infty}\frac{\sigma_l(r,\omega)}{r}\;Y^m_l(\theta,\varphi)\,e^{-im\varphi}\right] d\omega\,,
\end{equation}\\
where $Y^m_l(\theta,\varphi)$ are the spherical harmonics. In the usual spherical coordinates assumed in \eqref{eq:metric}, the scalar equation on a Schwarzschild background reads
\begin{equation}\label{eq:scalar_pert}
    r(r-\rh)\, \sigma''+\rh\, \sigma' +\left[ \frac{\omega^2r^3}{r-\rh} + \frac{12\alpha \rh^2}{r^4} - \frac{\rh}{r} -l(l+1) \right]\, \sigma = 0\,.
\end{equation}\\
As already mentioned, the $\beta$ term is absent since the Ricci invariant vanishes on a Schwarzschild background.
This equation can be recast into a Schr\"odinger type form by changing our coordinate from the radial one $r\in (0,\infty)$ to the tortoise one. The latter is defined for the general spherically symmetric background \eqref{eq:metric} by the equation
\begin{equation}
    dr_*= (A\,B)^{-1/2}dr\,.
\end{equation}
As is well known, in the case of the Schwarzschild background $r_*=r+2M\ln\left(r/{2M}-1\right)$, so then \eqref{eq:scalar_pert} can be recast into
\begin{equation}
    \label{eq:qnm_eq}
    \left(-\frac{d^2}{d r_*^2}+V_{\text{eff}}\right)\sigma\,=\,\omega^2\sigma\,,
\end{equation}
where the effective potential is given by
\begin{equation}
    V_{\text{eff}}=\left(1-\frac{2 M}{r}\right) \left[\frac{2 M}{r^3}-\frac{48 \alpha M^2}{r^6}+\frac{\ell  (\ell +1)}{r^2}\right] \, .
\end{equation}
The scalarization onset corresponds to the emergence of unstable solutions when one performs a scanning of the parameter space. Therefore we will look for purely imaginary modes $\omega\rightarrow i\omega$, in which case \eqref{eq:qnm_eq} yields
\begin{equation}
    \frac{d^2\sigma}{d r_*^2}\,=\,\left(V_{\text{eff}}+\omega^2\right)\sigma\,,
\end{equation}
We will be looking for bound solutions with energy $E=-\omega^2<0$ which correspond to the appearance of an instability with a growth parameter given by $\omega=\sqrt{-E}$.
In practise we vary the parameter $\alpha$ which changes the effective potential. Before doing that, we set $\omega=0$, which from continuity should correspond to a bound state too. The last step significantly simplifies our process. Then, we determine the values of the parameters (in this case just $\alpha$) for which $d\sigma/dr_*\rightarrow 0$ in the limit $r_*\rightarrow \infty$. In this case the effective potential (and hence the onset of scalarization) is controlled only by the rescaled mass $\hat M =M/\sqrt{\alpha}$, where $M$ is the mass of the black hole. GR solutions will become unstable for small values of $\hat M$, which correspond to large curvatures or large  $\alpha$ couplings. We search for such bound states while varying $\hat M$ and identify the threshold rescaled masses for different $n$ and $l$, namely $\hat{M}^{(n,l)}_{\text{th}}$, ($n,l=0,1,2,$ etc). This is shown in Fig.~\ref{fig:fig_Mth} and the results are presented in Table~\ref{tab:scalarization_thresholds}.
The mode associated with a threshold mass $\hat{M}^{(n)}_{\text{th}}$ has $n$ nodes. Hence, whenever $\hat M<\hat{M}^{(n)}_{\text{th}}$, GR black holes become unstable to a perturbation with $n$ nodes. Numerically, for $l=0$, these thresholds are $(n,\hat{M}^{(n)}_{\text{th}})\approx (0,1.179),\, (1,0.449),\, (2,0.277)$, etc.
Those along with more threshold results for different pairs $(n,l)$ are presented in Tab.~\ref{tab:scalarization_thresholds}.

\begin{table}[ht]
\centering
\begin{tabular}{ c c c c }
$\hat{M}_{\text{th}}^{(n,l)}$ & $l=0$ & $l=1$ & $l=2$ \\[1mm]
\hline\hline
$n=0$ & 1.179 & 0.692 & 0.494 \\  
$n=1$ & 0.449 & 0.363 & 0.302 \\ 
$n=2$ & 0.277 & 0.243 & 0.215 \\ 
\end{tabular}
\caption[Thresholds of scalarization]{Thresholds of scalarization for the first three modes and angular number $l=0,1,2$.}
\label{tab:scalarization_thresholds}
\end{table}

\section{Static, spherically symmetric black holes}
We will now derive the equations in the coordinate system we chose, namely \eqref{eq:metric}. The field equations can be cast as three coupled ordinary differential equations. The $(rr)$ component of the metric equations can be solved algebraically with respect to $B^{-1}$:
\begin{equation}\label{eq:grr}
    B^{-1}= \frac{-\mathcal{B}+\delta\sqrt{\mathcal{B}^2-4\mathcal{A}\,\mathcal{C}}}{2\mathcal{A}},\,\delta=\pm 1,
\end{equation}
where
\begin{align}
    \mathcal{A}=&\;4-\beta  \phi^2\, ,\\
    \begin{split}
        \mathcal{B}=
        &\; \beta  \phi ^2+A' (\beta  r^2 \phi \, \phi '-8 \alpha  \phi\,  \phi '+\beta  r \phi ^2-4 r)/A\\
        &+r^2 \phi '^2+4 \beta  r \phi\,  \phi '-4\, ,
    \end{split}
    \\
    \mathcal{C}=&\; 24 \alpha  A' \phi\, \phi '/A\, .
\end{align}
and a prime denotes differentiation with respect to the radial coordinate. By substituting \eqref{eq:grr} in the remaining field equations, we end up with a system of two coupled second order differential equations just like in Chapter~\ref{ch:Evasions}
\begin{align}
    \left(\frac{A'}{A}\right)'=&\; \tilde{A}\,(r,A'/A,\phi,\phi',\alpha,\beta)\label{eq:dif_1},\\
    \phi''=&\; \tilde{\phi}\,(r,A'/A,\phi,\phi',\alpha,\beta)\label{eq:dif_2}.
\end{align}

In order to search for black hole solutions, we assume the existence of a horizon, where $A,B\rightarrow 0$. In line with previous results for different models fashioning a coupling with $\GB$ (\textit{e.g}.~\cite{Kanti:1995vq,Sotiriou:2014pfa}), only $\delta=+1$ leads to black hole solutions.

\section{Near-horizon expansion}

As in Chapter~\ref{ch:Evasions}, near the horizon, one can perform the following expansion:
\begin{align}
    A(r\approx r_h)= & \,\alpha_1 (r-r_h)+\alpha_2 (r-r_h)^2+...\label{eq:exp_h1}\\
    B(r\approx r_h)= & \,\beta_1 (r-r_h)+\beta_2 (r-r_h)^2+...\label{eq:exp_h2}\\
    \phi(r\approx r_h)= & \,\phi_h +\phi_1 (r-r_h) +\phi_2 (r-r_h)^2+...\label{eq:exp_h3}
\end{align}
One can substitute these expressions in Eqs.~\eqref{eq:grr}, \eqref{eq:dif_1} and \eqref{eq:dif_2}, and obtain a near-horizon solution. In particular, $\phi''_\text{h}$ remains finite only provided that
\begin{equation}\label{eq:phi_h}
         \phi'(r_\text{h})=\phi_1= \big(a+\sqrt{\Delta}\big)/b,
\end{equation}
where the expressions for $a,\;b$ and $\Delta$ are as follows:
\begin{align}
    a =& \;24 \alpha  \beta  r_h \phi _h^2+r_h^3 \left(-3 \beta ^2 \phi _h^2+\beta  \phi _h^2-4\right)\, ,
    \\
    \begin{split}
        \Delta =&\; 9216 \alpha ^3 \beta  \phi _h^4+r_h^6 \left(3 \beta ^2 \phi _h^2-\beta  \phi _h^2+4\right)^2\\
        &-192 \alpha ^2 r_h^2 \phi _h^2 \left(9 \beta ^2 \phi _h^2-2 \beta  \phi _h^2+8\right)\, ,
    \end{split}
    \\
    \begin{split}
        b =&\; 2 \phi _h \left(8 \alpha -\beta  r_h^2\right) \big[24 \alpha  \beta  \phi _h^2+r_h^2 \left(-3 \beta ^2 \phi _h^2+\beta  \phi _h^2-4\right)\big]\\
        &/(\beta \phi_h^2-4)\, .
    \end{split}
\end{align}

Requiring that $\Delta\geq0$  defines a region on the $(r_\text{h},\phi_\text{h})$ space where regular black hole solutions with scalar hair can be found.

\section{Asymptotic expansion}

In order to analyze the asymptotic behaviour of the solutions, one can perform a suitable expansion, and solve the equations near spatial infinity imposing that $\phi$ vanishes there. This yields 
\begin{align}
    \begin{split}\label{eq:exp_f1}
        g_{tt}=&\; 1-2M\big/{r}+\beta \, Q^2\big/{4\,r^2} +\big(M Q^2-3 \beta  M Q^2\big)\big/{12\, r^3}\\
        &+\big(8 M^2 Q^2-28 \beta  M^2 Q^2-3 \beta ^3 Q^4+5 \beta ^2 Q^4-\beta  Q^4\big)\big/{48\, r^4}\\
        &+\big(288 M^3 Q^2
        -1040 \beta  M^3 Q^2+3072 \alpha  M Q^2-60 \beta ^3 M Q^4\\
        &+115 \beta ^2 M Q^4+10 \beta  M Q^4
        -9 M Q^4\big)\big/{960r^5}+\mathcal{O}\left(1/r^6\right) \, ,
    \end{split}\\[3.5mm]
    \begin{split}\label{eq:exp_f2}
        g_{rr}=&\; 1+2M\big/r+\big(16 M^2+2 \beta \, Q^2-Q^2\big)\big/{4\,r^2}+\big(32 M^3-5 M Q^2\\
        &+11 \beta  M Q^2\big)\big/{4\, r^3}+\big(488 \beta  M^2 Q^2-208 M^2 Q^2+768 M^4\\
        &-12 \beta ^3 Q^4
        +17 \beta ^2 Q^4-13 \beta  Q^4+3 Q^4\big)\big/{48\, r^4}+\big(6064 \beta  M^3 Q^2\\
        &-2464 M^3 Q^2+6144 M^5
        -1536 \alpha  M Q^2\-348 \beta ^3 M Q^4\\
        &+589 \beta ^2 M Q^4-442 \beta  M Q^4+97
        M Q^4\big)\big/{192\, r^5}
        +\mathcal{O}\left(1/r^6\right) \, ,
    \end{split}\\[3.5mm]
    \begin{split}\label{eq:exp_f3}
        \phi=&\; Q\big/r+M Q\big/r^2+\big(32 M^2 Q-3 \beta ^2 Q^3+2 \beta  Q^3-Q^3\big)\big/24\, r^3\\
        &+\big(48 M^3 Q-9 \beta ^2 M Q^3+9 \beta  M Q^3-4 M Q^3)\big/24\,r^4\\
        &+\big(2240 \beta  M^2 Q^3
        -1680 \beta ^2 M^2 Q^3-928 M^2 Q^3-4608 \alpha  M^2 Q\\
        &+6144 M^4 Q+117 \beta ^4 Q^5
        -144 \beta^3 Q^5+86 \beta ^2 Q^5-40 \beta  Q^5\\
        &+9 Q^5\big)\big/{1920 \, r^5}+\mathcal{O}\left(1/r^6\right) \, ,
    \end{split}
\end{align}
where as always in this thesis, $M$ is the ADM mass and $Q$ is the scalar charge. Equations \eqref{eq:exp_f1}-\eqref{eq:exp_f3} suggest, as one would expect, that the Ricci coupling dominates over the Gauss-Bonnet coupling at large radii. Specifically, the Ricci coupling appears at order $r^{-2}$, whereas the Gauss-Bonnet coupling appears initially at order $r^{-5}$.

\section{Numerical implementation}
\label{sec:Numerical-BHs}

The system of ordinary differential equations (ODEs) \eqref{eq:dif_1} and \eqref{eq:dif_2} can, in principle, be solved by starting from the horizon and integrating towards larger radii.   $\alpha$ and $\beta$ are theoretical parameters that are considered fixed. 
The values of $A'$, $A$, $\phi'$, and $\phi$ at $r=r_{h}$ appear to be ``initial data''. However, they are not all free to choose. $A(r_{h})$ is fixed by the condition $A(r_{h})=0$, {\em i.e.}~the fact that $r=r_{h}$ is a horizon. $A'/A$ has to diverge at $r=r_{h}$, or else $A$ will have a vanishing derivative on the horizon. Finally, $\phi'(r_{h})$, and $\phi(r_{h})$ are related by the regularity condition \eqref{eq:phi_h}.  
One also needs to fix $r_{h}$. The field equations are invariant under the global scaling symmetry $r\to\mu r$, $\alpha\to\mu^2\alpha$, where $\mu$ is a free parameter. We can make use of this symmetry to reduce the space of parameters that we have to explore. Practically, we can decide that the horizon is located at $r_{h}=1$; solutions with $r_{h}\neq1$ can later be obtained by a global scaling.

Hence, one can treat $\phi(r_{h})=\phi_{h}$ as the only free parameter. Integrating outwards, one will generically find a solution for arbitrary $\phi_{h}$. However, for given $\alpha$ and $\beta$, only one value of $\phi_{h}$ has the desired asymptotics, namely $\phi(r\to \infty)=\phi_\infty=0$. Imposing this condition (by a shooting method and to a desired precision) yields a unique solution. The global rescaling mentioned above turns this solution into a one-parameter family, that we can interpret as a family of black holes parametrized by their ADM mass $M$, for fixed couplings $\alpha$ and $\beta$. The scalar charge $Q$ is then determined as a function of $M$, $\alpha$ and $\beta$. 

A practical complication is that the regularity condition of Eq.~\eqref{eq:phi_h} cannot be imposed numerically with any reasonable accuracy. To circumvent this problem we start the numerical integration at $r\approx r_\text{h}[1 + \mathcal{O}(10^{-4})]$, similarly to what we did in Chapter~\ref{ch:Mass-Charge}, and we use the perturbative expansion in Eqs.~\eqref{eq:exp_h1}-\eqref{eq:exp_h3} to impose the regularity and propagate the data from the horizon to the starting point of the numerical integration. We typically integrate up to distances $r/r_\text{h}\approx 10^4$ and impose that $\phi$ vanishes there to a part in $10^{4}$.

In the next section, we will be using the scale-invariant masses and charges $\hat{M}$ and $\hat{Q}$ defined in \eqref{eq:scaled_quantities}, and we will assume that $\alpha>0$. 
Indeed, we will restrict our analysis to positive values of $\alpha$. As was explained in Chapter~\ref{ch:Evasions}, evading the no-hair theorem in spontaneous scalarization, where $f(\phi_\infty)=0$, requires $\alpha>0$ when $\beta=0$ and $\GB$ is positive, which is the case for a Schwarzschild black hole. Moreover, the Ricci coupling, controlled by $\beta$, does not contribute to linear perturbation theory around GR black holes.  It is hence unlikely that spontaneously scalarized spherically symmetric black hole solutions will exist for $\alpha<0$. It should be stressed, however, that the $\alpha<0$ case is  particularly interesting when studying rotating black holes \cite{Dima:2020yac} as explained in Chapter~\ref{ch:Evasions}.

\subsection{Domain of existence and \texorpdfstring{$n=0$}{} solutions}

\begin{figure}[ht]
  \centering
  \includegraphics[width=0.6\textwidth]{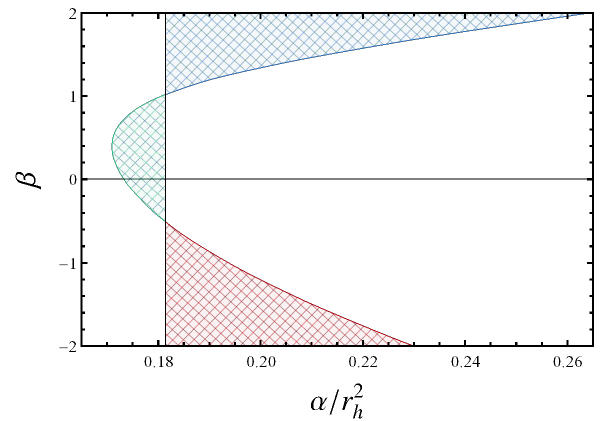}
  \caption[Parameter space and solution lines for scalarized black holes]{
  The shaded part corresponds to the region in the parameter space $(\beta,\alpha/r_h^2)$ where scalarized, and asymptotically flat black hole solutions exist.
  }
  \label{fig:domains}
\end{figure}

\begin{figure}[ht]
  \centering
  \includegraphics[width=0.49\textwidth]{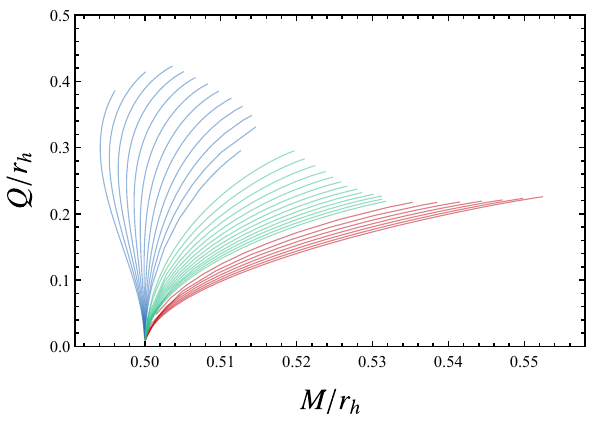}
  \hspace{0mm}
  \includegraphics[width=0.49\textwidth]{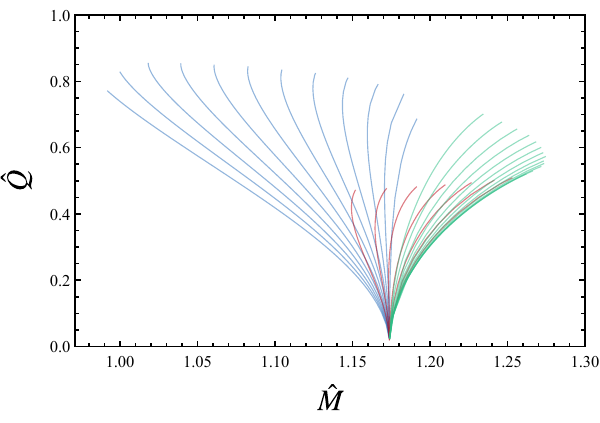}
  \caption[Parameter space and solution lines for scalarized black holes]{
  \textit{Left:} The mass and charge for scalarized black holes.
  The blue, green and red regions correspond to the upper, middle and lower regions of the left-panel plot respectively. The contours shown are of constant $\beta$ with separation of $\delta\beta=0.04,~0.01,~0.24$ in the blue, green and red regions respectively. \textit{Right:} Same but for mass and charge normalised with respect to $\alpha$.
  }
  \label{fig:QM}
\end{figure}

The domain of existence for black holes with non-zero scalar charge is non-trivial. In this subsection we present it for the fundamental branch ($n=0$).
To analyze the parameter space in which these solutions exist, we scan $(\alpha/r_h^2,\beta)$ for BHs with a non-trivial scalar field configuration that vanishes asymptotically.
For $\alpha=0$ the only solution that is regular at the horizon and asymptotically flat is the Schwarzschild BH~\cite{Sotiriou:2011dz}, while for $\beta=0$, the allowed values for $\alpha$ are in agreement with \cite{Doneva:2017bvd,Silva:2017uqg}

In Fig.~\ref{fig:domains} we show the solutions existence domain in the $(\alpha/r_h^2,\beta)$ plane. 
Scalarized solutions with the desired properties, \textit{i.e.}~regular everywhere and asymptotically vanishing, exist in the shaded regions. We can classify the space of solutions existing in three domains bounded by a seemingly vertical line given by the critical value for scalarization  $\alpha/r_h^2\approx0.18$ and a parabola-like curve defined by the condition $\Delta\geq 0$. To further analyze the solutions, in the left panel of Fig.~\ref{fig:QM}, we show the domain of existence in the $(M/r_h,Q/r_h)$ plane, which is suggestive of several properties of these solutions. 
First, it appears that the map $(\alpha/r_h^2,\beta) \mapsto (M/r_h,Q/r_h)$ is invertible offering a direct connection between the asymptotes of the solution and the underlying gravity model. 
The lines of constant $\beta$ approaching the vertical line in the left figure, merge at $(M/r_h,Q/r_h)=(0.5,0)$, which corresponds to the GR solution.
These same lines appear to be bounded from above as they approach the parabola in $(\alpha/r_h^2,\beta)$.

\begin{figure}[ht]
    \centering
    \includegraphics[width=.49\linewidth]{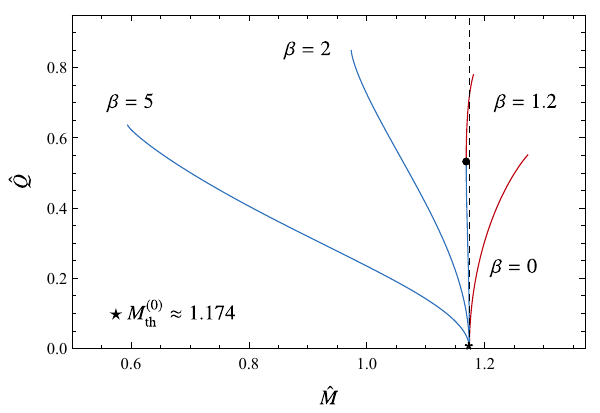}\hfill
    \includegraphics[width=.49\linewidth]{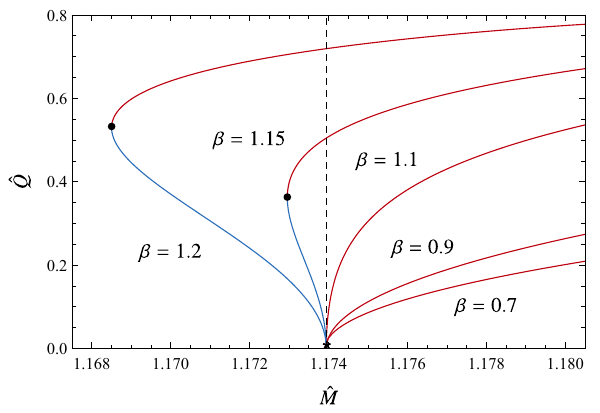}
    \caption[Charge-mass plots for scalarized sRGB black holes]{Charge-mass diagram for scalarized BHs with the Ricci term. The blue part of the curves corresponds to stable solutions while the red to unstable ones. The dotted vertical line represents $\hat{M}=\hat{M}_{\text{th}}$. \textit{Left}: Sample of solutions for different values of $\beta$. Solutions on the left part of the dotted vertical line on the diagram are expected to be stable against radial perturbations. \textit{Right:} Zoom on the stability threshold region. We see that for $\beta\approx1.15$ (possibly) stable solutions appear in the theory. For $\beta>\beta\approx 1.2$ (potentially) unstable solutions are no longer present.}
    \label{fig:scalarized_solutions}
\end{figure}

Now we turn our attention to Fig.~\ref{fig:scalarized_solutions}, which presents a few example lines for some characteristic values of $\beta$.
Schwarzschild BHs are radially unstable for $\hat{M}<\hat{M}_\text{th}^{(0,0)}\approx1.175$~\cite{Blazquez-Salcedo:2018jnn,Macedo:2019sem}.
It can be seen that $\hat{M}_\text{th}^{(0,0)}$ is the point where the curves converge in Fig.~\ref{fig:scalarized_solutions}. 
When $\hat{M}<\hat{M}_\text{th}^{(0,0)}$, scalar perturbations grow spontaneously and form a non-trivial scalar profile, and so are the energetically favourable solutions. 
We again stress that the final scalar field profile is determined by the full non-linear equations and not only the terms that trigger the instability.
In contrast, Schwarzschild  BHs are stable in the region $\hat{M}>\hat{M}_\text{th}^{(0,0)}$, and potential scalarized solutions would decay back to GR, as seen from purely energetic arguments. 
This reasoning can be confirmed by performing a radial stability analysis of the scalarized BH solutions. This is done in Chapter~\ref{ch:Stability}.

When $\beta$  is smaller than some critical value $\beta_{\text{crit}}\approx 1.15$, the charge-mass curve tilts to the right and all scalarized black holes have larger ADM masses than the GR mass instability threshold. Such scalarized black holes are unlikely to be produced dynamically. ADM mass is a measure of energy for the system. The fact that \textit{all} scalarized black holes for $\beta<\beta_{\text{crit}}$ have larger mass than \textit{all} GR black holes that are unstable implies that, if \textit{any} scalarized black hole is considered to be the end point of the tachyonic instability for a GR black hole, then this end state would have more energy than the initial state.

Based on the argument above, we conjecture that scalarized black holes are unstable for $\beta<\beta_{\text{crit}}$. Conversely, for $\beta>\beta_{\text{crit}}$ the ADM mass for scalarized black holes can be smaller than the GR counterparts and hence it is reasonable to expect that scalarized black holes are endpoints of the tachyonic instability\footnote{As we will see in Chapter~\ref{ch:Stability} this is not exactly the case even though it does capture the general trend. That is because very close to the GR mass two different types of solutions exist with one being stable and the other unstable}. These arguments are consistent with earlier results. In particular, it is already known that for $\beta=0$ scalarized black holes are radially unstable \cite{Blazquez-Salcedo:2018jnn}. Moreover, the general picture shown in Fig.~\ref{fig:scalarized_solutions} is very similar to the one presented in Ref.~\cite{Macedo:2019sem}. In that work, $\beta$ was vanishing and the $\phi^2 R$ term was absent, but a $\phi^4$ self-interaction had a similar effect. Analysis of radial stability did show in that case that stability was associated with whether the curves on the $\hat{Q}$-$\hat{M}$ plane tilt to the right or the left. In Chapter~\ref{ch:Stability} we will perform a proper stability analysis which will verify our conjecture. These considerations suggest strongly that the coupling between $\phi$ and the Ricci scalar, can have  a very interesting stabilizing effect for scalarized black holes, without having to resort to scalar self-interactions. 

Note that in some cases, when $\beta>\beta_{\text{crit}}$ and hence the $\hat Q$-$\hat M$ curve initially leans to the left, this same curve later turns towards the right. The points at which the curves turn right are marked by black dots in Fig.~\ref{fig:scalarized_solutions}. One expects configurations past the turning point to be unstable, as configurations of the same ADM mass and smaller scalar charge exist. 

\begin{figure}[ht]
\begin{center}
    \includegraphics[width=0.49\textwidth]{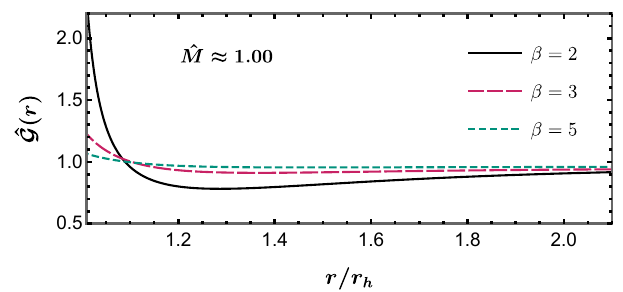}\hspace{0cm}
    \includegraphics[width=0.49\textwidth]{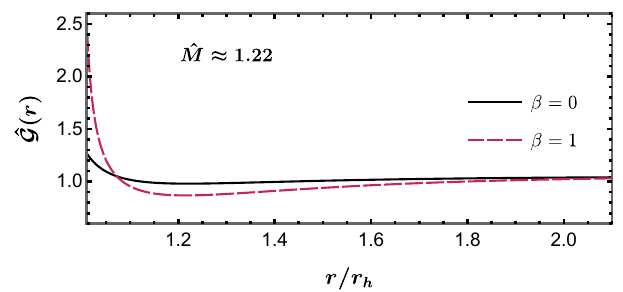}\\
    \includegraphics[width=0.49\textwidth]{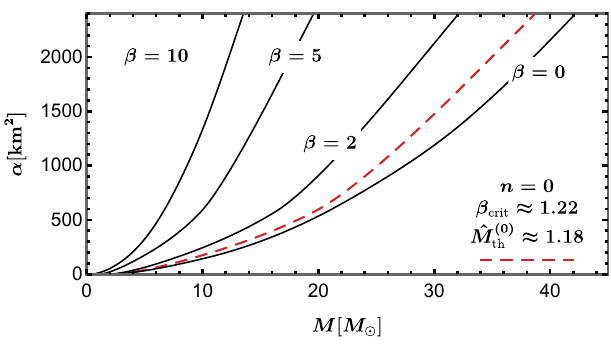}\hspace{0cm}
    \includegraphics[width=0.49\textwidth]{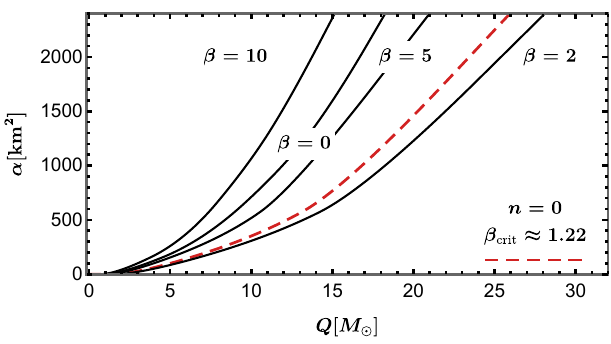}
\end{center}
\caption[Gauss-Bonnet couplings vs Mass and curvature vs distance for sRGB black holes]{\textit{Top left:} GB invariant as a function of the distance for different values of $\beta$ and $\hat{M}\approx 1$. \textit{Top right:} Same but for $\hat{M}\approx 1.22$.
\textit{Bottom left:} Domain of existence of $n=0$ scalarized black holes on the $\alpha$-$M$ plane. For a given $\beta$, solutions exist between the corresponding black line, and the dashed, red line. The latter coincides with the line where GR solutions of equal mass would become unstable. \textit{Bottom right:} Same but on the $\alpha$-$Q$ plane. Both panels can be obtained from an ``unfolding'' of Fig.~\ref{fig:scalarized_solutions}.}
\label{fig:fig_aMQdomain}
\end{figure}

As is clear from Fig.~\ref{fig:scalarized_solutions}, for $\beta>\beta_{\text{crit}}$, the normalized scalar charge $\hat Q$ increases as the normalized ADM mass $\hat M$ decreases, at least in the part of the curves up to the turning point (black dot), whereas for $\beta<\beta_{\text{crit}}$, the normalized scalar charge $\hat Q$ increases as the normalized ADM mass $\hat M$ increases. Interestingly, the dependence of the curvature near the horizon on the ADM mass turns out to be different in the two cases. For $\beta>\beta_{\text{crit}}$ scalarized black holes tend to have larger curvatures at the horizon when the ADM mass decreases, as is the case in GR, whereas for $\beta<\beta_{\text{crit}}$ the curvature on the horizon tends to increase as the mass (and scalar charge) increases. Hence, in both cases, the scalar charge seems to be controlled by the curvature. A way to see this is by comparing the curvature near the horizon for same mass $\hat{M}$ solutions but with different $\beta$. For $\beta>\beta_{\text{crit}}$ and fixed $\hat{M}$, increasing $\beta$ decreases the charge, as seen from the top panels of Fig.~\ref{fig:fig_aMQdomain}. The opposite holds for $\beta<\beta_{\text{crit}}$. This is equivalent to concluding that for all values of $\beta$, moving along the $\hat{M}$-$\hat{Q}$ curves for fixed $\beta$ towards larger charges means moving towards larger curvatures near the horizon.

In the bottom panel of Fig.~\ref{fig:fig_aMQdomain}, we show the domain of existence of scalarized black holes on the $\alpha$-$M$ and $\alpha$-$Q$ planes. As discussed, linear analysis showed distinct scalarization thresholds, the first (zero nodes) of which we denote with $\hat{M}^{(0,0)}_{\text{th}}\approx 1.179$. This threshold  is represented by the dashed, red line in Fig.~\ref{fig:fig_aMQdomain}. Note that $\hat{M}=\text{constant}$ (respectively $\hat{Q}=\text{constant}$) translates to a parabola in the $\alpha$-$M$ ($\alpha$-$Q$) plane. The rest of the curves correspond to the existence boundaries for various values of $\beta$. They are related to the horizon condition presented in Eq.~\eqref{eq:phi_h}. Solutions then exist everywhere between the red, dashed GR instability line and the plain, black existence line. Examining the plots reveals something rather interesting: the value of the Ricci coupling $\beta$ can affect the relative position of the existence line with respect to the instability parabola. This should not come as a surprise, based on the results presented in Fig.~\ref{fig:scalarized_solutions}, where $\beta$ has a similar effect on the relative position of the curve with respect to the threshold mass $\hat{M}^{(0,0)}_{\text{th}}$.

As mentioned earlier, we do not plan to consider the $\beta<0$ case in any detail as positive values appear to be better motivated. However, we can report the following based on a preliminary exploration. There is still a critical value of $\beta$, and for $\beta$ smaller than this value, scalarized black holes have smaller ADM masses than the GR instability threshold, together with scalar charges that tend to increase with decreasing mass. For $\beta$ larger than the critical value, the behaviour is reversed. Hence, the equivalent to Fig.~\ref{fig:scalarized_solutions} would be qualitatively similar for $\beta<0$.

\subsection{\texorpdfstring{$n=1,2$}{} nodes for the scalar profile}

\begin{figure}[ht]
\begin{center}
    \includegraphics[width=0.49\textwidth]{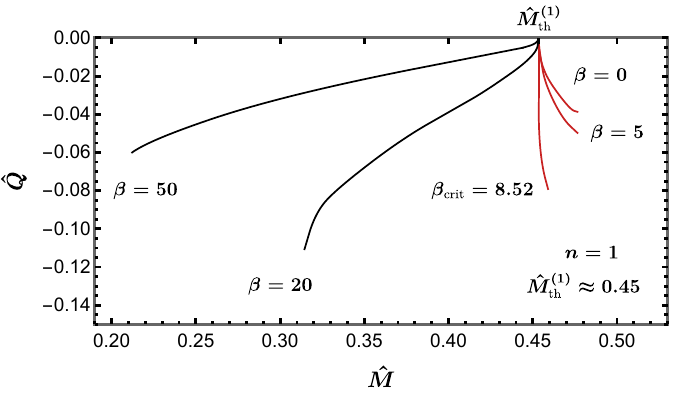}\hspace{0cm}
    \includegraphics[width=0.49\textwidth]{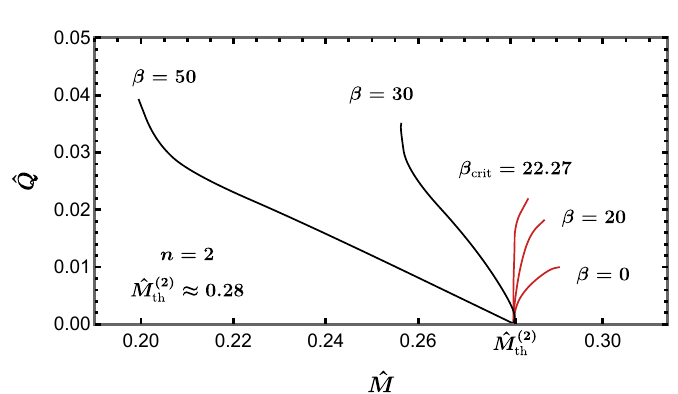}
\end{center}
\caption[Charge vs mass for the first and second overtone of scalarized sRGB black holes]{Normalized scalar charge versus normalized ADM mass for the solutions with $n=1$ (left panel) and 2 (right panel) nodes.}
\label{fig:fig_QM1}
\end{figure}

We now turn to solutions characterized by $n=1$ and $n=2$. For  $\beta>0$, the plot of the normalized charge versus the normalized mass is given in Fig.~\ref{fig:fig_QM1}.
A noticeable pattern is that, for $n=0$, the scalar charge is positive, while it is negative for $n=1$, positive for $n=2$, and so on. This is simply due to the fact that the scalar field has to approach 0 at spatial infinity from a different side, depending on the number of nodes. There is no deep significance in the sign of the charge, since the action \eqref{eq:ActionGeneric} possesses parity symmetry $\phi\to-\phi$, and the signs would have been flipped, had we chosen negative values of $\phi_\text{h}$ as initial conditions. Compared to the $n=0$ case, we see that the order of magnitude of the charge for all different values of $\beta$ is significantly smaller, and the range of masses for which we find scalarized solutions is strongly reduced. Once again there is a critical value of $\beta$ that separates right-leaning curves (likely unstable) from left leaning ones (likely stable).

\section{Discussion}
\label{sec:discussion}

We have considered the contribution that a coupling between the scalar, $\phi$, and the Ricci scalar, $R$, can have on black hole scalarization. We focused on static, spherically symmetric configurations. 

The $\beta \phi^2 R$ coupling is known not to affect the threshold of scalarization. However, our results show that it can  alter the domain of existence of scalarized black holes, significantly modify their properties, and control their scalar charge. Our results also strongly suggest that the strength of this coupling can have an impact on the stability of scalarized black holes. In particular, having $\beta$ be larger than some critical value, $\beta_\text{crit}$, is expected to resolve the stability problems for models that do not include the $\beta \phi^2 R$ coupling. We will investigate this issue in more detail in Chapter~\ref{ch:Stability}. 

We have mostly focused on positive values for $\beta$.  We did so for two reasons. First, as we showed in Chapter~\ref{ch:Cosmology}, including the $\beta \phi^2 R$ term in scalarization models and selecting a positive $\beta$ makes GR a cosmological attractor and allows one to have a consistent cosmological history, at least from the end of the inflationary era. The numerical values that we considered here for the couplings are similar to those used in Chapter~\ref{ch:Cosmology}. Second, for positive values of the Ricci coupling (and reasonably small values of the Gauss-Bonnet coupling), neutron stars do not scalarize \cite{Ventagli:2020rnx}. This allows one to evade the very tight binary pulsar constraints ({\textit e.g.} \cite{Freire:2012mg,Antoniadis:2013pzd,Shao:2017gwu}), related to energy losses due to dipolar emission of gravitational waves, without the need to add a bare mass to the scalar (and tune it appropriately). More discussion on this topic will follow in Chapter~\ref{ch:Neutron Stars}.

It is clear that the inclusion of the $\beta \phi^2 R$ coupling has multiple benefits in scalarization models. It is worth re-iterating that this coupling has lower mass dimensions than the $\alpha \phi^2 {\GB}$ coupling, which triggers scalarization at linear level. Moreover, unlike a bare mass term or scalar self-interactions, it allows the scalar to remain massless and free in flat space. Hence, the $\beta \phi^2 R$ coupling can be part of an interesting EFT that respects $\phi \to -\phi$ symmetry and in which shift symmetry can be broken only via the coupling to gravity (the complete EFT would potentially include more terms, such as  $R\phi^4$ and $G^{\mu\nu}\partial_\mu\phi\partial_\nu\phi$).

Gravitational wave observation of binaries that contain black holes would still be able to measure or constrain $\beta$ and $\alpha$. A detailed post-Newtonian analysis of the inspiral phase would be sufficient to provide some first constraints. Scalarization models in which the scalar charge is non-zero only below a mass threshold are also expected to be severely constrained by extreme mass ratio inspirals (EMRIs) observations by LISA: the supermassive black hole would be described by the Kerr metric, whereas the small black hole can carry a scalar charge. This is the ideal scenario to apply the considerations of Ref.~\cite{Maselli:2020zgv}.

It should be stressed that we only considered the case $\alpha>0$ throughout this paper, as this is a requirement for having scalarized black holes under the assumptions of staticity and spherical symmetry. However, it has been shown in Ref.~\cite{Dima:2020yac} that, for $\alpha<0$ (and $\beta=0$), scalarization can be triggered by rapid rotation. Indeed, some scalarized black hole have been found in this scenario in Refs.~\cite{Herdeiro:2020wei,Berti:2020kgk}. It would thus be very interesting to consider the effect of the $\beta \phi^2 R$ coupling for $\alpha<0$, \textit {i.e.}~in models where scalarization is induced by rotation.

It is likely that theoretical constraints on the value of the couplings in scalarization models will be imposed by the requirement that the initial value problem be well-posed in dynamical evolution scenarios where one expects the models to be good EFTs. Results in this direction have been obtained in Ref.~\cite{Ripley:2020vpk} for $\beta=0$. The inclusion of the coupling with the Ricci scalar is likely to affect the results quantitatively, and hence is an interesting prospect. We will come back to that in Chapter~\ref{ch:Stability}.

%% file: Chapters/sRGB_ns.tex
As we saw the onset of the tachyonic instability that triggers scalarization is controlled by linear terms (although Ref.~\cite{Doneva:2021tvn} also examined what happens if linear terms are absent from the potential) but eventually this instability is quenched by non-linearities, which control the end-state. In Chapters~\ref{ch:Cosmology} and \ref{ch:Black holes} we saw that including the Ricci term does seem to provide us with advantages (e.g. late-time attractor to GR in a cosmological scenario and potentially stable solutions). Additionally, it was shown in \cite{Ventagli:2020rnx} that the Ricci term can help in suppressing the scalarization of neutron stars, which would otherwise tend to place significant constraints. As we saw in Chapter~\ref{ch:Black holes}, even though the Ricci coupling does not affect the onset of black hole scalarization, it affects the properties of the scalarized solutions and, consequently, observables. For certain values of the Ricci coupling ---which happen to be consistent with the ones associated with a late-time attractor behaviour--- the presence of this operator is expected to render black holes radially stable, without the need to introduce self-interaction terms. This will be further examined in the next chapter.
For these reasons, it is of great interest to examine how the combination of Ricci and Gauss-Bonnet couplings affects neutron star properties as well.

The setup we need in order to study scalarized neutron stars in our theory \eqref{eq:ActionGeneric}, is different compared to the black hole scenario as we now assume the presence of matter. Therefore, the modified Einstein equations yield
\begin{equation}\label{eq:grav_eq}
    G_{\mu\nu}=\kappa T^{(\phi)}_{\mu\nu}+\kappa T^{M}_{\mu\nu}\,,
\end{equation}
where $T^{(\phi)}_{\mu\nu}$ is given in \eqref{eq:grav-BH} and comes from the variation of the $\phi$-dependent part of the action with respect to the metric, while
\begin{equation}
    T^{M}_{\mu\nu}=-\frac{2}{\sqrt{-g}}\frac{\delta S_{M}}{\delta g^{\mu\nu}}\, ,
\end{equation}
is the matter stress-energy tensor. We assume matter to be described by a perfect fluid with $T^{M}_{\mu\nu}=(\epsilon+p)u_\mu u_\nu+p\,g_{\mu\nu}$, where $\epsilon$, $p$ and $u_\mu$ are respectively the energy density, the pressure and the 4-velocity of the fluid. The pressure is  directly related to the energy density through the equation of state. The field equations then, for the spherically symmetric metric \eqref{eq:metric}, take the form of coupled ordinary differential equations for $A$, $B$, $\epsilon$ and $\phi$, as in the black hole case.

\section{Expansion for small radii}\label{Sec:Exp0}
Close to the center of the star, we can perform an analytic expansion of the form 
\begin{equation}\label{eq:smallr}
f(r)=\sum_{n=0}^\infty f_n r^n
\end{equation}
for the functions $\ln A$, $-\ln B$, $\epsilon$, $p$ and $\phi$.
Plugging these expansions in the field equations, we can solve order by order to determine the boundary conditions at the origin. At this point, there are essentially three quantities that one can freely fix: the central density $\epsilon_0$, the value of the scalar field at the center $\phi_0$, and the value of the time component of the metric at the center, determined by $A_0$. On the other hand, $B_0$ has to vanish in order to avoid a conical singularity at the center, while $p_0$ is directly related to $\epsilon_0$ by the equation of state. All higher order quantities $\{A_i, ...,\phi_i\}$, $i\geq 1$ can be determined in terms of the three quantities $\{\epsilon_0,A_0,\phi_0\}$. We will require that spacetime is asymptotically flat, with a trivial scalar field at spatial infinity, which fixes uniquely $A_0$ and $\phi_0$, or rather restricts $\phi_0$ to a discrete set of values, each corresponding to a different mode; technically, these values are found through a numerical shooting method. Therefore, for given parameters $\alpha$ and $\beta$, a solution is eventually fully determined by the central density $\epsilon_0$. Different choices of $\epsilon_0$ will translate into different masses.

We must underline the difference with the black hole case, studied in Chapter \ref{ch:Black holes}. For black holes, the equations are scale invariant up to a redefinition of the couplings. Practically, this means that it is enough to explore the full space of parameters $\alpha$ and $\beta$ for a \textit{fixed} mass. One can then deduce all solutions, of arbitrary mass, by an appropriate rescaling. For neutron stars this scaling symmetry is broken by the equation of state that relates $p$ and $\epsilon$. Therefore, one \textit{a priori} has to explore a 3-dimensional space of parameters ($\epsilon_0$, $\alpha$ and $\beta$) in the case of neutron stars. In order to keep this exploration tractable, as it was already done in \cite{Ventagli:2020rnx}, we will focus our study on a selection of central densities and equations of state. We pick these in order to cover very diverse solutions, typically corresponding to the lightest/heaviest observed stars in general relativity. We then explore a wide range of the $(\alpha,\beta)$ parameter space for these fixed densities and equations of state.

To complete this section, let us note that solving order by order the field equations for the higher order coefficients in the expansion \eqref{eq:smallr} does not always yield solutions. All first order coefficients in this expansion have to vanish; one can express $A_2$, $\epsilon_2$, $p_2$ and $\phi_2$ in terms of $B_2$; however, $B_2$ itself is determined by the following equation:
\begin{equation}\label{eq:Lambda2}
\begin{split}
& B_2^4(512\,\alpha^3\kappa\,\phi_0^2-256\,\alpha^3\beta\kappa^2\phi_0^4)
+B_2^3(512\,p_0\alpha^3\kappa^2\phi_0^2-64\,\alpha^2\beta\kappa\phi_0^2\\
&+32\,\alpha^2\beta^2\kappa^2\phi_0^4)
+B_2^2(12\,\alpha\beta^3\kappa^2\phi_0^4-24\,\alpha\beta^2\kappa\phi_0^2
-192\,p_0\alpha^2\beta\kappa^2\phi_0^2)\\
&+B_2\bigg(2\,\beta-\frac{16}{3}\alpha\epsilon_0\kappa-2\,\beta^2\kappa\,\phi_0^2
+3\,\beta^3\kappa\,\phi_0^2+24\,p_0\alpha\beta^2\kappa^2\phi_0^2\\
&+\frac{8}{3} \alpha\beta\epsilon_0\kappa^2\phi_0^2+\frac{16}{3}\alpha\beta^2\epsilon_0\kappa^2\phi_0^2+\frac{1}{2}\beta^3\kappa^2\phi_0^4
-\frac{3}{2}\beta^4\kappa^2\phi_0^4\bigg)\\
&-\frac{2}{3}\beta\epsilon_0\kappa+\frac{16}{9}\alpha\epsilon_0^2\kappa^2-p_0\beta^3\kappa^2\phi_0^2+\frac{1}{3}\beta^2\epsilon_0\kappa^2\phi_0^2-\frac{2}{3}\beta^3\epsilon_0\kappa^2\phi_0^2=0\, .
\end{split}
\end{equation}
Eq.~\eqref{eq:Lambda2} is a fourth order equation in $B_2$. Such an equation does not necessarily possess real solutions. Therefore, for any choice of parameters $(\alpha,\beta)$ and initial values $(\epsilon_0,\phi_0)$, we need to check that a real solution to Eq.~\eqref{eq:Lambda2} exists. In particular, we need to check this when implementing the shooting method that will allow us to find the values of $\phi_0$ such that the scalar field is trivial at spatial infinity. Such values might actually not exist in the domain where Eq.~\eqref{eq:Lambda2} possesses real solutions.
In practice, we make sure that each choice of parameters that we consider guarantees not only that Eq.~\eqref{eq:Lambda2} has a positive\footnote{An acceptable solution to Eq.~\eqref{eq:Lambda2} must be positive, otherwise $g_{rr}$ diverges at a finite radius, and consequently the pressure and the energy density diverge as well.} real solution, but that such a solution is connected to the GR one. We discard all other parameter combinations that do not respect such criteria.

We now analyze the asymptotic behaviour of the solutions at spatial infinity. This time, we expand the metric and scalar functions in inverse powers of $r$, and solve order by order.
We impose that the asymptotic value of the scalar field vanishes, that is $\phi(r\to\infty)\equiv\phi_\infty=0$, and that $A(r\to\infty)=0$. The asymptotic solution then reads
\begin{align}
\begin{split}\label{eq:Asymptotic1}
B=& \; 1-\frac{2M}{r}+\frac{1}{2}\frac{Q^2\kappa}{r^{2}}(1-2\,\beta\kappa)+\frac{1}{2}\frac{MQ^2\kappa}{r^{3}}(1-3\,\beta)\\
&+\frac{1}{12}\frac{Q^2\kappa}{r^{4}}\big[ M^2(8-28\,\beta)+Q^2\beta\kappa(1-5\,\beta+12\,\beta^2)\big]
\\
&+\frac{1}{48}\frac{MQ^2\kappa}{r^{5}}\big[ 768\,\alpha+8\,M^2(6-23\,\beta)-Q^2\kappa(1-18\,\beta+77\,\beta^2\\
&-156\beta^3)\big]+O(r^{-6}),
\end{split}
\\[2mm]
\begin{split}\label{eq:Asymptotic2}
A=& \; 1-\frac{2M}{r}+\frac{1}{2}\frac{Q^2\beta\kappa}{r^2}+\frac{1}{6}\frac{MQ^2\kappa}{r^3}(1-3\,\beta)+\frac{1}{r^4}\big[ 4\,M^4\\
&-\frac{1}{3}M^2Q^2\kappa(1+3\,\beta)
+\frac{1}{8}Q^4\beta^2\kappa^2 \big]
-\frac{1}{r^5}\big\{ 8\,M^5-\frac{1}{30}M^3Q^2\kappa(58\\
&-75\beta)-\frac{1}{80}M Q^2\kappa\big[ 512\,\alpha
-Q^2\kappa(3+10\,\beta-85\,\beta^2+60\,\beta^3) \big] \big\}\\
&+O(r^{-6}),
\end{split}
\\[2mm]
\begin{split}\label{eq:Asymptotic3}
\phi=&\; \frac{Q}{r}+\frac{MQ}{r^2}+\frac{1}{12}\frac{Q}{r^3}\big[ 16\,M^2-Q^2\kappa(1-2\,\beta+3\,\beta^2) \big]+\frac{1}{r^4}\big[ 2\,M^3Q\\
&-\frac{1}{12}MQ^3\kappa(4-9\,\beta+9\,\beta^2) \big]
+\frac{1}{480}\frac{Q}{r^5}\big\{Q^4\kappa^2(9-40\,\beta+86\,\beta^2\\
&-144\,\beta^3
+117\,\beta^4) -8M^2\big[ 144\,\alpha+Q^2\kappa(58-140\,\beta+105\,\beta^2) \big]\\
&+ 1536\,M^4 \big\}+O(r^{-6})\, ,
\end{split}
\end{align}\\[2mm]
where $M$ and $Q$ are free. Once again, we identify $M$ as the ADM mass and $Q$ as the scalar charge. As one can see from Eqs.~\eqref{eq:Asymptotic1}--\eqref{eq:Asymptotic3}, the contribution from the Ricci coupling dominates the asymptotic behaviour of the solutions over the Gauss-Bonnet coupling. Indeed, terms proportional to $\beta$ enter the expansion already at order $r^{-2}$, whereas $\alpha$-dependent terms arise only at order $r^{-5}$. This expansion is in fact entangled with the boundary conditions at the center of the star, as we already mentioned. For fixed parameters $\alpha$ and $\beta$, the freedom in $M$ directly relates to the freedom in the central density $\epsilon_0$. On the other hand, the fact that only discrete values of $\phi_0$ yield a vanishing scalar field at infinity means that the scalar profile is actually fixed once a central density (or a mass) is chosen. Therefore, $Q$ is fixed as a function of $M$, and does not constitute a free charge.

The scalar charge constitutes probably the most direct channel to test the theory through observations. Indeed, binaries of compact objects endowed with an asymmetric charge will emit dipolar radiation. This enhances the gravitational wave emission of such systems: in a Post-Newtonian (PN) expansion, dipolar radiation contributes to the energy flux at order -1PN with respect to the usual quadrupolar GR flux. Generically, this dipolar emission is controlled by the sensitivities of the compact objects, defined as\footnote{The factor of $1/\sqrt{4\pi}$ is added to match the standard definition of the sensitivity in the literature, where a different normalization for the scalar field is generally used.}
\begin{equation}
    \alpha_I=\dfrac{1}{\sqrt{4\pi}}\,\dfrac{\partial\text{ln}M_I}{\partial\phi_0},
\end{equation} 
$M_I$ being the mass of the component $I$, and $\phi_0$ the value of the scalar field at infinity. The observation of various binary pulsars, notably the PSR~J1738+0333 system, allows one to set the following constraint:
\begin{equation}
   | \alpha_A-\alpha_B|\lesssim2\times10^{-3},
      \label{eq:DEFbound}
\end{equation}
where $A$ and $B$ label the two components of the system \cite{Shao:2017gwu,Wex:2020ald}. We can then relate the sensitivity to the scalar charge $Q$, using the generic arguments of \cite{Damour:1992we}. We have
\begin{equation}
   Q_I=-\dfrac{1}{4\pi}\,\dfrac{\partial M_I}{\partial\phi_0}.
   \label{eq:DEFcharge}
\end{equation} 
If there is no accidental coincidence in the charge of the two components of the binary, Eqs.~\eqref{eq:DEFbound}-\eqref{eq:DEFcharge} translate as
\begin{equation}
\left|\dfrac{Q}{M}\right|\lesssim6\times10^{-4}
\label{eq:boundQ}
\end{equation}
for the solutions we consider. Only solutions satisfying this bound on the charge to mass ratio are relevant. It is however a non-trivial task to map this bound onto the parameters of the Lagrangian \eqref{eq:ActionGeneric}.

\section{Numerical implementation}

We solve the system of three differential equations for the three independent functions $A$, $\phi$ and $\epsilon$ by starting our integration from $r_0=10^{-5}~\text{km}$. We fix the parameters of the theory $\alpha$ and $\beta$, and the central density $\epsilon_0$, typically to values of order $10^{17}$~kg/m$^3$. Then, we give an initial guess for $\phi_0$, and determine boundary conditions as explained in Sec.~\ref{Sec:Exp0}. The integration will generically give a solution; however, we also demand that the scalar field vanishes at infinity, that is $\phi_\infty=0$. Only a discrete set of $\phi_0$ values will yield $\phi_\infty=0$. Each value corresponds to a different number of nodes of the scalar field in the radial direction, where by \textit{nodes} we refer to the number of times the scalar field radial profile crosses the horizontal axis. In practice, we integrate up to distances $r_\text{max}=300\, \text{km}$ and we implement a shooting method to select the solutions with $\phi_\infty=0$. Generally, we use Mathematica's built-in function FindRoot.

However, in some cases FindRoot fails to find the right solutions, even if one gives it a limited range $(\phi_{0,\:\text{min}},\phi_{0,\:\text{max}})$ where to look for. When this happens, we resort to  bisection instead. In this latter case, we require that $\phi(r_\text{max})/\phi_0 \leq 10^{-2}$. 

At each stage of the shooting method, we must check that Eq.~\eqref{eq:Lambda2} gives a real positive solution for $B_2$ that is connected to the GR solution. In some cases, we reach the limit of the region of the parameter space where these criteria are fulfilled before reaching $\phi_\infty=0$. When this is the case, there is no solution associated to the given choice of $\alpha$, $\beta$ and $\epsilon_0$. Note also that, given a set of $\alpha$, $\beta$ and $\epsilon_0$, there is a maximum number of nodes that the solution can have, consequently a maximum number of suitable choices of $\phi_0$ (typically up to three modes in the regions we explore). Solutions with more nodes are encountered only for higher values of the parameters $\alpha$ and $\beta$, or at higher curvatures (that is, at higher $\epsilon_0$).

Given a solution, we extract the value of the ADM mass $M$ and the scalar charge $Q$, as defined in the asymptotic expansion \eqref{eq:Asymptotic1}--\eqref{eq:Asymptotic3}. We have
\begin{equation}
    \begin{split}
    & M \approx -\left(\frac{1}{2}r^2\,B'\right) \bigg|_{r_\text{max}},\\
    & Q \approx -\left(r^2 \phi'\right)\big|_{r_\text{max}}.
    \end{split}
\end{equation}

\section{Existence regions of scalarized solutions}
\label{Sec:parameterSpace}

In this section, we will study the regions where scalarized solutions exist in the $(\alpha,\beta)$ parameter space.

\subsection{Light star with SLy EOS}
\label{sec:lightSLy}

\begin{figure}[ht]
\centering
\includegraphics[width=0.65\linewidth]{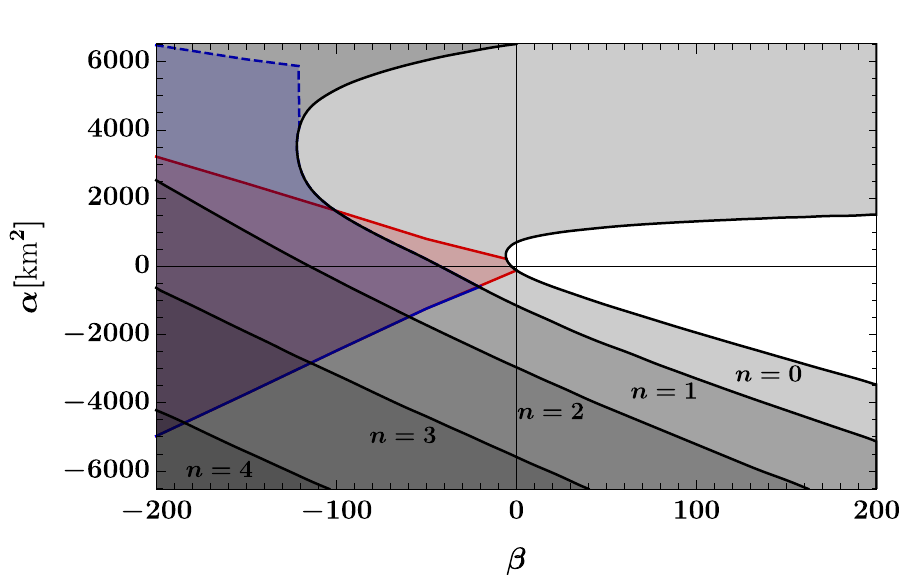}%
\caption[Parameter space for "light" scalarized sRGB neutron stars obeying the SLy equation of state]{Regions of existence of scalarized solutions in the $(\alpha,\beta)$ space, for the SLy EOS with $\epsilon_0=8.1\times 10^{17}~\text{kg}/\text{m}^3$. The red (respectively blue) region is the region where scalarized solutions with 0 (respectively 1) node exist. We superimposed the grey contours obtained in Ref.~\cite{Ventagli:2020rnx}, which represent the lines beyond which GR solutions with the same density are unstable to scalar perturbations with 0, 1, 2, \textit{etc} nodes. We see that the region where there exist scalarized solutions with $n$ nodes is included in the region where the GR solutions are unstable to scalar perturbations with $n$ nodes, but much smaller. The dashed boundary for the blue region corresponds to a breakdown of the integration inside the star. In GR, a star with this choice of $\epsilon_0$ and EOS has a light mass, $M_\text{GR}=1.12 M_\odot$.}
	\label{fig:Sly112}
\end{figure}

First, we consider a neutron star described by the SLy equation of state \cite{Haensel:2004nu}, with a central energy density of $\epsilon_0=8.1\times 10^{17}~\text{kg}/\text{m}^3$, so that its gravitational mass in GR is $M_{\text{GR}}=1.12 M_\odot$. The results are summarized in Fig.~\ref{fig:Sly112}, where we relate our new results to the previous study of the scalarization thresholds \cite{Ventagli:2020rnx}.
The white area corresponds to the region of the parameter space where the GR solution is stable. When cranking up the parameters $\alpha$ or $\beta$, a new unstable mode appears every time one crosses a black line. The first mode has 0 nodes, the second 1 node, \textit{etc}. We will refer to these black lines as \textit{instability lines}. Any point in the parameter space that lies within some grey region corresponds to a configuration where the GR solution is unstable. 
The red (respectively blue) area corresponds to the region where scalarized solutions with $n=0$ (respectively $n=1$) nodes exist. We do not include the equivalent regions for higher $n$, to not complicate further the analysis. The region where a scalarized solution does exist
is considerably reduced with respect to the region where the GR solution is unstable.

One of our main results is that the parameters $(\alpha,\beta)$ corresponding to the grey areas that are not covered by the colored regions must be excluded. Indeed, there, scalarized solutions do not exist while the GR solution itself is unstable. Therefore, neutron stars in these theories, when they reach a critical mass, will be affected by a tachyonic instability, but there does not exist a fixed point (a static scalarized solution) where the growth could halt. This would imply that neutron stars with this mass and EOS do not exist for the corresponding parameters  of the theory \eqref{eq:ActionGeneric}. Considering that the properties of the scalarized star are sensitive to nonlinearities, adding further nonlinear interaction terms to the action, \textit{e.g.} self-interactions in a scalar potential, as was proposed in \cite{Macedo:2019sem}, or non-linear terms in the coupling functions \cite{Doneva:2017bvd,Silva:2018qhn}, can potentially change this result.

In Fig.~\ref{fig:Sly112}, the regions where scalarized solutions exist are delimited by \textit{existence lines}, represented by a curve of the respective color. The plain lines correspond to boundaries beyond which it is no longer possible to find a value of $\phi_0$ that allows a suitable solution to Eq.~\eqref{eq:Lambda2}, while providing $\phi_\infty=0$. Beyond dashed lines, on the other hand, nothing special occurs at the center of the star, but the numerical integration breaks down at a finite radius inside the star. We do not know whether, when crossing these dashed lines, our integration is affected by numerical problems, or whether the divergence corresponds to an actual singularity of the solutions. 
It could be that this singularity emerges as an artifact of the method we employ. Indeed, in our analysis, we keep the central density $\epsilon_0$ fixed while pushing the couplings $\alpha$ and $\beta$ to larger and larger values. However, for each couple of parameters $(\alpha,\beta)$, there probably exists a maximal central density beyond which star solutions do not exist, or equivalently it becomes impossible to sustain such a high central density. The dashed line could correspond to this saturation, where we try to push all the parameters beyond values that can actually be sustained by the model.

A surprising feature, which is not visible in Fig.~\ref{fig:Sly112}, is that scalarized solutions always exist in a very narrow range along the instability lines. For example, when crossing the black instability line that delimitates the white region where the GR solution is stable, from the light-grey region where it is unstable against $n=0$ scalar perturbations, there exists a very narrow band (within the grey region) where scalarized solutions with zero nodes exist. We observed similar behaviours along each instability line, also in the scenarios discussed in the next paragraphs.

\subsection{Light star with MPA1 EOS}

\begin{figure}[ht]
\centering
\includegraphics[width=0.65\linewidth]{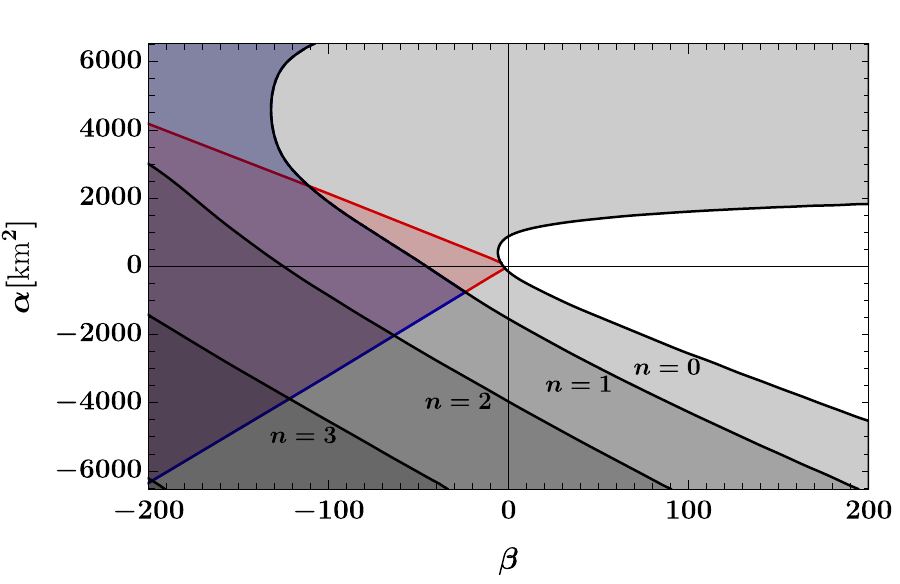}%
\caption[Parameter space for scalarized sRGB neutron stars obeying the MPA1 equation of state]{Regions of existence of scalarized solutions in the $(\alpha,\beta)$ space, for the MPA1 EOS with $\epsilon_0=6.3\times 10^{17}~\text{kg}/\text{m}^3$. The conventions are the same as in Fig.~\ref{fig:Sly112}.  In GR, a star with this choice of $\epsilon_0$ and EOS is again light, with $M_\text{GR}=1.12 M_\odot$.}
\label{fig:MPA1}
\end{figure}

We next consider a stellar model described by the MPA1 equation of state~\cite{Gungor:2011vq}. We choose a central energy density of $\epsilon_0=6.3\times 10^{17}\,\text{kg}/\text{m}^3$, such that it corresponds to the same GR mass as in the previous case, that is $M_{\text{GR}}=1.12 M_\odot$. We report the results in Fig.~\ref{fig:MPA1}.
As one can see, changing the EOS has only mild effects on the region of existence of scalarized solutions. The analysis of the parameter space is qualitatively the same as for the SLy EOS. The main difference is that, for the range of parameters we considered, no numerical divergences (associated with dashed lines) appear with the MPA1 EOS.

\subsection{Heavy star with SLy EOS}

\begin{figure}[ht]
\centering
\includegraphics[width=0.49\linewidth]{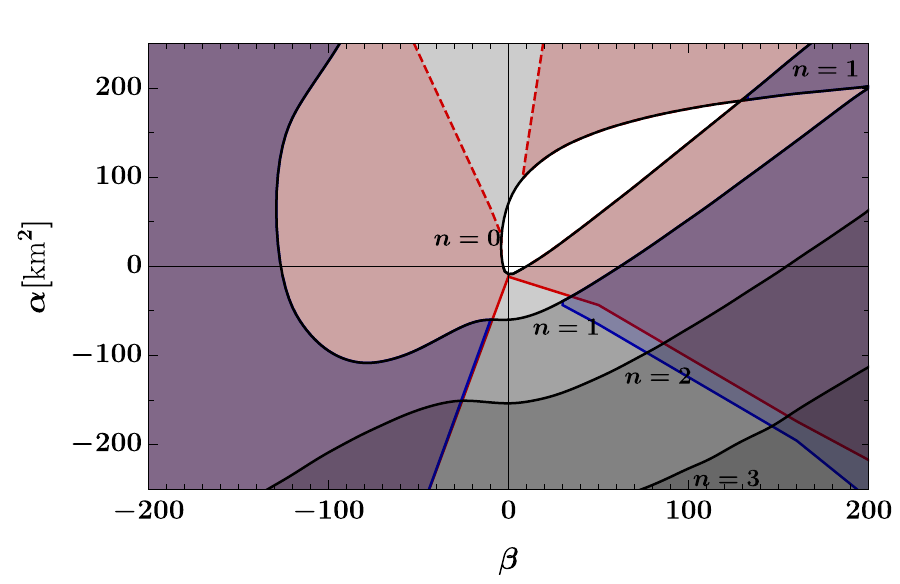}\hfill
\includegraphics[width=0.49\linewidth]{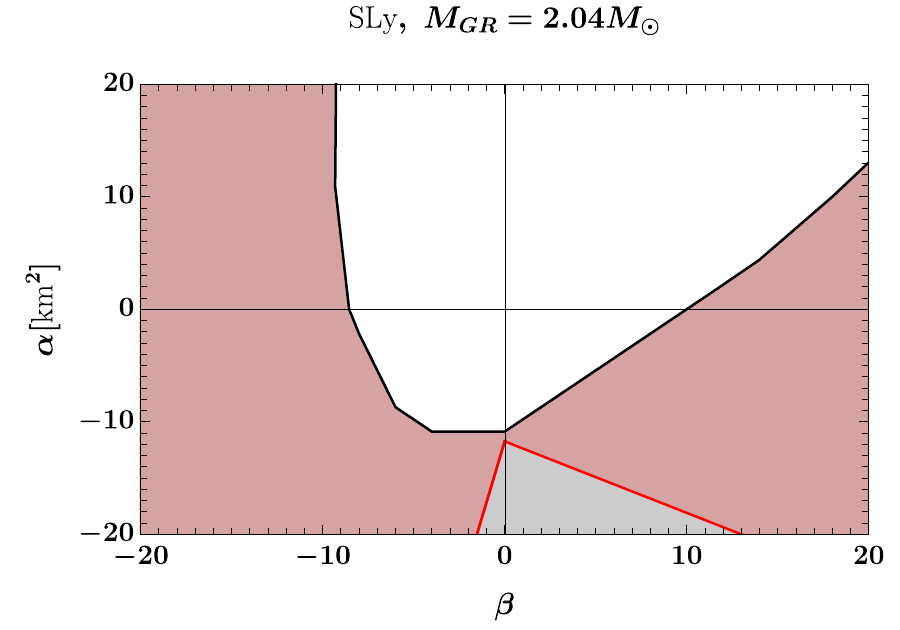}
\caption[Parameter space for "heavy" scalarized sRGB neutron stars obeying the SLy equation of state]{\textit{Left:} Regions of existence of scalarized solutions in the $(\alpha,\beta)$ space, for the SLy EOS with $\epsilon_0=3.4\times 10^{18}~\text{kg}/\text{m}^3$. The conventions are the same as in Fig.~\ref{fig:Sly112}. In GR, a star with this choice of $\epsilon_0$ and EOS is the heaviest possible, $M_\text{GR}=2.04 M_\odot$. \textit{Right:} This is simply a zoom of the left one.}
\label{fig:SLy204}
\end{figure}

Last, we consider a denser neutron star described by the SLy EOS, with $\epsilon_0=3.4\times 10^{18}\,\text{kg}/\text{m}^3$. It corresponds to an increased mass in GR of $M_{\text{GR}}=2.04 M_\odot$. The results are shown in Fig.~\ref{fig:SLy204}.
In this case, positive values of $\beta$ can also lead to scalarized solutions. Already in \cite{Mendes:2014ufa,Palenzuela:2015ima,Mendes:2016fby,Ventagli:2020rnx}, it was shown that, in GR, dense neutron stars have a negative Ricci scalar towards the center, which allows for scalarization to be triggered even when $\beta>0$. As before, a dashed line signals the appearance of divergences, which in this case show up already for the $n=0$ node.

In the right panel of Fig.~\ref{fig:SLy204}, we zoomed on the region of small couplings, in order to understand better what happens for natural values of the Ricci coupling $\beta$. In the absence of the Gauss-Bonnet coupling, scalarization can occur either if $\beta<-8.55$, or $\beta>11.5$. Let us concentrate on the $\beta>0$ scenario, which is motivated by Chapter \ref{ch:Cosmology}, where we showed that positive values of $\beta$ make GR a cosmological attractor. We remind that black hole scalarization (at least for non-rotating black holes) occurs for $\alpha>0$. Hence, we see that there exists an interesting region in the $\alpha>0,~\beta>0$ quadrant where even very compact stars do not scalarize, while black holes do. Such models can therefore \textit{a priori} pass all binary pulsar tests, while being testable with black hole observations. On the other hand, for $\beta \gtrsim 11.5$, the red region where GR solutions are replaced by scalarized solutions spreads very fast in the $\alpha$ direction, and one has to be careful, when considering black hole scalarization, that such models are not already excluded by neutron star observations.

\section{Mass and scalar charge of the \texorpdfstring{$\beta<0$}{} solutions}
\label{Sec:betaNeg}

\begin{figure}[!ht]
\centering
\includegraphics[width=0.49\linewidth]{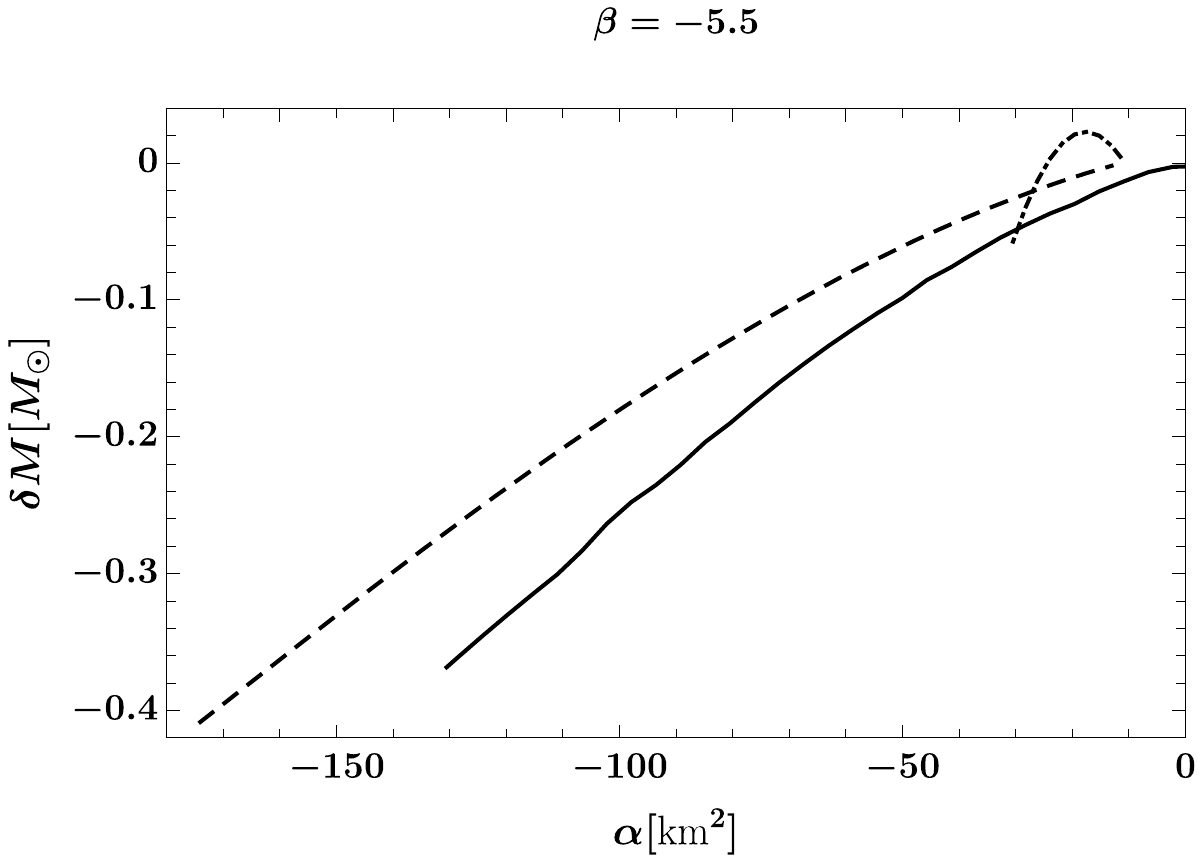}
\hfill
\includegraphics[width=0.49\linewidth]{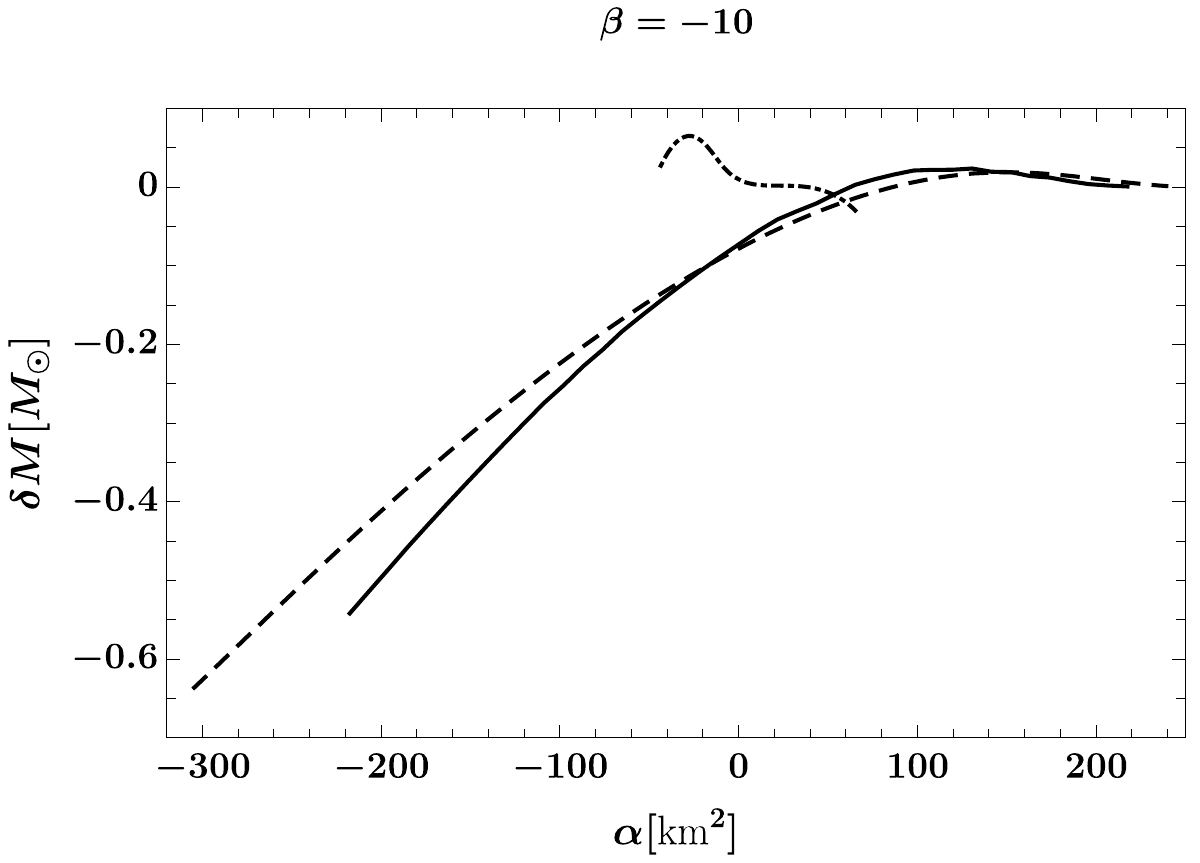}
\\
\includegraphics[width=0.49\linewidth]{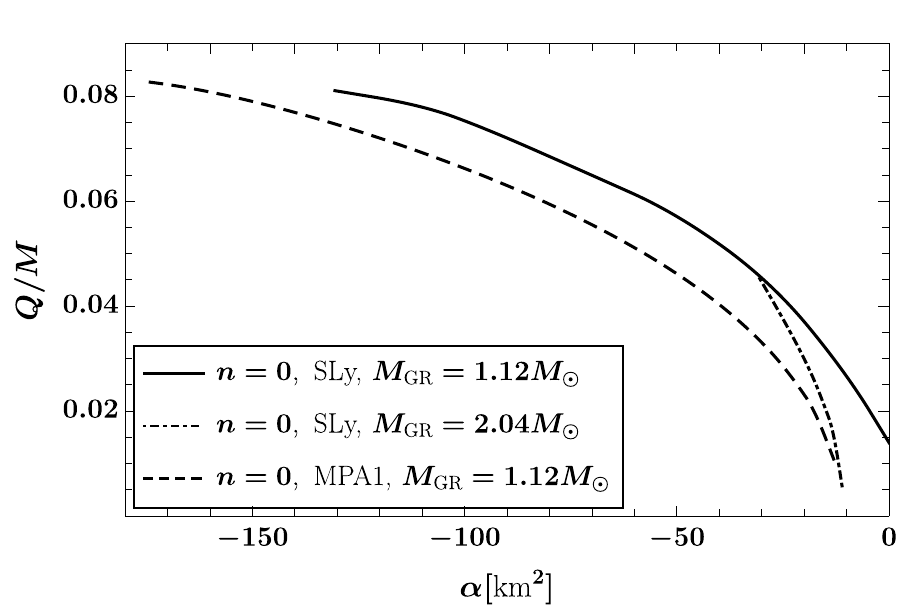}
\hfill
\includegraphics[width=0.49\linewidth]{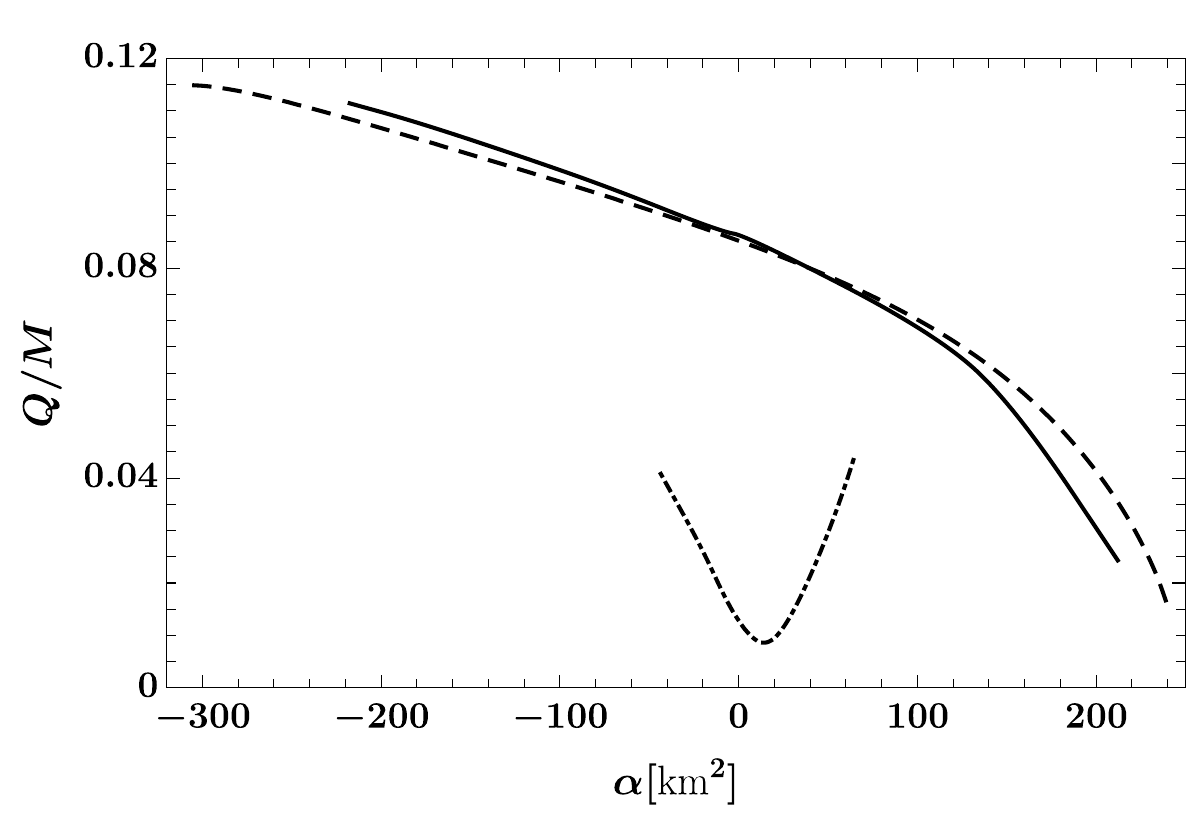}
\caption[Mass difference and scalar charge of scalarized sRGB neutron stars for negative Ricci couplings]{Mass difference and scalar charge of scalarized solutions for $\beta<0$. The two left (respectively right) panels show how these quantities evolve when varying $\alpha$ at fixed $\beta=-5.5$ (respectively $-10$). The scalar charge $Q$ (bottom panels) is normalized to the total mass of the solutions, $M$. For all curves, the mass difference $\delta M$ (upper panels) is computed with respect to a GR star with the same central density and EOS. Plain curves correspond to a GR mass of $1.12~M_\odot$, using the SLy EOS; dashed curves to the same GR mass, but the MPA1 EOS; and dotted-dashed curves to a GR mass of $2.04~M_\odot$, using the SLy EOS. In this region of the parameter space, only solutions with 0 nodes for the scalar field exist. A generic feature of lighter stars (plain and dashed curves), is that the charge decreases when $\alpha$ increases, \textit{a priori} offering a way to evade the stringent bound of Eq.~\eqref{eq:boundQ} when increasing $\alpha$. However, it is only for values of $\beta$ very close to the DEF threshold ($\beta=-5.5$) that we can obtain scalar charges compatible with observations.
}
\label{fig:smallCoup}
\end{figure}

We now focus on the scenario where $\beta<0$. This corresponds to the original situation studied by Damour and Esposito-Farèse. Typically, scalarized solutions with $\beta<0$ and $\alpha=0$ are extremely constrained by binary pulsar observations \cite{Freire:2012mg,Antoniadis:2013pzd,Shao:2017gwu}. A particular motivation to study solutions with $\beta<0$ is therefore to determine whether the addition of a non-zero Gauss-Bonnet coupling can improve their properties. We will consider three different choices of the Ricci coupling: $\beta=-5.5,-10$ and $-100$. The two first choices are relevant astrophysically: $\beta=-5.5$ is approximately the value where scalarization is triggered for small Gauss-Bonnet couplings, while $\beta=-10$ corresponds to a region where neutron stars are scalarized, but with rather small deviations with respect to GR. The third choice, $\beta=-100$, is certainly disfavored observationally, but it will allow us to illustrate an interesting behaviour concerning different scalar modes.

Let us start with the comparison between the cases $\beta=-5.5$ and $-10$. The results are summarized in Fig.~\ref{fig:smallCoup}.
This figure shows two properties of scalarized stars. First, the mass default (or excess) of scalarized stars with respect to GR stars with the same central density and EOS: 
$\delta M=M-M_{GR}$. 
Second, the scalar charge of the scalarized solutions, $Q$. We compare the results for the three different stellar models considered in Sec.~\ref{Sec:parameterSpace}, for the two values of $\beta$. 
All curves extend only over a finite range of $\alpha$. Indeed, passed a certain value of $\alpha$, we exit the red region on the $\beta<0$ side of Figs.~\ref{fig:Sly112}, \ref{fig:MPA1} and \ref{fig:SLy204} (moving vertically, since $\beta$ is fixed to $-5.5$ or $-10$). Scalarized solutions do not exist outside of this region. 

Figure \ref{fig:smallCoup} shows that the choice of EOS does not affect much the properties of the scalarized solutions.
However, increasing the density drastically modifies these properties. In particular, at higher densities, there exist solutions with $\delta M>0$. This can appear problematic at first. Indeed, one expects that, in a scalarization process, energy is stored in the scalar field distribution. Hence, the ADM mass, that constitutes a measure of the gravitational energy, should decrease in the process. 
However, we stress that we are not studying a dynamical process. Indeed, the stars for which we are computing the mass difference $\delta M$ have, by construction, the same central energy density $\epsilon_0$. In the scalarization process of a GR neutron star, the central energy density will not remain fixed. Hence, our results do not necessarily mean that a star will gain mass when undergoing scalarization.

Perhaps more interestingly for observations, Fig.~\ref{fig:smallCoup} also shows the behaviour of the scalar charge. For the light neutron stars, the scalar charge always decreases when $\alpha$ increases. Therefore, the constraint on the scalar charge, Eq.~\eqref{eq:boundQ}, disfavors the solutions with $\alpha<0$ with respect to standard DEF ($\alpha=0$) solutions. On the contrary, one could hope that a positive Gauss-Bonnet coupling could help evade these constraints even for $\beta<-5.5$, by quenching the charge. Effectively, there will be a direction in the $\alpha>0$ and $\beta<0$ quadrant where the effects of the two operators, Ricci and Gauss-Bonnet, combine to yield a small scalar charge.
This interesting possibility is moderated by what happens in the case of denser stars (dotted-dashed line in Fig.~\ref{fig:smallCoup}). For large negative values of the Ricci coupling ($\beta=-10$), the scalar charge does not have a monotonic behaviour with $\alpha$. In particular, as shown in the bottom-right panel of Fig.~\ref{fig:smallCoup}, $Q$ starts increasing for positive values of $\alpha$. Even at the point where $Q$ is minimal, its value ($Q/M\simeq8\times10^{-3}$) already exceeds the bound of Eq.~\eqref{eq:boundQ}. Therefore, it is only for values of $\beta$ that are very close to the DEF threshold $\beta\simeq-5.5$, that the addition of the Gauss-Bonnet coupling can help to reduce the scalar charge, and to pass the stringent binary pulsar tests.

\begin{figure}[!t]
\centering
\includegraphics[width=0.65\linewidth]{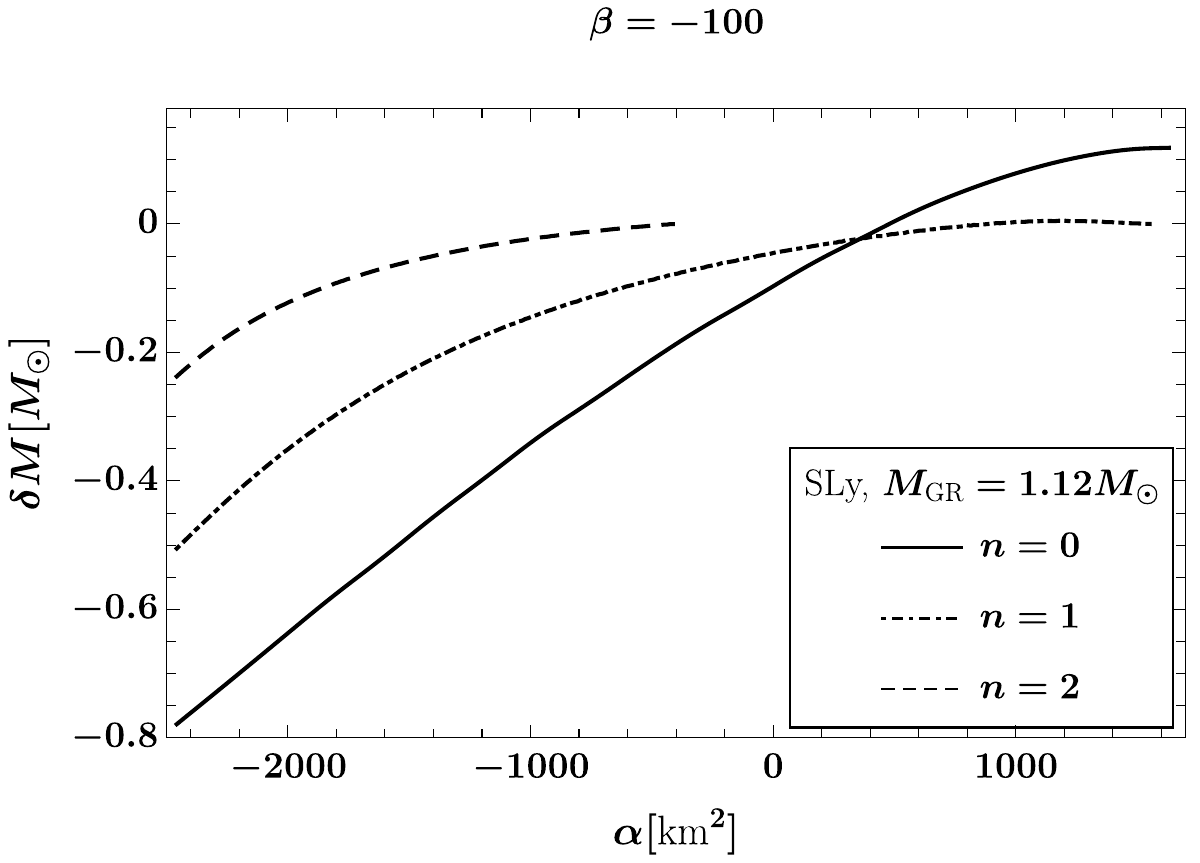}
\caption[Mass difference $\delta M$ vs $\alpha$ at $\beta=-100$ for the SLy EOS]{Mass difference $\delta M$ vs $\alpha$ at $\beta=-100$. The EOS considered here is the SLy one, with $\epsilon_0=8.1\times 10^{17}\,\text{kg}/\text{m}^3$, which in GR corresponds to $M_\text{GR}=1.12~M_\odot$. The color and dashing conventions is the same as in Fig.~\ref{fig:smallCoup}. We have more modes in this region of parameter space, that we represent as dotted-dashed (for $n=1$ node) and dashed (for $n=2$ nodes) curves. For $\alpha\gtrsim350\, \text{km}^2$, solutions with 1 node start having a smaller mass than solutions with 0 nodes, which can indicate that solutions with 1 node are more energetically favored.}
\label{fig:betaNeg100}
\end{figure}

To conclude the study of the $\beta<0$ region, we consider a significantly more negative Ricci coupling, namely $\beta=-100$. To illustrate what happens at these large negative values of $\beta$, it is enough to consider one scenario, for example the one of lighter neutron stars with the SLy EOS. For such negative values of $\beta$, there exist several scalarized solutions, with different number of nodes. We can then compare the mass difference of these solutions between each other. Figure \ref{fig:betaNeg100} shows that, for $\alpha>\alpha_\text{c}\simeq350\, \text{km}^2$, scalarized solutions with 1 node become lighter than scalarized solutions with 0 nodes.
This is a hint that, for $\alpha>\alpha_c$, the one node solution will be preferred energetically to the zero node solution. We cannot conclude definitively on this issue, as the ADM mass does not take into account the energy stored in the scalar distribution (which is non-zero for the two scalarized solutions). However, in the regime where this inversion happens, the mass difference with respect to GR, $\delta M$, is rather small. If our interpretation in terms of energetic preference is correct, the transition from a preferred solution with zero node to a solution with one node is interesting. Indeed, the scalarized solution with zero node is associated with the fundamental mode of the GR background instability. At the perturbative level, all the other modes of instability have higher energies. It would then be natural to expect that, at the non-linear level of scalarized solutions, this energy hierarchy is respected. This is the case up to $\alpha=\alpha_\text{c}$, but not anymore beyond. We will return to this in a while, where we provide a putative explanation for this inversion: that for $\alpha>\alpha_\text{c}$, the profile of the effective mass over the GR background tends to favor the growth of scalar field solutions with one node, rather than zero.

\section{Mass and scalar charge of the \texorpdfstring{$\beta>0$}{} solutions}\label{Sec:betaPos}

\begin{figure}[ht]
\includegraphics[width=0.47\linewidth]{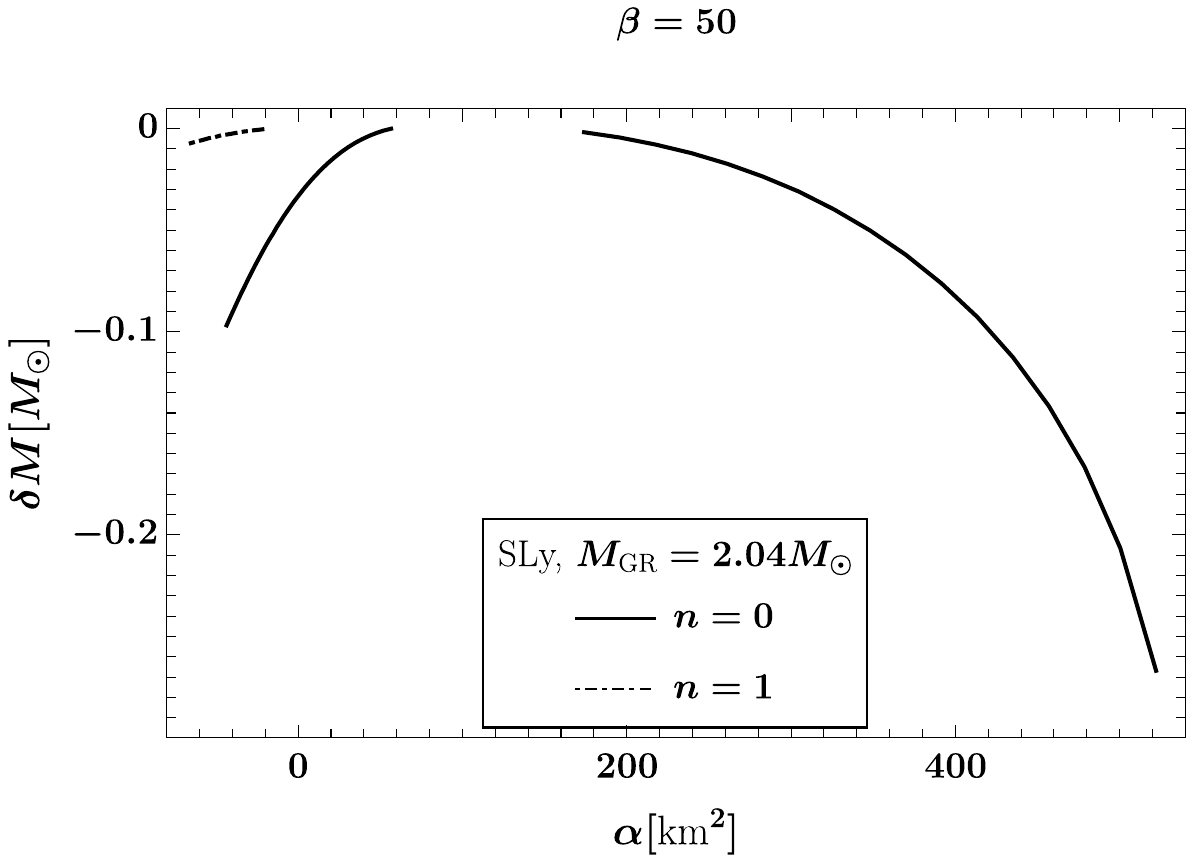}
\hfill
\includegraphics[width=0.47\linewidth]{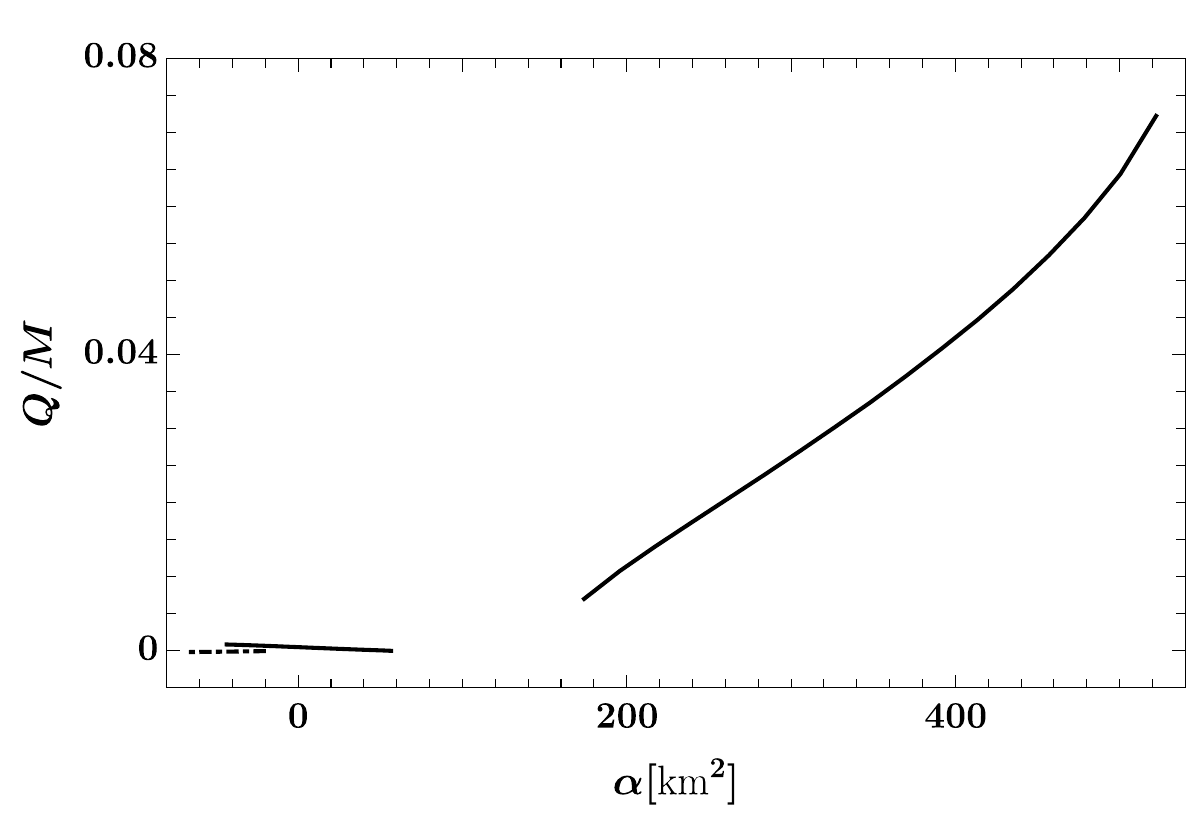}   
\caption[Mass difference and scalar charge of the scalarized neutron stars with positive Ricci couplings]{Mass difference and scalar charge of scalarized solutions for $\beta>0$ ($\beta=50$ here). Among the three neutron star scenarios that we considered throughout the paper, only the heavier star ($\epsilon_0=5.51 \times 10^{-3}$~kg/m$^3$, $M_\text{GR}=2.04~M_\odot$, SLy EOS) possesses some scalarized solutions in this region. The dashing convention is the same as in Fig.~\ref{fig:betaNeg100}. Solutions that correspond to the interval of $\alpha$ centered on 0 are interesting observationally, as they yield very small scalar charges, compatible with Eq.~\eqref{eq:boundQ}.}
\label{fig:50M204}
\end{figure}

We now consider the case of positive $\beta$. Such solutions are less constrained by observations than their $\beta<0$ counterparts. They are also very interesting from a cosmological perspective, where $\beta>0$ allows a consistent history throughout different epochs, as we showed in Chapter \ref{ch:Cosmology}. We have seen in Sec.~\ref{Sec:parameterSpace} that, among the three different possible neutron star configurations we focus on, only the denser one leads to scalarized solutions for $\beta>0$. In Fig.~\ref{fig:50M204}, we show the mass difference $\delta M$ and scalar charge $Q$ as functions of $\alpha$ when $\beta=50$.
Note that scalarized solutions with zero nodes exist over two disconnected ranges of $\alpha$ ($-44~\text{km}^2<\alpha<57~\text{km}^2$ and $174~\text{km}^2<\alpha<522~\text{km}^2$). In the gap, GR solutions are stable and no scalarized solutions exist. This is obvious from Fig.~\ref{fig:SLy204}, taking a cut along the vertical line $\beta=50$. 
    
Over the first interval, $\alpha$ is rather small and the scalarization process is dominated by the negative Ricci scalar. For strictly vanishing $\alpha$, the scalarization phenomenon with $\beta>0$ has already been examined in \cite{Mendes:2014ufa,Palenzuela:2015ima,Mendes:2016fby}. Here, we find that, in the interval of small values of $\alpha$, the scalar charges of the $n=0$ solutions (as well as of the $n=1$ solutions) are very small. Typically, $Q/M \simeq 10^{-4}-10^{-5}$, compatible with Eq.~\eqref{eq:boundQ}. Hence, all solutions  with $\beta>0$ and rather small values of $\alpha$ are interesting observationally: they display either no scalarization effects for neutron stars (for $\beta\lesssim 11.51$) or very mild scalar charges (for $\beta\gtrsim 11.51$). At the same time, they allow for a consistent cosmological history; finally, together with positive values of $\alpha$, they will generically give rise to black hole scalarization, as studied in detail in \cite{Antoniou:2021zoy}. In this region of parameter space, we can therefore hope to discover scalarization effects in the future gravitational-wave signals of binary black holes, that are either absent or suppressed in the case of neutron stars.
    
Over the second interval ($174~\text{km}^2<\alpha<522~\text{km}^2$), the contribution of the Gauss-Bonnet invariant tends to dominate, and the scalar charges are more significant, as one can immediately notice in Fig.~\ref{fig:SLy204}. Such setups are not compatible with Eq.~\eqref{eq:boundQ}, and therefore less interesting phenomenologically.

\section{Scalarized solutions in the proximity of the instability lines}\label{Sec:instabilityLines}

\begin{figure}[ht]
\centering
\includegraphics[width=0.47\linewidth]{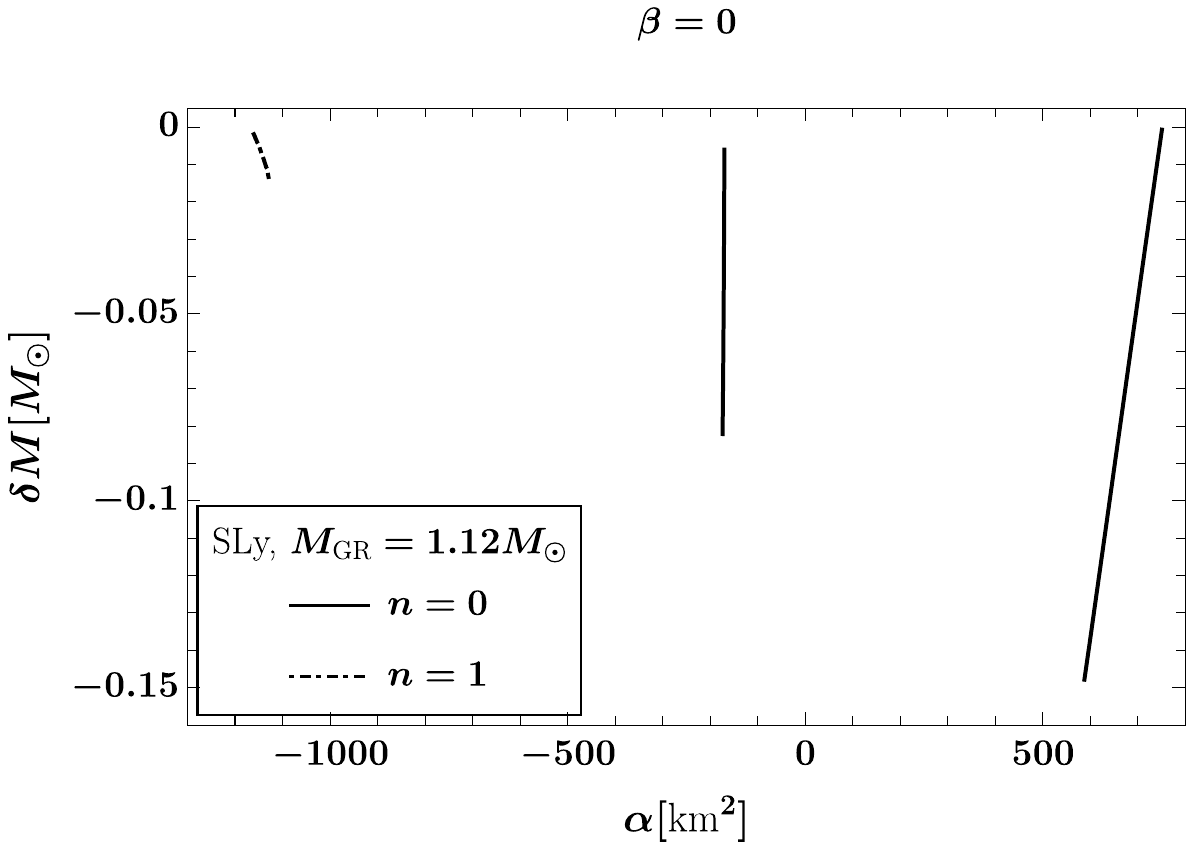}
\hfill
\includegraphics[width=0.47\linewidth]{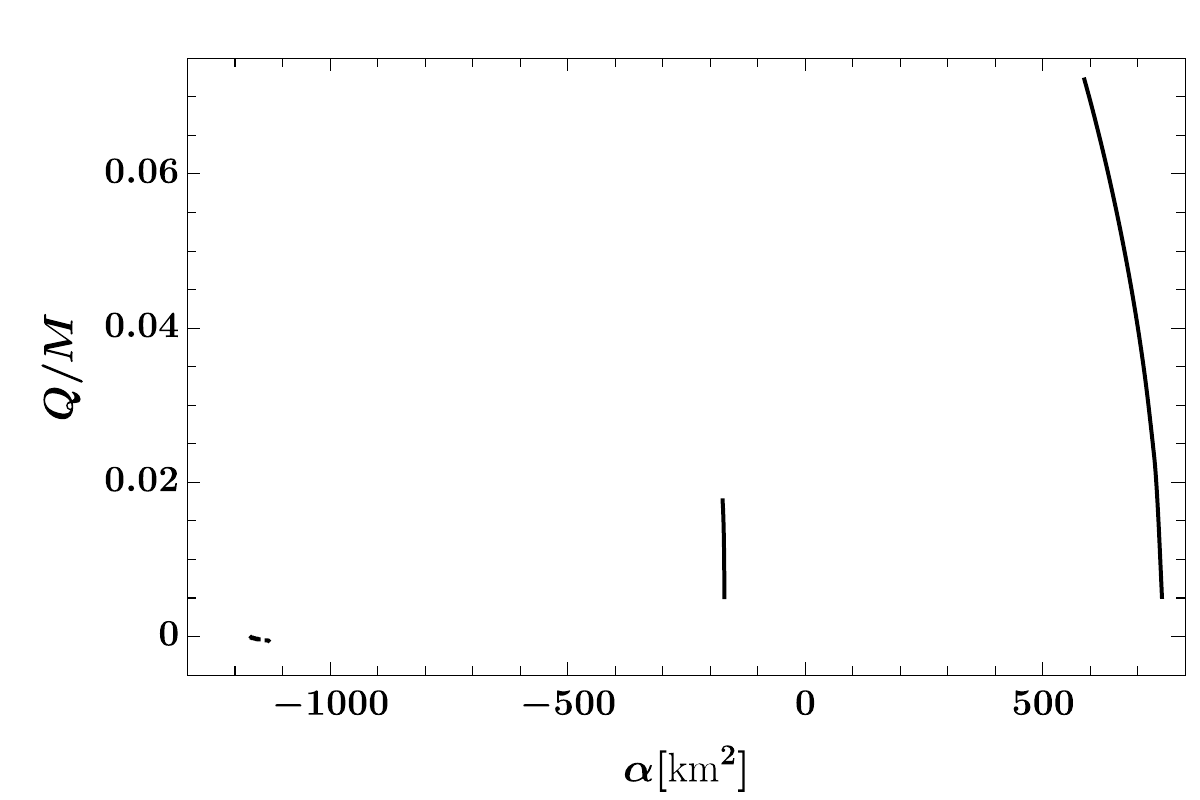}
\caption[Mass difference and scalar charge of the scalarized neutron stars along the instability lines for $\beta=0$]{Mass difference and scalar charge of the scalarized solutions along the instability lines, for $\beta=0$. The scenario considered here corresponds to $\epsilon_0=8.1\times 10^{17}~\text{kg}/\text{m}^3$ ($M_\text{GR}=1.12M_\odot$) together with the SLy EOS. Solutions with zero node acquire a significant charge and mass difference, and are apparently disconnected from GR when they appear while increasing $\alpha$ towards positive values. Solutions with $n=1$ nodes are very close to GR, with a small charge and mass difference. Since they extend only over a small range of $Q$ and $\delta M$, they are difficult to spot. They lie at the upper left (respectively lower left) of the top (respectively bottom) panel.
}
\label{fig:beta0}
\end{figure}

As we mentioned at the end of Sec.~\ref{sec:lightSLy}, a generic feature that is not observable in Figs.~\ref{fig:Sly112}, \ref{fig:MPA1} and \ref{fig:SLy204}, is that scalarized solutions are present in a tiny band close to each instability line.
Let us illustrate this with the light star model (with SLy EOS), that is the one which corresponds to Fig.~\ref{fig:Sly112}. For simplicity, we also restrict our study to solutions with $\beta=0$ (\textit{i.e.}, we take a cut along the vertical axis in Fig.~\ref{fig:Sly112}). The characteristics of the solutions are shown in Fig.~\ref{fig:beta0}.
Scalarized solutions with zero nodes (the ones lying close to the $n=0$ instability line of the GR solution) have a characteristic mass difference and scalar charge which is not particularly small. It is of the same order as for the solutions we previously examined (Figs.~\ref{fig:smallCoup}--\ref{fig:50M204}). They also exhibit a surprising behaviour: when increasing $\alpha$ progressively from 0 towards positive values, the mass and scalar charge suddenly deviate from GR, instead of being smoothly connected; further increasing $\alpha$, $\delta M$ and $Q$ then tend to decrease. This behaviour is significantly different from what we could observe in Figs.~\ref{fig:smallCoup}--\ref{fig:50M204}.
 
Solutions with more nodes ($n=1$, 2, 3...) exhibit a clear feature: they deviate very slightly from GR in terms of mass, and acquire only a small scalar charge (typically $\delta M < 10^{-2}$ and $Q/M < 10^{-4}$). We verified this behaviour for all higher nodes admitted; however, for simplicity, in Fig.~\ref{fig:beta0} we show only the case $n=1$.
This feature can be understood as follows; close to some instability line (on the unstable side), an unstable mode of the effective potential associated with the GR solution has just appeared. A very small deformation of the potential can therefore easily restore the equilibrium. This deformation can be caused by the back-reaction of the scalar onto the metric: the instability is triggered, the scalar field starts growing, but it immediately back-reacts on the potential, making it shallower and suppressing the instability. Clearly, such a behaviour can only happen close to instability lines, where a specific mode is on the edge of stability.

\section{The scalar profile of scalarized stars from GR solutions}\label{Sec:EffMass}

We will conclude this study by arguing that, already at the perturbative level of the GR solution, we can identify an influence on the profile of the scalar field in the fully scalarized solution. To this end, let us focus on the effective mass given in Eq.~\eqref{eq:scal_eq}, $ m_\text{eff}^2=\beta R/2-\alpha \GB$. This is a radially dependent quantity, and the scalar field is most likely to grow at radii where $m_\text{eff}^2$ is most negative. In particular, it is natural to expect that, if $m_\text{eff}^2$ has a minimum at $r=0$, this will favor a monotonic profile for the scalar field, and hence an $n=0$ type of solution. On the contrary, if $m_\text{eff}^2$ has a minimum at $r>0$, this favors a peaked profile for the scalar field, which is more common in $n\geq1$ solutions. Let us illustrate this with a concrete example. We will
consider the scenario that corresponds to $M_\text{GR}=1.12 M_\odot$, together with the SLy EOS, and two choices of $\beta$: $\beta=-10$ and $\beta=-100$. In the first case, only solutions with 0 nodes exist; in the second case, we can construct solutions with 0 or 1 node.

\begin{figure}[ht]
\centering
\includegraphics[width=0.47\linewidth]{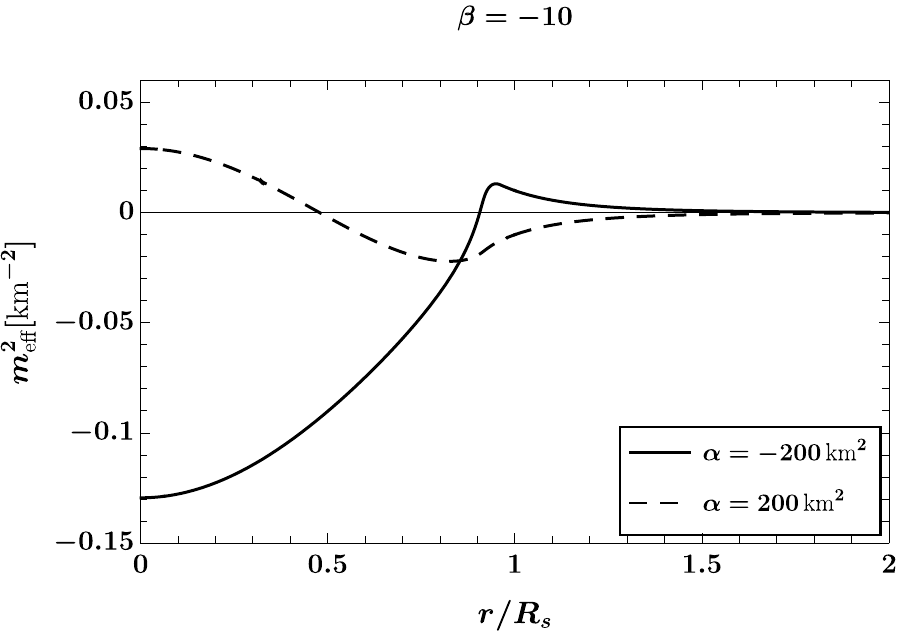}
\hfill
\includegraphics[width=0.47\linewidth]{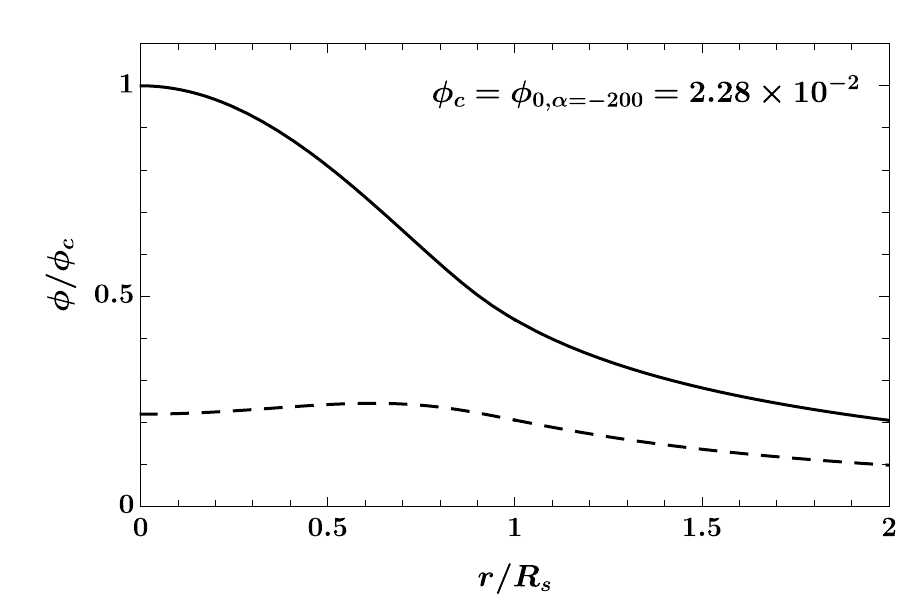}
\caption[Radial profiles of the effective mass and the scalar field]{\textit{Left Panel:} Radial profile of the effective mass squared over the GR background, using the SLy EOS and a central density $\epsilon_0=8.1\times 10^{17}\,\text{kg}/\text{m}^3$ (yielding $M_{\text{GR}}=1.12 M_{\odot}$), for $\beta=-10$ and $\alpha=\pm200$~km$^2$. \textit{Right panel:} Radial profile of the scalar field, this time in the fully scalarized solution with the same EOS, central density, and Lagrangian parameters. The radial coordinate is normalized by $R_\text{s}$, the radius of the star surface. In the lower panel, the scalar field is normalized to its central value for $\alpha=-200\, \text{km}^2$. When the minimum of $m_\text{eff}^2$ is shifted to $r>0$, so is the peak of $\phi$.}
\label{fig:EffMass2}
\end{figure}

We first focus on the case $\beta=-10$. The Ricci scalar is everywhere positive over the background we consider, with a maximum at $r=0$; hence, $\beta R$ contributes negatively to the squared mass, favouring the growth of the scalar field close to the center. The Gauss-Bonnet scalar, on the other hand, is negative in the central region of the star, and becomes positive towards the surface. Therefore, $-\alpha\mathscr{G}$ reinforces the effect of $\beta R$ if $\alpha<0$, while couterbalancing it if $\alpha>0$. This is illustrated in the left panel of Fig.~\ref{fig:EffMass2}.
The right panel shows the scalar profile of the fully scalarized solutions associated with the same parameters. In this range of parameters, only solutions with 0 nodes are allowed (as one can check in Fig.~\ref{fig:Sly112}); hence, pushing the minimum of $m_\text{eff}^2$ away from the center cannot favour $n=1$ solutions, which do not exist. Still, we notice that positive $\alpha$ values, which have the effect of displacing the minimum of $m_\text{eff}^2$ to $r>0$, also displace the peak of the scalar field to $r>0$. The peak of the scalar field is located approximately at the minimum of $m_\text{eff}^2$. Again, one must be careful in the comparison of the two panels, as one of them corresponds to a GR star while the other one corresponds to a scalarized star. However, our analysis seems to capture what happens during the transition from the GR to the scalarized branch.

\begin{figure}[!ht]
\centering
\includegraphics[width=0.49\linewidth]{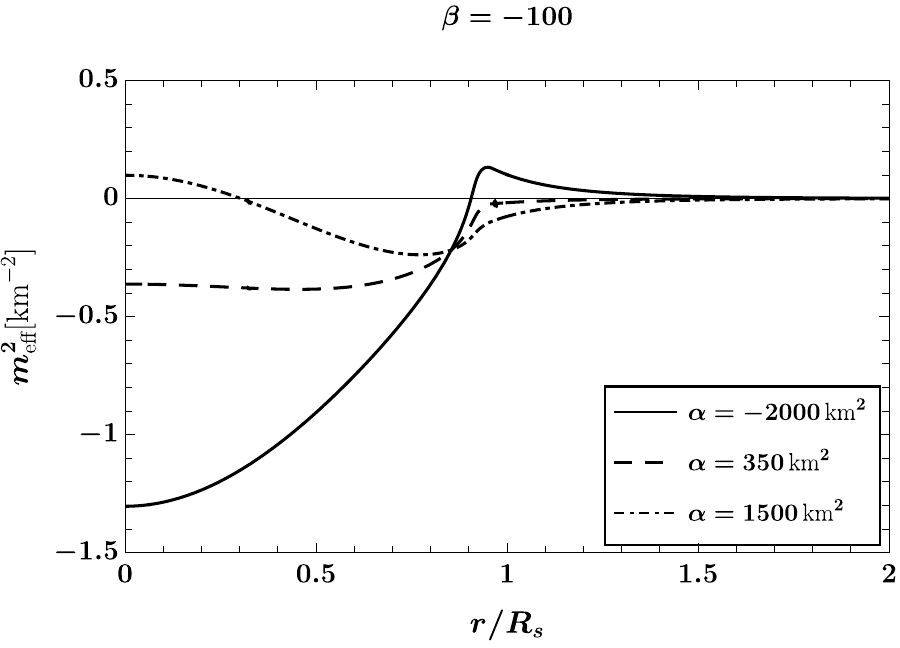}
\caption[Radial profile for the effective mass with zero and one nodes for the SLy EOS]{Radial profile of the effective mass squared over the GR background, using the SLy EOS and a central density $\epsilon_0=8.1\times 10^{17}\,\text{kg}/\text{m}^3$ (yielding $M_{\text{GR}}=1.12 M_{\odot}$), for $\beta=-100$ and $\alpha=-200$, 350 or 1500~km$^2$.}
\label{fig:EffMass3}
\end{figure}
\begin{figure}[!ht]
\centering
\includegraphics[width=0.49\linewidth]{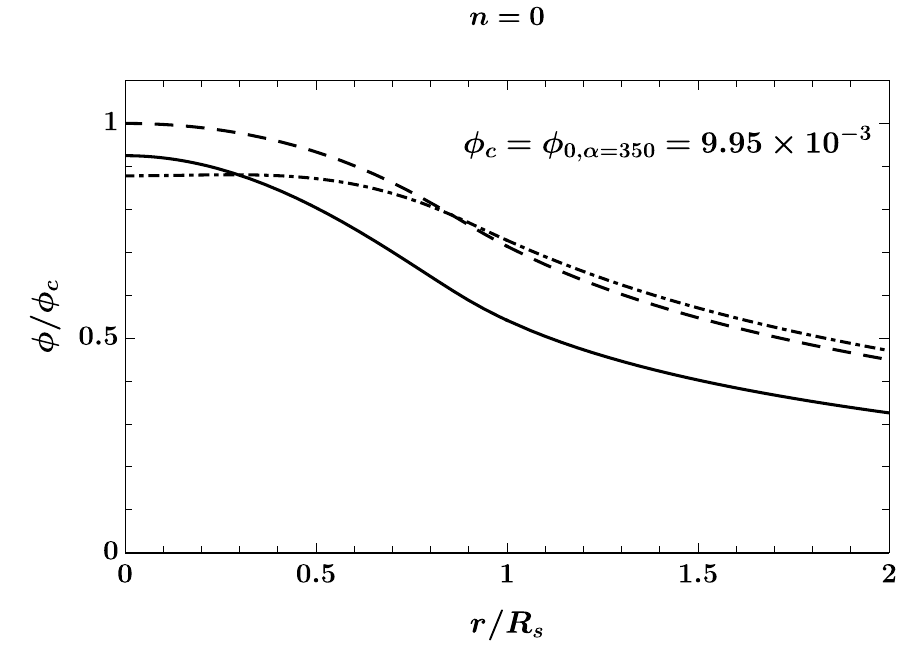}
\hfill
\includegraphics[width=0.49\linewidth]{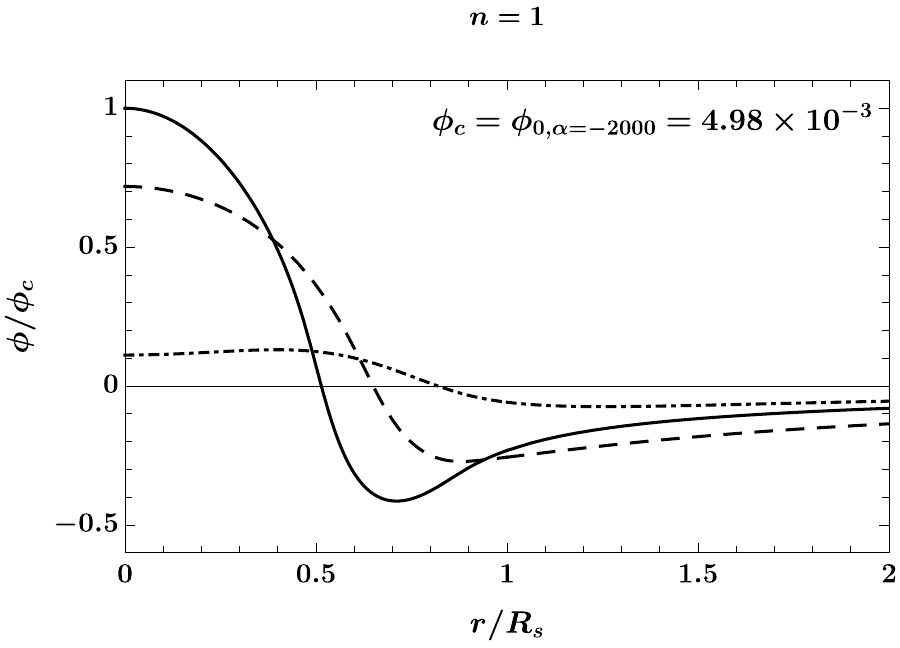}
\caption[Radial profile of the scalar field with $n=0,1$]{\textit{Left:} Radial profile of the scalar field solution with 0 nodes in the fully scalarized solution with the same EOS, central density, and Lagrangian parameters. The normalization is similar to the one of Fig.~\ref{fig:EffMass2}. When increasing $\alpha$, the minimum of $m_\text{eff}^2$ is progressively shifted from $r=0$ to a finite radius, alternatively favoring the growth of $n=0$ and $n=1$ solutions. \textit{Right:} Same but for the 1 node case.}
\label{fig:phi-r}
\end{figure}

To illustrate better the transition between $n=0$ and $n=1$ solutions, let us now consider the case $\beta=-100$. 
The qualitative discussion about the effect of $\beta R$ and $-\alpha\mathscr{G}$ over the effective mass is exactly the same as in the previous case. We will therefore consider again a large negative and a large positive value of $\alpha$, as well as an intermediate one: $\alpha=-2000, \,350$ and 1500~km$^2$. Note that the intermediate value corresponds to $\alpha_\text{c}$ in Sec.~\ref{Sec:betaNeg}, the critical value at which scalarized stars with $n=0$ node become more massive (and hence probably less stable) than those with $n=1$ node.
In Fig.~\ref{fig:EffMass3} we show the profile of the effective scalar mass. It behaves exactly as in the case $\beta=-10$, with a minimum at $r=0$ for negative values of $\alpha$, which is progressively shifted to larger radii when we increase $\alpha$. For the parameters we chose, this time, both solutions with zero and one nodes exist. In the left (respectively right)  panel of Fig.~\ref{fig:phi-r}, we show the $n=0$ (respectively $n=1$) solutions. In Sec.~\ref{Sec:betaNeg}, we stated that for $\alpha<\alpha_c$ we expected that the zero node solution will be energetically preferred over the one node solution, and vice-versa for $\alpha>\alpha_c$. The profiles of the effective mass squared give a complementary argument that strengthens this expectation. Indeed, for $\alpha=-2000\,\text{km}^2\ll\alpha_c$ the shape of $m_\text{eff}^2$ favours a scalar solution with a maximum at the center of the star, which decays monotonically with $r$, \textit{i.e.} a $n=0$ solution. For $\alpha=1500\,\text{km}^2\gg\alpha_c$, the tachyonic instability is still triggered inside the star, but away from the center. Thus, we expect that a solution with one node will be favoured. The transition between a minimum at $r=0$ and $r>0$ indeed seems to occur around $\alpha_\text{c}$. Even though it is beyond the scope of this work, a thermodynamical analysis of the solutions could potentially shed some more light on the aforementioned energetic considerations.

\section{Discussion}

In this chapter we performed a comprehensive study for the EsRGB model we introduced earlier in the thesis, in the neutron star scenario. We have identified the regions of the parameter space where solutions exist, considering three different stellar scenarios which correspond to different central densities and EOS. Although we have considered only a limited number of different central densities, we have selected the ones that correspond to the lowest/largest neutron star mass in GR, in order to cover very different setups. The regions where scalarized solutions exist are systematically smaller than the ones where the GR branch is tachyonically unstable. The complementary regions, where the GR solution is unstable while no scalarized solution exists, should be excluded.

We then investigated in detail the physical characteristics of the scalarized solutions. In general, large parameters ($|\beta|\gg 1$ or $|\alpha|\gg L^2$, where $L\simeq10$~km is the typical curvature scale) lead to scalar charges that would be in conflict with binary pulsar constraints. However, it is interesting to notice that solutions with $\beta>0$ and reasonably small $\alpha$ (typically $|\alpha|\lesssim 50$~km$^2$) lead either to stable GR configurations, or to scalarized stars with small charges. Remarkably, this is the region of the $(\alpha,\beta)$ parameter space for which GR is a cosmological attractor \cite{Antoniou:2020nax} and black holes scalarization can take place \cite{Antoniou:2021zoy}, as we explained in the two previous chapters. Therefore, it is possible to construct scalarization models that are consistent with current observations, while still having interesting strong field phenomenology. It is worth noting that future gravitational-wave observations have the potential to reach the precision required to measure small scalar charges for neutron stars.

We have also discovered that scalarized solutions systematically exist near the thresholds that delimit the stability of the GR solutions, and provided a putative explanation for this. Finally, we have shown that the profile of the effective mass at the GR level can foster the growth of certain modes, characterizing the scalar profile, with respect to others.

%% file: Chapters/stability_qnms.tex
In Chapter~\ref{ch:Black holes} we talked about how the effect that the scalar-Ricci coupling has on the black hole solution lines in the $\hat{Q}$-$\hat{M}$ plots, may be affecting the stability of the scalarized solutions.
We derived a threshold value for the scalar-$R$ coupling, associated with a change in the tilt of these curves and explained how in previous works this tilt has been associated with the radial stability of the solution \cite{Silva:2018qhn,Macedo:2019sem}. In this chapter we will further explore this by performing a stability analysis for the black holes derived in Chapter~\ref{ch:Black holes}.

Understanding how the various (self)interactions beyond the scalar-Gauss-Bonnet coupling affect the scalar profile of a scalarized compact object is essential from an observational perspective.
It has been shown that scalarized black holes are unstable under radial perturbations in the simplest, quadratic coupling scenario \cite{Blazquez-Salcedo:2018jnn}.
This issue can be overcome if one considers an additional quartic interaction in the Gauss-Bonnet coupling function, provided that the sign of the quartic coupling coefficient is opposite to the quadratic one \cite{Silva:2018qhn}. 
However, addressing the instability with quartic, or exponential couplings is not entirely appealing from an effective field theory (EFT) perspective. 
This is because these terms have a higher mass dimension than other terms that could contribute non-linearly, e.g. a simple $\phi^4$ self interaction.
It was, indeed, shown in Ref.~\cite{Macedo:2019sem} that including self interactions for the scalar can lead to radially stable scalarized solutions.

If we can actually demonstrate that the scalar-Ricci coupling can stabilize the solutions, we would confirm another important effect that this interaction poses in the general framework of scalar tensor theories.
This would come as an addition to the results of Chapters~\ref{ch:Cosmology} and \ref{ch:Neutron Stars}.
In this chapter, we perform a radial perturbation analysis for scalarized solutions and fully explore the role of the Ricci-scalar coupling in the stability of the solutions \cite{Antoniou:2022agj}. We also explore the interesting effect that this term has on the hyperbolicity of the problem.

\section{Thermodynamics and stability}
\begin{figure}[h]
\centering
\includegraphics[width=0.65\linewidth]{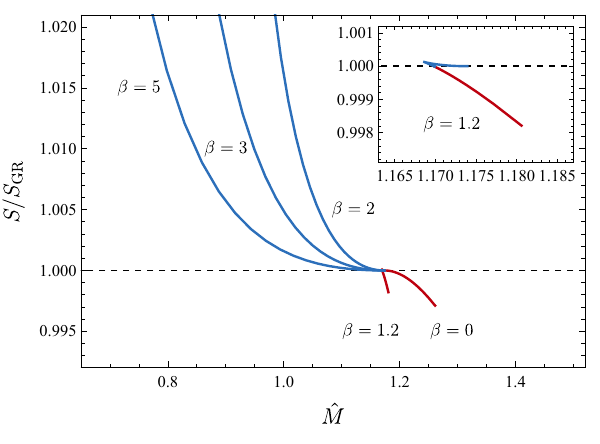}
\caption[Entropy of scalarized black holes in the sRGB model]{Ratio of the entropy of the scalarized solutions over the GR solutions with respect to $\hat{M}$, and for different values for the $\beta$ coupling. The inset corresponds to a zoom of the plot in order to depict clearly the case $\beta=1.2$.}
\label{fig:thermodynamics}
\end{figure}

Before proceeding to perturbing the black hole solutions we presented in Chapter~\ref{ch:Black holes} and numerically testing their stability, let us briefly examine the thermodynamical differences between these scalarized solutions and the GR black holes. In Appendix~\ref{ch:Appendix-Thermodynamics}, we derive the expression for the entropy if one allows for generic couplings of the scalar with the Ricci and Gauss-Bonnet invariants, which is given in Eq.~\eqref{eq:entropy}. Here it takes the following form
\begin{equation}
    S_h=\left[1-\frac{\beta\phi^2}{4}\right]\, \frac{A_h}{4} +2 \alpha \pi \phi^2\, ,
\end{equation}
where $A_h$ is the horizon area of the scalarized black hole. In Fig.~\ref{fig:thermodynamics} we plot the entropy of the scalarized black hole, normalized with the entropy of the GR black hole of the same mass, versus the normalized mass $\hat{M}$.

The results presented in  Fig.~\ref{fig:thermodynamics} are in agreement with the assumption we made in Chapter~\ref{ch:Black holes}, about left and right tilting curves. The blue lines in the entropy plots correspond to left tilting (parts of) curves, while the red ones to right tilting ones. For the blue (red) curves the scalarized black hole has a larger (smaller) entropy with respect to the GR solution indicating their stability (instbility). In Fig.~\ref{fig:scalarized_solutions} we saw that the curve corresponding to $\beta=1.2$, which is around the critical value, had both a left (blue) and a right (red) tilting part. This is reflected on the inset, zoomed plot in Fig.~\ref{fig:thermodynamics}. The blue part of the curve has an entropy ratio greater than one, indicating stable configurations, and the red part has a ratio smaller than one, correspondingly indicating unstable solutions.

\section{Radial perturbations}

After looking at the thermodynamical stability of the scalarized solutions, we proceed to verify our findings by employing a perturbative approach in order to investigate radial stability.
This not only allows us to further elucidate the timescale of the instability, but also to analyze possible modifications to the oscillatory spectrum of the BHs. 
It is worth noting that since a scalar degree of freedom is absent in GR, the radial modes contribute only to a shift in the mass in that case, and hence are not radiative in nature~\cite{Regge:1957td}.

We start by considering time-dependent radial perturbations of the metric tensor and scalar field over the static and spherically symmetric background \eqref{eq:metric}
\begin{align}
    ds^2=&-[A_0(r)+A_1(t,r)]dt^2+\frac{dr^2}{B_0(r)+ B_1(t,r)}+r^2d\Omega^2\, ,
    \\[2mm]
    \phi=&\;\phi_0(r)+\phi_1(t,r)\, ,
\end{align}
where $A_0$, $B_0$ and $\phi_0$ are the time-independent background solutions, while $A_1$, $B_2$ and $\phi_1$ are the time-dependent perturbations. We can then write down a system of equations for $(A_1,B_1,\phi_1)$, by substituting the metric and scalar perturbations in Eqs. \eqref{eq:grav-BH} and \eqref{eq:scal_eq}.
This system can be reduced to a second order partial differential equation system for $\phi_1$~\cite{Blazquez-Salcedo:2018jnn}. If we follow the process presented in \cite{Minamitsuji:2018xde}, the equations of motion can be written as
\begin{align}
    & \alpha _1 \phi _1''+\alpha _2 \phi _1'+\alpha _3 \phi _1+\alpha _4 B_1+\alpha _5 B_1'=0,\label{eq:radial_pert_tt}\\
    & \beta _1 \ddot{\phi}_1+\beta _2 \phi _1'+\beta _3 \phi _1+\beta _4 A_1+\beta _5 A_1'+\beta _6 B_1=0,\label{eq:radial_pert_rr}\\
    & \gamma _1 B_1+\gamma _2 \phi _1'+\gamma _3 \phi _1=0,\label{eq:radial_pert_tr}\\
    \begin{split}
    & c_1 \ddot{\phi}_1+c_2 \phi _1''+c_3 \phi _1'+c_4 \phi _1+c_5 A_1''+c_6 A_1'+A_1 c_7+c_8 \ddot{B}_1\\
    & +c_9 B_1'+c_{10} B_1 =0,\label{eq:radial_pert_thetatheta}
    \end{split}
    \\
    \begin{split}
    & d_1 \ddot{\phi}_1+d_2 \phi _1''+d_3 \phi _1'+d_4 \phi _1+d_5 A_1''+d_6 A_1'+A_1 d_7+d_8 \ddot{B}_1\\
    & +d_9 B_1'+d_{10} B_1=0\, .\label{eq:radial_pert_scalar}
    \end{split}
\end{align}
Then, we can solve \eqref{eq:radial_pert_tr} for
\begin{align}
    B_1=&-\frac{\gamma _2 \phi _1'+\gamma _3 \phi _1}{\gamma _1}\xrightarrow{\eqref{eq:radial_pert_rr}}\\
    A_1'=&-\frac{\beta _1 }{\beta _5}\ddot{\phi}_1+\frac{\left(\beta _6 \gamma _3-\beta _3 \gamma _1\right)}{\beta _5 \gamma _1}\phi _1+\frac{\left(\beta _6 \gamma _2-\beta _2 \gamma _1\right)}{\beta _5 \gamma _1}\phi _1'-\frac{A_1 \beta _4}{\beta _5}\, .
\end{align}
If we now multiply \eqref{eq:radial_pert_thetatheta} with $d_5$ and \eqref{eq:radial_pert_scalar} with $c_5$ and then subtract them, we can eliminate $A_1''$, and we reach the following equation
\begin{equation}
\begin{split}
        &\ddot{\phi}'_1\left(\frac{c_5 \gamma _2 d_8}{\gamma _1}-\frac{c_8 \gamma _2 d_5}{\gamma _1}\right) +\phi _1'' \bigg(c_2 d_5-\frac{c_9 \gamma _2 d_5}{\gamma _1}+\frac{c_5 \gamma _2 d_9}{\gamma _1}-c_5 d_2\bigg)\\
        &+\ddot{\phi}_1 \bigg[d_5 \left(-\frac{\beta _1 c_6}{\beta _5}-\frac{c_8 \gamma _3}{\gamma _1}+c_1\right)+c_5\left(\frac{\beta _1 d_6}{\beta _5}+\frac{\gamma _3 d_8}{\gamma _1}-d_1\right)\bigg]\\
        &+\phi _1' \bigg(\frac{\beta _6 c_6 \gamma _2 d_5}{\beta _5 \gamma _1}-\frac{\beta _6 c_5 \gamma _2 d_6}{\beta _5 \gamma_1}-\frac{\beta _2 c_6 d_5}{\beta _5}+\frac{\beta _2 c_5 d_6}{\beta _5}+\frac{c_9 \gamma _2 d_5 \gamma _1'}{\gamma_1^2}\\
        &-\frac{c_5 \gamma _2 d_9 \gamma _1'}{\gamma _1^2}-\frac{c_9 d_5 \gamma _2'}{\gamma _1}+\frac{c_5 d_9 \gamma_2'}{\gamma _1}-\frac{c_{10} \gamma _2 d_5}{\gamma _1}+\frac{c_5 \gamma _2 d_{10}}{\gamma _1}-\frac{c_9 \gamma _3 d_5}{\gamma _1}\\
        &+\frac{c_5 \gamma _3 d_9}{\gamma _1}-c_5 d_3+c_3 d_5\bigg)
        +\phi _1 \bigg(\frac{\beta _6 c_6 \gamma _3 d_5}{\beta _5 \gamma _1}-\frac{\beta _6 c_5 \gamma _3 d_6}{\beta _5 \gamma_1}-\frac{\beta _3 c_6 d_5}{\beta _5}\\
        &+\frac{\beta _3 c_5 d_6}{\beta _5}+\frac{c_9 \gamma _3 d_5 \gamma _1'}{\gamma_1^2}-\frac{c_5 \gamma _3 d_9 \gamma _1'}{\gamma _1^2}-\frac{c_9 d_5 \gamma _3'}{\gamma _1}+\frac{c_5 d_9 \gamma_3'}{\gamma _1}-\frac{c_{10} \gamma _3 d_5}{\gamma _1}\\
        &+\frac{c_5 \gamma _3 d_{10}}{\gamma _1}-c_5 d_4+c_4 d_5\bigg)+A_1 \bigg(-\frac{\beta _4 c_6 d_5}{\beta _5}+\frac{\beta _4 c_5 d_6}{\beta _5}\\
        &+c_7 d_5-c_5 d_7\bigg)=0\, .
\end{split}
\end{equation}
In the equation above for the metric element \eqref{eq:metric}, the coefficients of $\ddot{\phi}'_1$ and $A_1$ vanish, leaving us with the master equation for $\phi_1$:
%
%
\begin{equation}
   g(r)^2\frac{\partial^2\phi_1}{\partial t^2} -\frac{\partial^2\phi_1}{\partial r^2}+C(r)\frac{\partial\phi_1}{\partial r}+U(r)\phi_1=0,
   \label{eq:wavephi}
\end{equation}
where the coefficients depend only on the background solution.

When considering a quadratic coupling function between the scalar and the Gauss-Bonnet term, Ref.~\cite{Blazquez-Salcedo:2018jnn} pointed out that for some values of the coupling, the equation describing the perturbations is not hyperbolic. While this can hinder the investigation of linear stability as a Cauchy problem, we can still examine the mode structure of the spacetime by looking into the frequencies $\omega$.

To perform a mode analysis of the spacetime, we search for the natural frequencies of the system $\omega$, such that $\phi_1(t,r)=\phi_1(r)e^{-i\omega t}$. The perturbation equation~\eqref{eq:wavephi} can be manipulated to the more familiar Schr\"odinger form
\begin{equation}
\label{eq:pertEq}
    \left(-\frac{d^2}{d r_*^2}+V_{\text{eff}}\right)\psi\,=\,\omega^2\psi\, ,
\end{equation}
where $\phi_1(r)=F(r)\psi(r)$ and the tortoise coordinate is defined through $ dr_*=g(r)\, dr$. We also define
\begin{equation}
    \frac{2F'}{F}=\;C-\frac{g'}{g}\, ,
\end{equation}

\begin{equation}
\label{eq:potential}
    V_{\text{eff}}=\;\frac{1}{g^2}\bigg[U+\frac{C^2}{4}-\frac{C'}{2}-\frac{3\,g'^2}{4\, g^2}+\frac{g''}{2g}\bigg].
\end{equation}

The real part of the frequency $\omega$ describes the resonant modes of the system, \textit{i.e.} for which frequencies an initial perturbation would respond. 
The imaginary part of the frequency indicates the system's (modal) stability. 
For modes with a negative imaginary part, an initial perturbation decays exponentially, while when positive the perturbation grows and the system is rendered unstable. 
Generally, the effective potential identified from Eq.~\eqref{eq:potential} is useful when one attempts to verify the presence of unstable modes as a negative value for the integral of the effective potential with respect to the tortoise coordinate indicates the existence of unstable modes \cite{doi:10.1119/1.17935}, namely
\begin{equation}\label{eq:condition}
    \int_{-\infty}^{+\infty}V_{\text{eff}}(r_*)\,dr_*<0 \Rightarrow \; \textit{unstable modes}.
\end{equation}
It is worth noting that while the condition \eqref{eq:condition} is sufficient in indicating the presence of unstable modes, it is not a necessary condition. 

In what follows we will be making use of the compactified coordinate
\begin{equation}
    x=1-r_h/r,
\end{equation}
which maps all of spacetime, from the black hole horizon up to infinity, to a finite region, namely $x \in [0,1]$.
\begin{figure}
\centering
\includegraphics[width=0.47\linewidth]{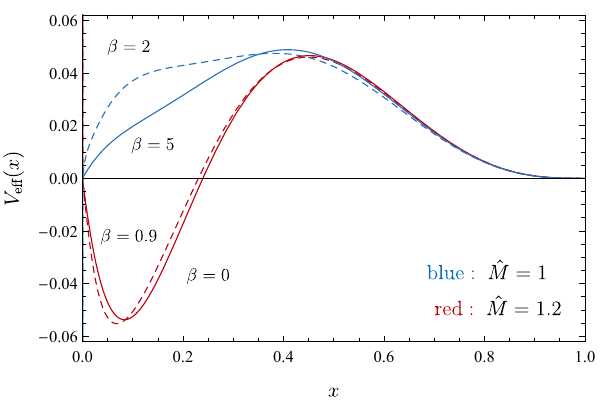}
\hfill
\includegraphics[width=0.47\linewidth]{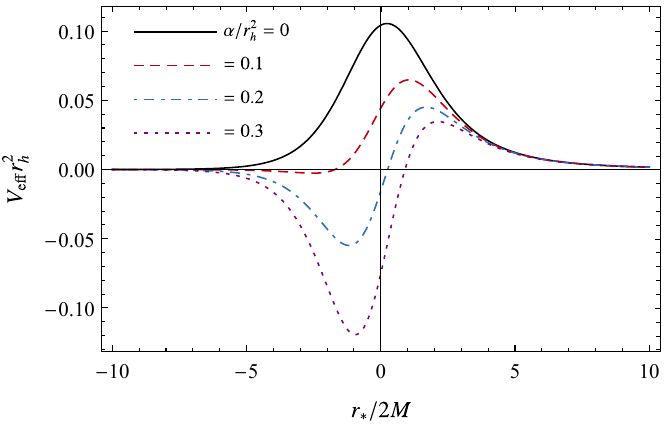}
\caption[Effective potential for radial perturbations in the non-decoupled numerical and the Schwarzschild background]{\textit{Left:} Plot of the effective potential for four different cases. The red lines correspond to right tilting curves with $\hat{M}=1.2$, while the blue ones to left-tilting curves with $\hat{M}=1$. In either case the dashed lines are characterized by a larger value of beta in comparison with the solid ones. \textit{Right:} Effective potential for radial perturbations in the Schwarzschild spacetime. The effective potential for the scalarized BH behaves similarly, as can be seen in Fig.~\ref{fig:potential}, presenting a minimum and a maximum as well.}
\label{fig:potential}
\end{figure}
In Fig.~\ref{fig:potential} we plot the effective potential for two random normalized masses $\hat{M}$ corresponding to the left and right parts of the left panel of Fig.~\ref{fig:scalarized_solutions}, \textit{i.e.} $\hat{M}=1$ and $\hat{M}=1.2$. 
We see that for $\hat{M}=1.2>\hat{M}_\text{th}$ the effective potential has a large negative region which is entirely absent for $\hat{M}=1<\hat{M}_\text{th}$ ($\hat{M}_\text{th}$ was defined in Chapter~\ref{ch:Black holes}). 
Taking into account \eqref{eq:condition}, we have a further indication suggesting the presence of unstable modes for $\hat{M}>\hat{M}_\text{th}$ and the absence of them for $\hat{M}<\hat{M}_\text{th}$.

To explore the modal structure of the spacetime, we have to impose proper boundary conditions in order to obtain the modes. These correspond to an ingoing wave at the horizon and an outgoing one at infinity
\begin{equation}
    \phi_1\xrightarrow[r_*\rightarrow -\infty]{x\rightarrow\, 0} e^{-i\omega r_*}\quad , \quad
    \phi_1\xrightarrow[r_*\rightarrow +\infty]{x\rightarrow\, 1} e^{+i\omega r_*}\, .
\end{equation}
We can see see that, for modes with $\omega_I>0$ (unstable), they simplify to $\phi_1(x\rightarrow 0,1)=0$.

To finish this section, let us mention that the equations describing the radial perturbations in the Schwarzschild spacetime can be directly obtained from~\eqref{eq:wavephi} by setting $\phi_0=0$ and $A=B=1-r_h/r$ which specify $g$, $C$ and $U$. 
The resulting scalar wave equation is given then by
\begin{equation}
    \frac{\partial^2\phi_1}{\partial t^2}-\frac{\partial^2\phi_1}{\partial r_*^2}+\left(1-\frac{r_h}{r}\right) \left[\frac{r_h}{r^3}-12 \frac{\alpha\, r_h^2}{r^6}
    +\frac{\ell  (\ell +1)}{r^2}\right]\phi_1=0\, ,
\label{eq:waveGR}
\end{equation}
where $dr_*=dr/(1-r/r_h)$ is the tortoise coordinate of the Schwarzschild spacetime. The effective potential is identified as
\begin{equation}
    V_{\text{eff}}^{(\text{d})}=\left(1-\frac{r_h}{r}\right)\left[\frac{r_h}{r^3} -\frac{12\, r_h^2\alpha}{r^6}+\frac{\ell  (\ell +1)}{r^2}\right],
\end{equation}
where the index \textit{d} stands for \textit{decoupling}. 
In the right panel of Fig.~\ref{fig:potential} we plot the potential in the decoupling limit for some values of $\alpha$, using the tortoise coordinate to improve visualization. 
Just as before, the equation describing radial perturbations can be used to access the stability properties of the Schwarzschild spacetime in sGB. 
From a direct integration of the potential, it is straightforward to see that \eqref{eq:condition} in this case yields $\alpha/r_h^2\gtrsim 0.208$.

\section{Numerical results}\label{sec:numerical_results}
In order to find the modes of scalarized BHs, we follow the direct integration method presented in previous works on the same subject~\cite{Blazquez-Salcedo:2018jnn,Silva:2018qhn}. 
We briefly summarize the method here.
After picking a value for $\omega$, we integrate Eq.~\eqref{eq:wavephi} from the horizon and infinity, using in-going and-outgoing waves as boundary conditions respectively\footnote{We could instead work with Eq.~\eqref{eq:pertEq}, but in our setup this would add an extra step, slowing down the integrations.}. 
In practice, the integration starts from finite values very close and very far away from the horizon's position, such that the potential is small. 
This changes the boundary conditions which are no longer purely in-going and out-going waves, but are rather given by the Taylor expansion of the field at the horizon and infinity.
Using the two separate solutions, we can demand that they are linearly dependent on a given frequency $\omega$. This is done by examining the Wronskian, given by
\begin{equation}
    W =\bigg[ \phi_1^{(-)}\frac{\partial \phi_1^{(+)}}{\partial r_*} - \phi_1^{(+)}\frac{\partial \phi_1^{(-)}}{\partial r_*}\bigg]\,.
\end{equation}
where $\phi_1^{(-)}$ represents the solution obtained by integrating from the horizon and $\phi_1^{(+)}$ the solution obtained from infinity. The Wronskian vanishes when the value of $\omega$ is a QNM frequency.
An alternative approach is to integrate from the horizon using the in-going boundary condition, up to a large distance $r_{\infty}/r_h$. 
Then one can decompose the solution at infinity onto in-going and out-going waves, and should the value of $\omega$ be a QNM frequency, the in-going amplitude is zero. 
For both methods, an initial choice for $\omega$ is made, and root-finding algorithms are employed to solve for the QNM frequency. 
This is usually called the shooting method, as one starts at one end, ``shooting'' for the value of $\omega$ for which the boundary condition is satisfied at the other end.

Both methods are suitable for finding quasinormal modes with large quality factors, i.e., with large $|\omega_r/\omega_i|$. This usually means that the fundamental mode is easily found. 
The reason is that the function describing radial perturbations grows exponentially in $r$, as $\sim e^{\omega_i r_*}$. 
The method also works remarkably well with unstable modes, as the perturbations decrease exponentially with $r$. 
As such, this method is reliable in identifying regions where BHs are linearly unstable.

Black hole solutions of the theory~\eqref{eq:ActionGeneric} were already studied to some extent in Chapter~\ref{ch:Black holes}.
In what follows, we present new insights considering the stability and modes of the Schwarzschild and EsRGB black holes.

\subsection{Radial oscillations and the existence of purely imaginary modes}

While the Schwarzschild BH is a solution in both GR and sGB theories, its dynamical response to perturbations can be completely different. In fact, it is precisely this difference that allows for spontaneous scalarized BHs, where the radial perturbation instabilities lead to the scalar hair. 
Here we investigate the radial mode structure of black holes in sGB, elucidating some major points considering Schwarzschild and scalarized solutions.

\begin{figure}[t]
\centering
\includegraphics[width=0.65\linewidth]{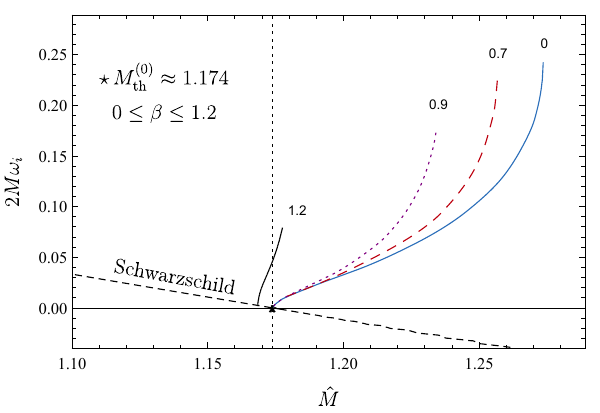}
\caption[Imaginary part of the unstable modes for scalarized black holes]{Imaginary part of the unstable modes for scalarized solutions in sGB considering the Ricci term. We see that as we increase the value of $\beta$, the instability decreases, having a critical value $\beta_\text{crit}< 1.2$ for which the $\hat{M}$-$\hat{Q}$ curves start tilting to the left and solutions becomes stable. We found no unstable modes for scalarized BHs with $\beta>\beta_\text{crit}$. The vertical dotted line marks the scalarization threshold and the dashed black line the Schwarzschild  mode.}
\label{fig:modes_scalarized}
\end{figure}

We will work with the black holes we presented in Sec.~\ref{sec:Numerical-BHs} of Chapter~\ref{ch:Black holes}.
Let us look into the solutions presented in Fig.~\ref{fig:scalarized_solutions}. 
As discussed there, we expect to find unstable modes for the region $\hat{M}>\hat{M}_c$, where the Schwarzschild BH is stable and energetically favourable. 
We performed a search for the modes using the shooting method. 
In Fig.~\ref{fig:modes_scalarized} we plot the unstable mode frequencies for the scalarized solutions considering different values for $\beta$, some of which have been presented in the right panel of Fig.~\ref{fig:scalarized_solutions}. These frequencies are purely imaginary. 
The imaginary part of the Schwarzschild fundamental mode is also plotted, and it is independent of our choice of $\beta$. 
We notice that all scalarized solutions with $\hat{M}>\hat{M}_\text{th}\approx1.175$ are unstable.

In agreement with the predictions made by observing Fig.~\ref{fig:scalarized_solutions}, for $\beta>\beta_\text{crit}\approx 1.15$, the behavior of the curves changes. 
We notice that the curve for which $\beta=1.2$ begins not from $\hat{M}_\text{th}$, but from some $\hat{M}\approx1.168<\hat{M}_\text{th}$. 
This means that the minimum BH mass for this parameter is no longer $\hat{M}_\text{th}$, but smaller.  
Once again, the reasoning for this can be understood qualitatively from purely energetic arguments. 
For $\beta=1.2$, we can have two scalarized solutions for a given mass $\hat{M}$, presenting different charges. 
Solutions with higher charges are unstable, decaying to the scalarized BH with a smaller charge. 
A similar feature was observed for the case of scalarized BHs with self-interaction~\cite{Macedo:2019sem}. 
Overall, the timescale of the instability $\tau=|\omega_I^{-1}|$ increases as $\beta$ increases, indicating the shift from unstable to stable solutions.

\subsection{Radial modes for the Schwarzschild-sGB spacetime}

Since Schwarzschild BHs are stable for $\hat M>\hat M_c$ and the dynamical response is different from GR, it is natural to investigate the impact of sGB terms on the Schwarzschild BH spectrum. The fundamental mode was already analyzed in Ref.~\cite{Macedo:2020tbm}, elucidating how the transition from stable to unstable modes occurs. Here we take an additional step, looking into the first and second overtones, as well as the first three values for the angular number $l$. The results of this subsection are independent of $\beta$ [cf. Eq.~\eqref{eq:waveGR}].

To find the modes for the Schwarzschild-sGB black hole we use three different techniques:
(i) the \textit{Continued Fraction} (CF) method, (see also Ref.~\cite{Macedo:2020tbm,Pani:2013pma}),
(ii) the Wentzel–Kramers–Brillouin (WKB) approximation,
and (iii) \textit{Direct Integration} (DI).
In general, the DI method has low accuracy for modes with low quality factors $|\omega_r/\omega_i|$.
For these modes, which can be expected at the onset of scalarization, we expect that the CF method is better suited in order to understand how the Schwarzschild spacetime becomes unstable. For completeness, however, we will attempt to do the same thing using WKB and DI in order to demonstrate the limitations of each technique.
The CF method relies on providing a semi-analytical approximation of the wave function through the Frobenius method~\cite{Leaver:1990zz}. In summary, the solution can be written in terms of coefficients that are computed by a recursive relation that takes the form of a continued fraction, justifying the name. Since the method requires the analytical form of the coefficients (at least for the expansion to be implemented), we focus mostly on the stability analysis of Schwarzschild BHs.  When applying the CF method we will be resorting to accuracy provided by about $N\sim2\times 10^4$ terms in the CF expansion.
In Appendix~\ref{ch:Appendix_qnms}, we discuss all three methods in more detail, and show how we can reduce any sequence with more than three terms, to a 3-term one that can always be solved.

\begin{figure}[t]
    \centering
    \includegraphics[width=0.47\textwidth]{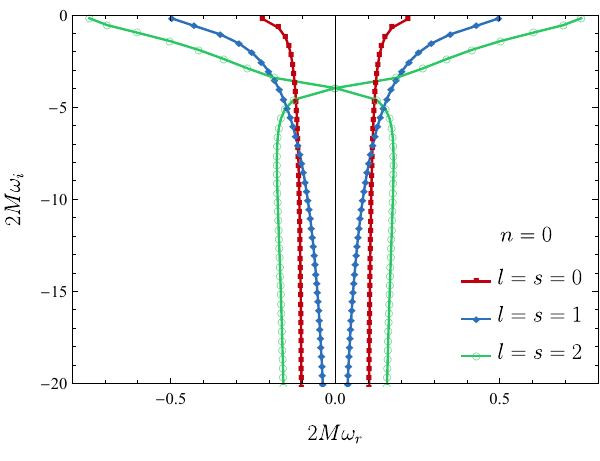}\hspace{5mm}
    \includegraphics[width=0.47\textwidth]{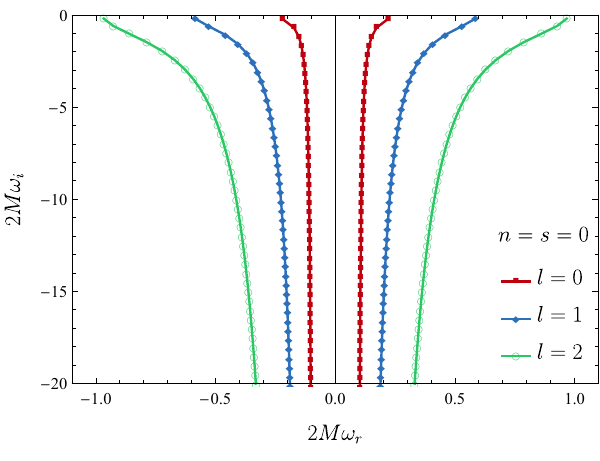}
    \caption[QNM frequencies for the Schwarzschild black hole]{\textit{Left:} QNM frequencies for the Schwarzschild black hole in the case $l=s$ and $n=0$. \textit{Right:} Same but in the case $n=s=0$ and $\ell=0,1,2$.}
    \label{fig:Scharzschild-QNMs}
\end{figure}

As a warm-up and since it is important in the context of this discussion we use our code to derive the modes for the GR Schwarzschild black hole. We use the CF method, which is presented in Appendix~\ref{ch:Appendix_qnms}, and we substitute
\begin{equation}
    \phi_l=\left(\frac{r}{\rh}-1\right)^{-i\rh\omega}\left(\frac{r}{\rh}\right)^{2i\rh\omega}e^{i\omega (r -\rh)}\sum_{n=0}^\infty a_n\left(1-\frac{\rh}{r}\right)^n\, ,
\label{eq:field-CF}
\end{equation}
in \eqref{eq:waveGR}, where now the effective potential is assumed to be
\begin{equation}
    V_{\text{eff}}=\left(1-\frac{r_h}{r}\right)\left[\frac{r_h}{r^3}+\frac{\ell  (\ell +1)}{r^2}+\frac{r_h(1-s^2)}{r^3}\right]\, ,
\end{equation}
with $s=0,1,2$ for a massless spin $0,1,2$ field \cite{Berti:2009kk}. As explained in Appendix~\ref{ch:Appendix_qnms} expression \eqref{eq:field-CF} contains the appropriate boundary conditions for the perturbations.
From this substitution we derive the following recurrence relation
\begin{equation}
\begin{split}
    &\alpha_n a_{n+1} + \beta_n a_n +\gamma_n a_{n-1} = 0\, ,\; n>0\, , \\
    &\alpha_0 a_{1} + \beta_0 a_0   = 0\, ,
\end{split}
\label{eq:3term-sequence}
\end{equation}
where the coefficients are given by
\begin{align}
\alpha_n = & (n+1)(n+1-2 i \rh \omega)\, , \\
\beta_n = &  -\rh^2\, l(l+1)-\rh^2 \left[2n(n+1)+1\right]+4i\rh\omega(2n+1)+8\rh^2\omega^2\, , \\
\gamma_n = & \left(n-2 i \omega  r_h\right)^2-s^2\, .
\end{align}
Recurrently solving this as described in  Appendix~\ref{ch:Appendix_qnms}, we find our results in agreement with the bibliography \cite{Berti:2009kk} and we present them in Fig.~\ref{fig:Scharzschild-QNMs}. On the left panel we show the QNMs for scalar, vector and gravitational perturbations with zero nodes, and on the right the QNMs for $\ell=0,1,2$ in the case of scalar perturbations with zero nodes.

\begin{figure}[t]
    \centering
    \begin{tabular}{ c c }
        \includegraphics[width=0.47\textwidth]{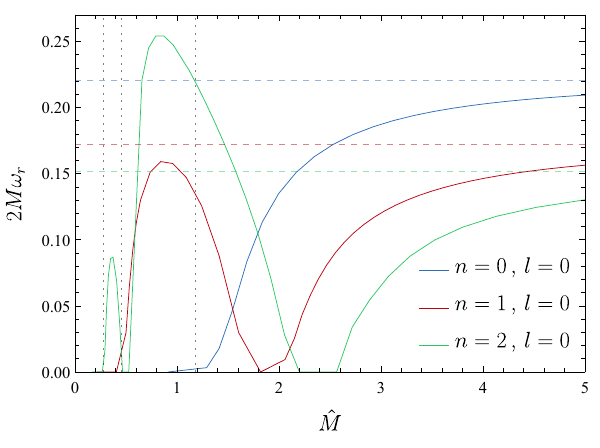} & \includegraphics[width=0.47\textwidth]{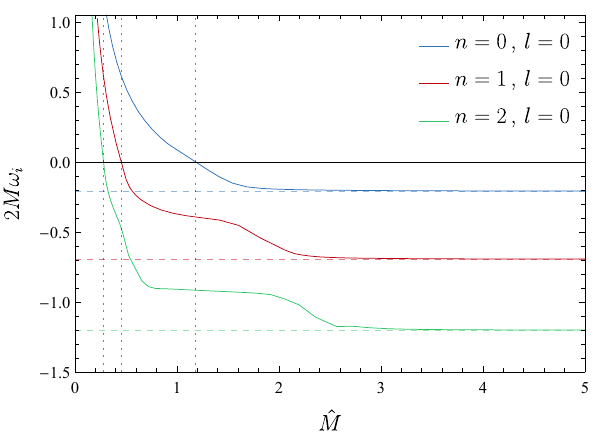} \\
        \includegraphics[width=0.47\textwidth]{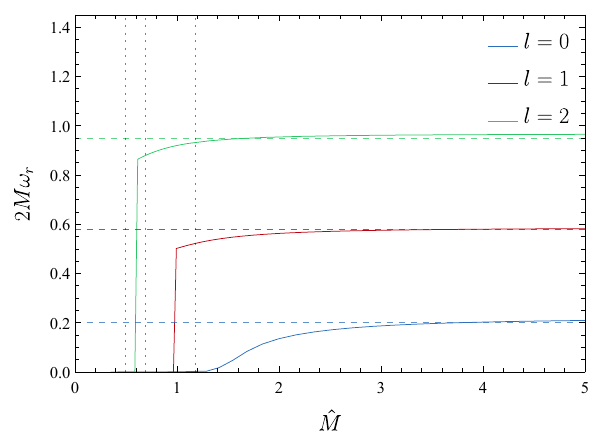} & \includegraphics[width=0.47\textwidth]{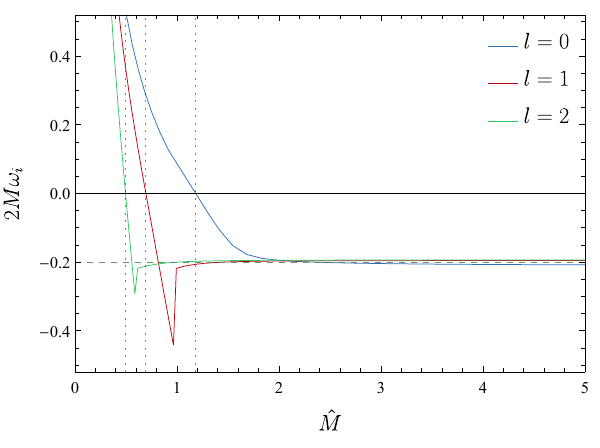}
    \end{tabular}
   \caption[Fundamental mode and first overtone of radial QNM frequencies in the "scalarized" Schwarzschild spacetime]{Fundamental mode and first overtone of radial QNM frequencies in the Schwarzschild spacetime. We see that the fundamental mode is responsible for the instability. The horizontal dotted lines represent the GR scalar modes and the vertical line, the thresholds for scalarization.}
\label{fig:sGB-CF}
\end{figure}

After this warm-up we move to the main part of this subsection which is the calculation of the modes in the decoupled Schwarzschild-sGB black hole.
First we employ the CF method.
Now, the additional GB coupling leads to the following 6-term sequence of recurrence relations
\begin{equation}
\begin{split}
    & \alpha_n a_{n+1} + \beta_n a_n +\gamma_n a_{n-1} + \delta_n a_{n-2} + \sigma_n a_{n-3} + \theta_n a_{n-4 }  = 0\, , \\
    & \alpha_3 a_{4} + \beta_3 a_3 +\gamma_3 a_{2} + \delta_3 a_{1} + \sigma_3 a_{0}   = 0\, , \\
    & \alpha_2 a_{3} + \beta_2 a_2 +\gamma_2 a_{1} + \delta_2 a_{0}   = 0\, , \\
    & \alpha_1 a_{2} + \beta_1 a_1 +\gamma_1 a_{0}    = 0\, , \\
    & \alpha_0 a_{1} + \beta_0 a_0   = 0\, ,
\end{split}
\end{equation}\\
where the coefficients are given by
\begin{align}
\alpha_n = & \rh^2 (n+1) (n+1 -2 i \rh \omega)\, , \\
\begin{split}
    \beta_n = & 12 \alpha -\rh^2\, l(l+1)-\rh^2 \left[2n(n+1)+1\right]+4i\rh^3\omega(2n+1)\\
    &+8\rh^4\omega^2\, , 
\end{split}
\\
\gamma_n = & -48 \alpha + r_h^2 \left(n-2 i \omega  r_h\right)^2\, , \\
\delta_n = & 72\alpha\, , \\
\sigma_n = & -48\alpha\, , \\
\theta_n= & 12\alpha\, .
\end{align}
Following the Gaussian elimintation step process explained in Appendix \ref{ch:Appendix_qnms}, we reduce the above to a 3-term sequence of the type \eqref{eq:3term-sequence}, which we then solve in the same manner as in the GR case.
In Fig.~\ref{fig:sGB-CF} we show the results.
The two top panels show the real and imaginary parts for the fundamendal mode and the first two overtones for $\ell=0$, while the lower two depict the same for the fundamendal mode with $\ell=0,1,2$.  The dotted vertical lines correspond to the scalarization thresholds for different values of $(n,l)$, while the horizontal dotted ones correspond to the Schwarzschild limits as presented in Fig.~\ref{fig:Scharzschild-QNMs}.

Let us first examine the two top panels. Starting from the rightmost part of the plots, we see that for $\hat{M}\gg 1$ (or equivalently $\alpha^2\ll M$) we recover the modes for a minimally coupled scalar field in GR, as expected. As we decrease $\hat{M}$, both the real and the imaginary part of the fundamental mode ($n=0$) approach zero monotonically at $\hat{M}=\hat{M}_c$. Beyond that point the frequency becomes purely imaginary and the corresponding mode unstable. For the first overtone ($n=1$), however, we notice that the real part seems to approach zero \textit{before} the instability occurs. This would imply that Schwarzschild BHs with $\hat{M}\approx 1.87$ in this theory have a purely imaginary mode, with the first one being $\omega r_h\approx -i0.55$. The real part of the first overtone goes to zero again when this mode becomes unstable (purely imaginary), in the region $\hat M<\hat M_c$. Going then to the second overtone ($n=2$) we find the real part of the mode vanishing twice before reaching the instability point, revealing two such purely imaginary modes. As a reminder the scalarization thresholds $\hat{M}^{(n,l)}_\text{th}$ for different overtones and angular numbers were given in tab.~\ref{tab:scalarization_thresholds}.
We can infer a general trend that overtones follow, according to which the $n$-mode has $n$ of these purely imaginary modes in a region of the parameter space where GR black holes are stable.

We highlight here that the usual CF method is not suitable to study purely imaginary modes~\cite{Cook:2016fge,Cook:2016ngj}. As the real part of the modes decreases, more terms in the CF expansion are needed in order to compute reliable values for the QNM frequencies. Typically as we said, near the purely imaginary mode, we consider $N=2\times 10^4$ or more terms in the CF expansion. Curiously, only recently this class of modes was computed with a reliable precision for the Kerr spacetime~\cite{Cook:2016fge,Cook:2016ngj}. We shall not attempt to generalize this method for sGB theories, but it would be interesting to further investigate the full spectrum of the Schwarzschild spacetime in the theory in light of such tools.

Purely imaginary modes for black holes are not an exclusive feature of sGB theories. As mentioned above, even within GR, the existence of such a class of modes has been known for quite some time. For instance, Schwarzschild black holes in GR have algebraically special modes that are purely imaginary~\cite{Chandrasekhar:1975nkd}, and many works have studied these modes for Kerr black holes. While the physical implications of these purely imaginary modes in black hole spacetimes are still poorly understood (they might be better understood in the context of AdS configurations \cite{Miranda:2005qx, Berti:2009kk, Konoplya:2011qq, Witczak-Krempa:2013xlz})), it is interesting seeing them arising in the overtones for Schwarzschild black holes in sGB. This feature, which seems to be an exception for GR black holes, seems to be common in black holes in sGB.

The lower two panels of Fig.~\ref{fig:Scharzschild-QNMs} show $\omega_r$ and $\omega_i$ for the fundamental mode while changing the angular number $\ell$. For large $\hat{M}$ all curves converge to the Schwarzschild limit as expected. For $\ell>1$ we observe an interesting behavior: close to the threshold $M_\text{th}^{(0,\ell)}$, both the real and imaginary parts of the frequency experience ``jumps''. The nature and reasoning behind these jumps is not clear and further exploration will follow in future work, eventually determining if their emergence is simply a result of the numerical techniques employed.

\begin{figure*}[t]
    \centering
    \includegraphics[width=0.47\textwidth]{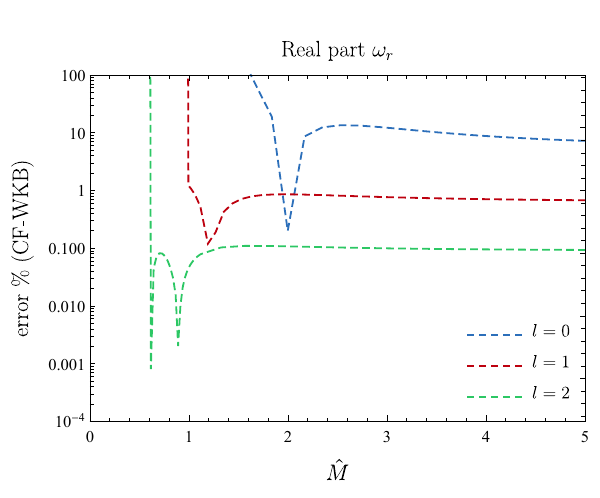}\hspace{2mm}
    \includegraphics[width=0.47\textwidth]{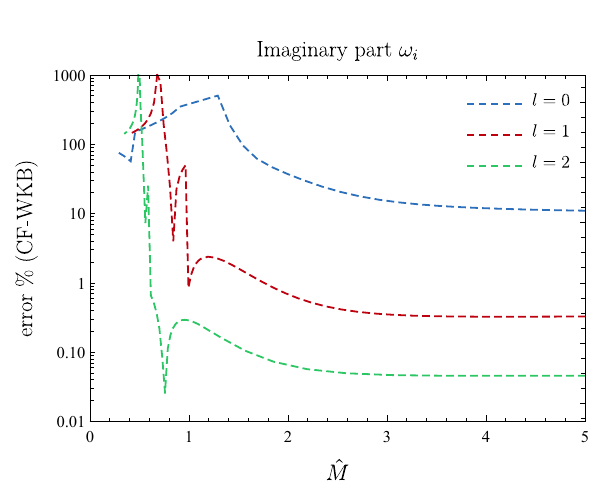}
    \caption{Relative error in the frequencies as those are calculated using the CF and the WKB method.}
    \label{fig:CF-WKB}
\end{figure*}

Next, we attempt to calculate these modes using the WKB method, which as described in Appendix~\ref{ch:Appendix_eom}, is not expected to yield trustworthy results near the thresholds where $\omega_r \rightarrow 0$. In Fig.~\ref{fig:CF-WKB} we present the relative error between the CF and WKB method. We see that especially close to the threshold WKB yields results that deviate significantly from the trustworthy results of the CF approach.
Finally, we also employed the DI method as described in the Appendix~\ref{ch:Appendix_eom}, and the results we produced were in good agreement with those from the CF method.

\subsection{The hyperbolic nature of the equations}

An important property of partial differential equations pertains to hyperbolicity. In physical theories, hyperbolic equations are necessary to describe the time evolution of initial data. In GR, where equations are quasilinear, proof of strong hyperbolicity establishes the well-posedness of the theory (see Ref.~\cite{Wald:1984rg} for a pedagogical discussion). In linear equations such as \eqref{eq:wavephi}, the situation is fairly straightforward. Eq. \eqref{eq:wavephi} is hyperbolic provided that $g^2(r)>0$. It has been shown that for the exponential coupling, the equation describing the radial perturbations is not hyperbolic for a variety of solutions~\cite{Blazquez-Salcedo:2018jnn}. Interestingly, even though the equations are not hyperbolic, we can still proceed to search for unstable modes. It was shown that for scalarized BHs in the exponential model with low mass, an unstable mode arises in the region where hyperbolicity is broken. Here we investigate the impact of the Ricci term on the hyperbolicity of the equation describing radial perturbations.

We start by analyzing the behavior of the coefficient $g(r)^2$ in the near-horizon regime. In this subsection, we shall replace the quadratic coupling with the Gauss-Bonnet term by a generic function of the form $\alpha \phi^2\to\alpha f(\phi)$, in order to compare our results with other works in the literature.
Note that the definition of the normalized charge and mass are unchanged under this substitution.
Using the expansion of the background near the event horizon we find that  the coefficient $g(r)^2$ appearing in Eq.~\eqref{eq:wavephi} behaves as
\begin{equation}
    g(r)^2(r-r_h)^2\approx  \frac{1}{2 a_1}\left(1+\frac{\sqrt{\delta}}{\lambda}\right),
\end{equation}
where the constants are defined as
\begin{align}
\begin{split}
    \delta=&\; 73728 \alpha ^3 \beta  \phi_h (\partial_\phi f)^3-768 \alpha ^2 \left[\beta  (9 \beta -2) \phi_h^2+8\right] (\partial_\phi f)^2\\
    &+\left[\beta  (3 \beta -1) \phi_h^2+4\right]^2,
\end{split}
    \\[2mm]
    \lambda=&\; \beta  \phi_h \left[(3 \beta -1) \phi_h-48 \alpha  (\partial_\phi f)\right]+4.
\end{align}
In order to investigate whether the Ricci term helps maintaining the hyperbolic nature of the equation, we look into large positive values of $\beta$ that, from our previous analysis, we know correspond to stable scalarized solutions. We find
\begin{equation}
    g(r)^2(r-r_h)^2\sim \frac{1}{a_1}\bigg[1+\frac{8\alpha}{\phi_h\beta}+\frac{8\alpha (\partial_\phi f)}{3\phi_h^2\beta^2}(\phi_h-24\alpha)+{\cal O}(\beta^{-3})\bigg],
    \label{eq:hyperbolic_c}
\end{equation}
which indicates that the Ricci term in the action acts \textit{in favor} of the hyperbolic character of the equation. Note, however, that in order to verify whether the equation maintains its hyperbolicity we need to properly solve the black hole solution and obtain the coefficient $g(r)^2$ numerically.
\begin{figure}[t]
\centering
\includegraphics[width=0.45\linewidth]{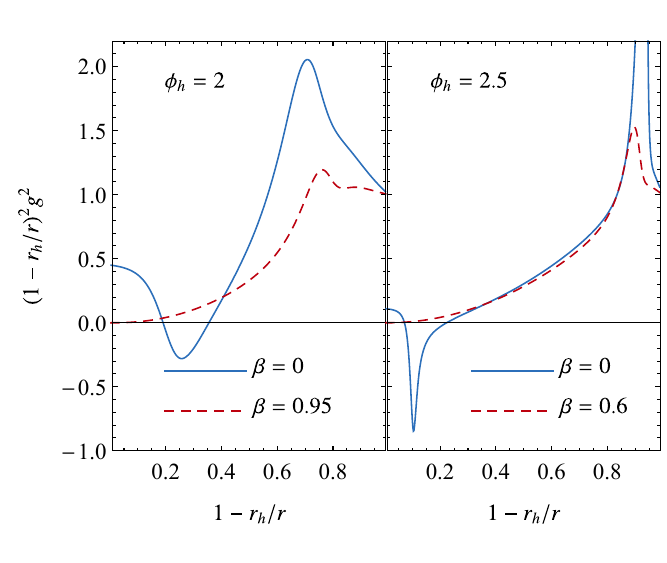}
\hfill
\includegraphics[width=0.54\linewidth]{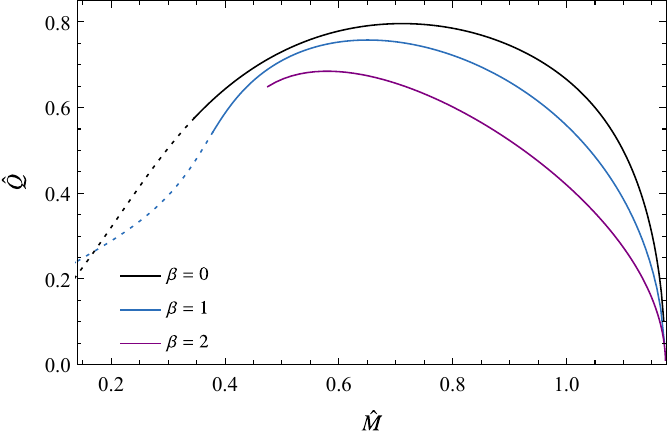}
\caption[Hyperbolicity healing with the scalar-Ricci coupling]{\textit{Left:} Respresentative solutions illustrating the effect of hyperbolicity healing from the Ricci term. The equations are not hyperbolic in the region $g^2<0$. The solid lines have $(\hat{M},\hat{Q})=(0.2677,0.4494)$ (left panel) and $(\hat{M},\hat{Q})=(0.0319, 0.110)$ (right panel). The dashed lines have $(\hat{M},\hat{Q})=(0.0915,0.195)$ (left panel) and $(\hat{M},\hat{Q})=(0.029, 0.068)$ (right panel). \textit{Right:} Normalised charge-mass diagram for scalarized black holes considering the exponential GB coupling and the Ricci term. The dotted lines indicate the region in which the pertubation equations are non-hyperbolic. For $\beta=2$ the perturbation equation is always hyperbolic.}
\label{fig:exponential}
\end{figure}
Now, let us consider a scenario that is known to break hyperbolicity and check the influence of the Ricci term. We consider the exponential model presented in Ref.~ \cite{Blazquez-Salcedo:2018jnn}, where 
\begin{equation}
f(\varphi)\propto [1-e^{-3\phi^2/2}]\, ,
\end{equation}
and look into solutions with a fixed $\phi_h$ and varying $\beta$. 
In Fig.~\ref{fig:exponential} we compare two scalarized solutions with $\phi_h=2$ and $2.5$. 
We observe that the radial domain in which the perturbation equation is non-hyperbolic decreases as $\beta$ increases, as predicted by Eq.~\eqref{eq:hyperbolic_c}.
Further, we observe that for some limiting value of $\beta$ the region with $g^2<0$ seems to vanish and the solution is hyperbolic for all $r>r_h$. 
We note, however, that as we approach this threshold the solutions for a given $\phi_h$ are increasingly hard to find, and beyond the threshold solutions cease to exist. 
It seems that while the $\beta$-term improves the hyperbolicity of the perturbation equation, it still is not enough to ensure it for all values of $(r,\phi_h)$ for a given $\beta$.

To further illustrate the hyperbolic properties of the equations as a function of the background solution, Fig.~\ref{fig:exponential} shows curves of constant $\beta$ and varying $\phi_h$ in the normalised charge-mass plane for the exponential model.
The dotted part of the curves corresponds to regions in which the radial perturbation equations are not hyperbolic. 
The fact that the $\beta = 2$ curve ends is indicative of our inability to find any solutions past this point. It seems that while the additional term helps with hyperbolicity, the parameter space of the solutions is truncated.

\section{Discussion}

In this chapter we have explored the influence of a coupling between a scalar field and the Ricci scalar on linear perturbations around scalarized black holes. 
This coupling, which is expected to be present in scalarization models based on EFT considerations, has already been shown in previous chapters to be crucial for observational viability: it can make GR a cosmological attractor, thereby providing the right conditions for compact objects without cosmological fine-tuning, and it can also suppress neutron star scalarization, thus removing binary pulsar constraints. 
It also affects the amount of scalar charge scalarized black holes can carry. 
We performed a radial stability analysis and have numerically shown that this same term renders scalarized black hole solutions stable, confirming the expectations we had, based on the work of Chapter~\ref{ch:Black holes}.
This happens for values of the coupling constants that are within the same range considered in the previous chapters.
Indeed, we have not found unstable modes for $\beta$ larger than some critical value, $\beta\approx 1.2$. 
Choosing a $\beta$ above this threshold is consistent with having a cosmological attractor and could quench neutron star scalarization for a range of $\alpha$ that still leads to black hole scalarization. It is worth noting that negative values of $\beta$ are also capable of stabilizing the solutions in a similar manner, but since they do not give rise to the cosmological attractor feature, they seem to be less interesting.

We have also performed a radial mode analysis in the Schwarzschild spacetime and looked for QNM modes. We were able to illustrate an interesting property: beyond the fundamental mode one can find purely imaginary modes in a region of the parameter space where Schwarzschild BHs are stable.
Finally, we analyzed the effects of the scalar-Ricci coupling on the hyperbolicity of the scalar perturbation equation, using the exponential GB coupling as an example. 
We demonstrated that it actually improves the hyperbolic nature of the problem, by reducing the region of the parameter space where hyperbolicity breaks down. 
This happens as the additional term changes the scalar field profile, as determined by the full non-linear field equations, which in turn changes the coefficients of the linear perturbation equations. It would be interesting to generalize the hyperbolicity analysis to more general perturbations and beyond the linear level, in order to check if the coupling with the Ricci scalar could have a positive effect when considering hair formation by collapse \cite{Ripley:2020vpk} or binary mergers \cite{East:2021bqk}.

%% file: Chapters/shadows.tex
In the last few years we have witnessed the detection of gravitational waves from the merging processes of stellar black holes \cite{Abbott:2016blz,TheLIGOScientific:2017qsa,Abbott:2020niy} but also the imaging observations of the supermassive black holes residing at the center of the M87 galaxy \cite{EventHorizonTelescope:2019dse,EventHorizonTelescope:2019uob,EventHorizonTelescope:2019jan,EventHorizonTelescope:2019ths,EventHorizonTelescope:2019pgp,EventHorizonTelescope:2019ggy,EventHorizonTelescope:2021bee,EventHorizonTelescope:2021srq} and of our own Galaxy  \cite{EventHorizonTelescope:2022xnr,EventHorizonTelescope:2022vjs,EventHorizonTelescope:2022wok,EventHorizonTelescope:2022exc,EventHorizonTelescope:2022urf,EventHorizonTelescope:2022xqj}.  These observations have been used extensively in the literature to probe the validity of General Relativity and to set limits and constraints on modified gravitational theories (see, for example,
\cite{Virbhadra:1999nm, Claudel:2000yi, Virbhadra:2002ju, Zakharov2005, Virbhadra:2007kw, Virbhadra:2008ws, Johannsen:2011dh, Zakharov2014, Psaltis:2015uza, Ezquiaga:2017ekz, Baker:2017hug, Creminelli:2017sry, Sakstein:2017xjx, Cunha:2018acu, Khodadi:2020jij, EventHorizonTelescope:2020qrl, Psaltis:2020ctj, Oikonomou:2020sij, Odintsov:2020sqy, Odintsov:2020zkl, Oikonomou:2021kql, EventHorizonTelescope:2021dqv, Volkel:2020xlc, Vagnozzi:2022moj}). Capturing the horizon-scale image of Sagittarius A$^*$ in particular, the supermassive black hole located in the center of our own Galaxy, presents a number of advantages. First, due to its proximity, the mass to distance ratio of Sagittarius A$^*$ is much more accurately determined than that of M87$^*$. In addition, Sagittarius A$^*$ has a much smaller mass than M87$^*$; this allows us to test a curvature scale which lies between the low curvature scale of the massive M87$^*$ black hole and the high curvature scale of stellar black holes.

The main feature in the horizon-scale images of the supermassive black holes is the bright photon ring which marks the boundary of a dark interior region, called the black-hole shadow \cite{Falcke:1999pj}. The bright ring is formed by photon trajectories originating from parts of the universe behind the black hole which are gravitationally lensed by its gravitational field and directed towards our line of sight. These photons have impact parameters slightly larger than the ones which lead to their capturing in bound, circular orbits around the black hole. The quantitative characteristics of the shadow can be calculated in the context of either GR or a modified theory of gravity and compared to the observed value, thus probing the validity of the theory in question.

In this chapter, we consider a set of modified gravitational theories with their common characteristic being the presence of a scalar field. This scalar field will be sourced by either gravitational terms, leading to induced or spontaneous scalarization, or gauge fields, leading to charged scalarized solutions \cite{Antoniou:2022dre}. The presence of the scalar field modifies the gravitational background as well as the geodesic structure of the spacetime including the photon trajectories and the size and shape of the black-hole shadow.
Employing the bounds on the deviation of the observed black-hole shadow of Sagittarius A$^*$ from that of the Schwarzschild solution
\footnote{Let us note that although we will make use of the bounds on the observed black-hole shadow from Sagittarius A$^*$ \cite{EventHorizonTelescope:2022xqj}, our analysis will cover also the corresponding bound from the M87$^*$ observation \cite{EventHorizonTelescope:2019dse,EventHorizonTelescope:2019uob,EventHorizonTelescope:2019jan,EventHorizonTelescope:2019ths,EventHorizonTelescope:2019pgp,EventHorizonTelescope:2019ggy,EventHorizonTelescope:2021bee,EventHorizonTelescope:2021srq,EventHorizonTelescope:2020qrl,EventHorizonTelescope:2021dqv}
as the latter is less stringent and thus easier to satisfy.},
as these were derived by the Event Horizon Telescope \cite{EventHorizonTelescope:2022xqj} in a mass-scale independent form, we will examine the validity of a number of scalar-tensor and tensor-scalar-vector theories. In particular, we will consider the Einstein-scalar-Gauss-Bonnet (EsGB) theory with three different forms of coupling function between the scalar field and the GB term, the EsRGB theory, and finally, the Einstein-Maxwell-scalar (EMs) theory with three different forms again of the coupling function between the scalar and the Maxwell fields. We demonstrate that the black-hole shadow bounds from Sagittarius A$^*$ can indeed impose restrictions on the parameter space or on the form of the coupling function of the scalar field in the aforementioned modified theories. However, the physical conclusions drawn depend very strongly on the particular EHT bound, or combination of EHT bounds, employed for this purpose. Thus, the use of individual bounds always allows amble parameter space where the majority of the modified theories considered are viable -- in certain cases, they are even favoured compared to General Relativity. In contrast, demanding that all EHT bounds are simultaneously satisfied significantly reduces the parameter space and, at times, eliminates it.

Before we start the main part of this chapter, let us point out that since we will only consider non-rotating and spherical scenarios, and since Sagittarius A$^*$ (and of course M87$^*$) is a supermassive black hole, we do not expect in principle any significantly strong constraints to be derived. We rather use the examples of Sagittarius A$^*$ and M87$^*$ as starting points motivating the study of the shadows produced in the context of different alternative gravitational theories. In other words, we will demonstrate the potential that the study of black hole and wormhole shadows has, in observationally confronting the validity of modified gravity. Therefore, some of the analysis that will follow will be using the observations regarding Sagittarius A$^*$ (and M87$^*$) merely as an ``observational compass''.

\section{The EHT bounds}
\label{sec:EHTbounds}

The Event Horizon Telescome (EHT) is a Very Long Baseline Interferometry (VLBI) array with Earth-scale coverage \cite{EventHorizonTelescope:2019dse,EventHorizonTelescope:2019uob,EventHorizonTelescope:2019jan,EventHorizonTelescope:2019ths,EventHorizonTelescope:2019pgp,EventHorizonTelescope:2019ggy,EventHorizonTelescope:2021bee,EventHorizonTelescope:2021srq}. It is observing the sky at 1.3 mm wavelength and has so far managed to provide the horizon-scale image of the two supermassive black holes located at the center of M87 galaxy and of our own Galaxy. The diameter $\hat d_m$ of the bright photon ring surrounding the inner dark area -- the most distinctive feature of these black-hole images -- may be used to test theoretical predictions of both GR and modified theories. As noted above, in  this work we will be using the horizon-scale image of Sagittarius A$^*$. Following \cite{EventHorizonTelescope:2022xqj}, one may write:
\begin{equation}
\hat d_m= \frac{\hat d_m}{d_{\text{sh}}}\,d_{\text{sh}} = \alpha_c\,d_{\text{sh}}= \alpha_c\, (1+\delta)\,d_\text{sh,th}\,.
\label{dm}
\end{equation}
The diameter $\hat d_m$ is the value of the diameter of the photon ring obtained by using imaging and model fitting to the Sagittarius A$^*$  data. The quantity $\alpha_c$ is a calibration factor which quantifies how accurately the ring diameter $\hat d_m$ tracks the shadow diameter $d_{\text{sh}}$. It encompasses both theoretical and potential measurement biases and thus may be written as 
\begin{equation}
\alpha_c= \alpha_1\,\alpha_2 \equiv \left(\frac{d_m}{d_{sh}}\right)\, \left(\frac{\hat d_m}{d_{m}}\right)\,.
\end{equation}
Specifically, $\alpha_1$ corresponds to the ratio of the true diameter of the peak brightness of the image (bright ring) $d_m$ over the diameter of the shadow $d_{sh}$. If $\alpha_1$ equals unity, the peak emission of the ring coincides with the shadow boundary.  Its value depends on the specific black-hole spacetime and the emissivity model in the surrounding plasma. A large number of time-dependent GRMHD simulations in Kerr spacetime as well as analytic plasma models in Kerr and non-Kerr metrics lead to small positive values $\alpha_1$, namely $\alpha_1 =1-1.2$. This result indicates that the radius of the brightest ring is always slightly larger than the black-hole shadow. 

The second calibration parameter $\alpha_2$ is the ratio between the inferred ring diameter $\hat d_m$ and its true value $d_{m}$. Three different imaging algorithms were used in the measurement of the ring diameter $\hat d_m$ denoted by \textit{eht-imaging}, \textit{SMILI} and \textit{DIFMAP}, respectively \cite{EventHorizonTelescope:2022xqj}. The ring diameter was also determined by fitting analytic models, and more specifically the \textit{mG-ring} model \cite{EventHorizonTelescope:2022exc}, to the visibility data. The three imaging methods led to a value of $\alpha_2$ close to unity, while the \textit{mG-ring} model allowed values of $\alpha_2$ in the range (1-1.3). 

Employing the above, the diameter of the boundary of the black-hole shadow may be written as $d_\text{sh}=\hat d_m/(\alpha_1\,\alpha_2)$. Then, Eq. (\ref{dm}) allows us to solve for the fractional deviation $\delta$ between the inferred shadow radius $r_{\text{sh,EHT}}$ and that of a theory-specific black hole  $r_{\text{sh,th}}$ \cite{EventHorizonTelescope:2022xqj}:
    \begin{equation}
    \delta=\frac{r_{\text{sh,EHT}}}{r_{\text{sh,th}}}-1\,.
    \label{delta}
    \end{equation}
 The above deviation parameter allows us to test the compatibility of the EHT measurements with GR or modified theories of gravity. The posterior over $\delta$ is obtained via the formula
    \begin{equation}
    \begin{split}
         P(\delta|\hat{d}\,)=C\, \int & d\alpha_1 \int d\alpha_2 \int d\theta_g\,\mathcal{L}[\hat{d}\,|\alpha_1,\alpha_2,\theta_g,\delta]\\
         & \times P(\alpha_1) P(\alpha_2) P(\theta_g) P(\delta)\,.
    \end{split}
    \label{post-delta}
    \end{equation}
In the above, $\theta_g=GM/Dc^2$ is a characteristic angular size set by the black-hole mass and physical distance. Then, $\mathcal{L}[\hat{d}\,|\alpha_1,\alpha_2,\theta_g,\delta]$ is the likelihood of measuring a ring diameter $\hat{d}$, and $P(\theta_g)$ is the prior in $\theta_g$. $P(\alpha_1)$ and $P(\alpha_2)$ are the distributions of the two calibration parameters and $C$ a normalization constant. 
\par
To obtain the characteristic angular size $\theta_g$ of Sagittarius A$^*$  one needs its mass and distance. Two different instruments, the Keck Observatory and the Very Large Telescope together with the interferometer GRAVITY (VLTI), were used to study the orbits of individual stars around Sagittarius A$^*$. The brightest star observed, S0-2, with a period of 16 years, has helped scientists to test relativistic effects such as gravitational redshift and the Schwarzschild precession \cite{GRAVITY:2018ofz, GRAVITY:2019zin, GRAVITY:2020gka, Do:2019txf} and to constrain alternative theories of gravity \cite{Hees:2019bmi, DellaMonica:2021xcf, deMartino:2021daj}. Its observation has also provided the most accurate so far measurements of the mass and distance of Sagittarius A$^*$. The Keck team found for the distance a value of $R=(7935\pm50\pm32)$\,pc  and for the black hole mass the value $M=(3.951 \pm 0.047)\times 10^6\,M_\odot$ \cite{Do:2019txf}. The VLTI team found correspondingly $R=(8277\pm9\pm33)$\,pc  and  $M=(4.297 \pm 0.012\pm0.040)\times 10^6\,M_\odot$. Therefore, two different priors for $\theta_g$ were derived, namely $\theta_g=4.92\pm0.03\pm0.01\,\mu as$ (Keck) and $\theta_g=5.125\pm0.009\pm0.020\,\mu as$ (VLTI). 

\begin{table}[t]
    \centering
    \begin{tabular}{ c c | c c c }
    
        & \multicolumn{4}{c}{Sgr $A^*$ estimates}\\[1mm]\hline\hline\\[-3.5mm]
        
        & &\hspace{\tabley} Deviation $\delta$ &\hspace{\tablex} 1-$\sigma$ bounds &\hspace{\tablex} 2-$\sigma$ bounds\\[1mm]\hline
        
        \parbox[ht]{3mm}{\multirow{3}{*}{\rotatebox[origin=c]{90}{\textit{\tiny eht-img}\hspace{4mm}}}} & {VLTI}\hspace{\tabley} &\hspace{\tabley} $-0.08^{+0.09}_{-0.09}$ &\hspace{\tablex} {$4.31\le \frac{r_{\text{sh}}}{M}\le 5.25$} &\hspace{\tablex} {$3.85\le \frac{r_{\text{sh}}}{M}\le 5.72$}\\[1mm]
        
        & Keck\hspace{\tabley} &\hspace{\tabley} $-0.04^{+0.09}_{-0.10}$ &\hspace{\tablex} {$4.47\le \frac{r_{\text{sh}}}{M}\le 5.46$} &\hspace{\tablex} {$3.95\le \frac{r_{\text{sh}}}{M}\le 5.92$}\\[1mm]

        & {\color{color1}Avg}\hspace{\tabley} &\hspace{\tabley} $-0.06^{+0.064}_{-0.067}$ &\hspace{\tablex} {\color{color1} $4.54\le \frac{r_{\text{sh}}}{M}\le 5.22$} &\hspace{\tablex} {\color{color1} $4.19\le \frac{r_{\text{sh}}}{M}\le 5.55$} \\[1mm]
        \hline
        
        \parbox[ht]{3mm}{\multirow{2}{*}{\rotatebox[origin=c]{90}{\textit{\tiny SMILI}\hspace{2mm}}}} & VLTI\hspace{\tabley} &\hspace{\tabley} $-0.10^{+0.12}_{-0.10}$ &\hspace{\tablex} $4.16\le \frac{r_{\text{sh}}}{M}\le 5.30$ &\hspace{\tablex} $3.64\le \frac{r_{\text{sh}}}{M}\le 5.92$\\[1mm]
        
        & Keck\hspace{\tabley} &\hspace{\tabley} $-0.06^{+0.13}_{-0.10}$ &\hspace{\tablex} $4.36\le \frac{r_{\text{sh}}}{M}\le 5.56$ &\hspace{\tablex} $3.85\le \frac{r_{\text{sh}}}{M}\le 6.24$\\[1mm]
        
        \hline
        
        \parbox[ht]{3mm}{\multirow{2}{*}{\rotatebox[origin=c]{90}{\textit{\tiny DIFMAP}\hspace{1mm}}}} & VLTI\hspace{\tabley} &\hspace{\tabley} $-0.12^{+0.10}_{-0.08}$ &\hspace{\tablex} $4.16\le \frac{r_{\text{sh}}}{M}\le 5.09$ &\hspace{\tablex} $3.74\le \frac{r_{\text{sh}}}{M}\le 5.61$\\[1mm]
        
        & Keck\hspace{\tabley} &\hspace{\tabley} $-0.08^{+0.09}_{-0.09}$ &\hspace{\tablex} $4.31\le \frac{r_{\text{sh}}}{M}\le 5.25$ &\hspace{\tablex} $3.85\le \frac{r_{\text{sh}}}{M}\le 5.72$\\[1mm]
        
        \hline
        
        \parbox[ht]{3mm}{\multirow{3}{*}{\rotatebox[origin=c]{90}{\textit{\tiny mG-ring}\hspace{3mm}}}} & VLTI\hspace{\tabley} &\hspace{\tabley} $-0.17^{+0.11}_{-0.10}$ &\hspace{\tablex} $3.79\le \frac{r_{\text{sh}}}{M}\le 4.88$ &\hspace{\tablex} $3.27\le \frac{r_{\text{sh}}}{M}\le 5.46$\\[1mm]
        
        & Keck\hspace{\tabley} &\hspace{\tabley} $-0.13^{+0.11}_{-0.11}$ &\hspace{\tablex} $3.95\le \frac{r_{\text{sh}}}{M}\le 5.09$ &\hspace{\tablex} $3.38\le \frac{r_{\text{sh}}}{M}\le 5.66$\\[1mm]
        
        & {\color{color2}Avg}\hspace{\tabley} &\hspace{\tabley} $-0.15^{+0.078}_{-0.074}$ &\hspace{\tablex} {\color{color2} $4.03\le \frac{r_{\text{sh}}}{M}\le 4.82$} &\hspace{\tablex} {\color{color2} $3.64\le \frac{r_{\text{sh}}}{M}\le 5.23$} \\[1mm]
        \hline
        
    \end{tabular}
\caption{Sagittarius A* bounds on the deviation parameter $\delta$. The colored bounds are the ones we use in the plots in the main part.}
\label{tab:bounds_sagA*}
\end{table}


\begin{table}[t]
    \centering
    \begin{tabular}{ r c c c }
        
        \multicolumn{4}{c}{M$87^*$ estimates}\\[1mm]\hline\hline\\[-3.5mm]
        
        &\hspace{\tablex} Deviation $\delta$ &\hspace{\tablex} 1-$\sigma$ bounds &\hspace{\tablex} 2-$\sigma$ bounds\\[1mm]\hline
        
        \hspace{5.3mm}EHT &\hspace{\tablex} $-0.01^{+0.17}_{-0.17}$ &\hspace{\tablex} $4.26\le \frac{r_{\text{sh}}}{M}\le 6.03$ &\hspace{\tablex} $3.38\le \frac{r_{\text{sh}}}{M}\le 6.91$
        \\[1mm]
        \hline
        
    \end{tabular}
\caption{M87* bounds on the deviation parameter $\delta$ \cite{EventHorizonTelescope:2021dqv}.}
\label{tab:bounds_M87*}
\end{table}

Employing these in Eq. (\ref{post-delta}), and assuming that the theory-specific solution considered in Eq. \eqref{delta} is the Schwarzschild solution, for which it holds that $r_{\text{sh,th}}=3\sqrt{3}\,GM/c^2=3 \sqrt{3}\,D\,\theta_{g}$, the corresponding values for the deviation parameter $\delta$, along with their errors, were derived in \cite{EventHorizonTelescope:2022xqj} and are displayed in the first column of Table I. We observe that the deviation $\delta$ always assumes negative values which means that the observed black-hole shadow is found to be smaller than the one predicted by GR for the Schwarzschild black hole. We also note that the value of $\delta$ derived by employing the measurements by VLTI is consistently more negative as compared to the one derived by Keck. The use of the specific algorithm for the image processing also affects the deviation parameter, with $\delta$ taking larger negative values as the {\textit{eht-imaging}} algorithm is gradually replaced by the {\textit{SMILI}}, the {\textit{DIFMAP}} or the {\textit{mG-ring}} algorithm. Finally, the value of $\delta$ is slightly modified by the type of simulations used in the calibration of $\alpha_1$; here we employ the values obtained using the GRMHD simulations as an indicative case. We note, however, that all values derived for $\delta$ by EHT \cite{EventHorizonTelescope:2022xqj} are consistent with each other independently of the specific telescope, image processing algorithm or type of simulation used. For completeness, in Table II we present the corresponding value for the deviation parameter $\delta$ as derived by the black-hole image of M87$^*$ \cite{EventHorizonTelescope:2021dqv}; we observe that the central value of $\delta$ is much closer to zero but the errors are larger, due the larger uncertainty in the measurement of the mass and distance of M87$^*$. 

The definition of $\delta$ via Eq. \eqref{delta} in conjunction with its values in the first column of Table I allows us to obtain the corresponding constraints on the dimensionless quantity $r_\text{sh}/M$ (for notational simplicity, henceforth we drop the subscript EHT from the quantity $r_\text{sh, EHT}$). The 1-$\sigma$ and 2-$\sigma$ bounds on $r_\text{sh}/M$ are displayed in the second and third column of Table I (and for completeness in the second and third column of Table II). We observe that, as expected, the constraints derived from Sagittarius A$^*$ are more stringent than the ones derived from M87$^*$: the allowed range of values in the former case is always narrower and this leads to a consistently smaller upper limit of $r_\text{sh}/M$. 

Here, we will use only two indicative sets of constraints, namely the ones obtained by using the \textit{eht-imaging} method and the \textit{mG-ring} analytic model, which lead to the smallest and largest $\delta$ (in absolute value), respectively. Moreover, in order to take a conservative stance, we will consider the Keck and VLTI values as independent and use their average value for $\delta$; these values together with the corresponding constraints on $r_\text{sh}/M$ are displayed in the two rows of Table I denoted by the word "Avg". In Sections IV-VI, these mass-scale independent constraints will be used to test the viability of compact solutions arising in the context of modified gravitational theories with a scalar degree of freedom, as potential candidates for Sagittarius A$^*$ and M87$^*$. Our analysis will pertain to current but also to future observed black-hole shadow images, and will act complementary to existing works placing bounds on the parameters of these modified gravitational theories. 

We would like to finish this section with the following comment. Throughout this work, we will focus on spherically-symmetric solutions obtained in the context of modified theories. It is for this reason that the theory-specific solution chosen above was the Schwarzschild solution and not the Kerr one. The rotation parameter and inclination angle of Sagittarius A$^*$ does affect the observed shadow radius. However, to our knowledge, at the moment there is no clear consensus on the value of these two parameters for Sagittarius A$^*$. In addition, it was found \cite{Vagnozzi:2022moj} that the shadow radius is affected very little by the rotation of the compact object, independently of the inclination angle. 
In fact, a recent study \cite{Fragione:2022oau} hints towards a rather small value of $a_*$, namely $a_* \leq 0.1$. In any case, it is estimated \cite{EventHorizonTelescope:2022xqj} that rotating black holes can have a shadow size which is smaller that that of a non-rotating black hole by up to 7.5\%. Therefore, considering the Schwarzschild solution as the theory-specific solution in our analysis seems to be a justified choice at the moment. In fact, due to the more compact geodesic structure of any rotating black hole compared to a non-rotating one, any "Schwarzschild" constraint applied in our analysis may be considered as the largest possible value for the corresponding "Kerr" one.

\section{Black Hole Shadow Radius}
Since the black hole spin affects the shadow feebly, as a first step, we investigate the shadow size for a static and spherically symmetric configuration \footnote{The notatation we use here is slightly different than what we have used so far \eqref{eq:metric}, but it is obvious that $g_{tt}\equiv A$ and $g_{rr}\equiv B^{-1}$ .}
\begin{equation}
    d s^2=g_{tt}\,d t^2+g_{rr}\,d r^2+r^2 d\Omega^2 \, .
\end{equation}
First let us locate the photon sphere for this background. To do that we consider the trajectory of a photon. Since spherical symmetry is assumed, we can consider, without loss of generality, the equatorial plane $\theta=\pi/2$. The killing vectors associated with the symmetries of this space time are $\xi_1^\mu=(1,0,0,0)$ and $\xi_2^\mu=(0,0,0,1)$. Then, following \cite{Psaltis:2007rv}, we can define the 4-momentum of a photon as $\tilde{k}=(k^t,k^r,k^\theta,k^\varphi)$, with $k^\theta=0$ (from symmetry arguments). Conservation of energy yields $E=-\xi_1^\mu k_\mu=-g_{tt}k^t$ and conservation of angular momentum $L=\xi_2^\mu k_\mu=r^2k^\varphi$. Morevover, $\tilde{k}^2=0$ fixes the $k^r$ component of the 4-momentum, so that we finally get:
\begin{align}
    \tilde{k}=\left( -\frac{E}{g_{tt}}\,,\sqrt{-\frac{E^2}{g_{tt}\,g_{rr}}-\frac{L^2}{g_{rr}\,r^2}}\,,0\,,\frac{L}{r^2} \right).
\end{align}
It is now straightforward to locate the radius for circular photon orbits by demanding $k^r=0$ and $dk^r/dr=0$. In terms of the impact parameter $b\equiv L/E$, these conditions yield
\begin{equation}
    b^2=-r^2/g_{tt}\,\bigg|_{r_{\text{ph}}}=\frac{r^3 \left(-g_{tt}\, g_{rr}'-g_{rr}\, g_{rr}'\right)}{g_{tt}^2 \left(r\, g_{rr}'+2 g_{rr}\right)}\,\bigg|_{r_{\text{ph}}},
\end{equation}
which can be easily solved to give the equation describing the photon radius
\begin{equation}
    \textit{Photon radius equation: }\quad r_{\text{ph}}=\frac{2g_{tt}}{g_{tt}'}\,\bigg|_{r_{\text{ph}}}\,.
\end{equation}
Our next step is to determine the shadow radius as observed by a far-away observer after lensing has been taken into account. For a null trajectory we can write $g_{\mu\nu}\dot{x}^\mu\dot{x}^\nu=0$, which in turn yields
\begin{equation}
    g_{rr}\left(\frac{\dot{r}}{\dot{\varphi}}\right)^2=-r^2-g_{tt}\left(\frac{\dot{t}}{\dot{\varphi}}\right)^2,
\end{equation}
where $E=-g_{tt}\dot{t}\;,\;L=r^2\dot{\varphi}$. Therefore we can equivalently solve for the radial deviation with respect to the polar angle
\begin{equation}
\label{eq:drdvarphi}
    \left(\frac{dr}{d\varphi}\right)^2=-\frac{r^2}{g_{rr}}\left( \frac{r^2}{g_{tt}\,b^2} + 1 \right),
\end{equation}
At the point of closest radial approach $r=r_0$, the equation above should vanish,
\begin{equation}
    \frac{1}{b^2}=-\frac{g_{tt}}{r^2}\,\bigg|_{r_0}\,.
\end{equation}
\begin{figure}[t]
\centering
\includegraphics[width=0.7\textwidth]{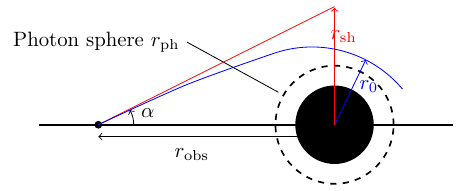}
\caption[Gravitational lensing in a black hole geometry]{Qualitative representation of a light ray reaching an observer at an angle $\alpha$, located at distance $r_{\text{\text{obs}}}$ from the point singularity. The blue line traces a light ray escaping from a closed orbit around the black hole to infinity. The red line aligns with the inferred angle of approach for the light ray to an asymptotic observer. The point of closest approach for the light ray with respect to the black hole is located at $r=r_0$. If $r_0=r_\text{ph}$ the light ray escapes the photon sphere. The shaded, circular area denotes the interior to the black-hole horizon, while the dashed, circular line corresponds to the location of the photon sphere.}
\label{fig:lensing}
\end{figure}
From FIG.~\ref{fig:lensing} we can easily deduce that
\begin{equation}
    \cot\alpha=\frac{\sqrt{g_{rr}}}{r}\frac{dr}{d\varphi}\,\bigg|_{r_{\text{obs}}}\, , \label{shadow_cot}
\end{equation}
and by substitution in \eqref{eq:drdvarphi} we can show that
\begin{equation}
    \sin^2\alpha=-\frac{g_{tt}\,b^2}{r^2}\,\bigg|_{r_{\text{obs}}}\,. \label{shadow_sin}
\end{equation}
Then, it is obvious that the angle for the shadow of the black hole is retrieved in the limit $r_0\rightarrow r_{\text{ph}}$. We assume that asymptotically far away the spacetime is flat, therefore, for a far-away observer $\sin\alpha\approx \alpha$ and $g_{tt}\rightarrow -1$, so $\alpha_{\text{sh}}=b_{\text{crit}}/r_{\text{obs}}$, where $b_{\text{crit}}$ is the impact parameter in the limit $r_0\rightarrow r_{\text{ph}}$ (cf. eq. (29) in \cite{Perlick:2021aok}). For $r_{\text{obs}}\gg r_{\text{ph}}$ we have $\alpha_{\text{sh}} \approx r_{\text{sh}}/r_{\text{obs}}$. We can, finally, deduce that
\begin{equation}
    r_{\text{sh}}=b_{\text{crit}}=\frac{r_{\text{ph}}}{\sqrt{-g_{tt}(r_{\text{ph}})}}\, .
\end{equation}

\section{A Wormhole Shadow}\label{sec:wormhole}

Wormholes constitute a different type of solution, the main characteristic of which is the presence of a throat connecting different regions of spacetime.
\begin{equation}
    ds^2= -e^{2v(l)}\,dt^2 + f(l)\,dl^2 + \left(l^2+l_0^2\right) \,(d\theta^2+ \sin^2 \theta\,d\varphi^2)\,.
\label{eq:wh-metric}
\end{equation}
The above line element can describe a wormhole geometry with the throat located at $l_0$.

\begin{figure}[!ht]
\centering
\vspace{-10mm}
    \includegraphics[width=0.5\textwidth]{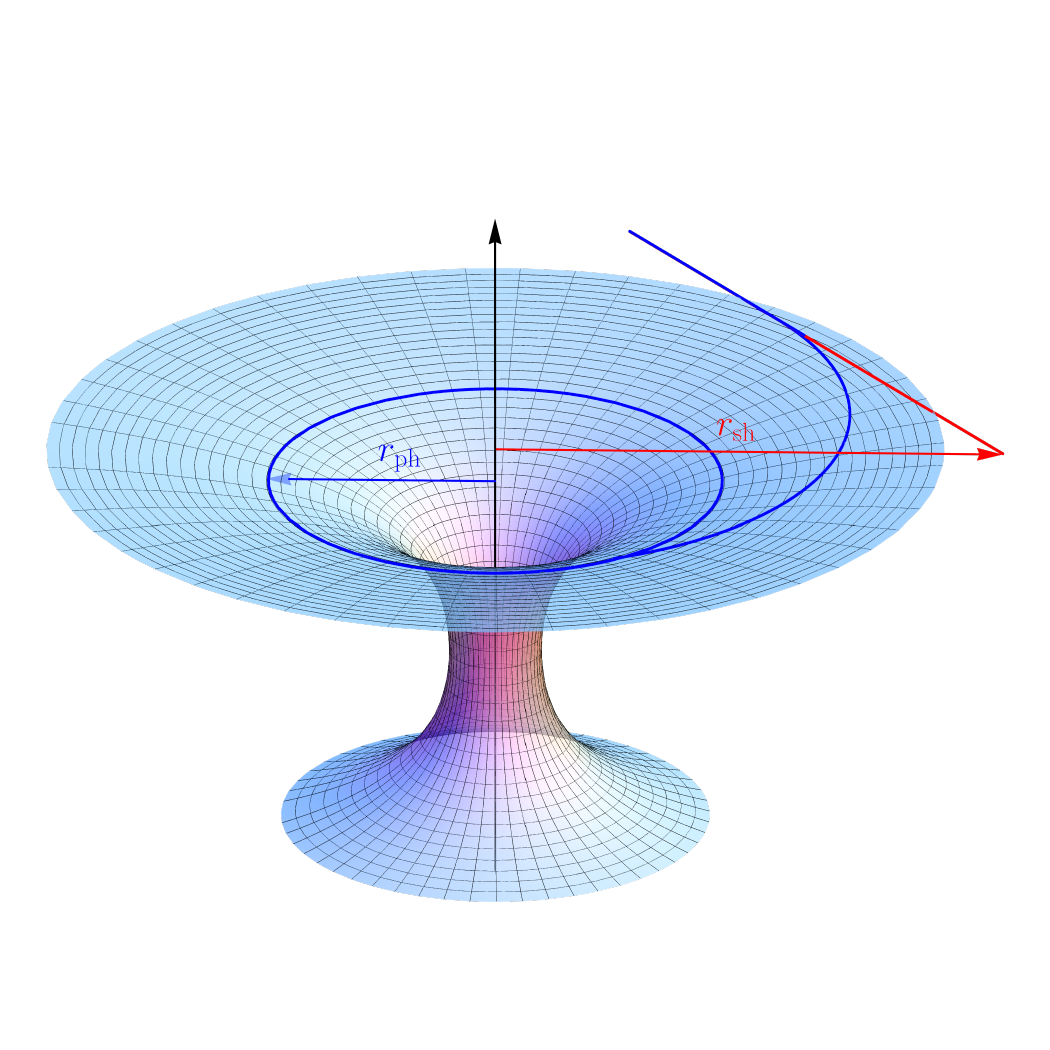}
\caption[Gravitational lensing in a wormhole geometry]{Here we show the embedding diagram depicting a finite radius throat along the vertical axis. The blue line traces a light ray escaping from the photon sphere to infinity, while the red straight line corresponds to the inferred line of approach to an asymptotic observer.}
\label{fig:wh_lensing}
\end{figure}
In the ansatz \eqref{eq:wh-metric} the conserved quantities derived from the geodesics Lagrangian are
\begin{equation}
    E=-g_{tt} k^t = e^{2v}\frac{dt}{d\lambda}\,, \quad L=g_{\varphi\varphi} k^\varphi = \left(l^2+l_0^2\right)\,\frac{d\varphi}{d\lambda}\,,
\end{equation}
As in the black hole-case we demand that for the photon sphere(s) $dk^l/dl=k^l=0$, which yields the following equation holding at the photon sphere(s):
\begin{equation}\label{eq:wh_photon_spheres}
    v'(l_{\text{ph}})=\frac{l_{\text{ph}}}{l_{\text{ph}}^2+l_0^2}\;\Rightarrow \; l_{\text{ph}}= \frac{1\pm\sqrt{1-4\, l_0^2\, v'^2_\text{ph}}}{2\, v'_\text{ph}}\,.
\end{equation}
For a null trajectory we find
\begin{equation}\label{eq:null_wh}
    \left(\frac{dl}{d\varphi}\right)^2=-\frac{\left(l^2+l_0^2\right)}{f}\left[1+\frac{\left(l^2+l_0^2\right)}{e^{2v}\,b^2}\right]
\end{equation}
At the point of closest approach $l=l_c$ we find the impact parameter
\begin{equation}
    b^2=\left(l_c^2+l_0^2\right)e^{-2v_c}\,.
\end{equation}
For the wormhole background \eqref{eq:wh-metric}, Eq. (\ref{shadow_cot}) for the lensing takes the form
\begin{equation}
    \cot \alpha=\sqrt{\frac{f(l)}{l^2+l_0^2}}\,\frac{dl}{d\varphi}\Bigg|_{l_\text{obs}}\,.
\end{equation}
\begin{figure}[!ht]
\centering
    \includegraphics[width=0.75\textwidth]{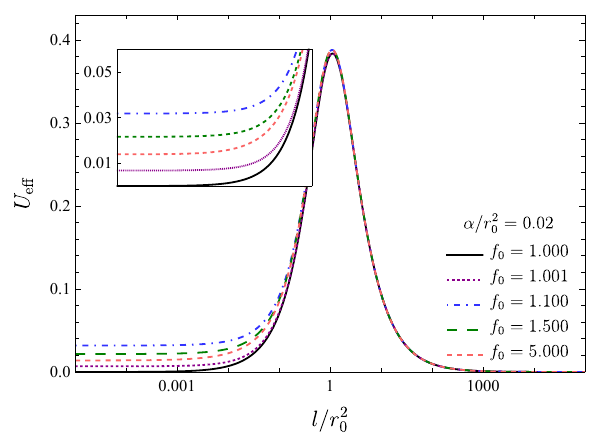}
\caption[Effective gravitational potential in the wormhole scenario]{Effective gravitational potential for the ansatz \eqref{eq:wh-metric}. We notice the existence of one maximum in each case corresponding to an unstable circular orbit.}
\label{fig:wh_gravitational_potential}
\end{figure}
If we now substitute this to \eqref{eq:null_wh} we find
\begin{equation}
    \sin^2\alpha=\frac{e^{2v}\,b^2}{l^2+l_0^2}\Bigg|_{l_\text{obs}}\,.
\end{equation}
For $l_{\text{obs}}\gg l_0,l_{\text{ph}}$, asymptotic flatness demands that $v\rightarrow 0$. The wormhole shadow is retrieved in the limit $l_c\rightarrow l_{\text{ph}},\,b\rightarrow b_{\text{crit}}$, $a_{\text{sh}}\approx b_{\text{crit}}/l_{\text{obs}}$. But also, from FIG.~\ref{fig:lensing}, $a_{\text{sh}}\approx (l_0+l_{\text{sh}})/l_{\text{obs}}$, so we can finally say that
\begin{equation}
    l_{\text{sh}}+l_0=b_{\text{crit}}=e^{-v(l_{\text{ph}})}\sqrt{\left(l_{\text{ph}}^2+l_0^2\right)}.
\end{equation}

To test the existence of one/two extrema of the gravitational potential, which are directly associated with the presence of photon spheres, we present the plots for the values of $f_0$ for which the shadow radius is calculated. The effective gravitational potential for photons in this ansatz is given by:
\begin{equation}
    e^{2v}f\left(\frac{dl}{d\lambda}\right)^2=(E+L\, U_{\text{eff}})(E-L\, U_{\text{eff}})\,
\end{equation}
where the effective gravitational potential is defined as
\begin{equation}
    U_{\text{eff}}=\frac{e^{v}}{\sqrt{l^2+l_0^2}}\, .
\end{equation}
By extremizing the potential we can find the two circular photon orbits, one being stable and one unstable. For the model we consider (EsGB), it turns out that for our solutions only one extremum can be found, due the positivity constraint of the quantity appearing under the square root in \eqref{eq:wh_photon_spheres}.

\section{The Einstein-Scalar-GB Theory} \label{sec:EsGB}

We initiate our analysis by considering the EsGB model.
As analyzed in previous chapters, the EsGB theory has produced a large number of solutions describing compact objects with interesting characteristics: black holes with scalar hair  
\cite{Campbell:1990, Campbell:1991, Maeda:1993ap, Kanti:1995vq, Kanti:1996gs, Kanti:1997br, Torii:1996, Guo:2008hf, Pani:2009wy, Pani:2011gy, Kleihaus:2011tg,  Yagi:2012ya, Sotiriou:2013qea, Sotiriou:2014pfa, Kleihaus:2015aje, Blazquez-Salcedo:2016enn, Antoniou:2017acq, Antoniou:2017hxj, Doneva:2017bvd, Silva:2017uqg, Brihaye:2017wln, Doneva:2018rou, Bakopoulos:2018nui, Witek:2018dmd, Minamitsuji:2018xde, Bakopoulos:2019fbx, Macedo:2019sem, Doneva:2019vuh, Zou:2019bpt, Cunha:2019dwb, Brihaye:2019dck}, traversable wormholes \cite{Bronnikov:1997gj, Kanti:2011jz, Kanti:2011yv, Mehdizadeh:2015jra} and particle-like solutions \cite{Brihaye:2017wln, Kleihaus:2019rbg,  Kleihaus:2020qwo}. Here, we will focus mainly on the first class of solutions, namely black holes, and examine their viability under the light of the mass-scale independent constraints coming from the measurement of the shadow radius of Sagittarius A*. For the sake of comparison, we will briefly discuss also the viability of the dilatonic wormhole solutions with a more detailed analysis of this type of compact objects, left for a future work.

\subsection{Black holes}

The presence of the GB term in the action \eqref{eq:action_GB}, which we re-write here for convenience:
\begin{equation*}
    S=\frac{1}{16\pi}\int{d^4x \sqrt{-g}\,\bigg\{ R+X+f(\phi)\,\GB\bigg\} }\,,
\end{equation*}
causes the evasion of the scalar no-hair theorems and leads to the emergence of a large number of scalarized solutions, as analyzed in Chapter~\ref{ch:Evasions}. In the context of the present analysis, we will consider spherically symmetric solutions that arise for three distinct coupling functions, namely for linear coupling (shift symmetry), quadratic coupling ($Z_2$ symmetry) and exponential coupling (dilatonic theory).

In principle, our solutions require the specification of three parameters beyond GR. We need first to specify the coupling constant $\alpha$ which quantifies the strength of the interaction between the Gauss-Bonnet curvature invariant and the scalar field; we also need two boundary conditions for the scalar field, since it obeys a second order differential equation. As explained in detail in previous chapters, regularity at the horizon imposes a condition for the derivative of the scalar at the horizon \eqref{eq:con-phi'}
This reduces the parameters from three to two, namely the field value at the horizon $\phi_h$ and the coupling strength $\alpha$. Let us make an important distinction between the solutions of this section and most of the black hole solutions we have seen in previous chapters. Here, we leave $\phi_h$ as an independent parameter, instead of fixing it by requiring the asymptotic vanishing of the scalar field. That is because in this section the black holes are not considered to be spontaneously scalarized. We will simply be referring to such black holes, as black holes with scalar hair.

In addition to the preceding constraint, we also need to limit the two dimensional plane $(\phi_h,\alpha)$ due to the requirement that the quantity under the square root in \eqref{eq:con-phi'} is positive definite. For this reason, in order to make the analysis easier to follow, we will trade the parameter $\alpha$ with $\mathcal{I}$ defined as follows
\begin{equation}
    \mathcal{I}\equiv\frac{\sqrt{96}}{r_h^2}(\partial_\phi f)_h\,.
\end{equation}
In this way, the parameter space we need to scan is $(\phi_h,\mathcal{I})$ with $-1<\mathcal{T}<1$ defined within clear boundaries. After the study of the complete parameter space, our results will be eventually expressed again in terms of $\alpha$ for clarity.

\subsubsection{Scenario I: \texorpdfstring{$\boldsymbol{f(\phi)=\alpha \,\phi(r)}$}{f1}}

For the case of the linear coupling, the two dimensional parameter space $(\phi_h,\mathcal{I})$ described above is reduced to one dimensional parameter space since the value of the field does not enter in the field equations as a result of the shift symmetry. This case is obviously equivalent to the theory we studied in Chapter~\ref{ch:Mass-Charge} if we set the couplings of the higher order operators to zero. In this case, the solutions are expected to form a line in the $(\alpha/M^2,r_{\rm sh}/M)$ plane that spans the $-1<\mathcal{I}<1$ parameter range.

\begin{figure}
    \centering
    \includegraphics[width=0.65\linewidth]{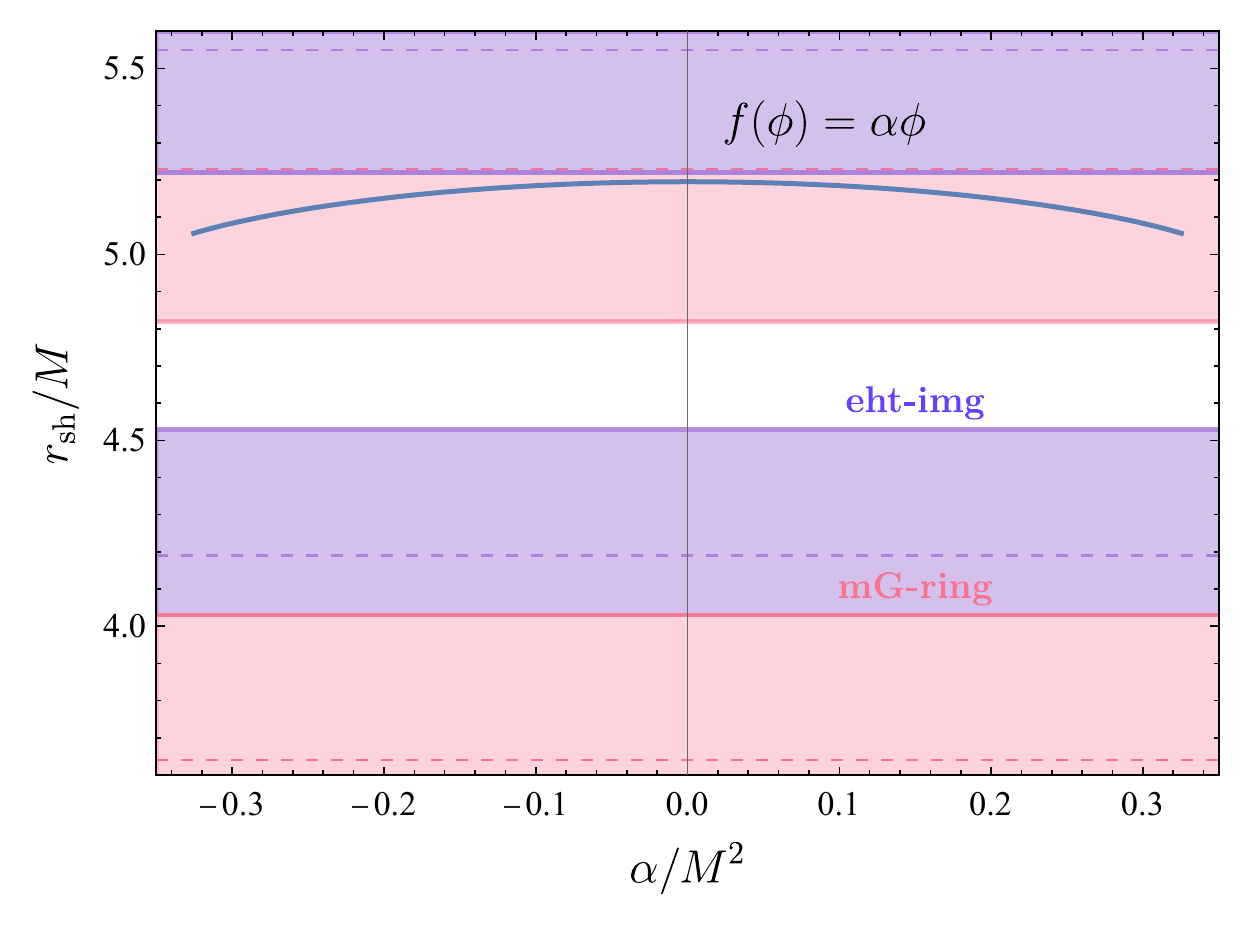}
    \caption{Shadow radius for EsGB theory with a linear coupling. The colored regions correspond to the parameter space of solutions excluded by the respective bounds presented in Tab.~\ref{tab:bounds_sagA*}.}
    \label{fig:linear}
\end{figure}

This is indeed the case as seen in Fig. \ref{fig:linear} where we depict the rescaled black-hole shadow $r_\text{sh}/M$ in terms of the dimensionless parameter $\alpha/M^2$ of the theory. We always choose positive values of $\phi$ so that the sign of the coupling parameter $\alpha$ directly reflects to the sign of $\mathcal{I}$. Here, we consider both positive and negative values for the coupling parameter, and present the complete family of solutions for the allowed range $-1<\mathcal{I}<1$. As $|\mathcal{I}| \rightarrow 1$, the minimum mass solutions -- a characteristic feature of the EsGB theory -- are reached and the solution lines terminate. We also note that the line of the solutions is mirror symmetric around the $\alpha=0$ line. This is expected since the field equations as well as the initial conditions are symmetric under the simultaneous exchange of the sign of $\dot{f}=\alpha$ and $(\phi', \phi'')$. This of course only holds for the linear coupling, for which $\ddot{f}=0$. 

We observe that the shadow radius $r_\text{sh}/M$ decreases as $\alpha/M^2$ increases. This is easily understood if we recall (see, for example \cite{Kanti:1995vq, Antoniou:2017acq, Antoniou:2017hxj, Antoniou:2019awm}) that the GB term causes a negative contribution to the total energy density of the theory and thus exerts a repulsive force. Therefore, if a black hole is to be created, any matter distribution needs to be compacted into a smaller area of spacetime compared to the case where the GB term is absent. As a result, the GB scalarized black holes have always a smaller horizon radius than e.g. the Schwarzschild black hole with the same mass \cite{Antoniou:2017acq, Antoniou:2017hxj, Antoniou:2019awm}. Since the whole geodesic structure gets more compact as $\alpha$ increases, the shadow radius will also get smaller. This decreasing trend of the solution line holds for both $\alpha>0$ and $\alpha <0$ since the GB contribution to the energy density is negative independently of the sign of $\alpha$. In fact, it is proportional to the combination $\phi_{h}' (\partial f)_h$, which is always negative according to Eq. \eqref{eq:con-phi'}. This holds also independently of the exact form of the coupling function $f(\phi)$, and thus we expect to see a similar behaviour for the other two forms of $f$. The generically smaller size of the shadow radius of any EsGB black-hole solution compared to the GR one brings to the foreground these type of modified theories since the EHT constraints \cite{EventHorizonTelescope:2022xqj} point to observed shadow radii which are always smaller than the Schwarzschild one.

Let us now focus on the constraints imposed on the shift symmetric theory by the mass-scale independent bounds depicted in Table~\ref{tab:bounds_sagA*}. As already explained, we will employ two of the derived bounds: the most `conservative' bound, the {\textit{eht-imaging}} one, which yields the smallest central value of the fractional deviation $\delta$, and the most `liberal' bound, the {\textit{mG-ring}} one, which allows for larger deviations from GR. The two solid, horizontal, blue lines denote the allowed 1-$\sigma$ range by the {\textit{eht-imaging}} bound, while the two solid, horizontal, red lines denote the corresponding range allowed by the {\textit{mG-ring}} bound (the blue and red horizontal, dashed lines denote the corresponding 2-$\sigma$ bounds). Likewise, the blue-shaded area is the one excluded by the {\textit{eht-imaging}} bound within 1-$\sigma$ accuracy and the red-shaded area the one excluded by the {\textit{mG-ring}} bound. The white area is the one which is allowed by both bounds.  

According to Fig. \ref{fig:linear}, the complete range of scalarized solutions in the shift symmetric EsGB theory is compatible with the {\textit{eht-imaging}} bound while it is altogether excluded by the {\textit{mG-ring}} bound within 1-$\sigma$! Our findings highlight in the best possible way the need to ``bridge the gap'' between the different EHT bounds as they lead to conflicting conclusions regarding the viability of certain solutions and, in a more general context, the physical relevance of their underlying theories.
We note that all solutions found, which are allowed by the {\textit{eht-imaging}} bound, satisfy also the recent experimental constraint on the dimensionless parameter $\alpha/M^2< 0.54$ \cite{Perkins:2021mhb} set on shift symmetric EsGB theories by the detection of gravitational waves from black-hole binaries. If, on the other hand, one takes a more conservative approach and demands that viable solutions should satisfy both of the EHT bounds within 1-$\sigma$ accuracy, one is forced to exclude the complete range of scalarized, shift-symmetric solutions as none of them falls in the optimum white area of the plot. All solutions are still allowed within 2-$\sigma$ accuracy.

\subsubsection{Scenario II: \texorpdfstring{$\boldsymbol{f(\phi)=\frac{\alpha}{2}\,\phi(r)^2}$}{}}

\begin{figure}
\centering
\includegraphics[width=0.65\linewidth]{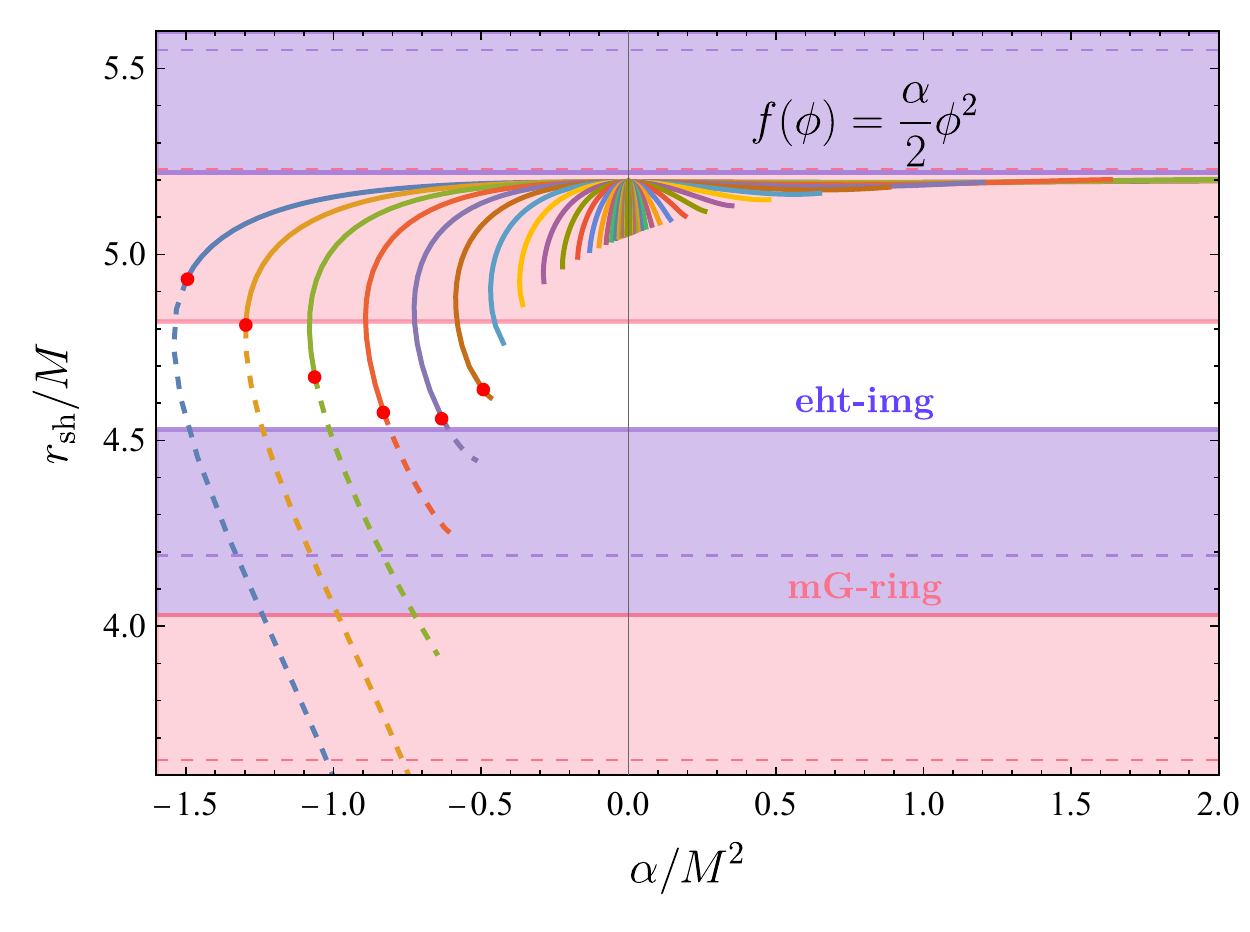}
\caption{Shadow radius for EsGB theory with a quadratic coupling. Each line corresponds to a different choice for $\phi_h$. The red dots denote the solutions beyond which the condition $\lim_{r \to r_h}T^{r\prime}_r(r)<0$ is exchanged for $\lim_{r \to r_h}T^{r\prime}_r(r)>0$ holds (from solid to dashed).}
\label{fig:quadratic}
\end{figure}

Unlike the linear case, the case of the quadratic coupling function necessitates searching along a two-dimensional parameter space due to the fact that the initial value of the field $\phi_h$ is physical. In order to facilitate the search, we select N=25 points equally spaced in the $\ln(\phi_h)$ space with $\phi_{h,{\rm min}}=0.1$ and $\phi_{h,{\rm max}}=100$. For each of these N points, i.e. for each choice of $\phi_h$, we plot a line that spans the entire parameter range $-1<\mathcal{T}<1$. The results are displayed in Fig.  \ref{fig:quadratic}. The red dots in the figure denote a transitioning point regarding the sign of $T^{r\prime}_r$ near the horizon, which will be discussed shortly.

The lines in Fig. \ref{fig:quadratic} denoting solutions with large $\phi_h$ are generally consolidated close to the vertical axis. In constrast, the smaller $\phi_h$ is, the more the lines spread out to larger values of $|\alpha/M^2|$. This is expected due to the definition of $\mathcal{T}$ which in this case takes the form
\begin{equation}
    \mathcal{T}=\frac{\sqrt{96}}{r_h^2}\alpha\,\phi_h\,.
\end{equation}
It is clear that in order to reach the values of $\mathcal{T}\approx \pm 1$, i.e. the limits of the range of $\mathcal{T}$, we need to choose an increasingly larger $\alpha$ in order to compensate for the smallness of $\phi_h$. This justifies the fact that the lines extend further and further away from the origin for small $\phi_h$ values.

Additionally, one may readily observe that the symmetry under the change in the sign of $\alpha$, present in the shift-symmetric case, is now broken. In fact, for negative values of the coupling constant $\alpha$, both the mass parameter $M$ and the shadow radius $r_\text{sh}$ are affected much more dramatically compared to the positive coupling case. This is manifest in Fig. \ref{fig:quadratic} where each solution line for $\alpha<0$ extends along a larger range of values of $r_\text{sh}/M$ compared to that of the $\alpha>0$ solutions. The fact that the former lines turn downward and to the right comes as a consequence of the dimensionless normalization we have applied to the axes. In addition to that, we have observed numerically that negative values of the coupling $\alpha$ lead to large and negative values of the scalar charge. The largeness of the charge and mass for values of the coupling deep into the negative regime is a generic consequence of the evolution of the field equations at intermediate scales between the horizon and infinity, and hence it is difficult to understand the origin of this effect by studying the asymptotic behavior of the solutions.

Moreover, it is interesting to note that, for some of the parameter space analyzed, one crosses the boundary beyond which one can obtain a solution with $\lim_{r \to r_h}T^{r\prime}_r(r)>0$. As explained in Chapter~\ref{ch:Evasions} the condition $\lim_{r \to r_h}T^{r\prime}_r(r)<0$ satisfied by the scalarized solutions found in \cite{Kanti:1995vq, Antoniou:2017acq, Antoniou:2017hxj} was employed to demonstrate the violation of the novel no-hair theorem \cite{Bekenstein:1995un}. Using the results of \cite{Papageorgiou:2022umj}, we can compute the boundary beyond which solutions with $\lim_{r \to r_h}T^{r\prime}_r(r)>0$ appear as follows
\begin{equation}
    \lim_{r \to r_h}T^{r\prime}_r(r)=0 \;\;\;\Rightarrow\;\;\; \alpha=-\frac{1}{4}+\frac{\mathcal{T}^2}{16}-\frac{2\sqrt{1-\mathcal{T}^2}}{9}\,.\label{eq:boundary}
\end{equation}
Simultaneously, due to the definition of $\mathcal{T}$, we can write
\begin{equation}
    \alpha= \frac{r_h^2}{4\sqrt{6}\phi_h}\mathcal{T} \,.
\end{equation}
The above result implies that depending on the choice of $\phi_h$ and $\mathcal{T}$, $\alpha$ can be above or below the boundary defined by (\ref{eq:boundary}). The points below the boundary, i.e. scalarized black-hole solutions with $\lim_{r \to r_h}T^{r\prime}_r(r)>0$, are denoted by dashed lines in Fig. \ref{fig:quadratic} and the transitioning points are marked by large red dots. We note that such solutions arise only in the case of negative coupling constant $\alpha$, and thus any analyses considering only positive $\alpha$ are bound to overlook them. 

In light of shadow observations of Sagittarius A$^*$, figure \ref{fig:quadratic} leads to similar conclusions regarding the validity of the quadratic, scalarized GB solutions with positive $\alpha$ to the ones found for the linear-coupling case: the {\textit{eht-imaging}} bound allows the complete range of solutions while the {\textit{mG-ring}} bound excludes all of them within 1-$\sigma$. No scalarized solutions with positive $\alpha$ fall in the white area. However, the situation is radically different for solutions with negative $\alpha$. There, as noted above, the lines of solutions with small or intermediate values of $\phi_h$ extend into the white area and thus survive all EHT bounds. These favoured solutions are characterised by either a positive or negative value of $\lim_{r \to r_h}T^{r\prime}_r(r)$.

\subsubsection{Scenario III: \texorpdfstring{$\boldsymbol{f(\phi)=\alpha\, e^{\gamma\,\phi(r)}}$}{}}

\begin{figure}
    \centering
    \includegraphics[width=0.49\linewidth]{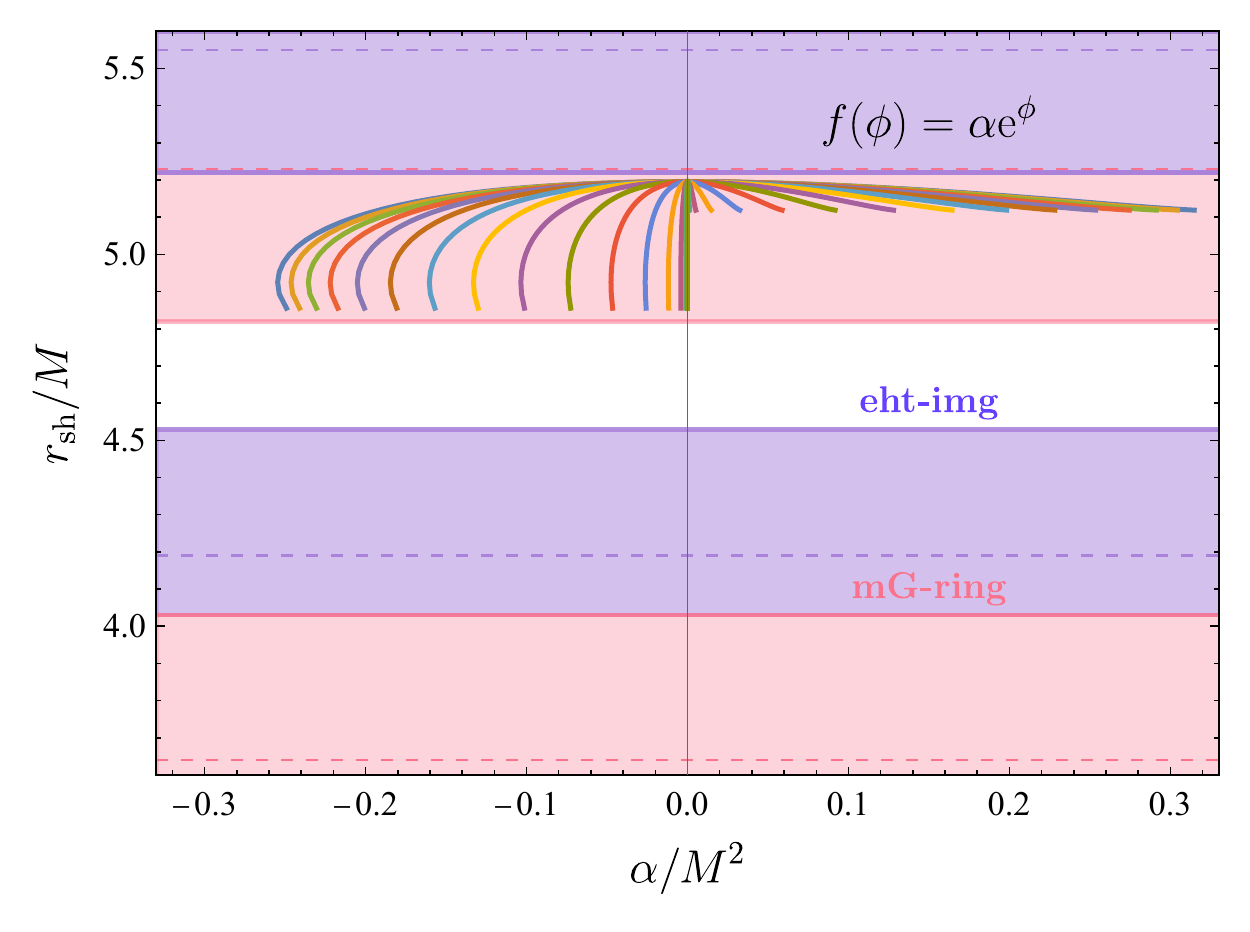}
    \hfill
    \includegraphics[width=0.49\linewidth]{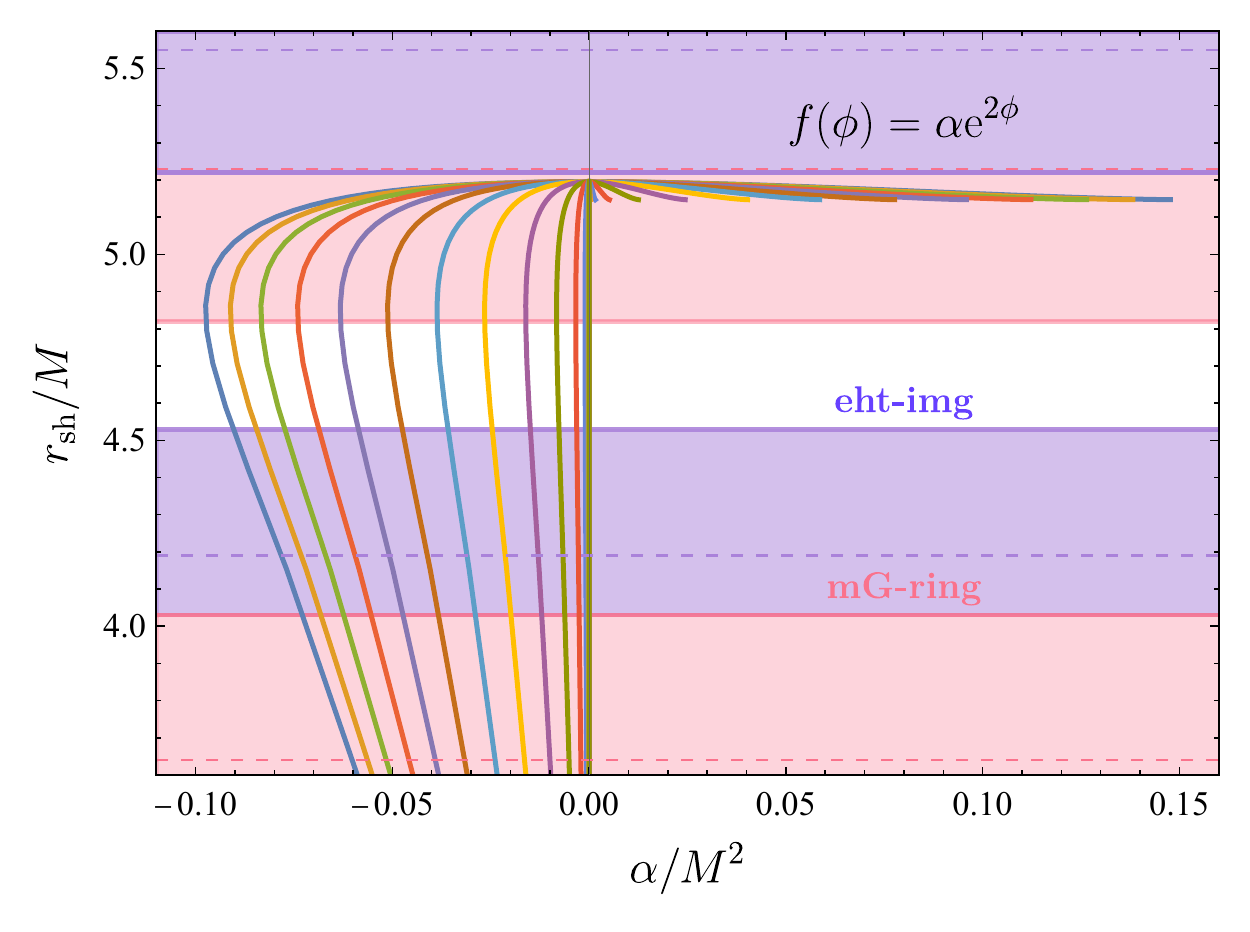}
    \caption[Black hole shadows for the EsGB model]{\textit{Left:} Shadow radius for EsGB theory with a dilatonic coupling with $\gamma=1$. Different lines lines correspond to different choiced for $\phi_h$. \textit{Right:} Same but for $\gamma=2$.}
    \label{fig:dilatonic1}
\end{figure}

For the dilatonic coupling, we need to scan a three dimensional parameter space since there is an additional parameter $\gamma$ that characterizes the coupling function. We follow the same procedure as before, and display the results for two distinct values of $\gamma=1,2$ in the left and right panels of Fig.  \ref{fig:dilatonic1}, respectively.

The subclass of solutions derived for positive values of the coupling parameter $\alpha$ exhibit the same profile, for both values of $\gamma$, as in the previous two cases: the whole range of solutions extend over a very restricted range of values of $r_\text{sh}/M$. As a result, they are all allowed by the {\textit{eht-imaging}} bound but they are all also excluded by {\textit{mG-ring}} bound within 1-$\sigma$. No positive-$\alpha$ solution manages to satisfy both bounds. In fact, all GB scalarized black holes derived for positive $\alpha$ demonstrate the same profile when it comes to their viability under the Sagittarius A* constraints independently of the particular form of the coupling function $f(\phi)$.  We note however that all these solutions satisfy the theoretical bound $\alpha/M^2 < 0.69$, for the existence of scalarized dilatonic black holes \cite{Pani:2009wy, Kanti:1995vq}, and the experimental bound $\alpha/M^2 < 0.54$ \cite{Perkins:2021mhb} (which although was derived for the shift symmetric case may apply also in the exponential case in the limit of small $\alpha$ as is the case here).

The situation however is different when we consider the solutions derived for negative values of the coupling constant $\alpha$. Considering also the behaviour observed in the previous two cases as well as the one depicted in Fig  \ref{fig:dilatonic1}, we conclude that this subclass of solutions is affected both by the form of the coupling function $f(\phi)$ and the particular values assumed for the parameters of the theory. In Fig. \ref{fig:dilatonic1}, we see that, for $\gamma=1$, none of the negative-$\alpha$ solutions manages to satisfy both EHT bounds despite the fact that they extend over a larger range of values of $r_\text{sh}/M$ compared to the positive $\alpha$ solutions. However, for $\gamma=2$, the solution lines manage to extend across the white optimum area and thus a subgroup of solutions, for a specific range of $\alpha$, are favoured by the EHT constraints and may be rendered viable. 
We note that the only way to cross into the regime where the the condition $\lim_{r \to r_h}T^{r\prime}_r(r)>0$ for the dilatonic coupling holds, is to increase the value of $\gamma$ even further. However, this yields a less observationally motivated theory. Another important observation is that for the dilatonic coupling the ratio $r_{\rm sh}/M$ depends on $\gamma$ but not on $\alpha$ and $\phi_h$ simultaneously - this is reflected in the fact that all solution lines corresponding to different values of $\phi_h$ terminate at the same horizontal line in Fig.  \ref{fig:dilatonic1}. This is due to the presence of a symmetry in the Lagrangian that allows us to absorb any change in the value of the scalar field on the horizon $\phi_h$ into a redefinition of the coupling constant $\alpha$ \cite{Kanti:1995vq}, namely
\begin{equation}
    \phi_h \rightarrow \phi_h + \phi_*\,, \quad \alpha \rightarrow \alpha \,e^{-\gamma \phi_*}\,.
\end{equation}
As a result, the parameter space reduces from a 3-dimensional to a 2-dimensional one, explaining why in Fig.~\ref{fig:dilatonic1}, for a set value of $\gamma$ the minimum of each curve corresponds to the shame ration of shadow over mass.

\subsection{Wormholes}

In the context of the EsGB theory, traversable wormhole solutions have been discovered for a variety of scalar-GB couplings, featuring single or double-throat geometries \cite{Kanti:2011jz,Kanti:2011yv,Antoniou:2019awm}.
More on wormhole solutions discovered within Horndeski theory can be found in Appendix~\ref{ch:Appendix_othersol}.
Exploring the shadows of these solutions in depth is beyond the scope of this work and is left for future analysis. Here, however, we will present the results for one characteristic example in order to demonstrate the potential of our analysis as a tool to observationally distinguish wormhole from black-hole solutions.

\begin{figure}[h]
    \centering
    \includegraphics[width=0.65\linewidth]{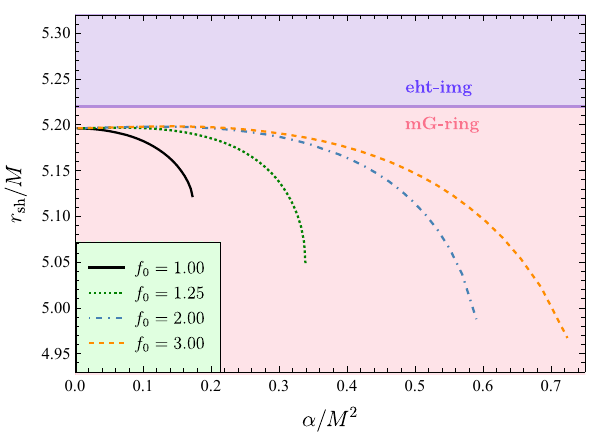}
    \caption[Wormhole shadows in EsGB]{Wormhole solutions in EsGB theory with coupling function $f(\phi)=\alpha e^{-\phi}$, for $f_0=\{1,1.25,1.5,2,3\}$.}
    \label{fig:wh_shadows}
\end{figure}

The case we consider here is the first one historically studied \cite{Kanti:2011jz,Kanti:2011yv} and involves an exponential coupling function of the form $f(\phi)=\alpha e^{-\gamma\phi}$ with $\gamma=1$. Single throat solutions are then discovered if one assumes the line element given in ansatz 1. In accordance with the black-hole scenario, a regularity for the scalar field's derivative on the throat is derived
\begin{equation}
    \phi_0'^2=\frac{f_0(f_0-1)}{2\alpha e^{-\phi_0}\left[f_0-2(f_0-1)\tfrac{\alpha}{l_0^2}e^{-\phi_0} \right]}\, ,
\end{equation}
where $f_0$ and $\phi_0$ are the values of $f$ and $\phi$ evaluated at the throat. For simplicity here, we chose $\phi_0$ so that asymptotically the field vanishes. Additionally, the value of the other metric function $v_0$ at the throat is chosen so that an asymptotically flat spacetime is recovered. We are left therefore with one free parameter, i.e. $f_0$, in addition to the coupling one.

In the limit $f_0\rightarrow 1$, the redshift function $v_0$ tends to larger negative values and a horizon emerges, thus yielding the relevant black-hole solutions in this theory. This allows us to directly compare the shadow radii between black holes and wormholes, and as an example we chose the case with $\gamma=1$. The results are presented in Fig.~\ref{fig:wh_shadows}, where we see that $f_0$ has non-trivial consequences both on the shadow radius and on the mass range of the solutions. Specifically, it appears that as we increase $f_0$ the mass range can also increase significantly. In terms of the shadow radius, we see that all solutions -including the black hole- presented lay within the averaged 1-$\sigma$ {\textit{eht-imaging}} bounds presented in Table~\ref{tab:bounds_sagA*}. On the other hand, all solutions are excluded within 1-$\sigma$ if one chooses to consider the averaged {\textit{mG-ring}} estimates. Once again, no solution exists that satisfies both bounds.

\section{Curvature-induced spontaneous scalarization}

We now write the general action allowing for spontaneously scalarized solutions to emerge, in the following form:
\begin{equation}
\begin{split}
    S=\frac{1}{2\kappa}\int d^4x\sqrt{-g}\;\bigg[& R -\frac{1}{2}\nabla_\alpha \phi \nabla^\alpha \phi
    +h(\phi)R + f(\phi)\GB +V(\phi)\bigg]\, .
\end{split}
\end{equation}

\subsection{Minimal model}

Here, we consider the minimal model associated with spontaneous scalarization identified in \cite{Andreou:2019ikc} and explored in \cite{Ventagli:2020rnx,Ventagli:2021ubn} as well as in earlier chapters, and the coupling functions are defined as
\begin{equation}
    h(\phi)=-\frac{\beta}{4}\phi^2\quad,\quad f(\phi)=\frac{\alpha}{2} \phi^2\,.
\end{equation}

\begin{figure}[t]
    \centering
    \includegraphics[width=0.49\linewidth]{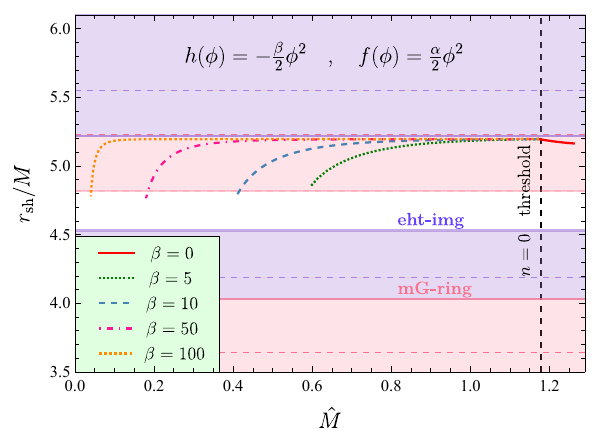}
    \hfill
    \includegraphics[width=0.49\linewidth]{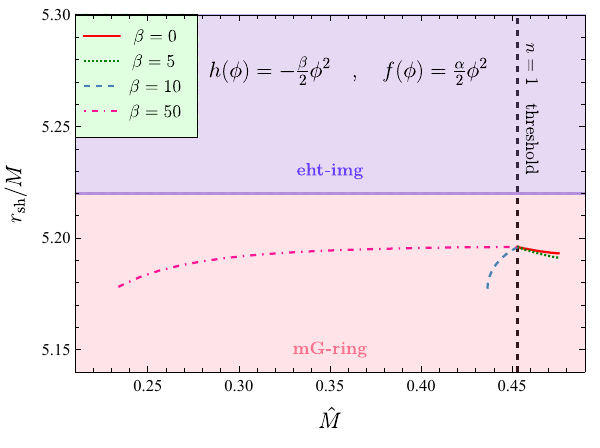}
    \caption[Shadow radius for the minimal scalarization model]{\textit{Left:} Shadow radius of the fundamental mode $(n=0)$ for spontaneously scalarized black holes in the EsRGB theory with quadratic couplings between the scalar field and curvature. The values of the $\phi\text{-}R$ coupling for the lines plotted are $\beta=0,5,10,50,100$. At the same time the $\phi\text{-}\GB$ coupling spans all the allowed values for which spontaneously scalarized solutions are retrieved. \textit{Right:} Same as left panel but for the first overtone $n=1$. The $\beta=100$ case is not presented here for illustrative purposes as it extends to values of $\hat{M}$ that are much smaller than the rest.}
    \label{fig:minimal_scalarization}
\end{figure}

To this end, in Fig.~\ref{fig:minimal_scalarization} we present the shadow radius for spontaneously scalarizeed black-hole solutions derived for different values of the scalar-Ricci coupling constant $\beta$.
We will only consider $\beta>0$ motivated from the analysis of Chapters~\ref{ch:Cosmology}, \ref{ch:Neutron Stars}, \ref{ch:Stability}.
In Chapter~\ref{ch:Stability} in particular we saw that we can distinguish between three regions for scalarized black holes: (I) for $\beta\lessapprox 1$ solutions are unstable, (II) for $1 \lessapprox \beta\lessapprox 1.2$ the solution curves have both a stable and an unstable part (effectively yielding one stable and one unstable solution for any $\hat{M}$), and (III) for $\beta > \beta_{\text{crit}}\approx 1.2$ all solutions are stable. Moreover, as we saw $\mathcal{O}(\beta)\sim 1$ with $\beta>0$ is consistent with a GR cosmological attractor, and can suppress neutron star scalarization.

However, here we aim at conducting a comprehensive study and thus, in Fig.~\ref{fig:minimal_scalarization}, we present the results for the radius of the black-hole shadow in the minimal model for a variety of values of $\beta$, namely $\beta=\{0,5,10,50,100\}$. The left panel depicts the solutions for the fundamental mode ($n=0$). Here, the case $\beta=0$, shown with a solid red line, corresponds to the radially unstable EsGB scalarization model. The solutions in this case lie to the right of the scalarization threshold at $\hat{M}_{\text{th}}^{(0,0)}\approx 1.179$. The rest of the curves shown correspond to values of $\beta$ that are larger than the critical value and therefore to stable configurations. The right panel shows the solutions for the first overtone ($n=1$). Here, only the solutions with $\beta \gtrapprox 10$, which lie to the left of the threshold scalarization value of $\hat{M}_{\text{th}}^{(0,0)}\approx 0.453$, are stable. In both plots, the horizontal axis depicts the value of the dimensionless parameter $\hat M$. The vertical axis showing the shadow radius $r_\text{sh}$ of the black hole is also properly re-scaled in terms of the mass $M$ so that the results are independent of the black-hole mass under consideration.

We readily observe that significant deviations from GR appear in the value of the shadow radius especially towards the lower mass limit of each curve. This is to be expected since it is for the lightest black holes that the curvature is stronger and the effect of both the GB and additional Ricci term becomes increasingly more important. As in the EsGB theory, the quadratic GB term leads to black holes with a more compact geodesic structure, compared to the Schwarzschild solution with the same mass, with the radius of the black-hole shadow following along and taking smaller values, too. The more conservative {\it{eht-imaging}} bound allows all of the solutions to 1-$\sigma$ accuracy, whereas the {\it{mG-ring}} bound, which favours larger deviations from GR, excludes almost all of the solutions to 1-$\sigma$ accuracy. The only solutions allowed are the ones towards the bottom tip of the curves for the fundamental modes. Considering the mG-ring 2-$\sigma$ bounds however all solutions are allowed. 

Therefore, if future observations of horizon-scale images of much lighter black holes are made with the same error bounds, scalarized black-hole solutions would be either favoured or even admitted as the only possible choice compared to the GR solution. Focusing on the character of Sagittarius A$^*$, though, spontaneous scalarization may not be a viable option: all stable solutions arise in the regime $\hat M < 1.2$, which translates to $0.7 < \alpha/M^2$. If, in addition, we focus on the subclass of solutions which survive both the {\it{eht-imaging}} and the {\it{mG-ring}} bounds, these emerge for $\beta\gtrapprox 7$ in the regime $\hat M < 0.5$ or for $4 < \alpha/M^2 $. At the moment, there are no bounds on the dimensionless parameter $\alpha/M^2$ derived in the context of the EsRGB theory. However, if we take the  theoretical bound $\alpha/M^2 < 0.69$, for the existence of dilatonic black holes \cite{Pani:2009wy, Kanti:1995vq}, or the experimental bound $\alpha/M^2 < 0.54$ for shift symmetric solutions \cite{Perkins:2021mhb} as indicative values, we see that the aforementioned range significantly surpasses the latter ones. A more detailed study dedicated to the EsRGB theory needs to be performed before concluding whether Sagittarius A* is a spontaneously scalarised black-hole solution.

\subsection{Quartic sGB coupling}
Here we examine a variation of the EsGB model (without the Ricci coupling) that has been shown to yield stable black-hole solutions under certain assumptions \cite{Silva:2018qhn}
\begin{equation}
    h(\phi)=0\quad,\quad f(\phi)=\frac{\alpha}{2} \phi^2+\frac{\zeta}{4}\phi^4.
\end{equation}
For sufficiently negative values of the ratio $\zeta/\alpha \lessapprox -0.7$, black holes in this theory do get stabilized. Considering positive ratios, on the other hand, produces solutions that are unstable. As in the minimal model, there is a particular range of negative values for the ratio $\zeta/\alpha$ for which both stable and unstable solutions emerge.

\begin{figure}[t]
    \centering
    \includegraphics[width=0.65\linewidth]{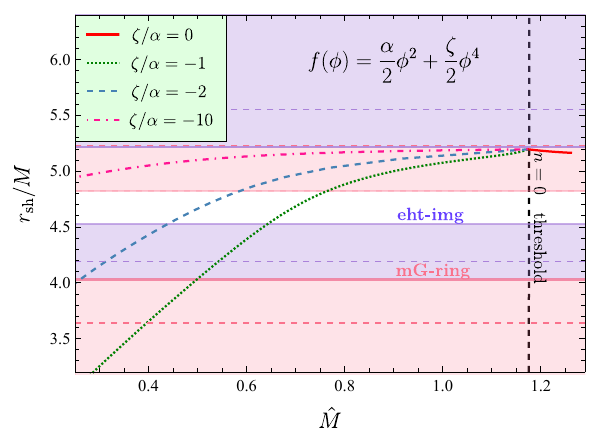}
    \caption[Shadow radius for the quartic $\phi\text{-}\GB$ coupling]{Shadow radius of the fundamental modes $(n=0)$ for spontaneously scalarized black holes in EsGB theory with a quartic $\phi\text{-}\GB$ coupling, for different ratios $\alpha/\zeta=\{0,-1,-2,-10\}$.}
    \label{fig:quartic_scalarization}
\end{figure}

One of the reasons why this model is particularly interesting relates to the fact that even for small values of the quartic coupling, the minimum mass can in principle be pushed to very small values, contrary to the minimal model presented in the last subsection. This feature has evaded attention in other works and is of significant importance as it allows us to probe a much larger range of masses. A consequence of this large mass range is an equally large range in the shadow radii as can be seen in  Fig.~\ref{fig:quartic_scalarization}. It is important to mention that the minimal mass for any $\zeta/\alpha \lessapprox -0.7$ seems to have the potential to be arbitrarily pushed to small values.

Employing the mass-scale independent bounds of Table~\ref{tab:bounds_sagA*}, we may draw a number of useful conclusions. To start with, solutions with fairly large, negative values of $\zeta/\alpha$, i.e. $\zeta/\alpha \simeq -10$, seem to be excluded by the {\textit{mG-ring}} bound, at least in the intermediate and larger mass regime.
For less negative values of $\zeta/\alpha$ the region allowed by the bounds from Table~\ref{tab:bounds_sagA*}
is pushed to intermediate masses. In general, for some fixed $\zeta/\alpha$, solutions with large masses tend to be disfavoured by the {\textit{mG-ring}} bound while small-mass solutions are excluded by the {\textit{eht-imaging}} bound, and this holds independently of the value of that ratio.

We note that, in this case, the solutions which are allowed by the existing bounds of Table~\ref{tab:bounds_sagA*} emerge for $\hat M < 0.85$ or for $1.4 < \alpha/M^2 $. This is an improvement since the lower bound on $\alpha/M^2$ is now much closer to the indicative theoretical and experimental bounds mentioned earlier. Again, in the absence of a bound on $\alpha/M^2$ specifically for the quartic EsGB model, we cannot conclusively state whether Sagittarius A* can be a spontaneously scalarized solution arising in the framework of this model.

\section{The Einstein-Maxwell-scalar Theory}
\label{sec:EMs}

In the black-hole scenario there exists a wider class of theories that also includes Einstein-Maxwell-scalar (EMs) models as spontaneous scalarization frameworks \cite{Herdeiro:2018wub}. The EMs model describes a scalar field non-minimally coupled to Maxwell’s tensor, while being minimally coupled to gravity. It has been shown that under certain assumptions black-hole solutions appear to spontaneously scalarize \cite{Herdeiro:2018wub,Fernandes:2019rez,Blazquez-Salcedo:2020nhs}. For small values of the charge to mass ratio $q$, these solutions have been demonstrated to be the endpoints of a dynamical evolution of unstable Reissner–Nordstr\"om (RN) solutions with the same $q$ within numerical error, while for larger values dynamical scalarization decreases its value. The action functional describing the EMs theory is given by
\begin{equation}
\label{eq:Action_EMS}
\begin{split}
    S=\frac{1}{2\kappa}\int d^4x\sqrt{-g}\;\bigg[& R -\frac{1}{2}\nabla_\alpha \phi \nabla^\alpha \phi
    +f(\phi)F_{\mu\nu}F^{\mu\nu}\bigg]\,.
\end{split}
\end{equation}
The theory we consider here admits the RN solution which is scalar free. To accommodate this we require that asymptotically our theory must approach the RN solution, which translates to $\phi\rightarrow 0$ and $f(\phi)\rightarrow -1$, as $r\rightarrow \infty$. 

The Einstein, Maxwell and scalar field equations are produced by variation with respect to the metric tensor, the scalar field, and the electromagnetic tensor respectively, and they read
\begin{align}
    &G_{\mu\nu}=T_{\mu\nu}\,,\label{eq:EMS_1}\\
    &\Box\phi+\big[\partial_\phi f(\phi)\big]F_{\mu\nu}F^{\mu\nu}=0\,,\label{eq:EMS_2}\\
    &\partial_\mu\left(\sqrt{-g}\,f(\phi)\,F^{\mu\nu}\right)=0\,,\label{eq:EMS_3}
\end{align}
where the energy-momentum tensor contains contributions from the scalar and electromagnetic field in the following way
\begin{equation}
\begin{split}
    T_{\mu\nu}=&-\frac{1}{4}g_{\mu\nu}(\nabla\phi)^2+\frac{1}{2}\nabla_\mu\phi\nabla_\nu\phi\\
    &+f(\phi)\left[\frac{1}{2}g_{\mu\nu}F_{\mu\nu}F^{\mu\nu}-2g^{\rho\sigma}F_{\mu\rho}F_{\nu\sigma}\right]\,.
\end{split}
\end{equation}
The explicit form of the field equations \eqref{eq:EMS_1}-\eqref{eq:EMS_3} for a spherically-symmetric line element can be found in Appendix~\ref{ch:Appendix_eom}.

As in the curvature-induced  scalarization scenario, for the model to be continuously connected to GR, the property $(\partial_\phi f)\big|_{\phi_0}=0$ should be satisfied for some $\phi_0$. The coupling functions which we will consider here satisfy the aforementioned properties and are given by
\begin{align}
    f_e(\phi)=&-e^{-\alpha\phi^2},\label{f1_Max}\\
    f_q(\phi)=&-1+\alpha\phi^2,\label{f2_Max}\\
    f_h(\phi)=&-\cosh\left(\sqrt{-2\alpha}\phi\right),
    \label{f3_Max}
\end{align}
where the coupling constant $\alpha$ is negative. In this case, by taking perturbations of the scalar equation around a RN background, we find that the requirement for the emergence of a tachyonic instability is equivalent to the condition $(\partial_{\phi\phi} f)\big|_{\phi_0} F^2>0$. Here, we consider a purely electric field, namely:
\begin{equation}
    A_\mu dx^\mu=V(r)\,dt \Rightarrow F_{\mu\nu}F^{\mu\nu}<0,
\end{equation}
which in turn requires $(\partial_{\phi\phi} f)\big|_{\phi_0} < 0$. If we also integrate by parts, a second condition is derived, namely $\phi(\partial_\phi f)\big|_{\phi_0} <0$.

\begin{figure}[!ht]
    \centering
    \begin{tabular}{c c}
    \includegraphics[width=0.49\linewidth]{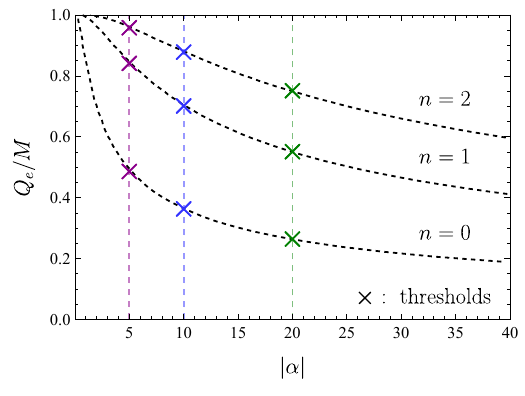}\hfill & \includegraphics[width=0.49\textwidth]{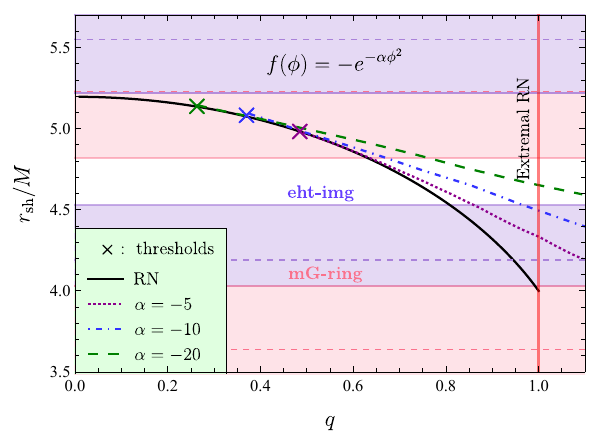}\\
    \includegraphics[width=0.49\textwidth]{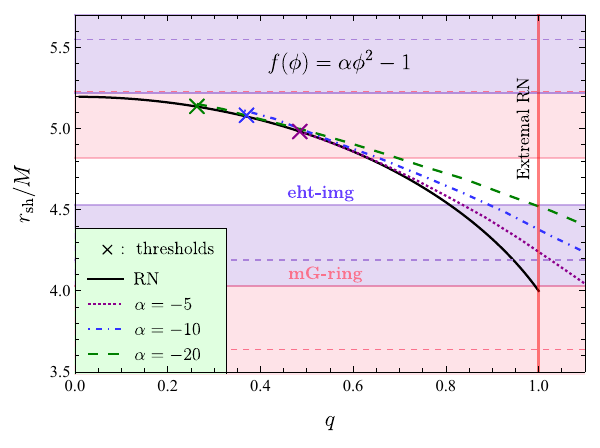}\hfill & \includegraphics[width=0.49\linewidth]{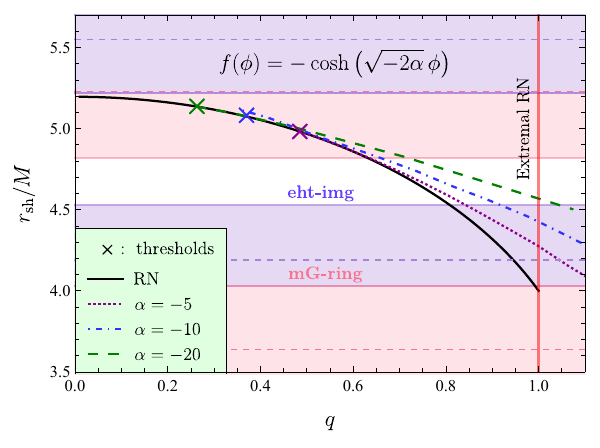} 
    \end{tabular}
    \caption[Shadows for the Einstein-Maxwell-scalar model]{\textit{Top Left}: Onset of scalarization for different overtone numbers. The threshold does not depend on the coupling function. \textit{Top Right}: Shadow radius for the fundamental mode for spontaneously scalarized EMS black holes with an exponential coupling function $f(\phi)=-e^{-\alpha\phi^2}$, for an s-EM coupling with values $\alpha=\{-5,-10,-20\}$. The solid line corresponds to the GR limit (RN). \textit{Bottom Left}: Same as top right but for a quadratic coupling function $f(\phi)=\alpha\phi^2-1$. \textit{Bottom right}: Same as top right but for a hyperbolic coupling function of the form $f(\phi)=-\cosh(\sqrt{-2\alpha}\phi)$.}
    \label{fig:EMS}
\end{figure}

In order to demonstrate the dependence of the shadow radius on the parameters of the theory, we fix $\alpha$ to different negative values and allow for our code to scan the parameter space for the values of $q\equiv Q_e/M$, where $Q_e$ is the electric charge, for which scalarized solutions exist. The existence line for scalarization is presented in the top left panel of Fig.~\ref{fig:EMS}. To create this plot, we examine the linear stability of scalar perturbations around the RN background. We decompose the field perturbation as we did in the curvature induced calarization scenario in Chapter~\ref{ch:Black holes} and we follow the same procedure.
Following this method, we determine the scalarization thresholds for the first three modes, \textit{i.e.} for $n=0$, $n=1$ and $n=2$. This yields the minimum value of $|\alpha|$ for a fixed value of $q$ for which we expect spontaneous scalarization to occur. This value appears to be increasing as one increases the overtone number $n$. It is worth-pointing out that since the threshold of scalarization corresponds to small values of $\phi$, it is independent of our choices of the coupling function accounting for the fact that all of them become identical for small $\phi$.

In the remaining three panels of Fig.~\ref{fig:EMS}, we present the rescaled black-hole shadow $r_{\text{sh}}/M$ in terms of $q$ for the three coupling functions given in Eqs. \eqref{f1_Max}-\eqref{f3_Max}. The scalarized solutions depicted refer to the fundamental mode of the scalar field with $n=0$ and $\ell=0$. The black solid line in each of the three plots corresponds to the shadow radius for RN black holes with different parameters $q$. The value of it can be found analytically to be
\begin{equation}
    \frac{r_{\text{sh}}}{M}=\frac{\sqrt{9-8 q^2}+3}{\sqrt{2+{\left(\sqrt{9-8 q^2}-3 \right)}/{\left(2 q^2\right)}}}\,.
\end{equation}
The coloured lines correspond to solutions with a  different value for the EMS coupling, namely $\alpha=\{-5,-10,-20\}$. The "$\times$" symbol appearing in each coloured line corresponds to the scalarization threshold for each $\alpha$. For all three choices of the coupling function we observe similar results: First, the extremality limit can be exceeded for scalarized solutions, i.e. solutions with $q>1$ emerge. Second, the charge range appears to increase the more we increase the absolute value of the coupling parameter. This confirms the results appearing in \cite{Herdeiro:2018wub,Fernandes:2019rez,Blazquez-Salcedo:2020nhs}. 

The latter result effectively means that a larger domain in the parameter space of $q$ allows for solutions with a shadow radius lying within the desired bounds. Indeed, as we observe from the three plots of Fig.~\ref{fig:EMS} for the three different forms of the coupling function, an increase in $|\alpha|$ decreases the slope of each solution line and thus increases the range of solutions which fall in the white area. These solutions satisfy again all bounds of Table~\ref{tab:bounds_sagA*} coming from the Sagittarius A* constraints. The more `conservative' bound, the {\textit{eht-imaging}} one, clearly favours solutions with small and up to intermediate values of $q$. On the other hand, the more `liberal' bound, the {\textit{mG-ring}} one, tends to favour solutions with intermediate and large values of the charge parameter including the ones beyond RN extremality. According to these results, charged scalarized solutions can be viable candidates for future-observed black holes. However, on the average, they are expected to possess a significant $q$ parameter. This does not seem to be the case with Sagittarius A* for which a very strict upper bound of $q \leq 8.6 \times 10^{-11}$ has been derived \cite{Zajacek:2018ycb, Zajacek:2018vsj}.

\section{Discussion}
\label{sec:conclusions}

The recent publication of black hole images by the EHT collaboration gave rise to a novel way to probe the near horizon regime of black holes that is a valuable and complimentary way to test deviations from GR. The data available by the EHT display a bright ring of emission which surrounds a dark depression that is roughly the size of the black hole shadow. In order to connect the size of the bright ring to the underlying shadow, one has to use the mass-to-distance ratio which for the supermassive black hole in the center of our galaxy Sagittarius A$^*$ is much more accurately known compared to the previously available M87$^*$ due to the proximity of Sagittarius A$^*$ to Earth. For this reason, the bounds presented in the recent EHT publication \cite{EventHorizonTelescope:2022xqj} are the strongest to date regarding black hole metric deviations from GR in the near horizon regime, derived from black hole imaging. In this chapter, motivated by the observational derivation of these bounds, we explored shadows for a number of selected theories of modified gravity whose overarching theme is that they predict the existence of black holes bestowed with non-trivial scalar field profiles.

As there is no clear consensus yet on the spin parameter of Sagittarius A*, we limited our analysis to the spherically symmetric case. For this particular case, the deviation of the black hole metric from the Schwarzschild scenario is quantified by the fractional deviation $\delta$ whose bounds were announced in \cite{EventHorizonTelescope:2022xqj} and were presented here in Table \ref{tab:bounds_sagA*}. Among the various choices displayed in the Table, we settled with displaying the results of the image-domain feature extraction procedure {\textit{eht-imaging}} and the fitting to the analytic model {\textit{mG-ring}}. Our choices were motivated by the fact that these two constraints represent two very  distinct methodologies. In addition, they lie at the two extremes of the spectrum of possible results, with the {\textit{eht-imaging}} constraints being the most conservative ones allowing only for small deviations from GR and the {\textit{mG-ring}} constraints being the most liberal ones favouring much larger deviations from GR.

Regarding the theories under consideration, we first focused on the EsGB theory. Our focus in section \ref{sec:EsGB} was to study black holes with non-trivial scalar hair that are regular from the horizon to infinity and for several different choices for the scalar coupling function. We found that, for the linear coupling, the parameter space of the theory cannot be constrained by the EHT observations since the entire range of solutions are either all allowed by the {\textit{eht-imaging}} constraint or excluded by the {\textit{mG-ring}} constraints within 1-$\sigma$ accuracy. However, for the quadratic and exponential couplings, we found a distinctly different behaviour of the solutions with positive and negative coupling parameter. The solutions derived for a positive coupling exhibit the same behaviour as in the linear coupling case with the whole set of solutions being allowed by the former EHT constraint and excluded by the latter. On the other hand, solutions with a negative coupling extend over a larger part of the parameter space and may thus be more effectively constrained by the EHT bounds. In these cases, subclasses of solutions that satisfy all EHT constraints within 1-$\sigma$ accuracy could be determined.
We also find that special solutions for which the energy momentum tensor component $T_r^r(r)$ can have a local maximum from the horizon to infinity can only occur for the quadratic coupling in a way that is consistent with the EHT results. In the context of this theory, we also highlighted differences in the shadows between black-holes and wormholes. However, a detailed analysis featuring wormhole solutions is left for a future work.

Subsequently, we turned our attention to spontaneous scalarization which for the most part has been on the center of this thesis. We considered two different scenarios; in the first one scalarization is associated with the compactness of the object in accordance with Chapters~ \ref{ch:Cosmology}, \ref{ch:Black holes}, and \ref{ch:Neutron Stars}.
In this case we examined in detail the effects on the shadow radius from the couplings of a scalar field with curvature invariants (Ricci and GB). We saw that in principle the EHT can place constraints on the theory depending on the choices of the coupling parameters under examination. For the minimal EsRGB model we saw that there exists a small region in the parameter space of solutions that satisfies even the tightest combinations of the EHT bounds presented in Table~\ref{tab:bounds_sagA*}. If we also allow for higher order operator corrections in the EsGB coupling, then the allowed parameter space widens due to the fact that the minimal black-hole mass in this case is pushed towards zero.

Finally, in section \ref{sec:EMs} we studied scalarization as a result of a non-minimal coupling of a scalar field with the Maxwell tensor. Compared to the RN scenario, we were able to demonstrate that scalarized EMS black holes allow for agreement with the EHT bounds for a broader range of electric charges. Additionally, solutions are retrieved beyond the GR extremality limit with shadow radii within the desired bounds. 

As we have underlined earlier, the scope of the work presented in this chapter was rather theoretical than observational. We used the constraints derived from the images we currently have at hand only as reference points, in order to study differences between black hole (and wormhole) shadows pertaining to the theories discussed in this thesis.
Looking to the future, the Next Generation EHT (ngEHT) project will provide us with significantly sharper images of the shadow of supermassive black holes such as M87$^*$ and SgrA$^*$ and also possibly real time video of the evolution of the accretion disk around the black hole horizon. This will usher a whole new era in fundamental physics in the strong gravity regime while giving birth to a whole new field: imaging and time resolution of black holes on horizon scales \cite{Blackburn:2019bly,2021AAS...23722101D}. It remains to be seen if the preference for a smaller black hole shadow than the one predicted in the Schwarzschild case will persist in the next generation of experiments, and what implications on the viability of modified theories this will entail.

%% file: Chapters/conclusions.tex
\section{Summary}

Despite the numerous successes of GR, both mathematical arguments (\textit{e.g.} singularity issue, non-renormalizability) and observational considerations (\textit{e.g} dark energy) constitute it incomplete.
Over the decades, a vast effort has been made in order to modify GR, in pursuit of a correct description of gravity at all scales.
The simplest way to extend GR is to increase the number of degrees of freedom by one, in the form of a scalar field. Scalar tensor theories have been considered as viable candidates for gravity from the 1950s onward, and have been used to tackle issues traditionally emerging in standard GR, as we explained in the introductory Chapter~\ref{ch:Introduction}. There, we briefly discussed the roadmap of modified gravitational theories and focused particularly on the Horndeski model, the most general theory involving the metric and a scalar field while yielding second order equations of motion. We then presented the most well-known no scalar hair theorems, which prevent black holes from acquiring nontrivial scalar hair, while once again paying particular tribute to the no-hair theorem governing the generalized Galileon (shift-symmetric Horndeski).

In Chapter~\ref{ch:Evasions} we proceeded to explain how the no-hair theorem for the Galileon may be evaded if not all of its assumptions are satisfied. In this shift-symmetric scenario, we explained how the presence of a nonminimal coupling between the scalar field and the Gauss-Bonnet invariant has the potential to source nontrivial scalar hair in curved spacetime, and yield second order equations of motion. By allowing for more complicated (non shift-symmetric) couplings with the Gauss-Bonnet term (EsGB theory), we then presented a larger family of hairy black holes within the Horndeski framework. We explained how numerical solutions may be found in this theory, and we demonstrated that for numerical black hole solutions to be found, some existence conditions need to be imposed at the black-hole horizon. These conditions yield a minimum black hole size depending on the coupling parameter of the sGB term. Based on these numerical solutions we explicitly showed how the old and novel no-hair theorems formulated by Bekenstein are evaded.
In the last part of this chapter we introduced the concept of spontaneous scalarization of compact objects. In this process compact objects, including black holes but also neutron stars experience a \textit{phase transition} away from the GR solution and toward a scalarized one, when some particular threshold is exceeded. The threshold can be for instance curvature or spin related, even though rotating configurations were not studied in this thesis.

The main goal in Chapter~\ref{ch:Mass-Charge} was to study in depth the aforementioned existence conditions and deduce the various effects higher order terms in the Lagrangian ---that do not directly contribute to the evasion of the no-hair theorems, may have on the properties of the hairy solutions--- For that reason we resorted to the somewhat simpler shift symmetric model. We found that these additional terms have in principle important effects on the properties of the hairy solutions including their mass and scalar charge. We also showed that even though the additional couplings do not strongly alter the minimum black hole mass (at least for values that are reasonable from an EFT perspective) they yield nontrivial differences between solutions that exist near the minimum mass limit. Perhaps the most interesting conclusion from this analysis is that we cannot resort to these additional higher order terms in order to retrieve significantly charged and heavy black holes.

For the next few chapters, namely Chapters~\ref{ch:Cosmology} to \ref{ch:Stability}, we focused on a theory predicting spontaneously scalarized solutions. Overall, it had been demonstrated that scalar-Gauss-Bonnet models fail to be properly incorporated within a cosmological framework, since they predict divergences in the late universe.
In Chapter~\ref{ch:Cosmology} we proposed a model (EsRGB), where except for the nonminimal coupling of the scalar with the Gauss-Bonnet invariant ---which is the one that sources the scalar hair--- we considered another nonminimal interaction in the form of a scalar-Ricci coupling.
We showed that for reasonably small and positive values of the scalar-Ricci coupling $\beta$, a late time cosmological attractor is retrieved. A shortcoming of our theory is found in its validity during the early stages of the universe.
That is because one of our main assumptions, namely the subdomincance of the scalar energy density, breaks down for high enough redshift ($z\approx 10^{12})$.
This means that our theory eventually needs to be embedded in a UV complete one that is consistent with inflation, and that it can only be considered as an effective limit at later times.

In Chapter~\ref{ch:Black holes} we studied scalarized black holes in the EsRGB model, and we demonstrated the dependence of their properties on the exact synergy between the scalar-Ricci and scalar-Gauss-Bonnet couplings, $\beta$ and $\alpha$ respectively.
First we determined the parameter space $(\beta,\alpha)$ for spontaneously scalarized black holes to exist, and by performing a perturbative analysis on a GR background we derived the scalarization thresholds for the first few modes and angular numbers.
In the numerical solutions we derived, we saw that for values of $\beta$ consistent with a late time attractor, there exist strong indications that beyond a critical value, black holes get stabilized. This comes in contradiction with the EsGB theory which ignores the scalar-Ricci interaction.
Next, in Chapter~\ref{ch:Neutron Stars} we continued the study of compact objects in the EsRGB theory, by focusing on neutron stars. We explored the parameter space $(\beta,\alpha)$ where spontaneously scalarized neutron stars emerge in three different scenarios, spanning values from the lowest to the largest neutron star masses appearing in GR, while resorting to two different EOS.
We found that the parameter space for scalarized solutions is significantly smaller than the one determined by the onset of the scalarization, and we investigated extensively the properties of the solutions appearing in this region.
One of our main results pertained to the fact that for values of $\beta$ consistent with the cosmological attractor behaviour (and small $\alpha$), neutron star scalarization is suppressed and the solutions that manage to emerge carry small charges.
This way our theory evades the very strong binary pulsar constraints, by allowing spontaneously scalarized black holes to emerge in a region of the parameter space where neutron star scalarization is suppressed.

In Chapter~\ref{ch:Stability} we returned to the issue regarding the stability of scalarized black holes in EsRGB.
The entropy of the EsRGB black holes was shown to be larger that their GR counterpart for values of $\beta$ beyond the critical value already derived in Chapter~\ref{ch:Black holes}, indicating therefore their thermodynamical stability.
We then performed a radial stability analysis of the solutions, decidedly confirming that black holes get stabilized for $\beta$ larger than its critical value.
We also looked into the QNMs of the Schwarzschild spacetime in order to explicitly show the transition from GR to scalarized solutions, and we found that beyond the fundemental mode, purely imaginary modes may be found in a region of the parameter space where Schwarzschild black holes are stable.
We finally demonstrated the effects of the sRGB synergy in the hyperbolicity of the scalar perturbations. Even though this constitutes a preliminary analysis, we found that in principle, this synergy reduces the region of space where hyperbolicity breaks down.

Finally, in Chapter~\ref{ch:Shadows} we focused on a different aspect of compact object solutions with scalar fields, namely their shadow properties, and we considered various theories to that extent.
We performed this analysis in light of the recent images of the two supermassive black holes M87$^*$ and Sagittarius A$^*$ captured by the EHT. Our main goal, however, was to use the images as a guide, in order to explore differences concerning the shadows of black holes (and wormholes) with scalar fields.
We began with EsGB black holes and wormholes. Overall, the bounds set by the current shadow observations by EHT, do not place siginificant constraints to the theory with positive scalar-Gauss-Bonnet couplings. For negative couplings on the other hand, the parameter space of solutions may extend deep into the forbidden regions set by EHT. Importantly, we showed that wormhole solutions present larger shadows compared with black holes in the same theory.
We then, moved to spontaneously scalarized black holes in the quadratic EsGB, in EsGB with quartic scalar-Gauss-Bonnet couplings, in EsRGB, and finally in the EMs model where scalarization is induced by a scalar-Maxwell coupling.
We noticed significant differences between the shadows corresponding to the different choices of the coupling parameters especially in the EMs model, where the extra coupling allows for more solutions within the allowed bounds (electrically charged GR black holes are very constraint by EHT).

\section{Future perspectives}

As we explained in Chapter~\ref{ch:Cosmology}, the EsRGB model ceases to be a trustworthy theory at very high redshifts and perhaps needs to be emdeded in a UV complete theory consistent with very early universe cosmology and inflation, in order to yield reliable predictions. In this respect, it may be worthwhile to examine the effect higher order terms (in a manner similar to that followed in Chapter~\ref{ch:Mass-Charge}) may have, and whether they can push the validity of the subdominance hypothesis to larger redshift.

In terms of the scalarized black hole solutions we found in Chapter~\ref{ch:Black holes} in the EsRGB theory, we should stress that we only considered the case $\alpha>0$, as this is a requirement for having scalarized black holes under the assumptions of staticity and spherical symmetry. However, as we mentioned in Chapter~\ref{ch:Evasions} it has been shown in Ref.~\cite{Dima:2020yac} that, for $\alpha<0$ (and $\beta=0$), scalarization can be triggered by rapid rotation, and some scalarized black holes have been found in this scenario in Refs.~\cite{Herdeiro:2020wei,Berti:2020kgk}. It would thus be very interesting to consider the effect of the $\beta \phi^2 R$ coupling for $\alpha<0$, {\textit{i.e.}}~in models where scalarization is induced by rotation. 

In terms of the neutron star solutions in Chapter~\ref{ch:Neutron Stars}, an obvious continuation of the work presented there, is the stability analysis of the scalarized solutions.
Moreover, since rotation is known to have important effects on black hole scalarization, by either quenching it or triggering it, it is interesting to see what its impact is on scalarized neutron stars in different models. The effect of rotation on neutron star scalarization was investigated in the framework of the DEF model \cite{Doneva:2013qva}, and it would be interesting to extend this analysis to coupled Ricci/Gauss-Bonnet couplings, or pure Gauss-Bonnet ones.

It will also be interesting to combine the bounds coming from neutron star and black hole observations with the theoretical constraints that relate to the requirement that scalarization models have a well-posed initial value problem \cite{Ripley:2020vpk}. So far, the combined theory with both Ricci and Gauss-Bonnet couplings has not been studied in detail from the initial value problem perspective.
In Chapter~\ref{ch:Stability} we explored the impact of the Ricci coupling on the hyperbolicity of scalar perturbations and deduced that it actually improves it.
It would be interesting to generalize the hyperbolicity analysis to more general perturbations and beyond the linear level, in order to check if the coupling with the Ricci scalar could have a positive effect when considering hair formation by collapse \cite{Ripley:2020vpk} or binary mergers \cite{East:2021bqk}.

It is also worth pointing out that, gravitational wave observations could potentially place constraints in the case of the scalar-Ricci-Gauss-Bonnet synergy. Approaches similar to the one followed in \cite{Perkins:2021mhb} can be extended in the EsRGB model (even though the analysis becomes more complicated in non shift-symmetric theories), but even a Post-Newtonian analysis of the inspiral could place some preliminary constraints (see for example \cite{Julie:2019sab, Julie:2022huo}).

With regard to the study of QNMs in EsRGB, but also more generally in the Horndeski Lagrangian a proper analysis of their spectrum could be performed.
A significant challenge in this direction pertains to the fact that most of the black hole solutions with scalar hair found are numerical, and hence some of the usual QNM techniques are tough to apply.
From an observational point of view however, potential deviations of the no-hair theorems are expected to yield imprints on the QNM spectrum, and therefore their analysis is critical.

Finally, regarding the shadows of compact objects with scalar fields, an interesting work to follow would be the analytical study of wormhole shadows in the theories we examined here. Such a study would point out whether shadow observations could potentially be used in the future to distinguish black hole solutions from wormholes.

%% file: Chapters/appendix_thermodynamics.tex
In this appendix, we calculate the thermodynamical properties, which can reveal interesting characteristics of the scalarized solutions, namely their temperature and entropy. The first quantity may be easily derived by using the following definition \cite{York:1984wp,Gibbons:1994ff}
\begin{align}
    T=\frac{k_h}{2\pi}=\frac{1}{4\pi}\,\left(\frac{1}{\sqrt{|g_{tt} g_{rr}|}}\,
    \left|\frac{dg_{tt}}{dr}\right|\right)_{r_h}=\frac{\sqrt{a_1 b_1}}{4\pi}\,,
\label{eq:Temp-def}
\end{align}
that relates the black-hole temperature $T$ to its surface gravity $k_h$. The above formula is valid for spherically-symmetric black holes in theories that may contain also higher-derivative terms such as the GB term. The final expression of the temperature in Eq. \eqref{eq:Temp-def} is derived by employing the near-horizon asymptotic forms \eqref{eq:A-rh}-\eqref{eq:B-rh} of the metric functions.

The entropy of the black hole may be calculated by using the Euclidean approach in which the entropy is given by the relation \cite{Gibbons:1976ue}
\begin{equation}
    S_h=\beta\left[\frac{\partial (\beta F)}{\partial \beta} -F\right],
\label{eq:entropy-def}
\end{equation}
where $F=I_E/\beta$ is the Helmholtz free-energy of the system given in terms of the Euclidean version of the action $I_E$, and $\beta=1/(k_B T)$. The above formula has been used in the literature to determine the entropy of the asymptotically-flat coloured GB black holes \cite{Kanti:1995vq,Kanti:1997br} and of the family of novel black-hole solutions found in \cite{Antoniou:2017acq} for different forms of the GB coupling function.

Alternatively, one may employ the Noether current approach developed in \cite{Wald:1993nt} to calculate the entropy of a black hole. In this, the Noether current of the theory under diffeomorphisms is determined, with the Noether charge on the horizon being identified with the entropy of the black hole. In \cite{Iyer:1994ys}, the following formula was finally derived for the entropy
\begin{equation}
    S=-2\pi \oint{d^2x \sqrt{h_{(2)}}\left(\frac{\partial \mathcal{L}}{\partial R_{abcd}}\right)_\mathcal{H}\hat{\epsilon}_{ab}\,\hat{\epsilon}_{cd}}\,,
\label{eq:entropy_AdS}
\end{equation}
where $\mathcal{L}$ is the Lagrangian of the theory, $\hat{\epsilon}_{ab}$ the binormal to the horizon surface $\mathcal{H}$, and $h_{(2)}$ the 2-dimensional projected metric on  $\mathcal{H}$. The equivalence of the two approaches has been demonstrated in \cite{Dutta:2006vs}, in particular in the context of theories that contain higher-derivative terms such as the GB term.  Here, we will use the Noether current approach to calculate the entropy of the black holes as it leads faster to the desired result. 

To this end, we need to calculate the derivatives of the scalar gravitational quantities, appearing
in the Lagrangian of our theory with respect to the Riemann tensor.
Let us here consider the general theory with Ricci and Gauss-Bonnet couplings relevant for Chapters~\ref{ch:Black holes}, \ref{ch:Neutron Stars}, and \ref{ch:Stability},
\begin{equation}
    S=\frac{1}{2\kappa}\int d^4x\sqrt{-g}\;\bigg[R -\frac{1}{2}\nabla_\alpha \phi \nabla^\alpha \phi
    +h(\phi)R + f(\phi)\GB +V(\phi)\bigg]\, .
\label{eq:sRGB}
\end{equation}
Then, substituting in Eq. \eqref{eq:entropy_AdS}, we obtain
\begin{equation}
\begin{split}
    S=&-\frac{1}{8}\oint d^2x \sqrt{h_{(2)}}\bigg\{
    \frac{1+h(\phi)}{2}\left(g^{ac}g^{bd}-g^{bc}g^{ad}\right)\\
    & +f(\phi)\Big[2R^{abcd}+-2\left(g^{ac}R^{bd}-g^{bc}R^{ad}-g^{ad}R^{bc}+g^{bd}R^{ac}\right)\\
    & +R\left(g^{ac}g^{bd}-g^{bc}g^{ad}\right)\Big]\bigg\}_\mathcal{H}\hat{\epsilon}_{ab}\,\hat{\epsilon}_{cd}
    \,.
\end{split}
\label{eq:entropy_1}
\end{equation}
The first term inside the curly brackets of the above expression comes from the variation
of the Einstein-Hilbert term and leads to:
\begin{equation}
    S_1=-\frac{1}{16}\oint{d^2x\sqrt{h_{(2)}}\,
    \big[1+h(\phi)\big]\left(\hat{\epsilon}_{ab}\,
    \hat{\epsilon}^{\,ab}-\hat{\epsilon}_{ab}\,\hat{\epsilon}^{\,ba}\right)}.
\end{equation}
We recall that $\hat{\epsilon}_{ab}$ is antisymmetric, and, in addition, satisfies 
$\hat{\epsilon}_{ab}\,\hat \epsilon^{\,ab}=-2$. Therefore, we easily obtain the result
\begin{equation}
    S_1=\big[1+h(\phi)\big]\, \frac{A_{\mathcal{H}}}{4}. \label{eq:S1}
\end{equation}
where $A_{\mathcal{H}}=4\pi r_h^2$ is the horizon surface. The remaining terms in Eq. \eqref{eq:entropy_1} are all proportional to the coupling function $f(\phi)$ and follow from the variation of the GB term. To facilitate the calculation, we notice that, on the horizon surface, the binormal vector is written as:
$\hat{\epsilon}_{ab}=\sqrt{-g_{00}\,g_{11}}\big|_{\mathcal{H}} \left(\delta^0_a\delta^1_b-\delta^1_a\delta^0_b\right)$. This means that we may alternatively write:
\begin{equation}
    \left(\frac{\partial \mathcal{L}}{\partial R_{abcd}}\right)_\mathcal{H}\hat{\epsilon}_{ab}\,\hat{\epsilon}_{cd}=4g_{00}\,g_{11}\big|_{\mathcal{H}}\left(\frac{\partial \mathcal{L}}{\partial R_{0101}}\right)_\mathcal{H}.
\end{equation}
Therefore, the terms proportional to $f(\phi)$ may be written as
\begin{equation}
\begin{split}
    S_2=-\frac{1}{2}\,f(\phi)&\,g_{00}\,g_{11}\big|_{\mathcal{H}}\,\oint d^2x \sqrt{h_{(2)}}\,\bigg[2R^{0101} +g^{00}g^{11}R\\[2mm]
    & -2\left(g^{00}R^{11}-g^{10}R^{01}-g^{01}R^{10}+g^{11}R^{00}\right)\bigg]_{\mathcal{H}}\,.
\end{split}
\label{eq:S2}
\end{equation}
To evaluate the above integral, we will employ the near-horizon asymptotic solution \eqref{eq:A-rh}-\eqref{eq:phi-rh} for the metric functions and scalar field. Substituting in Eq. \eqref{eq:S2}, we straightforwardly find 
\begin{equation}
    S_2=\frac{f(\phi_h)A_{\mathcal{H}}}{r_h^2}= 4\pi f(\phi_h)\, .
\label{eq:S2-final}
\end{equation}
Combining the expressions \eqref{eq:S1} and \eqref{eq:S2-final}, we finally derive the result
\begin{equation}
    S_h=\big[1+h(\phi_h)\big]\, \frac{A_h}{4} +4 \pi f(\phi_h)\,.
\label{eq:entropy}
\end{equation}
The above describes the entropy of a GB black hole arising in the context of the theory \eqref{eq:sRGB}, with a general coupling function $f(\phi)$ between the scalar field and the GB term.

%% file: Chapters/appendix_other_solutions.tex
In this appendix let us briefly discuss other types of solutions appearing in the context of scalar-tensor theories with couplings between the scalar and the GB invariant.

\section{Solutions with cosmological constant}
One interesting class of solutions was recovered in the presence of a cosmological constant \cite{Bakopoulos:2018nui}.
\begin{equation}
S=\frac{1}{16\pi}\int{d^4x \sqrt{-g}\left[R+X+f(\phi)\,\GB- 2\Lambda\right]}.
\end{equation}
The procedure for finding these regular solutions bares a lot of similarities with the steps we followed in the case of asymptotically flat solutions.
Since the uniform distribution of energy associated with the cosmological constant permeates the whole spacetime, we expected $\Lambda$ to have an effect on both the near-horizon and far-field asymptotic solutions. Indeed, analytical calculations in the small-$r$ regime reveal that the  cosmological constant modifies the constraint that determines the value of $\phi'_h$ for which a regular, black-hole horizon forms. In addition, such a horizon is indeed formed, for either positive or negative $\Lambda$ and for all choices of the coupling function $f(\phi)$. The behaviour of the solution in the far-field regime depends strongly on the sign of the cosmological constant. For $\Lambda>0$, a second horizon, the cosmological one, is expected to form at a distance $r_c>r_h$, whereas for $\Lambda<0$, an Anti-de Sitter type of solution is sought for at asymptotic infinity. Both types of solutions are analytically shown to be admitted by the set of our field equations at the limit of large distances, thus opening the way for the construction of complete black-hole solutions with an (Anti)-de Sitter asymptotic behaviour.

\begin{figure}[ht]
    \centering
    \includegraphics[width=0.49\textwidth]{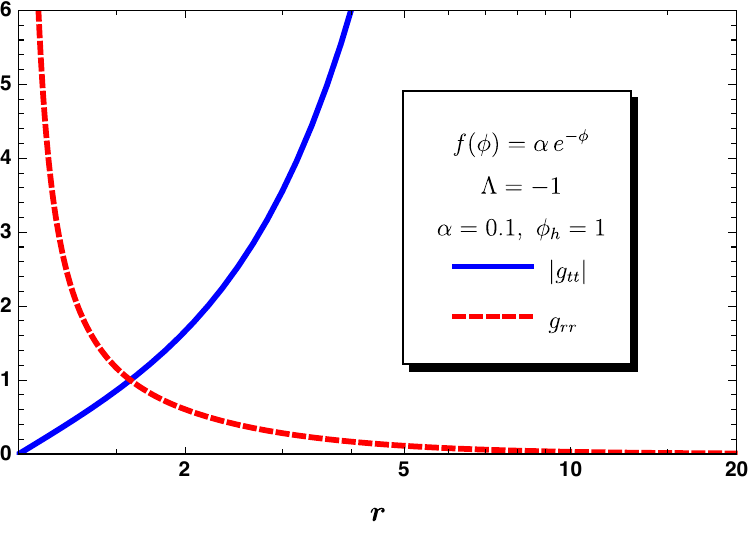}\hfill
    \includegraphics[width=0.49\textwidth]{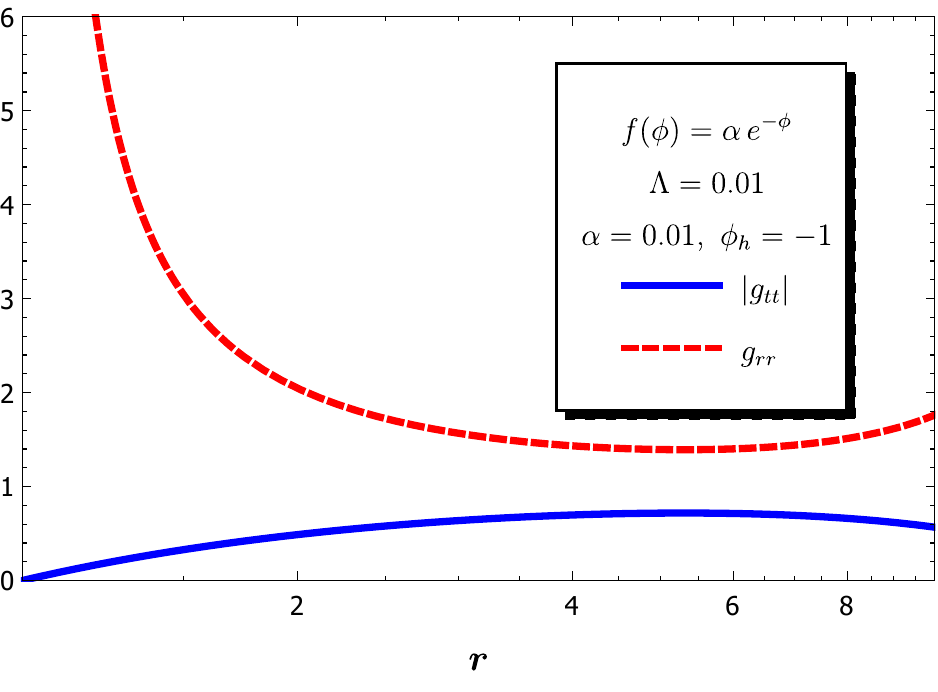}\hspace{2mm}
    \caption[Metric elements for EsGB-$\Lambda$]{\textit{Left:} The metric components $|g_{tt}|$ and $g_{rr}$ in terms of the radial coordinate $r$, for $f(\phi)=\alpha e^{-\phi}$.
    \textit{Right:} Same as left but for a positive cosmological constant. The plots are taken from \cite{Bakopoulos:2018nui}.}
    \label{fig:metric_Lambda}
\end{figure}

GB black-hole solutions with a negative cosmological constant smoothly merge with the SAdS ones, in the large mass limit. As in the asymptotically-flat case, it is the small-mass range that provides the characteristic features for the GB solutions. These solutions have a modified dependence of both their temperature and horizon area on their mass compared to the SAdS solution. Another characteristic is also the minimum horizon, or minimum mass, that all our GB solutions possess due to the existence condition inequality, which in this case is given by
\begin{equation}
    256 \Lambda  \dot f^4_c \left(\Lambda  r_c^2-6\right) +
    32 r_c^2 \dot f^2_c \left(2\Lambda  r_c^2-3\right)+r_c^6 \geq0 \,.
\end{equation} 

Contrary to the asymptotically flat case the form the field acquires asymptotically does not contain a $1/r$ fall-off but instead a logarithmic contribution, $\sim\ln r$.
The absence of a $(1/r)$-term in the expression of the scalar field at large distances excludes the presence of a scalar charge, even a secondary one. The coefficient $d_1$ in front of the logarithmic term in the expression of $\phi$ can give us information on how much the large-distance behaviour of the scalar field deviates from the power-law, valid in the asymptotically-flat case. It turns our that this deviation is stronger for GB black holes with a small mass whereas the more massive ones have a $d_1$ coefficient that tends to zero.

\section{Wormholes}

A particularly interesting property emerging in the EdGB solutions is the
presence of regions with negative {\sl effective} energy density  -- this
is due to the presence of the higher-curvature GB term and is therefore
of purely gravitational nature \cite{Kanti:1995vq,Kanti:2011jz}. Consequently, the EdGB theory has been shown to allow for Lorentzian, traversable wormhole solutions without the need for exotic matter \cite{Kanti:2011jz,Kanti:2011yv}. It is tempting to conjecture that the more general EsGB theories should also allow for traversable wormhole solutions. Indeed, traversable wormholes require violation of the energy conditions \cite{Morris:1988cz,Visser:1995cc}. But whereas in General Relativity this violation is typically achieved by a phantom field \cite{Ellis:1973yv,Ellis:1979bh,Bronnikov:1973fh,Kodama:1978dw,Kleihaus:2014dla,Chew:2016epf}, in EdGB theories it is the effective stress-energy tensor  that allows for this violation \cite{Kanti:2011jz,Kanti:2011yv}.

The metric element initially considered in the case of EdGB wormholes was the following:
\begin{equation}
    ds^2= -e^{2v(l)}\,dt^2 + f(l)\,dl^2 + \left(l^2+l_0^2\right) \,(d\theta^2+ \sin^2 \theta\,d\varphi^2)\,,
\end{equation}
and allowed for single-throat wormhole geometries. By changing the metric element to the following one:
\begin{equation}
    ds^2
    = -e^{f_0(\eta)}dt^2+e^{f_1(\eta)}\left\{d\eta^2
    +\left(\eta^2+\eta_0^2\right)\left(d\theta^2+\sin^2\theta d\varphi^2 \right)\right\}\,,
\end{equation}
a richer family of wormhole solutions emerges in EsGB with single and double throat geometries.

\begin{figure}[ht]
    \centering
    \includegraphics[width=0.47\textwidth]{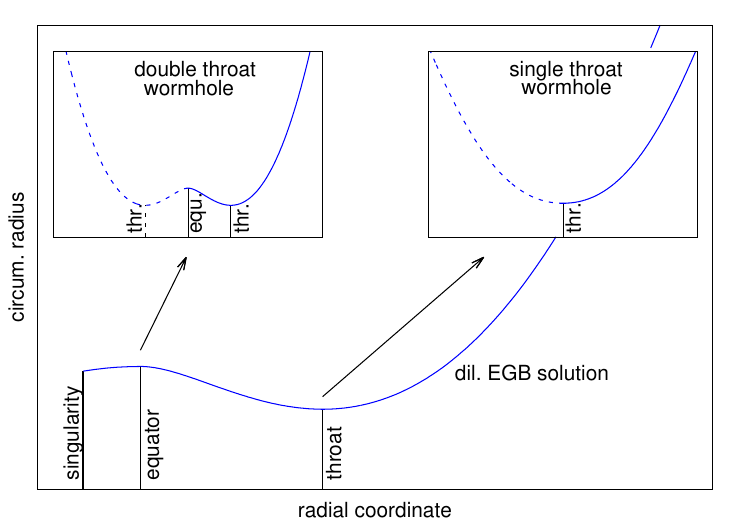}\hspace{5mm}
    \includegraphics[width=0.47\textwidth]{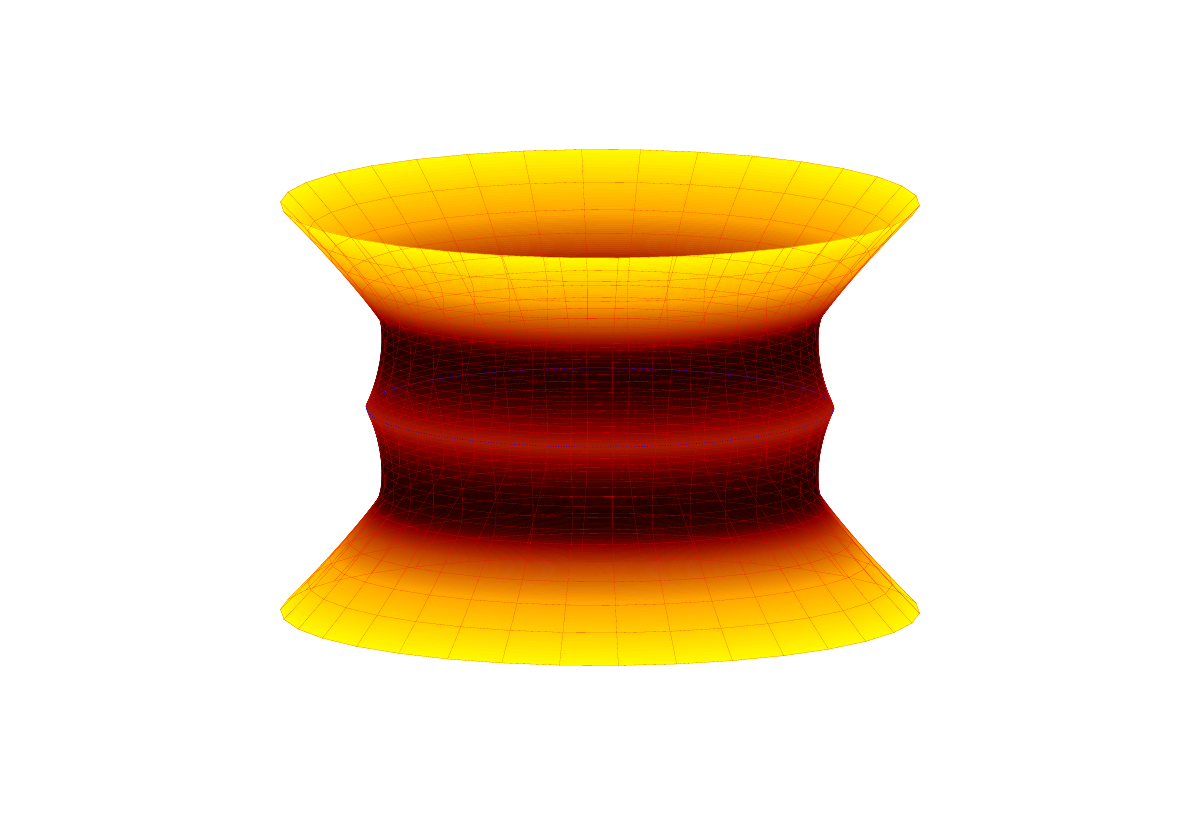}\hspace{2mm}
    \caption[Wormholes in EsGB]{\textit{Left:} Schematic picture for the construction of double-throat and single-throat wormhole solutions. \textit{Right:} The embedded equatorial plane is shown for the double-throat wormhole with $\alpha/\eta_0^2=0.35$ and $Q/M=0.886$, for $f(\phi)=0.25\, e^{-\phi}$. The plots are taken from \cite{Antoniou:2019awm}.}
    \label{fig:whs_EsGB}
\end{figure}

A traversable wormhole solution is characterised by the absence of horizons or singularities.
In order to ensure that this is the case for the wormhole solutions, the following
regular expansions for the metric functions and scalar field, are considered near the throat/equator $\eta=0$
\begin{align}
    e^{f_0}=&a_0\,(1+a_1 \eta +a_2 \eta^2+...)\,,\label{appf0}\\[1mm]
    e^{f_1}=&b_0\,(1+b_1 \eta +b_2 \eta^2+...)\,,\label{appf1}\\[1mm]
    \phi=&\phi_0+\phi_1 \eta +\phi_2 \eta^2+...\,.\label{appff}
\end{align}
The Lorentzian signature of spacetime demands that both parameters $a_0$ and $b_0$ must be positive;
in addition, they should be finite and non-vanishing. The emergence of an extremum in the circumferential radius $R_c$ dictates that $f_1'(0)=0$;
this leads to the result $b_1=0$. The $(\eta\eta)$ component of the Einstein equations yields near the throat a constraint equation
\begin{equation}\label{eq10}
\left[\left(\eta_0^2\phi'^2 + 4\right) e^{f_1} - 8 f_0'\phi' \dot{F}\right]_{\eta=0}= 0\,.
\end{equation}
similar to the existence conditions emerging at the black hole horizon which we encountered repeatedly in this thesis.

%% file: Chapters/appendix_eom.tex
In this appendix we present the various equations of motion discussed in the main part of this thesis, and other lengthy expressions that we omitted.

\section{Scalar equation in the Horndeski Lagrangian}
In Chapter~\ref{ch:Introduction} we introduced the Horndeski Lagrangian and equations of motion.
The equations for the full Lagrangian are given in \eqref{eq:Horndeski-eom} but the exact expressions for the terms we used there were omitted. Since, the scalar equation of motion is used extensively in Chapter~\ref{ch:Mass-Charge}, here we present the terms related with the scalar equation as they were derived in \cite{Kobayashi:2011nu}.

The ``source terms'' are given by
\begin{align}
P_\phi^2=&G_{2\phi}\, ,
\\
P_\phi^3=&\nabla_\mu G_{3\phi}\nabla^\mu\phi,
\\
P_\phi^4=&G_{4\phi}R+G_{4\phi X}\left[(\Box\phi)^2-(\nabla_\mu\nabla_\nu\phi)^2\right]\, ,
\\
\begin{split}
    P_\phi^5=&-\nabla_\mu G_{5\phi }G^{\mu\nu} \nabla_\nu\phi -
    \frac{1}{6}G_{5\phi X}
    \big[
    (\Box\phi)^3-3\Box\phi (\nabla_\mu\nabla_\nu \phi)^2\\
    &+2(\nabla_\mu\nabla_\nu \phi)^3
    \big]\, ,
\end{split}
\end{align}
while the terms contributing to the shift-symmetric current are given by
\begin{align}
J_\mu^2=&-{\mathcal L}_{2X}\nabla_\mu\phi\, ,
\\
J_\mu^3=&-{\mathcal L}_{3X}
\nabla_\mu\phi+ G_{3X}\nabla_\mu X+2G_{3\phi}\nabla_\mu\phi\, ,
\\
\begin{split}
    J_\mu^4=&-{\mathcal L}_{4X}\nabla_\mu\phi+
    2G_{4X}R_{\mu\nu}\nabla^\nu\phi
    -2G_{4XX}(\Box\phi\nabla_\mu X
    \\
    &-\nabla^\nu X\nabla_\mu\nabla_\nu\phi)
    -2G_{4\phi X}\left(\Box\phi\nabla_\mu \phi +\nabla_\mu X\right)\, ,
\end{split}
\\
\begin{split}
    J_\mu^5=&
    -{\mathcal L}_{5X}\nabla_\mu\phi-2G_{5\phi }G_{\mu\nu}\nabla^\nu\phi
    -G_{5X}\big[G_{\mu\nu}\nabla^\nu X+
    R_{\mu\nu}\Box\phi \nabla^\nu\phi
    \\
    &-R_{\nu\lambda}\nabla^\nu\phi\nabla^\lambda\nabla_\mu\phi
    -R_{\alpha\mu\beta\nu}\nabla^\nu\phi\nabla^\alpha\nabla^\beta\phi
    \big]
    \\
    &+G_{5XX}\bigg\{
    \frac{1}{2}\nabla_\mu X\left[(\Box\phi)^2-(\nabla_\alpha\nabla_\beta\phi)^2\right]
    -\nabla_\nu X(\Box\phi \nabla_\mu\nabla^\nu\phi
    \\
    &-\nabla_\alpha\nabla_\mu\phi
    \nabla^\alpha\nabla^\nu\phi)
    \bigg\}
    +G_{5\phi X}\bigg\{
    \frac{1}{2}\nabla_\mu\phi\big[(\Box\phi)^2-(\nabla_\alpha\nabla_\beta\phi)^2\big]
    \\
    &+\Box\phi\nabla_\mu X-\nabla^\nu X\nabla_\nu\nabla_\mu \phi
    \bigg\}\, .
\end{split}
\end{align}

\section{High-order shift symmetric model}
In this section we present the equations of motion related to the theory we considered in Chapter~\ref{ch:Mass-Charge}. Varying action \eqref{eq:action} with respect to the scalar field yields the following scalar equation
\begin{align}\label{eq:scalar}
    \Box\phi=\,&+\alpha \GB+2\gamma G^{\mu \nu } \nabla_{\nu }\nabla_{\mu }\phi+\kappa(\nabla\phi)^2\Box\phi+2\kappa \nabla^\mu\phi\nabla^\nu\phi\nabla_{\mu}\nabla_{\nu}\phi\\ \nonumber
    &-\sigma \nabla_{\mu }\,\Box\phi \nabla^{\mu}\phi -  \sigma\,(\Box\phi)^2 + \sigma\nabla^{\mu }\phi \,\Box\,\nabla_{\mu }\phi +\sigma(\nabla_{\mu}\nabla_{\nu }\phi)^2,
\end{align}
while varying with respect to the metric yields
\begin{equation}
\begin{split}
    G_{\mu\nu}=
    &+\frac{1}{2}\nabla_\mu\phi\nabla_\nu\phi-\frac{1}{4}g_{\mu\nu}(\nabla\phi)^2\\
    &-\frac{\alpha}{2 g}g_{\mu(\rho}g_{\sigma)\nu}\epsilon^{\kappa\rho\alpha\beta}\epsilon^{\sigma\gamma\lambda\tau}{R}_{\lambda\tau\alpha\beta}\nabla_{\gamma}\nabla_{\kappa}\phi\\
    &-\frac{\gamma}{2}  {R} \nabla_{\mu }\phi \nabla_{\nu }\phi +\gamma \nabla_{\nu }\nabla_{\mu }\phi \,\Box\phi- \gamma G_{\nu \rho } \nabla_{\mu }\phi \nabla^{\rho }
    \phi\\
    &-\gamma G_{\mu \rho } \nabla_{\nu }\phi \nabla^{\rho}\phi+\frac{\gamma}{2} {R}_{\mu \nu } (\nabla_{\mu }\phi)^2 - \gamma \nabla_{\rho }\nabla_{\nu }\phi \nabla^{\rho }\nabla_{\mu }\phi\\
    &-\frac{\gamma}{2} g_{\mu \nu } (\Box\phi)^2 +  \frac{\gamma}{2} G_{\rho \sigma } g_{\mu \nu } \nabla^{\rho 
    }\phi \nabla^{\sigma }\phi+\frac{\gamma}{2} g_{\mu \nu } {R}_{\rho \sigma } \nabla^{\rho }\phi \nabla^{\sigma }\phi\\
    &-\gamma 
    {R}_{\mu \rho \nu \sigma } \nabla^{\rho }\phi \nabla^{\sigma
    }\phi +  \frac{\gamma}{2} g_{\mu \nu } (\nabla_{\sigma }\nabla_{\rho }\phi)^2\\
    &+\frac{\sigma}{2} \nabla_{\mu}\phi \nabla_{\nu}\phi \;\Box\phi - \sigma\nabla_{\rho }\nabla_{(\mu}\phi\nabla_{\nu)}\phi \nabla^{\rho }\phi\\
    &+ \frac{\sigma}{2} g_{\mu\nu} \nabla^{\rho }\phi \nabla_{\sigma }\nabla_{\rho }\phi \nabla^{\sigma }\phi\\
    &-\frac{\kappa}{2}(\nabla\phi)^2\nabla_\mu\phi\nabla_\nu\phi+\frac{\kappa}{8}g_{\mu\nu}(\nabla\phi)^2(\nabla\phi)^2.\\
\end{split}
\end{equation}

\section{Equations motion in EsRG}
\subsection{Black holes}
Here we present the equations used to derive the numerical static and spherically symmetric black hole solutions in Chapter~\ref{ch:Black holes} that were also used in Chapter~\ref{ch:Stability}.
\begin{align}
    \begin{split}
        (t,t):\quad & B \big\{\beta  \phi ^2+2 \phi  \big[\phi '' \big(-8 \alpha +8 \alpha  B+\beta  r^2\big)+2 \beta  r \phi '\big]\\
        &+\phi'^2\big[-16 \alpha +16 \alpha  B+(2 \beta -1) r^2\big]-4\big\}+4=0\\
        &-\beta \, \phi ^2+B' \big[\phi \, \phi ' \big(-8 \alpha +24 \alpha B+\beta  r^2\big)+\beta  r \phi ^2\\
        &-4 r\big]=0\, ,
    \end{split}\\[2mm]
    \begin{split}
        (r,r):\quad & B \phi  \, \phi ' \big[A' \big(-8 \alpha +24 \alpha  B+\beta  r^2\big)+4 A \beta  r\big]\\
        &+\big(\beta  \phi ^2-4\big) \big[B r A'+A (B-1)\big]+A B r^2 \phi'^2=0\, ,
    \end{split}\\[2mm]
    \begin{split}
        (\theta,\theta):\quad & A \,A' \big\{B' \big[48 \alpha  B \phi  \phi'+r \big(\beta  \phi ^2-4\big)\big]+2 B \big(\beta  \phi ^2+16 \alpha  B \phi'^2\\
        &+2 \beta  r \phi  \phi'-4\big)\big\}+2 A^2 B' \big[\beta  \phi  \big(2 r \phi '+\phi \big)-4\big]\\
        &+8 A B \phi\,  \phi'' \big(4 \alpha  B A'+A \beta  r\big)+2 A B A'' \big[16 \alpha B \phi \, \phi '\\
        &+r \big(\beta  \phi ^2-4\big)\big]-B A'^2 \big[16 \alpha  B \phi \,\phi'+r \big(\beta  \phi ^2-4\big)\big]\\
        &+4 A^2 B \phi' \big[2 \beta  \phi +(2\beta -1) r \phi'\big]=0\, ,
    \end{split}
\end{align}
The background equation for the scalar field reads
\begin{equation}
    \begin{split}
        (\phi):\quad  &\phi  \big\{A A' \big[B' \big(24\alpha B -8\alpha+\beta  r^2\big)+4 \beta  B r\big]+4 A^2 \beta  \big(r B'+B\\
        &-1\big)-B A'^2 \big(8 \alpha  (B-1)+\beta  r^2\big)\big\}+2 A r \phi ' \big(B r A'+A r B'\\
        &+4 A B\big)+2 A B \phi  A'' \big[8\alpha (B-1)+\beta  r^2\big]+4 A^2 B r^2 \phi ''=0\, .
    \end{split}
\end{equation}

\subsection{Neutron stars}

Here we report the field equations solved in Chapter~\ref{ch:Neutron Stars} for a static and spherically symmetric spacetime and with matter described as a perfect fluid
\begin{align}
\begin{split}
    (t,t):\quad & B^{-2}\big(\beta  \kappa  \phi ^2+2 \kappa  r^2 \epsilon -2)
    +B^{-1}\big(8 \alpha  \kappa  \phi  B' B^{-1} \phi '+16 \alpha  \kappa  \phi '^2\\
    &+16 \alpha  \kappa  \phi  \phi ''-\beta  \kappa  \phi ^2-\beta\kappa  r^2 \phi  B' B^{-1} \phi '-2 \beta  \kappa  r^2 \phi '^2\\
    &-2 \beta  \kappa  r^2 \phi  \phi ''+\kappa  r^2 \phi'^2-\beta\kappa  r \phi ^2 B' B^{-1}-4 \beta \kappa  r \phi\phi'\\
    &+2 r B' B^{-1}\big)+2-24 \alpha  \kappa  \phi  B' B^{-1} \phi'-16 \alpha  \kappa  \phi '^2\\
    &-16 \alpha  \kappa  \phi  \phi''=0\, ,
\end{split}\\[2mm]
\begin{split}
    (r,r):\quad & B^{-2} \big(\beta  \kappa  \phi ^2-2 \kappa  p r^2-2\big)
    +B \big(8 \alpha  \kappa  \phi  A' A^{-1} \phi'-4 \beta  \kappa  r \phi  \phi '\\
    &+2-\beta  \kappa  r^2 \phi A' A^{-1} \phi '-\beta  \kappa  r \phi ^2 \Gamma'+2 r A' A^{-1}-\beta  \kappa  \phi ^2\\
    &-\kappa  r^2 \phi '^2\big)-24 \alpha  \kappa  \phi A' A^{-1} \phi'=0\, ,
\end{split}\\[2mm]
\begin{split}
    (em):\quad &2p'+(\epsilon+p)\Gamma'=0\, .
\end{split}
\end{align}
The equation for the scalar field reads
\begin{equation}
    \begin{split}
    (\phi):\quad & 4 \beta   \phi\,B^{-2} +B^{-1}\big[8 \alpha\phi  A' A^{-1} B' B^{-1}+8 \alpha   \phi  \big(A' A^{-1}\big)^2\\
    &+16 \alpha   \phi  \big(A'A^{-1}\big)'-4 \beta  \phi
    -\beta   r^2 \phi  A'A^{-1} B' B^{-1}\\
    &-\beta   r^2 \phi  \big(A'A^{-1}\big)^2-2 \beta   r^2 \phi  \big(A'A^{-1}\big)'-2   r^2 A' A^{-1} \phi '\\
    &-2  r^2 B' B^{-1} \phi '-4   r^2 \phi ''-4 \beta   r \phi  \Gamma '-4 \beta   r \phi  B' B^{-1}-8  r \phi '\big]\\
    &-24 \alpha   \phi  A' A^{-1} B' B^{-1}-8 \alpha   \phi \big[(A'A^{-1})^2\\
    &+16 \alpha  \phi  \big(A'A^{-1}\big)'\big]=0\, .
\end{split}
\end{equation}

\section{Einstein-Maxwell-scalar model}

Here we derive the equations of motion used to solve for scalarized black holes in the context of the EMs theory, pertaining to Section~\ref{sec:EMs} of Chapter~\ref{ch:Shadows}
\begin{equation}
\label{eq:Action_1}
\begin{split}
    S=\frac{1}{2\kappa}\int d^4x\sqrt{-g}\;\bigg[& R -\frac{1}{2}\nabla_\alpha \phi \nabla^\alpha \phi
    +f(\phi)F_{\mu\nu}F^{\mu\nu}\bigg]\,.
\end{split}
\end{equation}
Variation the action \eqref{eq:Action_1} with respect to the metric tensor, the electromagnetic tensor and the scalar field, yields
\begin{align}
    &G_{\mu\nu}=T_{\mu\nu}\,,\\
    &\Box\phi+\dot{f}(\phi)F_{\mu\nu}F^{\mu\nu}=0\,,\\
    &\partial_\mu\left(\sqrt{-g}\,f(\phi)\,F^{\mu\nu}\right)=0\,,
\end{align}
where the energy-momentum tensor contains scalar and electromagnetic contributions as follows
\begin{equation}
\begin{split}
    T_{\mu\nu}=&-\frac{1}{4}g_{\mu\nu}(\nabla\phi)^2+\frac{1}{2}\nabla_\mu\phi\nabla_\nu\phi\\
    &+f(\phi)\left[\frac{1}{2}g_{\mu\nu}F_{\mu\nu}F^{\mu\nu}-2g^{\rho\sigma}F_{\mu\rho}F_{\nu\sigma}\right]\,.
\end{split}
\end{equation}
For the EMS scalarization model discussed in, we use the following metric ansatz (in order to be consistent with \cite{}):
\begin{equation}
    ds^2=-N(r)e^{-2\delta(r)}dt^2+N(r)^{-1}dr^2+r^2 d\Omega^2\,
\end{equation}
where $N(r)=1-2m(r)/r$, with $m(r)$ being the Misner-Sharp mass \cite{}. Then, the Einstein ($tt$ and $rr$), scalar and electromagnetic equations \eqref{eq:EMS_1}-\eqref{eq:EMS_3} yield:
\begin{align}
    (t,t):\quad & m' -\frac{1}{8} r^2 \left(1-\frac{2 m}{r}\right) \phi'^2 +\frac{1}{2} e^{2 \delta } r^2 f\, V'^2 = 0\, , \\
    (r,r):\;& 4 \delta '+r \phi'^2 = 0\, , \\
    \begin{split}
       (\phi):\quad & 4 r\, (r-2 m) \phi''+r^2 (2 m-r)\phi'^3-8 e^{2 \delta } r^2 V'^2 \dot{f}\\
       & \hspace{-3mm}+4 \left[e^{2 \delta } r^3 f V'^2+r \delta' (2 m-r)-2 m+2 r\right]\phi' = 0\, ,
   \end{split}\\
   (em):\quad & r^2 f\, V'-e^{-\delta } Q_e = 0\, .
\end{align}
It is then straightforward to solve with respect to $m''$ and $\phi''$, which leaves with a system of ordinary differential equations that can be integrated. The appropriate boundary conditions are found by taking the near-horizon expansions of the functions $m,\,\phi,\,\delta,\,V$.

\section{Perturbative coefficients}
\label{sec:appendix_perturbative}
Here we present the coefficients appearing in the $\tilde{a}$-expansion of the scalar charge and the ADM mass. In order to find the perturbative expressions we solve the perturbed field and scalar equations. To that extent, we emply the asymptotic expansions introduced in Eqs~\ref{eq:pertA}-\eqref{eq:pertphi}. To \textbf{first order} with respect to $\Tilde{\alpha}$, the equations read
\begin{align}
    (tt):\quad &(r-2 m) B_1'+B_1=0,\\[2mm]
    (rr):\quad&(2 m-r) A_1'+B_1=0,\\[2mm]
    (\phi):\quad&r^5 (2 m-r) \phi _1''+2 r^4 (m-r) \phi_1'-48 m^2=0\, .
\end{align}
To \textbf{second order} with respect to $\Tilde{\alpha}$, the field equations yield
\begin{align}
    \begin{split}
    (tt):\quad &m^2 r^{10} (r-2 m) B_2'+m^2 r^{10} B_2+2304 \gamma  m^6-384 \gamma  m^3 r^3\\[2mm]
    &-52 m^2 r^6-8 m^3 r^5-16 m^4 r^4+736 m^5 r^3-2 m r^7\\[2mm]
    &+12 \gamma  r^6-r^8=0\, ,
    \end{split}\\[5mm]
    \begin{split}
    (rr):\quad &\left(4 m^2+2 m r+r^2\right) \big[8 m^3 \left(4\gamma +5 r^2\right)-16 m^2 r^3-4 r^3 \gamma\\[2mm]
    &+r^5\big]+m^2 r^9 (2 m-r) A_2'+m^2 r^9 B_2=0\, ,
    \end{split}\\[5mm]
    \begin{split}
    (\phi):\quad &40 m^2 r^4 \sigma -672 m^3 r^3 \sigma -224 m^4 r^2 \sigma -2 m^2 r^{11} \phi _2'\\[2mm]
    &+2 m^3 r^{10} \phi _2'-m^2 r^{12} \phi _2''+2 m^3 r^{11} \phi _2''-512 m^5 r \sigma\\[2mm]
    &+3456 m^6 \sigma +16 m r^5 \sigma +24 r^6 \sigma=0\, .
    \end{split}
\end{align}
Higher order equations are very lengthy but we have calculated them up to $\mathcal{O}(\tilde{a}^5)$. We can then solve for the coefficients appearing in the expressions \eqref{eq:Qa}-\eqref{eq:Ma} for the charge and mass expansions, in an attempt to verify how the higher order terms appearing in Lagrangian \eqref{eq:action} enter the eperturbative expressions. The charge coefficients are given by
\begin{equation}
        \begin{split}
            Q_1=&\;\frac{2}{m},\;Q_3=-\frac{1}{60 m^5},\;Q_4=-\frac{689\sigma }{36960 m^9},\\[5mm]
            Q_5=&\;\frac{11051 \kappa }{720720 m^{11}}-\frac{268867 \gamma^2}{16336320 m^{13}}-\frac{84317 \gamma }{180180 m^{11}}\\[2mm]
            &-\frac{4609603 \sigma ^2}{130690560 m^{13}}-\frac{118549}{158400 m^9}\, .
            \end{split}
\end{equation}
We notice that the $\sigma$-term contributes to the 4th perturbative order, while the $\gamma$ and $\sigma$ terms enter the coefficients at 5th order.
The mass coefficients yield
\begin{equation}
        \begin{split}
            M_2=&\;\frac{49}{40 m^3},\; M_3=\;\frac{18107 \sigma }{73920 m^7},\\[5mm]
            M_4=&\;\frac{244007 \gamma }{360360 m^9}-\frac{635421 \gamma ^2}{10890880 m^{11}}+\frac{11838611 \sigma ^2}{87127040 m^{11}}+\frac{408253}{246400 m^7}\\[2mm]
            &+\frac{130309 \kappa }{1441440 m^9},\\[5mm]
            M_5=&\;\frac{995527207 \gamma  \sigma }{1241560320 m^{13}}-\frac{210006269 \gamma ^2 \sigma }{2595989760 m^{15}}+\frac{56711635 \kappa  \sigma }{496624128 m^{13}}\\[2mm]
            &+\frac{12276069473 \sigma ^3}{119189790720 m^{15}}+\frac{84509327587 \sigma }{50315865600 m^{11}}\, .
        \end{split}
\end{equation}
Now, we notice that $\sigma$ enters at 3rd order, while $\sigma$ and $\kappa$ enter at the 4th-order coefficient of the mass.
In Eqs. \eqref{eq:phi_pert}-\eqref{eq:GB_pert} we presented the general form of the expansions for the scalar field and the GB invariant up to third order in $\tilde{\alpha}$, which were used in Fig.~\ref{fig:perturbative_treatment}.
In the following equations we give the analytic expressions for the coefficients appearing in these expansions.
For $\phi$, in accordance with the results for the mass and charge, the higher-order term to enter the perturbative expressions at the lowest order is the $\sigma$ one, with the $\gamma$ and $\kappa$ following. For $\GB$ the $\gamma$ contribution is already noticed in 2nd order, while $\sigma$ contributes at 3rd order.
Let us also point out that setting $\gamma=\sigma=\kappa=0$ we retrieve the perturbative result of \cite{Sotiriou:2014pfa}.
\begin{align}
\begin{split}
    \phi_0 = \,
    & 0\, ,
\end{split}\\[5mm]
\begin{split}
    \phi_1= \, 
    & \frac{2 \left(3 r^2+4 m^2+3 r\right)}{3 m r^3}\, ,
\end{split}\\[5mm]
\begin{split}
    \phi_2 = \,
    & \frac{2 \sigma  \left(21 r^5-224 m^5-84 m^4 r-24 m^3 r^2+84 m^2 r^3+42 m r^4\right)}{21 m^2 r^9}\, ,
\end{split}\\[5mm]
\begin{split}
    \phi_3 = \,
    & \frac{1}{{30 m^4}}\bigg[\frac{18432 m^9 \sigma ^2}{r^{15}}+\frac{7680 m^8 \left(8 \gamma ^2+3 \sigma ^2\right)}{7 r^{14}}\\
    & +\frac{15360 m^7 \left(8 \gamma ^2-\sigma ^2\right)}{13 r^{13}}+\frac{240 m (2 \gamma +\kappa )-952 m^3}{5 r^5}\\
    &+\frac{48 m^2 \left(-10 \gamma ^2+58 m^4-5 m^2 (\gamma -\kappa )+15 \sigma ^2\right)}{r^8}\\
    & +\frac{320 m^6 \left(24 \gamma ^2+8 m^2 (7 \gamma
    -\kappa )-33 \sigma ^2\right)}{r^{12}}\\
    & +\frac{96 m^4 \left(-8 \gamma ^2+24 m^2 (3 \gamma -\kappa )-5 \sigma ^2\right)}{r^{10}}\\
    &+\frac{32 m^2 \left(15 (2 \gamma +\kappa )-82 m^2\right)}{3 r^6}+\\
    & +\frac{160 m \left(-6 \gamma ^2+35
    m^4+18 \kappa  m^2+12 \sigma ^2\right)}{7 r^7}+\frac{66 m^2}{r^4}+\frac{526 m}{3 r^3}\\
    & +\frac{3840 m^5 \left(4 \gamma ^2+8 m^2 (4 \gamma -\kappa )-9 \sigma ^2\right)}{11 r^{11}}-\frac{1}{2 m r}+\frac{73}{r^2}\\
    & +\frac{160 m^3 \left(-72 \gamma ^2+424 m^4-24 m^2 (2 \gamma +\kappa )+99 \sigma ^2\right)}{9 r^9}\bigg] \, ,
\end{split}
\end{align}

\begin{align}
\begin{split}
    \GB_1 = \;
    & \frac{48 m^2}{r^6} \, ,
\end{split}\\[5mm]
\begin{split}
    \GB_2 = \;
    & \frac{79872 \gamma  m^5}{r^{15}}+\frac{14336 \gamma m^4}{r^{14}}+\frac{6656 \gamma  m^3}{r^{13}}-\frac{12288 \gamma    m^2}{r^{12}}\\
    &+\frac{53760 m^4}{r^{12}}-\frac{4096 m^3}{5 r^{11}}-\frac{448 m^2}{r^{10}}+\frac{588}{5 m^2 r^6}-\frac{1408 \gamma  m}{r^{11}}\\
    &+\frac{384 \gamma }{m r^9}-\frac{4608 m}{r^9}-\frac{64}{m r^7}-\frac{640 \gamma }{r^{10}}-\frac{32}{r^8} \, ,
\end{split}\\[5mm]
\begin{split}
    \GB_3 = \;
    &\frac{106496 \gamma  m^5 \sigma }{r^{19}}-\frac{5603328 \gamma  m^7 \sigma }{r^{21}}-\frac{98304 \gamma  m^6 \sigma }{r^{20}}\\
    &+\frac{1892352 \gamma  m^4 \sigma }{r^{18}}+\frac{129024 \gamma  m^3 \sigma }{r^{17}}+\frac{28672 \gamma  m^2 \sigma }{r^{16}}\\
    &-\frac{2174976 m^6 \sigma}{r^{18}}+\frac{3981312 m^5 \sigma}{11 r^{17}}+\frac{841728 m^4 \sigma }{5 r^{16}}\\
    &+\frac{507904 m^3 \sigma }{r^{15}}-\frac{38016 m^2 \sigma }{r^{14}}+\frac{18107 \sigma}{770 m^6 r^6}-\frac{181248 \gamma  m \sigma }{r^{15}}\\
    &+\frac{4608 \gamma  \sigma }{m^2 r^{12}}-\frac{4096 \gamma \sigma }{m r^{13}}-\frac{122880 m \sigma}{7 r^{13}}-\frac{12288 \gamma\sigma}{r^{14}}\\
    &-\frac{27648 \sigma }{r^{12}} \, .
\end{split}
\end{align}

%% file: Chapters/appendix_qnms.tex
Solving the eigenvalue problem for the equations of the type presented in \eqref{eq:qnm_eq}, requires the consideration of appropriate boundary conditions and the application of problem-specific numerical techniques. In this appendix we present the boundary conditions and three different ways of solving for the radial quasinormal modes in Chapter~\ref{ch:Stability}. For a detailed overview of QBN techniques one may refer to \cite{Berti:2009kk,Pani:2013pma}.

The boundary conditions are those of an ingoing wave at the horizon, and of an outgoing one at infinity. Physically we don't expect anything to cross the black-hole horizon outwards, or incoming waves from infinity
\begin{equation}\label{eq:boundaries}
    \begin{split}
        &\sigma_l\xrightarrow[r_*\rightarrow -\infty]{r\rightarrow\, \rh} e^{-i\omega r_*}\\
        &\sigma_l\xrightarrow[r_*\rightarrow +\infty]{r\rightarrow\, \rh} e^{+i\omega r_*}
    \end{split}\;,\quad \text{where}\quad dr_*=(A\,B)^{-1/2}dr.
\end{equation}\\
At this point let us consider a Schwarzschild background, so that $r_*=r+2M\ln\left(r/{2M}-1\right)$. Then the boundary conditions yield
\begin{equation}
\begin{split}
    & \sigma_l\xrightarrow[r_*\rightarrow -\infty]{r\rightarrow\, \rh} e^{-i\omega r_*}\sim \left(\frac{r}{2M}-1\right)^{-2iM\omega}\, ,\quad
    \\
    & \sigma_l\xrightarrow[r_*\rightarrow +\infty]{r\rightarrow\, \rh} e^{+i\omega r_*}\sim e^{i\omega r}\left(\frac{r}{2M}\right)^{2iM\omega}\, ,
\end{split}
\label{eq:boundary-conditions}
\end{equation}

As mentioned above, in linearized theory, scalarization manifests itself as a tachyonic instability around a GR solution.  Linearizing around a Schwarzschild background and neglecting backreaction,  we can recast the scalar equation \eqref{eq:scal_eq} into the following form:

\begin{equation}\label{eq:sigma}
    -\frac{\partial^2 \sigma}{\partial t^2}+\frac{\partial^2 \sigma}{\partial r_*^2}= V_{\text{eff}}\,\sigma,
\end{equation}
where the scalar field is decomposed into spherical harmonics, $\delta \phi = \epsilon\, \sigma(r,t) Y^m_\ell (\theta,\varphi)/r$, $d r= d r_* \sqrt{A B}$, and 
\begin{equation}
    V_{\text{eff}}=A \left[\frac{B}{2r}\left(\frac{A'}{A}+\frac{B'}{B}\right)-\alpha\, \GB \right].
\end{equation}

In the next subsections we present three techniques we can follow in order to solve the differential equation \eqref{eq:qnm_eq} with the boundary conditions we just introduced.

\section{Continued fraction method}

For a scalar field satisfying the boundary conditions we discussed, we may assume the following form
\begin{equation}
    \phi_l=\left(\frac{r}{\rh}-1\right)^{-i\rh\omega}\left(\frac{r}{\rh}\right)^{2i\rh\omega}e^{i\omega (r -\rh)}\sum_{n=0}^\infty a_n\left(1-\frac{\rh}{r}\right)^n\, ,
\end{equation}
which assumes the appropriate form in the boundaries, in accordance with Eqs.~\eqref{eq:boundary-conditions}.
Substituting this expression into \eqref{eq:qnm_eq} yields a recurrence relation for the coefficients $a_n$. These relations are satisfied for $\omega$ being a QNM frequency. In order to find the $n$-th QNM frequency we need to consider the $n$-th inversion of the recurrence relations. We will be reducing our equations to a three-term recurrence relation where we consider the ladder operator $R^+_n$, since that is a problem which we can always solve.
As a matter of fact, especially for low quality factors $\omega_r/\omega_i$ this method works better than WKB or even DI. Consider the following recurrence relations
\begin{equation}\label{eq:3-term}
\begin{split}
    &\alpha_n a_{n+1} + \beta_n a_n +\gamma_n a_{n-1} = 0\, ,\; n>0\, , \\
    &\alpha_0 a_{1} + \beta_0 a_0   = 0\, . 
\end{split}
\end{equation}
We now define a ladder operator $R^+$ such as
\begin{equation}
    R^+_n\,a_n=a_{n+1},
\end{equation}
so from \eqref{eq:3-term}, after we substitute $n\rightarrow n+1$ we have the ladder operator satisfying the following recurrence relations
\begin{equation}
    R^+_n=-\frac{\gamma_{n+1}}{\beta_{n+1}+\alpha_{n+1} R^+_{n+1}}\;\; ,\quad R^+_n=-\frac{\gamma_{n}+\beta_{n}R_{n-1}}{\alpha_{n} R^+_{n-1}}\, ,
\end{equation}
In order for the series to converge, we demand that
\begin{equation}
    \lim_{n\rightarrow\infty}R_n=0.
\end{equation}
In practice, we start from some very high $n=N$, so that $R_{n+1}=0$, and we recurrently solve for $R_n$ from top to bottom $N\rightarrow n$, where $n$ corresponds to the $n$-th QNM frequency. We then solve recurrently for $R_n$ from bottom to top $0\rightarrow n$, where $R_0=-\beta_0/\alpha_0$.
\begin{equation}
\begin{split}
    &-\left[{\gamma_n+\beta_n\left(\ldots-\frac{\gamma_1-\frac{\beta_1\beta_0}{\alpha_0}}{-\frac{\alpha_1\beta_0}{\alpha_0}}\right)}\right]\bigg/
    \left[{\alpha_n\left(\ldots-\frac{\gamma_1-\frac{\beta_1\beta_0}{\alpha_0}}{-\frac{\alpha_1\beta_0}{\alpha_0}}\right)}\right]\\
    & =-\frac{\gamma_{n+1}}{\beta_{n+1}+\alpha_{n+1}\left(\ldots -\frac{\gamma_{N-1}}{\beta_{N-1}+\frac{\alpha_{N-1}\gamma_N}{\beta_N}}\right)}\, .
\end{split}
\end{equation}
The continued fractionmethod has been sucessfully adopted in the context of decoupled EsGB in\cite{Macedo:2020tbm}.

\subsection{Six-term sequence elimination steps}
We will now show how to reduce the 6-term sequence of recurrence relations appearing in Chapter~\ref{ch:Stability} to a 3-term one.
Consider the following
\begin{equation}
\begin{split}
    & \alpha_n a_{n+1} + \beta_n a_n +\gamma_n a_{n-1} + \delta_n a_{n-2} + \sigma_n a_{n-3} + \theta_n a_{n-4 }  = 0 \\
    & \alpha_3 a_{4} + \beta_3 a_3 +\gamma_3 a_{2} + \delta_3 a_{1} + \sigma_3 a_{0}   = 0 \\
    & \alpha_2 a_{3} + \beta_2 a_2 +\gamma_2 a_{1} + \delta_2 a_{0}   = 0 \\
    & \alpha_1 a_{2} + \beta_1 a_1 +\gamma_1 a_{0}    = 0 \\
    & \alpha_0 a_{1} + \beta_0 a_0   = 0
\end{split}
\end{equation}
In order to bring the above into a 3-term sequence that can always be solved, we first eliminate $\theta_n$:
\begin{equation}
\begin{split}
    & \alpha_n^{(1)} a_{n+1} + \beta_n^{(1)} a_n +\gamma_n^{(1)} a_{n-1} + \delta_n^{(1)} a_{n-2} + \sigma_n^{(1)} a_{n-3} = 0 \\
    & \alpha_2^{(1)} a_{3} + \beta_2^{(1)} a_2 +\gamma_2^{(1)} a_{1} + \delta_2^{(1)} a_{0} = 0 \\
    & \alpha_1^{(1)} a_{2} + \beta_1^{(1)} a_1 +\gamma_1^{(1)} a_{0} = 0 \\
    & \alpha_0^{(1)} a_{1} + \beta_0^{(1)} a_0 = 0 
\end{split}
\end{equation}
The coefficients of the first level are determined as
\begin{equation}
    \begin{split}
        \boldsymbol{n\ge 4:}\quad&
        \alpha_n^{(1)} = \alpha_n^{(0)}\;,\quad
        \beta_n^{(1)}  = \beta_n^{(0)} - \frac{\theta_n^{(0)}}{\sigma_{n-1}^{(1)}} \alpha_{n-1}^{(1)}\;,\\
        &\gamma_n^{(1)}  =   \gamma_n^{(0)} - \frac{\theta_n^{(0)}}{\sigma_{n-1}^{(1)}} \beta_{n-1}^{(1)}\;,\quad \delta_n^{(1)} =     \delta_n^{(0)} - \frac{\theta_n^{(0)}}{\sigma_{n-1}^{(1)}} \gamma_{n-1}^{(1)}\;,\\
        &\sigma_n^{(1)} =  \sigma_n^{(0)} - \frac{\theta_n^{(0)}}{\sigma_{n-1}^{(1)}} \delta_{n-1}^{(1)}\;,\\[2mm]
        \boldsymbol{n<4:}\quad&\beta_n^{(1)}=\beta_n^{(0)}\;,\quad \gamma_n^{(1)}=\gamma_n^{(0)}\;,\quad \delta_n^{(1)}=\delta_n^{(0)}\;,\quad \sigma_n^{(1)}=\sigma_n^{(0)}
    \end{split}
\end{equation}
Then we eliminate $\sigma_n$
\begin{equation}
\begin{split}
    &\alpha_n^{(2)} a_{n+1} + \beta_n^{(2)} a_n +\gamma_n^{(2)} a_{n-1} + \delta_n^{(2)} a_{n-2} = 0 \, ,\\
    &\alpha_1^{(2)} a_{2} + \beta_1^{(2)} a_1 +\gamma_1^{(2)} a_{0}    = 0 \, ,\\
    &\alpha_0^{(2)} a_{1} + \beta_0^{(2)} a_0   = 0\, . 
\end{split}
\end{equation}
The coefficients of the second level are determined as
\begin{equation}
    \begin{split}
        \boldsymbol{n\ge 3:}\quad&
        \alpha_n^{(2)} = \alpha_n^{(1)} = \alpha_n^{(0)}\;,\quad
        \beta_n^{(2)}  = \beta_n^{(1)} - \frac{\sigma_n^{(1)}}{\delta_{n-1}^{(2)}} \alpha_{n-1}^{(2)}\;,\\
        &\gamma_n^{(2)}  =   \gamma_n^{(1)} - \frac{\sigma_n^{(1)}}{\delta_{n-1}^{(2)}} \beta_{n-1}^{(2)}\;,\quad\delta_n^{(2)} = \delta_n^{(1)} - \frac{\sigma_n^{(1)}}{\delta_{n-1}^{(2)}} \gamma_{n-1}^{(2)}\;,\\[2mm]
        \boldsymbol{n<3:}\quad&\beta_n^{(2)}=\beta_n^{(0)}\;,\quad \gamma_n^{(2)}=\gamma_n^{(0)}\;,\quad \delta_n^{(2)}=\delta_n^{(0)}\, .
    \end{split}
\end{equation}
Finally, we eliminate $\delta_n$:
\begin{equation}
\begin{split}
    &\alpha_n^{(3)} a_{n+1} + \beta_n^{(3)} a_n +\gamma_n^{(3)} a_{n-1} = 0 \\
    &\alpha_0^{(3)} a_{1} + \beta_0^{(3)} a_0   = 0\, .
\end{split}
\end{equation}
The coefficients of the third level are determined as
\begin{equation}
    \begin{split}
        \boldsymbol{n\ge 2:}\quad&
        \alpha_n^{(3)}=\alpha_n^{(2)} = \alpha_n^{(1)} = \alpha_n^{(0)}\;,\quad
        \beta_n^{(3)}  = \beta_n^{(2)} - \frac{\delta_n^{(2)}}{\gamma_{n-1}^{(3)}} \alpha_{n-1}^{(3)}\;,\\
        &\gamma_n^{(3)} = \gamma_n^{(2)} - \frac{\delta_n^{(2)}}{\gamma_{n-1}^{(3)}} \gamma_{n-1}^{(3)}\;,\\[2mm]
        \boldsymbol{n<2:}\quad& \beta_n^{(3)}=\beta_n^{(0)}\;,\quad \gamma_n^{(3)}=\gamma_n\, .
    \end{split}
\end{equation}
%

\section{WKB method}

%
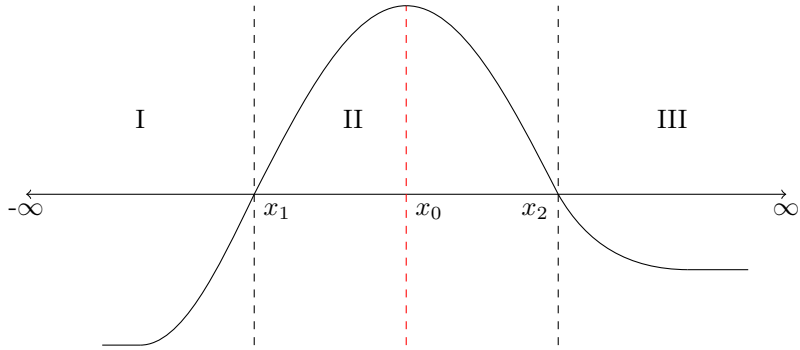
\begin{figure}[ht]
\centering
\begin{tikzpicture}
  \draw [<->] (-5,0) node[below] {-$\infty$} -- (-2,0) node[below right] {$x_1$} -- (0,0) node[below right] {$x_0$} -- (2,0) node[below left] {$x_2$} -- (5,0) node[below] {$\infty$};
  
  \node at (-3.5,1) {I};
  \node at (-0.7,1) {II};
  \node at (3.5,1) {III};
  
  \draw[dashed] (-2,-2) -- (-2,2.5);
  \draw[dashed] (2,-2) -- (2,2.5);
  \draw[dashed,red] (0,-2) -- (0,2.5);
  
  \draw (-4,-2) -- (-3.5,-2);
  \draw (-3.5,-2) cos (-2,0);
  \draw (-2,0) sin (0,2.5);
  \draw (0,2.5) cos (2,0);
  
  \draw (2,0) to[bend right] (3.7,-1);
  \draw (3.7,-1) -- (4.5,-1);
  
\end{tikzpicture}
\caption{Qualitative graph of a generic potential $V(r)$ with two turning points, located at $r_*^{(1)}$ and $r_*^{(2)}$ and a maximum at $r_*^{(0)}$.}
\label{fig:WKB_generic_potential}
\end{figure}

The WKB method, which is named after G.
Wentzel, H. Kramers and L. Brillouin, provides a useful way of approximating the solution to linear differential equations such us the one we consider below
\begin{equation} \label{General equation for Q}
	\frac{d^2 \psi(x)}{dx^2} + Q(x) \psi(x)=0\,, \quad Q(x) \neq 0,
\end{equation}
where $x$ is a tortoise coordinate spanning all space from $-\infty$ to $+\infty$. The function $Q(x)$ approaches some constant asymptotically (not necessarily the same at $\pm \infty$), therefore the solutions there are wave-like. In regions I and II we write the solution as an exponential power series of the form
\begin{equation}
    \psi(x) \approx \exp \left[ \frac{1}{\epsilon} \sum_{n=0}^{\infty} \epsilon^n S_n(x) \right]\,, \quad \epsilon \ll 1,
\end{equation}
in order to get the solution in the \textit{physical optics} approximation. The solutions, then, take the form
\begin{equation}
    \psi_I\sim \,\frac{1}{Q^{1/4}}\exp\left[\pm i\int_{x_2}^x \sqrt{Q(s)} \, ds \right]\,,\;\psi_{III}\sim \,\frac{1}{Q^{1/4}}\exp\left[\pm i\int_x^{x_1} \sqrt{Q(s)} \, ds \right]\,.
\end{equation}

The WKB method is based on the fact that the function $Q\equiv \omega^2-V_{\text{eff}}$ necessarily has two turning points, and the procedure involves relating different WKB solutions across a matching region with limits being the turning points, where $\omega^2=V(x)$, \textit{i.e.} $Q(x_1)=Q(x_2)=0$. The technique works best when the classical turning points are close, \textit{i.e.} when $\omega^2 \sim
V_{\rm max}$, where $V_{\rm max}$ is the peak of the potential. Then we can expand the function $Q$ around the extremum of the potential $x_0$ up to some arbitrary order
\begin{equation}
    Q\sim
    Q_0+\sum_{n=2}^{\text{order}}Q_0^{(n)}\frac{(x-x_0)^n}{n!}\,,\quad \text{where}\quad Q_0^{n}=\frac{d^2Q^{(n)}}{d x^2}\bigg|_{x_0}
\end{equation}
This is done so that we can retrieve the following approximate differential equation in region II
\begin{equation}
    \frac{d^2\psi}{dx^2}+\left
    [Q_0+\sum_{n=2}^{\text{order}}Q_0^{(n)}\frac{(x-x_0)^n}{n!} \right ]\psi=0 \,.
\end{equation}
The procedure followed involves matching the solutions over all regions while respecting the appropriate boundary conditions \eqref{eq:boundaries}, and leads to a simple equation for the n-th overtone solution of the QNM frequency. For an early application of this method to black holes see \cite{1985ApJ...291L..33S}. This method has been applied to 3rd order \cite{Iyer:1986nq,Iyer:1986np} and has been extended up to 6th order \cite{Konoplya:2003ii}.

Since the WKB method requires that $Q\equiv \omega^2-V_{\text{eff}}$ has two turning points close to one another, we expect it to yield better results for higher excitations (large $n,l$). From the CF approach in Chapter~\ref{ch:Stability}, we saw that as we approach the scalarization thresholds from large masses, both the real and the imaginary part of the frequency goes to zero. We expect the WKB method to break down the closer we get to the thresholds.

\section{Direct integration}

In order to apply the direct integration method we need to determine the boundary conditions at the horizon and at asymptotic infinity. Near the horizon and at infinity we assume
\begin{align}
    \phi_\text{near} = & \, e^{-i\omega r_*}\sum_n h_n(r-\rh)^n\approx (r-\rh)^{-i\omega \rh}\sum_n h_n(r-\rh)^n\,,\\
    \phi_\text{far} = & \, e^{i\omega r_*}\sum_n \frac{g_n}{r^n}\approx e^{i\omega r}\left(\frac{r}{\rh}\right)^{i\omega\rh}\sum_n \frac{g_n}{r^n}\,,
\end{align}
where $r_*=r+\rh\ln\left(r/\rh-1\right)$ is the tortoise coordinate. For the calculations done in Chapter~\ref{ch:Stability}, we have calculated the coefficients $h_n,\,g_n$ up to 10th order. Here we present them up to 2nd order
\begin{align}
    h_1 = & \; h_0\,\frac{i \left[\left(\ell^2+\ell+1\right) r_h^2-12 \alpha \right]}{r_h^3 \left(2 \omega  r_h+i\right)}\\
    \begin{split}
        h_2 = & \; h_0\big(144 \alpha ^2-24 \alpha  \left(\ell^2+\ell-1\right) r_h^2+2 i \left(2 \ell^2+2 \ell+3\right) \omega  r_h^5\,,\\
        & +\ell \left(\ell^3+2\ell^2+3\ell+2\right) r_h^4-144 i \alpha  \omega  r_h^3\big)\\
        &
        \big/\big[4 r_h^6 \left(\omega  r_h+i\right) \left(2 \omega  r_h+i\right)\big]\,,
    \end{split}\\
    g_1 = & \; g_0\,\frac{i \ell (\ell+1)}{2 \omega }\,,\\
    g_2 = & \; g_0 \,\frac{\left[\ell \left(-\ell^3-2\ell^2+\ell+2\right)+2 i \omega  r_h\right]}{8 \omega ^2}\,.
\end{align}
We then perform two integrations: the first one from the horizon to some intermediate matching point, imposing the condition of purely in-going waves at the horizon. The second one is done from infinity to the matching point, imposing the condition of purely out-going waves asymptotically far away from the horizon. We finally demand that the Wronskian of the two solutions vanishes at the matching point.